\documentclass[12pt]{article}
\def\int {\intop \limits}

\def\fnote#1{\footnote}
\begin{document}
\newcommand{\dst}[1]{\displaystyle{#1}}
\newcommand{\barl}{\begin{array}{rl}}
\newcommand{\ball}{\begin{array}{ll}}
\newcommand{\ear}{\end{array}}
\newcommand{\barc}{\begin{array}{c}}
\newcommand{\sne}[1]{\displaystyle{\sum _{#1} }}
\newcommand{\sn}[1]{\displaystyle{\sum ^{\infty }_{#1} }}
\newcommand{\ini}[1]{\displaystyle{\int ^{\infty }_{#1}}}
\newcommand{\myi}[2]{\displaystyle{\int ^{#1}_{#2}}}
\newcommand{\inn}{\displaystyle{\int }}
\newcommand{\be}{\begin{equation}}
\newcommand{\ee}{\end{equation}}
\newcommand{\aq}[1]{\label{#1}}
\renewcommand \theequation{\thesection.\arabic{equation}}

\vspace*{4.0cm}
\centerline{\Large {\bf Concept of Formation Length }}
\vskip .25cm
\centerline{\Large {\bf in Radiation Theory}}
\vskip .5cm
\centerline{\large{\bf V. N. Baier and V. M. Katkov}}
\centerline{Budker Institute of Nuclear Physics,
 630090 Novosibirsk, Russia}
\vskip 0.5cm {\bf Contents}
\begin{enumerate}
\item Introduction
\renewcommand{\theenumii}{\arabic{enumii}}
\begin{enumerate}
\item General outlook
\item Processes in medium
\item Radiation in presence of an external field
\end{enumerate}
\item The effect of a medium in an infinitely thick target
\begin{enumerate}
\item The Landau-Pomeranchuk-Migdal (LPM) effect for radiation
\item Integral characteristics of bremsstrahlung
\item The LPM effect for pair creation
\item Anomalous magnetic moment of the electron in a medium
\item Propagation of high-energy photon in a medium in presence
of an external field
\item An influence of multiple scattering on coherent scattering of photon
\item The polarization of a medium and bremsstrahlung process
\end{enumerate}
\item Impact of a medium for a target of finite thickness
\begin{enumerate}
\item General consideration
\item Boundary effect for a thick target
\item A thin target
\item A target of intermediate thickness
\item Multi-photon effects in energy loss spectra
\item Radiation from a structured target
\item Comparison of the results obtained in different papers
\item A qualitative behavior of the spectral intensity of radiation
\item Experimental investigation of the LPM effect
\item Discussion of theory and experiment
\end{enumerate}
\item Effects in colliding beams
\begin{enumerate}
\item Mechanisms of radiation
\item Beam-size effect in bremsstrahlung
\item Coherent radiation
\item Mechanisms of $e^+e^-$ pair creation
\end{enumerate}
\item Appendix
\end{enumerate}

\begin{abstract}
The features of electromagnetic processes are considered which
connected with finite size of space region in which final
particles (photon, electron-positron pair) are formed. The
longitudinal dimension of the region is known as the formation
length. If some external agent is acting on an electron while
traveling this distance the emission process can be disrupted.
There are different agents: multiple scattering of projectile,
polarization of a medium, action of external fields, etc. The
theory of radiation under influence of the multiple scattering,
the Landau-Pomeranchuk-Migdal (LPM) effect, is presented. The
probability of radiation is calculated with an accuracy up to
"next to leading logarithm" and with the Coulomb corrections
taken into account. The integral characteristics of
bremsstrahlung are given, it is shown that the effective
radiation length increases due to the LPM effect at high
energy. The LPM effect for pair creation is
also presented. The multiple scattering influences also on
radiative corrections in a medium (and an external field too)
including the anomalous magnetic moment of an electron and the
polarization tensor as well as coherent scattering of a photon in
a Coulomb field. The polarization of a medium alters the
radiation probability in soft part of spectrum. Specific features
of radiation from a target of finite thickness include: the
boundary photon emission, interference effects for thin target,
multi-photon radiation. The theory predictions are compared with 
experimental data obtained at SLAC and CERN SPS. 
For electron-positron colliding beams following items
are discussed: the separation of coherent and incoherent 
mechanisms of radiation, the beam-size effect
in bremsstrahlung, coherent radiation and mechanisms of
electron-positron creation.

\end{abstract}

\newpage
\section{Introduction}

\subsection{General outlook}

The electromagnetic interaction is of the local nature, e.g. the
interaction Hamiltonian in quantum electrodynamics is
\begin{equation}
H_{int}(x)=e\overline{\psi}(x)\gamma_{\mu}\psi(x) A^{\mu}(x),
\label{1}
\end{equation}
i.e. the current in a point $x$ interacts with an electromagnetic
field $A^{\mu}(x)$ in the same point. However the process of
photon radiation takes place not in one point but in some domain
of space-time, in which a photon (an emitted wave in the classical
language) is originating. The longitudinal dimension of this domain
is called {\it the formation (coherence) length}. It is evident
that the minimal size of the formation length is the wave length of the
emitted photon. One can estimate the formation length from the phase factor
entering the expression for the probability of photon emission.
In the quasiclassical operator method (see Eq.(\ref{a1}), Appendix A)
this factor is (in this paper we employ units such that $\hbar=c=1$)
\begin{equation}
\exp \left[-\frac{i\varepsilon}{\varepsilon'}
\left(\omega t -{\bf k r}(t)\right) \right],
\label{2}
\end{equation}
where $\varepsilon (\varepsilon')$ is the energy of the initial (final)
electron, $\omega$ is the photon energy, $\varepsilon'=\varepsilon-\omega$,
${\bf k}$ is the wave vector of the photon, ${\bf r}(t)$ is the particle
location on a classical trajectory, $t$ is the time. The main contribution
into the integral over time in the general expression for the probability
of radiation (\ref{a1}) gives the region
\begin{equation}
\frac{\varepsilon}{\varepsilon'}
\left(\omega t -{\bf k r}(t)\right) \sim 1,
\label{3}
\end{equation}
Taking into account that
\begin{equation}
{\bf k}=\sqrt{{\cal E}(\omega)}\omega {\bf n},\quad {\bf r}(t) = {\bf v} t,
\quad {\cal E}(\omega)=1-\frac{\omega_0^2}{\omega^2},\quad \omega_0^2=
\frac{4\pi \alpha n_e}{m}, \quad {\bf n v} =v\cos \vartheta,
\label{4}
\end{equation}
where ${\cal E}(\omega)$ is the dielectric constant, $\omega_0$
is the plasma frequency, $n_e$ is the electron density, we have
from Eq.(\ref{3}) for $\vartheta \ll 1$
\begin{equation}
t \sim t_f=l_f(\omega)=\frac{2\varepsilon \varepsilon'}{\omega m^2
\displaystyle{\left(1+
\gamma^2\vartheta^2+\frac{\omega_p^2}{\omega^2} \right)}},\quad
\omega_p=\omega_0\gamma.
\label{5}
\end{equation}
In this paper we consider the processes at high energies, i.e. the
Lorentz factor $\gamma=\varepsilon/m \gg 1$. Then from
Eq.(\ref{5}) follows:
\begin{enumerate}
\item ultrarelativistic particle radiates into the narrow cone
with the vertex angle $\vartheta \leq 1/\gamma$ along the  momentum of the
initial particle, the contribution of larger angles is suppressed because
of shortening of the formation length;
\item the effect of the polarization of a medium described by the
dielectric constant ${\cal E}(\omega)$ manifest itself
for soft photons only when $\omega \leq \omega_b$, since on the Earth
$\omega_0 < 100$~eV we have $\omega_b < \varepsilon \omega_0/m< 2 \cdot
10^{-4} \varepsilon$;
\item  when $\vartheta \leq 1/\gamma,~ \omega_0 \gamma/\omega \ll 1$
the formation length is of the order
\begin{equation}
l_f \simeq l_{f0}(\omega)=\frac{2\varepsilon \varepsilon'}{\omega
m^2}. \label{6}
\end{equation}
\end{enumerate}
In the classical case $\omega \ll \varepsilon$
one has $l_{f0}=2\gamma^2/\omega$.
So for ultrarelativistic particles the formation length extends
substantially. For example for a $\varepsilon=25$~GeV electron
emitting a $\omega=100$~MeV photon, $l_{f0}=10\mu$m, i.e.$\sim 10^5$
interatomic distances.

The formation length is important in many electromagnetic processes
including creation of electron-positron pair, magnetic bremsstrahlung,
transition radiation, $\check{{\rm C}}$erenkov radiation. There is also
a set of applications involving other forces.

There is a number of physical interpretation of the formation length.
\begin{itemize}
\item The momentum transfer ${\bf q}$ to a
nucleus in the process of the photon
emission by the particle with the momentum ${\bf p}$
\begin{equation}
{\bf q} = {\bf p} - {\bf p}' - {\bf k},
\label{7}
\end{equation}
takes the minimal value when it is longitudinal
$q_{\parallel}=q_{min}=p-p'-k$, and in this case
$q_{min}=1/l_{f0}= (\hbar/l_{f0})$, i.e. from the uncertainty
principle follows that if the minimal momentum transfer is small,
then the formation length is large.
\item For creation of the electron-positron pair the formation length
is the distance required for the final state particles to separate
enough that they act as independent particles.
\item This is the distance over which the amplitudes of several
interactions can add coherently to the total probability.
\item The transverse size of the region, where the radiation is
formed $\varrho_{\perp}$ is determined by the minimal transverse
momentum transfer $q_{\perp}^{min}$:
$\varrho_{\perp}=1/q_{\perp}^{min}=(\hbar/q_{\perp}^{min})$.
Although usually $q_{\perp} \gg q_{\parallel}$ and this dimension
is much smaller, we consider below the situation when the
constraint of $q_{\perp}$ is the main effect.
\end{itemize}
The formation length concept is important because within it
the amplitudes can add coherently if a charged particle is moving freely.
In the opposite case, when some perturbation is acting on the particle
within the formation length,
breakdown of the coherence happens and radiation will be suppressed.

\subsection{Processes in a medium}

Landau and Pomeranchuk were the first to show that if the formation
length of the bremsstrahlung becomes comparable to the distance over which
the multiple scattering becomes important, the bremsstrahlung will be
suppressed \cite{1}. They considered radiation of soft photons.
Migdal \cite{2}, \cite{3} developed a quantitative
theory of this phenomenon. Now the common name is the Landau- Pomeranchuk
-Migdal (LPM) effect.

Let us estimate a disturbance of the emission process
due to a multiple scattering. As it is known, the mean angle
of the multiple scattering at some length $l_{f0}$ is
\begin{equation}
\vartheta_s = \sqrt{\vartheta_s^2} =
\frac{\varepsilon_s}{\varepsilon}
\sqrt{\frac{l_{f0}}{2L_{rad}}},\quad
\varepsilon_s=m\sqrt{\frac{4\pi}{\alpha}}=21.2~{\rm MeV},
\label{8}
\end{equation}
where $\alpha=e^2=1/137$, $L_{rad}$ is the radiation length. Since we are
interesting in influence of the multiple scattering on the radiation
process we put here the formation length (\ref{6}). One can expect
that when $\vartheta_s \geq 1/\gamma$ this influence will be substantial.
From this inequality we have
\begin{equation}
\frac{\varepsilon'}{\varepsilon_{LP}} \geq
\frac{\omega}{\varepsilon}, \quad \varepsilon_{LP}=\frac{m^4
L_{rad}}{\varepsilon_s^2}, \label{9}
\end{equation}
here $\varepsilon_{LP}$ is the characteristic energy scale, for
which the multiple scattering will influence the radiation
process for the whole spectrum. It was introduced in \cite{1} and
denoted by $E_0=\varepsilon_{LP}$. For tungsten we have
$\varepsilon_{LP}=2.7$~TeV, and similar values for all the heavy
elements. For the light elements the energy $\varepsilon_{LP}$ is
much larger. When $\varepsilon \ll \varepsilon_{LP}$ the multiple
scattering will influence emission of soft photons only with
energy
\begin{equation}
\omega \leq \omega_{LP}=\frac{\varepsilon^2}{\varepsilon_{LP}},
\label{10}
\end{equation}
e.g. for particles with energy $\varepsilon=25$~GeV one has
$\omega_{LP} \simeq 230$~MeV. Of course, we estimate here the
criterion only. For description of the effect one has to calculate
the probability of the bremsstrahlung taking into account multiple
scattering.

Side by side with the multiple scattering of an emitting electron
one has to take into account also the influence of the medium on
the radiated electromagnetic field. As it was shown above
Eq.(\ref{5}), this effect leads also to suppression of the soft
photon emission (the dielectric suppression, the Ter-Mikaelian
effect, see in \cite{4}).

A clear qualitative analysis of different mechanisms
of suppression is presented in early papers
\cite{5},\cite{6}. More simple derivation of the Migdal's results is given
in \cite{7}.

The next step in a quantitative theory of the
Landau-Pomeranchuk-Migdal (LPM) effect was made in \cite{8}. This
theory is based on the quasiclassical operator method in QED
developed by the authors (see \cite{7a}, \cite{7}, \cite{9},
\cite{16a}). One of the basic equations (obtained with the use of
kinetic equations describing the motion of an electron in the
medium in the presence of an external field) is the
Schr\"{o}dinger equation in an external field with imaginary
potential (Eq.(3.3) of \cite{8}; see (\ref{a21}) in Appendix A).

New activity with the theory of the LPM effect is connected with
a very successful series of experiments \cite{10} - \cite{E2}
performed at SLAC recently. In these experiments the cross section
of the bremsstrahlung of soft photons with energy from 200~keV to
500~MeV from electrons with energy 8~GeV and 25~GeV was measured
for a variety in materials with an accuracy of the order of a few
percent. Both the LPM and the dielectric suppression was observed
and investigated. These experiments were the challenge for the
theory since in all the published before papers calculations are
performed to logarithmic accuracy which is not enough for a
description of the new experiments. The contribution of the
Coulomb corrections (at least for heavy elements) is larger then
experimental errors and these corrections should be taken into
account. In papers \cite{13}-\cite{19} the problem was
investigated using different approaches.

Very recently the LPM was studied at CERN \cite{HU} using iridium
target and electrons with energy 287~GeV, 207~GeV and 149~GeV.
For iridium according to Eqs.(\ref{9}) and (\ref{10})
$\varepsilon_{LP}$=2.2 TeV, $\omega_{LP} \simeq$ 37~GeV for the
highest energy and the LPM suppression of bremsstrahlung was
observed in much wider interval of photon energies than at SLAC.

The crossing process for the bremsstrahlung is the pair creation by a
photon. The created particles in a medium
can undergo the multiple scattering.
It should be emphasized that for the bremsstrahlung the formation length
(\ref{6}) increases strongly if $\omega \ll \varepsilon$. Just because
of this the LPM effect was investigated at SLAC at a relatively
low energy. For the pair creation
the formation length
\begin{equation}
l_{p}=\frac{2\varepsilon(\omega-\varepsilon)}{m^2\omega}
\label{10a}
\end{equation}
attains maximum at $\varepsilon=\omega/2$ and this maximum is
$l_{p,max}=(\omega/2m)\lambda_c$. Because of this even for heavy
elements the effect of
multiple scattering on pair creation becomes noticeable at photon energies
$\omega \geq 10$~TeV \cite{L5}.
Starting from these energies one has to take into account
the influence of a medium on the pair creation
and on the bremsstrahlung hard part of the spectrum in electromagnetic
showers being created by the cosmic ray particles of the ultrahigh energies.
These effects can be quite significant in the electromagnetic calorimeters
operating in the detectors on the colliders in TeV range.

The nonlinear effects of QED are due to the interaction of a photon with
electron-positron field. These processes are the photon-photon
scattering,  the coherent photon scattering,
the photon splitting into two photons, and
the coalescence of two photons into photon in the Coulomb field.

The process of creation of the electron-positron pair by a photon
connected with the
coherent photon scattering \cite{19b}. In the quasiclassical
approximation the amplitude $M$ of the coherent photon scattering
is described by diagram where the electron-positron pair is
created by the initial photon with 4-momentum $k_1~(\omega,{\bf
k}_1)$ and then annihilate into the final photon with 4-momentum
$k_2$. For high energy photon $\omega \gg m$ this process occurs
over a rather long distance, known as the time of life of the
virtual state
\begin{equation}
l_f=\frac{\omega}{2q_c^2},
\label{1.1}
\end{equation}
where $q_c \geq m$ is the characteristic transverse momentum of the process.
When the virtual electron (or positron)
is moving in a medium it scatters on atoms. The mean square of
momentum transfer to the electron from a medium on the distance
$l_f$ is
\begin{equation}
q_s^2=4\pi Z^2\alpha^2n_aL(q_c)l_f,\quad L(q_c)=\ln \frac{q_c^2}{q_{min}^2},
\quad q_{min}^2=a^{-2}+\Delta^2+\frac{m^4}{\omega^2},
\label{1.2}
\end{equation}
where $Z$ is the charge of nucleus,
$n_a$ is the number density of atoms in the medium,
$\Delta$ is the photon momentum transfer ($\Delta=|{\bf k}_2-{\bf k}_1|
\ll q_c$),
$a$ is the screening radius of atom.

The coherent photon scattering amplitude $M$
can be obtained from general formulas for probabilities of
electromagnetic processes in the frame of
the quasiclassical operator method (see e.g.
\cite{9}). It can be estimated as
\begin{equation}
M \sim \frac{\alpha \omega}{2\pi l_f n_a}\frac{q_s^2}{q_c^2}=
\frac{\alpha}{\pi n_a} q_s^2.
\label{1.3}
\end{equation}
We use the normalization
condition Im$M=\omega \sigma_p$ for the case $\Delta=0$,
where $\sigma_p$ is the total cross section of pair creation by a photon.

In the case of small momentum transfer $q_s \equiv \sqrt{q_s^2} \ll m$
we have in the region of small $\Delta \ll m$
\begin{equation}
q_c^2=m^2,\quad M \sim \frac{2Z^2 \alpha^3 \omega}{m^2}
\ln \frac{m^2}{a^{-2}+\Delta^2+\frac{m^4}{\omega^2}}.
\label{1.4}
\end{equation}

At an ultrahigh energy it is possible that $q_s \gg m$. In this
case the characteristic momentum transfer is $q_c$, the
combination $q_c^2/m^2$ enters in the formation length $L_f$
(\ref{5}). The self-consistency condition is
\begin{equation}
q_c^2=q_s^2=\frac{2\pi \omega Z^2 \alpha^2 n_a
L(q_c)}{q_c^2},\quad q_c^2=\varepsilon^2\vartheta_c^2 \label{1.5}
\end{equation}
where $L(q_c)$ is defined in Eq.(\ref{1.2}). So using
Eq.(\ref{1.3}) one gets for the estimate of the coherent photon
scattering amplitude $M$ (the influence of the multiple
scattering manifests itself at the high photon energies such that
$m^2 a/\omega \ll 1$)
\begin{equation}
M \sim \frac{2Z^2\alpha^3 \omega}{\Delta_s^2}
\sqrt{\ln \frac{\Delta_s^2}{a^{-2}+\Delta^2}},\quad
\Delta_s^2=\sqrt{2\pi \omega Z^2 \alpha^2 n_a} \gg \Delta^2.
\label{1.6}
\end{equation}

\subsection{Radiation in the presence of an external field}

The authors have shown \cite{20} that an external electromagnetic
field can also modify the bremsstrahlung process. We consider the
effect of an external field on the photon emission in
electron-electron(positron) collision. This effect is associated
as well with a reduction of the formation length of a photon
(either real or virtual) due to relatively large turn of particle
velocity over this length, and the corresponding increase in the
vertex angle of the radiation cone. If a photon of energy $\Omega$
is emitted by an electron (positron) at an angle $\vartheta$ to
its velocity, the formation length of such a photon will be
according to Eq.(\ref{5}) (we neglect the polarization of a
medium)
\begin{equation}
l_f(\Omega) = \frac{2(\varepsilon-\Omega)}{\varepsilon \Omega
\vartheta_c^2} = \frac{2}{\varepsilon u_{\Omega} \vartheta_c^2},
\label{11}
\end{equation}
where $u_{\Omega} \equiv \Omega/(\varepsilon-\Omega),~ q_c$
defined in Eq.(\ref{1.5}). The characteristic emission angle in
weak field is $\vartheta_c = 1/\gamma$, and we can neglect the
influence of the external field if
\begin{equation}
|\dot{\bf v}|l_f(\Omega) = \frac{eHl_f(\Omega)}{\varepsilon} \ll
\frac{1}{\gamma}. \label{12}
\end{equation}
Substituting Eq.(\ref{11}) into Eq.(\ref{12}), we have the
criterion for a field to be weak
\begin{equation}
\gamma \frac{eH}{\varepsilon} \frac{2\gamma^2}{\varepsilon
u_{\Omega}} =\frac{2\chi}{u_{\Omega}} \ll 1. \label{13}
\end{equation}
Here we introduced the important parameter $\chi$, which
characterizes the quantum effects in an external field, when
$\chi \ll 1$ we are in the classical domain and with $\chi \geq
1$ we are already well inside the quantum domain
\begin{equation}
\mbox{\boldmath$\chi$} =\frac{{\bf F}\gamma}{F_0} \quad
\chi=|\mbox{\boldmath$\chi$}|, \quad {\bf F}={\bf
E}_{\perp}+({\bf v}\times {\bf H}),
\label{14}
\end{equation}
where ${\bf E} ({\bf H})$ is an electric (magnetic) field,
Eq.(\ref{14}) contains the transverse with respect to particle
velocity ${\bf v}$ component of fields, $F_0$ is one of
\begin{equation}
 E_0=\frac{m^2}{e}=\left(\frac{m^2c^3}{e\hbar} \right) =1.32
\cdot 10^{16} {\rm V/cm},~{\rm or}~
 H_0=\frac{m^2}{e}=\left(\frac{m^2c^3}{e\hbar} \right)
=4.41 \cdot 10^{13} {\rm Oe}.
\label{15}
\end{equation}
In strong fields, where $\chi/u \gg 1$, with characteristic
radiation angles $\vartheta \gg 1/\gamma$, the effective emission
angle $\vartheta_{ef}$ is determined by a self-consistency
argument: the deviation angle of the particle in the field over
the formation length must not exceed $\vartheta_{c}$, i.e.
\begin{eqnarray}
&& |\dot{\bf v}|l_f(\Omega, \vartheta_{c}) = \vartheta_{c}, \quad
\vartheta_{c} =
\frac{1}{\gamma}\left(\frac{\chi}{u_{\Omega}}\right)^{1/3}=
\left(\frac{2eH}{\varepsilon^2 u_{\Omega}} \right)^{1/3}
\nonumber \\
&& l_f(\Omega, \vartheta_{c}) = \frac{\gamma}{mu_{\Omega}}
\left(\frac{u_{\Omega}}{2\chi} \right)^{2/3} =
\left(\frac{\varepsilon}{4u_{\Omega} e^2 H^2} \right)^{1/3}.
\label{16}
\end{eqnarray}
It can be seen from Eq.(\ref{16}) that when $\chi/u \gg 1$,
neither the characteristic emission angle nor the photon
formation length depends on the mass of the radiating particle. A
parameter characterizing the effect of an external field on the
radiation process was derived in \cite{8}.

It follows from Eq.(\ref{16}) that, for $\chi/u_{\Omega} \gg 1$,
the characteristic photon formation length is reduced by factor
$(\chi/u_{\Omega})^{2/3}$, and the emission angle is increased by
factor $(\chi/u_{\Omega})^{1/3}$. Since the relevant parameter
$\chi/u_{\Omega}$ depends on the frequency $\Omega$ (via
$u_{\Omega}$), this effect is manifested earlier for soft photons.
Because of this a specific situation arises in bremsstrahlung in
a collision of an electron and a positron(an electron) in
colliding-beam experiment. The point is that the external factors
act differently on the radiating particle and on the recoil
particle. For the radiating particle the criterion of influence
of external factors is the same in scattering by a nucleus as in
collision of particles. For the recoil particles the effect turns
out to be enhanced by $\gamma^2$ times (in the laboratory frame).
This is due to the fact that the main contribution to the
bremsstrahlung cross section give the virtual photons emitted by
recoil particle with very low energy
\begin{equation}
\displaystyle{\Omega \simeq q_0 = \frac{\omega m^2}{4\varepsilon
\varepsilon'}}, \label{17}
\end{equation}
where $\omega$ is the frequency of the real photon. The formation
length of the virtual photon is
\begin{equation}
\displaystyle{l_v=l_f(q_0)=
\frac{8\varepsilon^3(\varepsilon-\omega)}{m^4\omega}}, \label{18}
\end{equation}
for example, for $\varepsilon=2~$GeV and $\omega=10~$MeV the
formation length is enormous: $l_v \simeq 4 \cdot 10^3$cm. This is
the reason why the field has a significant effect on virtual
photon emission in $e^{-}e^{+}$ collisions even for moderate
field and at relatively low particle energies, when the parameter
$\displaystyle{\frac{\chi (\varepsilon-\omega)}{\omega} \equiv
\frac{\chi}{u}}$ is small while the value
$\displaystyle{\frac{\chi(\varepsilon-q_0)}{q_0} \simeq 4\gamma^2
\frac{\chi}{u}}$ is large. In that event, the picture of emission
of a bremsstrahlung photon (radiation vertex) does not change
directly, but a significant change takes place in the virtual
photon spectrum at momentum transfers $|q| \leq
q_{min}(4\gamma^2\chi/u)^{1/3}$, increasing the lower bound of
the effective momentum transfer and resulting in a corresponding
decrease of the cross section. The bremsstrahlung cross section
under these conditions was derived to logarithmic accuracy in
\cite{20} using the equivalent photon method. In \cite{21} the
problem was solved to relativistic accuracy (terms $O(1/\gamma)$
were neglected).

It follows from Eq.(\ref{13}), and was demonstrated in \cite{8}
that in the case $\chi/u \geq 1$ there is a change of the
radiation vertex describing absorption of a virtual photon and
emission of a real one. Since for $\chi/u \gg 1$ the formation
length of a real photon $l_{\omega}$ falls off as
$(u/\chi)^{2/3}$ (see Eq.(\ref{16})), the bremsstrahlung cross
section falls off in just the same extent.

\section{The effect of a medium in an infinitely thick target}

\setcounter{equation}{0}

\subsection{The LPM effect for radiation}

We consider first the case where the formation length is much
shorter than the thickness of a target $l (l_f \ll l)$. The
derivation of basic formulas for probability of radiation under
influence of multiple scattering is given in Appendix A. The
spectral distribution of the probability of radiation per unit
time in this case is given by Eq.(\ref{a19}) and functions
$\varphi_{\mu}(\varphi_0, \mbox{\boldmath$\varphi$})$ satisfy the
Eq.(\ref{a21}). Introducing new variables
\begin{equation}
t=\frac{a}{2}\tau,\quad \mbox{\boldmath$\varrho$} =
\sqrt{\frac{a}{b}}{\bf x} = \frac{1}{\gamma}{\bf x},
\label{2.7}
\end{equation}
where the quantities $a$ and $b$ are defined in
Eqs.(\ref{a13}),(\ref{a14}), we obtain for the spectral
distribution of the probability of radiation per unit time
\begin{equation}
\frac{dW}{d\omega}=\frac{2\alpha}{\gamma^2} {\rm Re} \int_{0}^{\infty} dt
e^{-it}\left[R_1
\varphi_0(0, t) + R_2
{\bf p} \mbox{\boldmath$\varphi$}(0, t)\right],
\label{2.8}
\end{equation}
where $\displaystyle{R_1=\frac{\omega^2}{\varepsilon
\varepsilon'},\quad R_2
=\frac{\varepsilon}{\varepsilon'}+\frac{\varepsilon'}{\varepsilon}}$,
and the functions $\varphi_{\mu}=\varphi_{\mu}(\varphi_{0},
\mbox{\boldmath$\varphi$})$ now satisfy the equation
\begin{eqnarray}
&&\displaystyle{i\frac{\partial \varphi_{\mu}}{\partial t}={\cal
H} \varphi_{\mu},\quad {\cal H}={\bf p}^2 -
iV(\mbox{\boldmath$\varrho$}),\quad {\bf p}= -i
\mbox{\boldmath$\nabla$}_{\mbox{\boldmath$\varrho$}},\quad
V(\mbox{\boldmath$\varrho$})=-Q\mbox{\boldmath$\varrho$}^2
\Big(\ln \gamma^2 \vartheta_1^2}
\nonumber \\
&&\displaystyle{+\ln \frac{\mbox{\boldmath$\varrho$}^2}{4}+2C-1\Big),\quad
Q=\frac{2\pi n_a Z^2 \alpha^2 \varepsilon \varepsilon'}{m^4 \omega},\quad
C=0.577216...}
\label{2.9}
\end{eqnarray}
with the initial conditions $\varphi_0(\mbox{\boldmath$\varrho$},
0)= \delta(\mbox{\boldmath$\varrho$}), \quad
\mbox{\boldmath$\varphi$}(\mbox{\boldmath$\varrho$}, 0) ={\bf p}
\delta(\mbox{\boldmath$\varrho$})$. Here $\varepsilon,
\varepsilon', \omega$ are defined in Eq.(\ref{2}) and $Z, n_a$ are
defined in Eq.(\ref{1.2}). The functions $\varphi_0$ and
$\mbox{\boldmath$\varphi$}$ in Eq.(\ref{2.8}) are rescaled
according with the initial conditions (factors $1/\gamma^2$ and
$1/\gamma^3$, correspondingly). Note, that it is implied that in
formula (\ref{2.8}) subtraction at $V=0$ is made.

The potential $V(\mbox{\boldmath$\varrho$})$ Eq.(\ref{2.9})
corresponds to consideration of scattering in the Born
approximation. The difference of exact (as a function of
$Z\alpha$)  potential $V(\mbox{\boldmath$\varrho$})$ and taken in
the Born approximation is computed in Appendix A of \cite{13}.
The potential $V(\mbox{\boldmath$\varrho$})$ with the Coulomb
corrections taken into account is
\begin{eqnarray}
&& \displaystyle{V(\mbox{\boldmath$\varrho$})=-Q\mbox{\boldmath$\varrho$}^2
\Big(\ln \gamma^2 \vartheta_1^2
+\ln \frac{\mbox{\boldmath$\varrho$}^2}{4}+2C-1+2f(Z\alpha)\Big)}
\nonumber \\
&& \displaystyle{=-Q\mbox{\boldmath$\varrho$}^2
\Big(\ln \gamma^2 \vartheta_2^2
+\ln \frac{\mbox{\boldmath$\varrho$}^2}{4}+2C\Big)},
\label{2.9a}
\end{eqnarray}
where $\vartheta_2=\vartheta_1 \exp (f-1/2)$, the function $f=f(Z\alpha)$
is
\begin{equation}
f(\xi)={\rm Re}\left[\psi(1+i\xi)-\psi(1) \right]=
\xi^2\sum_{n=1}^{\infty}\frac{1}{n(n^2+\xi^2)},
\label{2.9b}
\end{equation}
where $\psi(\xi)$ is the
logarithmic derivative of the gamma function.

In above formulas $\mbox{\boldmath$\varrho$}$ is space of the
impact parameters measured in the Compton wavelengths $\lambda_c$,
which is conjugate to space of the transverse momentum
transfers measured in the electron mass $m$.

An operator form of a solution of Eq. (\ref{2.9}) is
\begin{eqnarray}
&&\varphi_0(\mbox{\boldmath$\varrho$}, t) = \exp(-i{\cal H}t)
\varphi_0(\mbox{\boldmath$\varrho$}, 0)=
<\mbox{\boldmath$\varrho$}| \exp(-i{\cal H}t) |0>,\quad {\cal
H}={\bf p}^2-iV(\mbox{\boldmath$\varrho$}),
\nonumber \\
&&\mbox{\boldmath$\varphi$}(\mbox{\boldmath$\varrho$}, t)=
\exp(-i{\cal H}t) {\bf p} \varphi_0(\mbox{\boldmath$\varrho$},0)=
<\mbox{\boldmath$\varrho$}| \exp(-i{\cal H}t){\bf p} |0>,
\label{2.10}
\end{eqnarray}
where we introduce the following Dirac state vectors:
$|\mbox{\boldmath$\varrho$}>$ is the state vector of coordinate
$\mbox{\boldmath$\varrho$}$, and
$<\mbox{\boldmath$\varrho$}|0>=\delta(\mbox{\boldmath$\varrho$})$.
Substituting Eq.(\ref{2.10}) into Eq.(\ref{2.8}) and taking
integral over $t$ we obtain for the spectral distribution of the
probability of radiation
\begin{equation}
\frac{dW}{d\omega}=\frac{2\alpha}{\gamma^2} {\rm Im} T,\quad T= <0|
R_1 \left(G^{-1}-G_0^{-1} \right)
+ R_2 {\bf p} \left(G^{-1}-G_0^{-1} \right) {\bf p} |0>,
\label{2.11}
\end{equation}
where
\begin{equation}
G={\bf p}^2+1-iV,\quad G_0={\bf p}^2+1.
\label{2.12}
\end{equation}
Here and below we consider an expression $<0|...|0>$ as a limit:
${\rm lim}~{\bf x} \rightarrow 0,
\newline {\rm lim}~{\bf x'} \rightarrow 0$ of
$<{\bf x}|...|{\bf x'}>$.

Let us estimate the effective impact parameters $\varrho_c$ which
give the main contribution into the radiation probability. Since
the characteristic values of $\varrho_c$ can be found
straightforwardly by calculation of Eq.(\ref{2.11}), we the
estimate characteristic angles $\vartheta_c$ connected with
$\varrho_c$ by an equality $\varrho_c=1/(\gamma \vartheta_c)$.
The mean square scattering angle of a particle on the formation
length of a photon $l_f$ Eq.(\ref{5}) has the form
\begin{equation}
\vartheta_s^2=\frac{4\pi Z^2 \alpha^2}{\varepsilon^2}n_a l_f
\ln \frac{\zeta}{\gamma^2 \vartheta_1^2}=
\frac{4Q}{\gamma^2\zeta}
\ln \frac{\zeta}{\gamma^2 \vartheta_1^2},
\label{2.13}
\end{equation}
where $\zeta=1+\gamma^2\vartheta^2$, we neglect here the polarization
of a medium.
When $\vartheta_s^2 \ll 1/\gamma^2$ the contribution in the probability
of radiation gives a region where $\zeta \sim 1 (\vartheta_c=1/\gamma)$,
in this case $\varrho_c=1$. When $\vartheta_s \gg 1/\gamma$
the characteristic angle of radiation is determined by  self-consistency
arguments (compare with Eq.(\ref{16})):
\begin{equation}
\vartheta_s^2 \simeq \vartheta_c^2 \simeq \frac{\zeta_c}{\gamma^2}=
\frac{4Q}{\zeta_c \gamma^2} \ln \frac{\zeta_c}{\gamma^2\vartheta_1^2},\quad
\frac{4Q}{\zeta_c^2} \ln \frac{\zeta_c}{\gamma^2\vartheta_1^2}=1,\quad
4Q\varrho_c^4 \ln \frac{1}{\gamma^2\vartheta_1^2 \varrho_c^2}=1.
\label{2.14}
\end{equation}
It should be noted that when the characteristic impact parameter
$\varrho_c$ becomes smaller than the radius of nucleus $R_n$,
the potential $V(\mbox{\boldmath$\varrho$})$ acquires an oscillator form
(see Appendix B, Eq.(B.3) in \cite{13})
\begin{equation}
V(\mbox{\boldmath$\varrho$})=Q\mbox{\boldmath$\varrho$}^2
\left(\ln \frac{a_s^2}{R_n^2}-0.041 \right),\quad a_s=0.81 a_B
Z^{-1/3},
\label{2.15}
\end{equation}
here $a_s$ is the screening radius, $a_B$ is the Bohr radius.

Allowing for the estimates Eq.(\ref{2.14}) we present the
potential $V(\mbox{\boldmath$\varrho$})$ Eq.(\ref{2.9}) in the
following form
\begin{eqnarray}
&&\displaystyle{V(\mbox{\boldmath$\varrho$})=V_c(\mbox{\boldmath$\varrho$})+
v(\mbox{\boldmath$\varrho$}),\quad V_c(\mbox{\boldmath$\varrho$})=
q\mbox{\boldmath$\varrho$}^2,\quad q=QL_c,\quad L_c \equiv
L(\varrho_c)=\ln \frac{1}{\gamma^2\vartheta_2^2\varrho_c^2}},
\nonumber \\
&&\displaystyle{L_1 \equiv L(1)=\ln
\frac{1}{\gamma^2\vartheta_2^2}= \ln
\frac{a_{s2}^2}{\lambda_c^2}, \quad v(\mbox{\boldmath$\varrho$})=
-\frac{q\mbox{\boldmath$\varrho$}^2}{L_c} \left(2C+\ln
\frac{\mbox{\boldmath$\varrho$}^2}{4\varrho_c^2} \right)},
\label{2.16}
\end{eqnarray}
here $a_{s2}=a_s \exp (-f+1/2)$. So, one can to redefine the
parameters $a_s$ and $\vartheta_1$ to include the Coulomb
corrections. The inclusion of the Coulomb corrections
($f(Z\alpha)$ and -1) into $\ln \vartheta_2^2$ diminishes
effectively the correction $v(\mbox{\boldmath$\varrho$})$ to the
potential $V_c(\mbox{\boldmath$\varrho$})$. In accordance with
such division of the potential we present the propagators in Eq.
(\ref{2.11}) as
\begin{equation}
G^{-1}-G_0^{-1}=G^{-1}-G_c^{-1} + G_c^{-1}-G_0^{-1}
\label{2.17}
\end{equation}
where
\[
G_c={\bf p}^2+1-iV_c,\quad G={\bf p}^2+1-iV_c-iv
\]
This representation of the propagator $G^{-1}$ permits one to
expand it over the "perturbation" $v$. Indeed, with an increase of
$q$ the relative value of the perturbation is diminished
$\displaystyle{(\frac{v}{V_c} \sim \frac{1}{L_c})}$ since the
effective impact parameters diminishes and, correspondingly, the
value of logarithm $L_c$ in Eq.(\ref{2.16}) increases. The maximal
value of $L_c$ is determined by the size of a nucleus $R_n$
\begin{equation}
L_{max}=\ln \frac{a_{s2}^2}{R_n^2} \simeq 2
\ln \frac{a_{s2}^2}{\lambda_c^2}
\equiv 2 L_1,\quad L_1=2(\ln (183Z^{-1/3})-f(Z\alpha)),
\label{2.18}
\end{equation}
The value $L_1$ is the important parameter of the radiation theory.

The matrix elements of the operator $G_c^{-1}$ can be calculated
explicitly. The exponential parametrization of the propagator is
\begin{equation}
\displaystyle{G_c^{-1}=i\int_{0}^{\infty}dt e^{-it} \exp (-i{\cal
H}_ct),\quad {\cal H}_c={\bf p}^2 - iq
\mbox{\boldmath$\varrho$}^2} \label{2.19}
\end{equation}
The matrix elements of the operator $\displaystyle{\exp (-i{\cal
H}_ct)}$ has the form (details see in \cite{13})
\begin{eqnarray}
&& <\mbox{\boldmath$\varrho$}_1|\exp(-i{\cal
H}_ct)|\mbox{\boldmath$\varrho$}_2> \equiv
K_c(\mbox{\boldmath$\varrho$}_1, \mbox{\boldmath$\varrho$}_2, t),
\nonumber \\
&& K_c(\mbox{\boldmath$\varrho$}_1, \mbox{\boldmath$\varrho$}_2, t)=
\frac{\nu}{4\pi i \sinh \nu t} \exp \left\{ \frac{i\nu}{4}
\left[ (\mbox{\boldmath$\varrho$}_1^2+\mbox{\boldmath$\varrho$}_2^2)
\coth \nu t - \frac{2}{\sinh \nu t}
\mbox{\boldmath$\varrho$}_1\mbox{\boldmath$\varrho$}_2\right] \right\},
\label{2.24}
\end{eqnarray}
where $\nu=2\sqrt{iq}$ (see Eq.(\ref{2.16})).

Substituting formulas (\ref{2.19}) and (\ref{2.24}) in the
expression for the spectral distribution of the probability of
radiation (\ref{2.11}) we have
\begin{eqnarray}
&&\displaystyle{\frac{dW_c}{d\omega}=
\frac{\alpha}{2\pi \gamma^2} {\rm Im}~\Phi (\nu)},
\nonumber \\
&&\displaystyle{\Phi(\nu)=\nu\int_{0}^{\infty} dt e^{-it}\left[R_1
\left(\frac{1}{\sinh z}-\frac{1}{z}\right)-i\nu R_2
\left( \frac{1}{\sinh^2z}- \frac{1}{z^2}\right) \right]}
\nonumber \\
&&=R_1\left(\ln p -\psi\left(p+\frac{1}{2} \right) \right)
+R_2\left( \psi\left(p \right)-\ln p+\frac{1}{2p} \right)
\label{2.25}
\end{eqnarray}
where $z=\nu t,~p=i/(2\nu)$, let us remind that $\psi(x)$ is the
logarithmic derivative of the gamma function (see
Eq.(\ref{2.9b})). If we substitute in Eq.(\ref{2.25}) $t
\rightarrow t \exp (-i\pi/4)$ the probability will be transformed
into the form containing the real functions only. Then it can be
written as
\begin{equation}
\frac{dW_c}{d\omega}=\frac{\alpha |\nu|^2}{12\gamma^2}\left[R_1 G(s)
+2R_2 \phi(s)\right],
\label{2.41}
\end{equation}
where
\begin{eqnarray}
&& G(s)=12s^2\left[\frac{\pi}{4}-\int_{0}^{\infty}\exp (-sx)\frac{\sin (sx)}
{\sinh x} dx\right],
\nonumber \\
&& \phi(s)=6s^2\left[\int_{0}^{\infty}\exp (-sx)
\sin (sx) \coth x dx - \frac{\pi}{4}\right],
\label{2.42}
\end{eqnarray}
are the functions introduced by Migdal (see Eqs.(46), (47) in \cite{2}),
note that parameter $\displaystyle{s=\frac{1}{\sqrt{2}|\nu|}}$
in Eq.(\ref{2.42}) is two times larger than that used by Migdal.
If in Eqs.(\ref{2.41}), (\ref{2.42}) one omits the Coulomb correction,
then the probability (\ref{2.41}) coincides formally with the
probability calculated by Migdal (see Eq.(49) in \cite{2}).

Expanding the expression $G^{-1}-G_c^{-1}$ over powers of $v$ we
have
\begin{equation}
G^{-1}-G_c^{-1}=G_c^{-1}(iv)G_c^{-1}+G_c^{-1}(iv)G_c^{-1}(iv)G_c^{-1}
+...
\label{2.26}
\end{equation}

In accordance with Eqs.(\ref{2.17}) and (\ref{2.26}) we present
the probability of radiation in the form
\begin{equation}
\frac{dW}{d\omega}=\frac{dW_c}{d\omega}+\frac{dW_1}{d\omega}+
\frac{dW_2}{d\omega}+...
\label{2.27}
\end{equation}
At $Q \geq 1$ the expansion
(\ref{2.26}) is a series over powers of $\displaystyle{\frac{1}{L}}$.
It is important that variation of the parameter $\varrho_c$ by a factor
order of 1 has an influence on the dropped terms
in (\ref{2.26}) only.
The probability of radiation $\displaystyle{\frac{dW_c}{d\omega}}$
is defined by Eq.(\ref{2.25}).

The term $\displaystyle{\frac{dW_1}{d\omega}}$ in Eq.(\ref{2.27})
corresponds to the term linear in $v$ in Eq.(\ref{2.26}). The
explicit formula for the first correction to the probability of
radiation \cite{13} is
\begin{eqnarray}
&&\displaystyle{\frac{dW_1}{d\omega}=-\frac{\alpha}{4\pi \gamma^2
L_c} {\rm Im}~F(\nu);\quad -{\rm
Im}~F(\nu)=D_1(\nu_0)R_1+\frac{1}{s}D_2(\nu_0)R_2;\quad
s=\frac{1}{\sqrt{2}\nu_0}},
\nonumber \\
&&\displaystyle{D_1(\nu_0)=\int_{0}^{\infty}\frac{dz e^{-sz}}{\sinh^2z}
\left[d(z)\sin sz+\frac{\pi}{4}g(z)\cos sz \right],\quad
D_2(\nu_0)=\int_{0}^{\infty}\frac{dz e^{-sz}}{\sinh^3 z}}
\nonumber \\
&&\displaystyle{\times \left\{\left[d(z)-
\frac{1}{2}g(z)\right]\left(\sin sz+\cos sz\right)
+\frac{\pi}{4}g(z)\left(\cos sz - \sin sz\right)\right\}},
\nonumber \\
&&\displaystyle{d(z)=(\ln \nu_0 \vartheta(1-\nu_0)-\ln \sinh z
-C)g(z) -2G(z) \cosh z}, \label{2.30}
\end{eqnarray}
where
\begin{eqnarray}
&&g(z)=z \cosh z-\sinh z,\quad G(z)=\int_{0}^{z}(1-y\coth y)dy
\nonumber \\
&&=z-\frac{z^2}{2}-\frac{\pi^2}{12}- z\ln \left(1-e^{-2z} \right)
+\frac{1}{2}{\rm Li}_2 \left(e^{-2z} \right),
\label{2.31a}
\end{eqnarray}
here ${\rm Li}_2 \left(x \right)$ is the Euler dilogarithm; and
\begin{equation}
\nu_0^2 \equiv |\nu|^2=4q= 4 QL(\varrho_c)=
\frac{8\pi n_a Z^2\alpha^2 \varepsilon \varepsilon'}{m^4 \omega}L(\varrho_c),
\label{2.32a}
\end{equation}
The functions $G(s)$ and $\phi(s)$ are given in Fig. 1 and the
functions $D_1(\nu_0)$ and $D_2(\nu_0)$ are shown in Fig. 2.

As it was said above (see Eqs.(\ref{2.14}), (\ref{2.18})),
$\varrho_c=1$ at
\begin{equation}
|\nu^2|=\nu_1^2=4QL_1 \leq 1.
\label{2.30a}
\end{equation}
The logarithmic functions $L_c \equiv L(\varrho_c)$ and $L_1$ are
defined in Eqs.(\ref{2.16}) and (\ref{2.18}). If the parameter
$|\nu|
> 1$, the value of $\varrho_c$ is defined from the Eq.
(\ref{2.14}), where $\vartheta_1 \rightarrow \vartheta_2$, up to
$\varrho_c=R_n/\lambda_c$.  So, we have two representation of
$|\nu|$ depending on $\varrho_c$: at $\varrho_c = 1$ it is
$|\nu|=\nu_1$ and at $\varrho_c \leq 1$ it is $|\nu|=\nu_0$. The
mentioned parameters can be presented in the following form
\begin{eqnarray}
&& \nu^2=i\nu_0^2,\quad \nu_0^2=|\nu|^2 \simeq \nu_1^2\left(1+
\frac{\ln \nu_1}{L_1}\vartheta(\nu_1-1) \right),\quad
\nu_1^2=\frac{\varepsilon}{\varepsilon_e}\frac{1-x}{x},\quad
x=\frac{\omega}{\varepsilon},
\nonumber \\
&&\varepsilon_e=m\left(8\pi Z^2 \alpha^2 n_a \lambda_c^3 L_1
\right)^{-1}, \quad L_c \simeq L_1 \left(1+ \frac{\ln
\nu_1}{L_1}\vartheta(\nu_1-1) \right)
\label{2.30b}
\end{eqnarray}
It should be noted that in the logarithmic approximation the
parameter $\varrho_c$ entering into the parameter $\nu$ is
defined up to the factor $\sim 1$. However, we calculated the
next term of the decomposition over
$v(\mbox{\boldmath$\varrho$})$ (an accuracy up to the "next to
leading logarithm") and this permitted to obtain the result which
is independent of the parameter $\varrho_c$ (in terms $\propto
1/L$). Our definition of the parameter $\varrho_c$ minimizes
corrections to Eq.(\ref{2.25}) practically for all values of the
parameter $\varrho_c$.

The approximate solution of Eq.(\ref{2.14}) given in
Eq.(\ref{2.30b}) has quite good numerical accuracy:it is $\sim
2~\%$ at $\nu_1=100$ and $\sim 4.5~\%$ at $\nu_1=1000$. The LPM
effect manifests itself when
\begin{equation}
\nu_1(x_c)=1,\quad
x_c=\frac{\omega_c}{\varepsilon}=\frac{\varepsilon}{\varepsilon_e+\varepsilon}.
\label{2.31b}
\end{equation}
So, the characteristic energy $\varepsilon_e$ is the energy starting from
which the multiple scattering distorts the whole
spectrum of radiation including its hard part. If the radiation
length $L_{rad}$ is taken within the logarithmic
approximation the value $\varepsilon_e$ coincides with $\varepsilon_{LP}$
Eq.(\ref{9}). The difference between these values is
\[
\frac{\varepsilon_e-\varepsilon_{LP}}{\varepsilon_e}=\frac{1}{9L_1} < 2\%
\].

The formulas derived in \cite{13},\cite{L5} and written down
above are valid for any energy. In Fig. 3 the spectral radiation
intensity in gold ($\varepsilon_e=2.5$~TeV) is shown for
different energies of the initial electron. In the case when
$\varepsilon \ll \varepsilon_e$ ($\varepsilon=25~$GeV and
$\varepsilon=250~$GeV) the LPM suppression is seen in the soft
part of the spectrum only for $x \leq x_c \simeq
\varepsilon/\varepsilon_e \ll 1$ while in the region $\varepsilon
\geq \varepsilon_e$ ($\varepsilon=2.5~$TeV and
$\varepsilon=25~$TeV) where $x_c \sim 1$ the LPM effect is
significant  for any $x$. For relatively low energies
$\varepsilon=25~$GeV and $\varepsilon=8~$GeV used in famous SLAC
experiment \cite{13}, \cite{E2} we have analyzed the soft part of
spectrum, including all the accompanying effects: the boundary
photon emission, the multiphoton radiation and influence of the
polarization of the medium (see below). The perfect agreement of
the theory and data was achieved in the whole interval of measured
photon energies (200~keV$ \leq \omega \leq$500~MeV), see below
and the corresponding figures in \cite{13},\cite{14},\cite{15}. It
should be pointed out that both the correction term with $F(\nu)$
and the Coulomb corrections have to be taken into account for this
agreement.

When a scattering is weak ($\nu_1 \ll 1$), the main contribution
in Eq.(\ref{2.30}) gives a region where $z \ll 1$. Then
\begin{eqnarray}
&&\displaystyle{-{\rm Im}~F(\nu)=\frac{1}{9}{\rm
Im}~\nu^2\left(R_2- R_1 \right),\quad L_c \rightarrow L_1},
\nonumber \\
&&\displaystyle{\Phi(\nu) \simeq \frac{\nu^2}{6}\left(R_1+2R_2 \right),
\quad (|\nu| \ll 1)}
\label{2.33}
\end{eqnarray}
Combining the results obtained in Eq.(\ref{2.33}) we obtain the
spectral distribution of the probability of radiation in the case
when scattering is weak $(|\nu| \ll 1)$
\begin{eqnarray}
&&\displaystyle{\frac{dW}{d\omega}=\frac{dW_c}{d\omega}+\frac{dW_1}{d\omega}=
\frac{\alpha}{2\pi \gamma^2} {\rm
Im}~\left[\Phi(\nu)-\frac{1}{2L_1}F(\nu) \right]}
\nonumber \\
&&\displaystyle{=\frac{\alpha}{2\pi \gamma^2}\frac{2Q}{3}\left[R_1\left(
L_1-\frac{1}{3} \right)+2R_2\left(L_1+\frac{1}{6} \right) \right]}
\nonumber \\
&&\displaystyle{=\frac{4Z^2\alpha^3n_a}{3m^2\omega}\Bigg[\frac{\omega^2}
{\varepsilon^2}
\left(\ln \left(183Z^{-1/3} \right)-\frac{1}{6} -f(Z\alpha)\right)}
\nonumber \\
&&\displaystyle{+2\left(1+\frac{\varepsilon'^2}{\varepsilon^2} \right)
\left(\ln \left(183Z^{-1/3} \right)+\frac{1}{12}-f(Z\alpha) \right) \Bigg]},
\label{2.35}
\end{eqnarray}
where $L_1$ is defined in Eq.(\ref{2.18}). This expression
coincide with the known Bethe-Maximon formula for the probability
of bremsstrahlung from high-energy electrons in the case of
complete screening (if one neglects the contribution of atomic
electrons) written down within power accuracy (omitted terms are
of the order of powers of $\displaystyle{\frac{1}{\gamma}}$) with
the Coulomb corrections, see e.g. Eq.(18.30) in \cite{7}, or
Eq.(3.83) in \cite{22c}.

At $\nu_0 \gg 1$ the function ${\rm Im}~F(\nu)$ Eq.(\ref{2.30})
has the form
\begin{equation}
\displaystyle{-{\rm Im}~F(\nu)=\frac{\pi}{4}R_1+\frac{\nu_0}{\sqrt{2}}
\left(\ln 2-C+\frac{\pi}{4} \right)R_2}.
\label{2.38}
\end{equation}
Under the same conditions ($\nu_0 \gg 1$) the function ${\rm
Im}~\Phi(\nu)$ Eq.(\ref{2.25}) is
\begin{equation}
\displaystyle{{\rm Im}~\Phi(\nu)=\frac{\pi}{4}R_1+\frac{\nu_0}{\sqrt{2}}R_2
\rightarrow \frac{\nu_0}{\sqrt{2}}R_2}.
\label{2.39}
\end{equation}
So, in the region where the LPM effect is strong the probability
Eq.(\ref{2.25}) can written as
\begin{equation}
\frac{dW_c}{d\omega}=\frac{\alpha R_2}{\varepsilon^2}
\left(\frac{Z^2\alpha^2 \varepsilon \varepsilon' n_a}
{\pi \omega} L(\varrho_c)\right)^{1/2}.
\label{2.39a}
\end{equation}
This means that in this limit the emission probability is proportional to
the square root of the density. This fact was pointed out by Migdal \cite{2},
the expression (\ref{2.39a}) coincides formally with the
probability calculated by Migdal (see Eq.(52) in \cite{2}) with the same
remarks which were made after Eq.(\ref{2.42}).

Thus, at $\nu_0 \gg 1$ the relative contribution of the first correction
$\displaystyle{\frac{dW_1}{d\omega}}$ is defined by
\begin{equation}
\displaystyle{r=\frac{dW_1}{dW_c}=\frac{1}{2L(\varrho_c)}
\left(\ln 2-C+\frac{\pi}{4} \right) \simeq \frac{0.451}{L(\varrho_c)}},
\label{2.40}
\end{equation}
where $\displaystyle{L(\varrho_c)=\ln \frac{a_{s2}^2}{\lambda_c^2
\varrho_c^2}}$. In this expression the value $r$ with the
accuracy up to terms $\sim 1/L_c^2$ doesn't depend on the photon
energy:$L_c \simeq L_1+\ln(\varepsilon/\varepsilon_e)/2$. Hence
we can find the correction to the total probability at
$\varepsilon \gg \varepsilon_e$. The maximal value of the
correction is attained at $\varepsilon \sim 10\varepsilon_e$, it
is $\sim 6\%$ for heavy elements.

In the above analysis we did not consider an inelastic scattering of a
projectile on atomic electrons. The potential
$V_e(\mbox{\boldmath$\varrho$})$ connected with this process
can be found from formula (\ref{2.9a}) by substitution
$Z^2 \rightarrow Z, \vartheta_1 \rightarrow \vartheta_e=0.153 \vartheta_1$
(an analysis of an inelastic scattering on atomic electrons as well as
the parameter $\vartheta_e$ can be found in \cite{22}).
The summary potential including both an elastic and an inelastic scattering
is
\begin{eqnarray}
&& \displaystyle{V(\mbox{\boldmath$\varrho$})+V_e(\mbox{\boldmath$\varrho$})
=-Q(1+\frac{1}{Z})\mbox{\boldmath$\varrho$}^2
\Big[\ln \gamma^2 \vartheta_2^2
+\ln \frac{\mbox{\boldmath$\varrho$}^2}{4}+2C +
\frac{1}{Z+1}\left(\ln \frac{\vartheta_e^2}{\vartheta_1^2}-2f\right)\Big]}
\nonumber \\
&& \displaystyle{=-Q_{ef}\mbox{\boldmath$\varrho$}^2
\Big(\ln \gamma^2 \vartheta_{ef}^2
+\ln \frac{\mbox{\boldmath$\varrho$}^2}{4}+2C \Big)},
\label{40a}
\end{eqnarray}
where
\[
\displaystyle{Q_{ef}=Q(1+\frac{1}{Z}),\quad \vartheta_{ef}
= \vartheta_1 \exp \left[\frac{1}{1+Z}\left(Zf(\alpha Z)-1.88
\right)-\frac{1}{2} \right]}.
\]

\subsection{Integral characteristics of bremsstrahlung}

Integrating Eq.(\ref{2.25}) over $x=\omega/\varepsilon$ we obtain
the total intensity of radiation in the logarithmic approximation
\begin{eqnarray}
&&\frac{I}{\varepsilon}L_{rad}^0=2\frac{\varepsilon_e}{\varepsilon}
{\rm
Im}\Bigg[\int_{0}^{1}\frac{dx}{g}\sqrt{\frac{x}{1-x}}(2(1-x)+x^2)
\nonumber \\
&&+ \int_{0}^{1}\frac{x^3dx}{1-x}
\left(\psi(p+1)-\psi\left(p+\frac{1}{2}\right) \right)
+2\int_{0}^{1}xdx\left(\psi\left(p+1\right) -\ln p\right)\Bigg],
\label{2.1i}
\end{eqnarray}
where
\[
p=\frac{g\eta}{2},\quad \eta=\sqrt{\frac{x}{1-x}},\quad g=\exp
\left(i\frac{\pi}{4}\right)
\sqrt{\frac{L_1}{L_c}\frac{\varepsilon_e}{\epsilon}},
\]
$L_{rad}^0$ is the radiation length in the logarithmic
approximation. The relative energy loss of electron per unit
time in terms of the Bethe-Maximon radiation length $L_{rad}^0$:
$\displaystyle{\frac{I}{\varepsilon}L_{rad}^0}$ in gold is given
in Fig. 4 (curve 1), it reduces by 10\% (15\% and 25\%) at
$\varepsilon \simeq 700$~GeV ($\varepsilon \simeq 1.4$~TeV and
$\varepsilon \simeq 3.8$~TeV) respectively, and it cuts in half
at $\omega \simeq 26$~TeV. This increase of effective radiation
length can be important in electromagnetic calorimeters operating
in detectors on colliders in TeV range as well as in analysis of
high-energy cosmic rays. The relative energy loss of electron 
per unit time in terms of the exact Bethe-Maximon radiation 
length $L_{rad}$ in iridium and lead calculated taking into 
account the correction terms Eq.(\ref{2.30}) is shown in 
Fig.19 (see Sec.3.10.2). 

The spectral distribution of bremsstrahlung intensity and the
spectral distribution over energy of created electron (positron)
as well as the reduction of energy loss and the photon conversion
cross section was calculated by Klein  \cite{E2} using the Migdal
\cite{2} formulas. As will be explained below (see Sec. 3.7) we
use more accurate procedure of fine tuning and because of this
our calculation in logarithmic approximation differs from Migdal
one. We calculated also the correction term and include the
Coulomb corrections. For this reason the results shown here in
Figs.3-4 are more precise than given in \cite{E2}.

In the case $\varepsilon \ll \varepsilon_e$ the correction term
to the total intensity of radiation (or to the radiation length)
can be written as \cite{L5}
\begin{eqnarray}
&&\frac{I}{\varepsilon}=\frac{\alpha m^2}{4\pi \varepsilon_e}
\left(1+\frac{1}{9L_1}-\frac{4\pi}{15}\frac{\varepsilon}{\varepsilon_e}
\right) \simeq L_{rad}^{-1}\left(1
-\frac{4\pi}{15}\frac{\varepsilon}{\varepsilon_e}\right),
\nonumber \\
&& \frac{1}{L_{rad}}= \frac{1}{L_{rad}^0}
\left(1+\frac{1}{9L_1} \right),\quad \frac{1}{L_{rad}^0}=\frac{2Z^2\alpha^3n_a L_1}{m^2}
\label{2.2i}
\end{eqnarray}

In the opposite case $\varepsilon \gg \varepsilon_e$ we have
\cite{L5}
\begin{equation}
\frac{I}{\varepsilon L_{rad}} \simeq \frac{5}{2}
\sqrt{\frac{\varepsilon_e}{\varepsilon}}
\left[1-2.37\sqrt{\frac{\varepsilon_e}{\varepsilon}} +4.57
\frac{\varepsilon_e}{\varepsilon} +\frac{1}{4L_1} \left(\ln
\frac{\varepsilon}{\varepsilon_e} -0.3455 \right)\right]
\label{2.3i}
\end{equation}
Although the coefficients in the last expression are rather large
at two first terms of the decomposition over
$\sqrt{\varepsilon_e/\varepsilon}$ this formula has the accuracy
of the order of 10\% at $\varepsilon \sim 10 \varepsilon_e$.

Analogous expression for the integral probability of radiation
has the form
\begin{equation}
W=\frac{11\pi Z^2 \alpha^3 n_a}{2\sqrt{2} m^2}
\sqrt{\frac{\varepsilon_e}{\varepsilon}}L_1
\left[1-1.23\sqrt{\frac{\varepsilon_e}{\varepsilon}} +1.645
\frac{\varepsilon_e}{\varepsilon} +\frac{1}{4L_1} \left(\ln
\frac{\varepsilon}{\varepsilon_e} +2.53 \right)\right].
\label{2.4i}
\end{equation}

Ratio of the main terms of Eqs.(\ref{2.3i}) and (\ref{2.4i}) gives
the mean energy of radiated photon
\begin{equation}
\bar{\omega}=\frac{9}{22}\varepsilon \simeq 0.409 \varepsilon.
\label{2.5i}
\end{equation}

\subsection{The LPM effect for pair creation}

The probability of the pair creation by a photon can be obtained
from the probability of the bremsstrahlung with help of the substitution law:
\begin{equation}
\omega^2d\omega \rightarrow \varepsilon^2 d\varepsilon,\quad
\omega \rightarrow -\omega,\quad \varepsilon \rightarrow -\varepsilon,
\label{2.1p}
\end{equation}
where $\omega$ is the photon energy, $\varepsilon$ is the energy
of the particle. Making this substitution in Eq.(\ref{2.11}) we
obtain the spectral distribution of the pair creation probability
\cite{L5} (over the energy of the created electron $\varepsilon$)
\begin{equation}
\frac{dW_p}{d\varepsilon}=\frac{2\alpha m^2}{\varepsilon \varepsilon'}
{\rm Im}\left<0|s_1\left(G^{-1}-G_0^{-1}\right)+
s_2 {\bf p}\left(G^{-1}-G_0^{-1}\right){\bf p}|0\right>,
\label{2.2p}
\end{equation}
where
\begin{equation}
s_1=1,\quad s_2=\frac{\varepsilon^2+\varepsilon'^2}
{\omega^2},\quad\varepsilon'=\omega-\varepsilon,
\label{2.3p}
\end{equation}
the other notations are given in Eqs.(\ref{2.12}), (\ref{2.9}),
(\ref{2.9a}), (\ref{2.9b}).

Just as in the analysis of the LPM effect for radiation we present
the potential Eq.(\ref{2.9a}) in the form $
V(\mbox{\boldmath$\varrho$})= V_c(\mbox{\boldmath$\varrho$})
+v(\mbox{\boldmath$\varrho$})$ and will proceed as we did in
Eqs.(\ref{2.16})- (\ref{2.24}). Substituting the expression
(\ref{2.24}) in the formula for the spectral distribution of the
pair creation probability (\ref{2.2p}) we have
\begin{eqnarray}
&&\displaystyle{\frac{dW_p^c}{d\varepsilon}=
\frac{\alpha m^2}{2\pi \varepsilon \varepsilon'} {\rm Im}~\Phi_p (\nu)},
\nonumber \\
&&\displaystyle{\Phi_p(\nu)=\nu\int_{0}^{\infty} dt e^{-it}\left[s_1
\left(\frac{1}{\sinh z}-\frac{1}{z}\right)-i\nu s_2
\left( \frac{1}{\sinh^2z}- \frac{1}{z^2}\right) \right]}
\nonumber \\
&&=s_1\left(\ln p-\psi\left(p+\frac{1}{2}\right) \right)
+s_2\left(\psi (p) -\ln p+\frac{1}{2p}\right),
\label{2.10p}
\end{eqnarray}
where we used the same notations as in Eq.(\ref{2.25}). This
formula gives the spectral distribution of the pair creation
probability in the logarithmic approximation which was used also
by Migdal \cite{3}. It should be noted that the parameter
$\varrho_c$ entering into the parameter $\nu$ (see
Eqs.(\ref{2.24}), (\ref{2.16}) and (\ref{2.14})) is defined up to
the factor $\sim 1$, what is inherent in the logarithmic
approximation. However, below we will calculate the next term of
the decomposition over $v(\mbox{\boldmath$\varrho$})$ (an accuracy
up to the "next to leading logarithm") and this permits to obtain
the result which is independent of the parameter $\varrho_c$. It
will be shown that the definition of the parameter $\varrho_c$
minimizes corrections to Eq.(\ref{2.10p}) practically for all
values of the parameter $\varrho_c$. It should be emphasized also
that here the Coulomb corrections are included into the parameter
$\nu$ in contrast to \cite{3}.

Substituting the expansion (\ref{2.26}) in Eq.(\ref{2.2p}) we
obtain the decomposition of the probability of the pair creation:
\begin{equation}
\frac{dW_p}{d\varepsilon}=\frac{dW_p^c}{d\varepsilon}+
\frac{dW_p^1}{d\varepsilon}+\frac{dW_p^2}{d\varepsilon}+...
\label{2.12p}
\end{equation}
The probability of pair creation
$\displaystyle{\frac{dW_p^c}{d\varepsilon}}$ is defined by
Eq.(\ref{2.10p}).The term
$\displaystyle{\frac{dW_p^1}{d\varepsilon}}$ in Eq.(\ref{2.12p})
corresponds to the first term (linear in $v$) in Eq.(\ref{2.26}).
The explicit expression for the first correction to the pair
creation probability is \cite{L5}
\begin{eqnarray}
&& \frac{dW_p^1}{d\varepsilon}= -\frac{\alpha m^2}{4\pi
\varepsilon \varepsilon' L_c} {\rm Im}~F_p(\nu);\quad F_p(\nu)=
\int_{0}^{\infty}\frac{dz e^{-it}}{\sinh^2z}
\left[s_1f_1(z)-2is_2f_2(z) \right],
\nonumber \\
&& f_1(z)=\left(\ln \varrho_c^2+\ln \frac{\nu}{i}
-\ln \sinh z-C\right)g(z) - 2\cosh z G(z),
\nonumber \\
&& f_2(z) = \frac{\nu}{\sinh z}
\left(f_1(z)-\frac{g(z)}{2} \right), \quad t=t_1+t_2,
\label{2.15p}
\end{eqnarray}
where the functions $g(z)$ and $G(z)$ are defined in
Eq.(\ref{2.31a}), $L_c$ is defined in Eq.(\ref{2.16}). Use of the
last representation of function $G(z)$ simplifies the numerical
calculation. As in radiation case the parameter $\nu$ has two
representations: Eqs.(\ref{2.32a}) and (\ref{2.30a}).

When the scattering of created particles is weak ($\nu_1 \ll 1$),
the main contribution in Eqs.(\ref{2.10p}) and (\ref{2.15p}) gives
the region where $z \ll 1$ and one can expand the functions of
$z$ (compare with Eq.(\ref{2.33})). As result we  obtain the
spectral distribution of the pair creation probability for $|\nu|
\ll 1$:
\begin{eqnarray}
&&\displaystyle{\frac{dW_p}{d\varepsilon}=
\frac{dW_p^c}{d\varepsilon}+\frac{dW_p^1}{d\varepsilon}=
\frac{\alpha m^2}{2\pi \varepsilon \varepsilon'} {\rm
Im}~\left[\Phi_p(\nu)-\frac{1}{2L_c}F_p(\nu) \right]}
\nonumber \\
&& \displaystyle{=\frac{4Z^2\alpha^3n_a}{3m^2\omega}\Bigg\{
\left(\ln \left(183Z^{-1/3} \right) -f(Z\alpha)\right)
\left(1-\frac{31\nu_1^4}{21} \right)-\frac{1}{6}}
\nonumber \\
&& \displaystyle{+2\frac{\varepsilon^2+\varepsilon'^2}{\omega^2}
\left[\left(\ln \left(183Z^{-1/3} \right)-
f(Z\alpha)\right)\left(1-\frac{16\nu_1^4}{21} \right)+\frac{1}{12}
\right] \Bigg\}},
\label{2.22p}
\end{eqnarray}
where $L_1$ is defined in Eq.(\ref{2.16}). Integrating
Eq.(\ref{2.22p}) over $\varepsilon$ we obtain
\begin{equation}
W_p=\frac{28Z^2\alpha^3n_a}{9m^2}\left[\left(\ln(183 Z^{-1/3})-f(Z\alpha)
\right)\left(1-\frac{3312}{2401} \frac{\omega^2}{\omega_e^2} \right)-
\frac{1}{42} \right],
\label{2.23p}
\end{equation}
where
\begin{equation}
\omega_e=m\left(2\pi Z^2 \alpha^2 n_a \lambda_c^3 L_1 \right)^{-1}
\label{2.23ap}
\end{equation}
Note that $\omega_e$ is four times larger than
$\varepsilon_e$~Eq.(\ref{2.30b}),~ in gold it is
$\omega_e=10.5~$TeV. This is just the value of photon energy
starting with the LPM effect becomes essential for the pair
creation process in heavy elements. If one omits here the terms
$\propto \nu_1^4$ and $\propto (\omega/\omega_e)^2$ these
expressions coincide with the known Bethe-Maximon formula for the
probability of pair creation by a high-energy photon in the case
of complete screening (if one neglects the contribution of atomic
electrons) written down within power accuracy (omitted terms are
of the order of powers of $\displaystyle{\frac{m}{\omega}}$) with
the Coulomb corrections, see e.g. Eqs.(19.4) and (19.17) in
\cite{7}.

The pair creation spectral probability $dW/dx$ vs
$x=\varepsilon/\omega$ in gold is shown in Fig. 5 for different
energies. It is seen that for $\omega=2.5~$TeV which below
$\omega_e$ the difference with the Bethe-Maximon probability is
rather small. When $\omega > \omega_e$ there is significant
difference with the Bethe-Maximon spectrum increasing with
$\omega$ growth. In Fig. 5 are shown the curves (thin lines
2,3,4) obtained in logarithmic approximation
$dW_p^c/d\varepsilon$ Eq.(\ref{2.10p}), the first correction to
the spectral probability $dW_p^1/d\varepsilon$ Eq.(\ref{2.15p}),
curves $c2, c3, c4$ and the sum of these two contributions:
curves $T1, T2, T3, T4$. It should be noted that for our
definition of the parameter $\varrho_c$ Eq.(\ref{2.14}) the
corrections are not exceed 6\% of the main term. The corrections
are maximal for $\nu_0 \sim 3$.

Integrating Eq.(\ref{2.10p}) over $y=\varepsilon/\omega$ we obtain
the total probability of pair creation in the logarithmic
approximation
\begin{eqnarray}
&&\frac{W_p^c}{W_{p0}^{BM}}=\frac{9}{14}\frac{\omega_e}{\omega}
{\rm Im}\int_{0}^{1}\frac{dy}{y(1-y)}\Bigg[
\left(\ln p-\psi\left(p+\frac{1}{2}\right) \right)
\nonumber \\
&&+\left(1-2y+2y^2 \right)\left(\psi\left(p\right)
-\ln p+\frac{1}{2p}\right)\Bigg],
\label{2.23bp}
\end{eqnarray}
where
\[
p=\frac{bs}{4},\quad s=\frac{1}{\sqrt{y(1-y)}},\quad
b=\exp \left(i\frac{\pi}{4}\right)
\sqrt{\frac{L_1}{L_c}\frac{\omega_e}{\omega}},
\]
$W_{p0}^{BH}$ is the Bethe-Maximon probability of pair photoproduction
in the logarithmic approximation. The total probability of pair creation
$W_p^c$ in gold is given in Fig. 4 (curve 2),it reduced by 10\%
at $\omega \simeq 9$~TeV and it cuts in half at
$\omega \simeq 130$~TeV.

At $\nu_0 \gg 1$ the function $F_p(\nu)$ (see Eq.(\ref{2.15p}))
has the form
\begin{equation}
\displaystyle{-{\rm Im}~F_p(\nu)=\frac{\pi}{4}(s_1-s_2)+\frac{\nu_0}{\sqrt{2}}
\left(\ln 2-C+\frac{\pi}{4} \right)s_2},
\label{2.25p}
\end{equation}
Under the same conditions ($\nu_0 \gg 1$) the function ${\rm
Im}~\Phi_p(\nu)$ Eq.(\ref{2.10p}) is
\begin{equation}
\displaystyle{{\rm Im}~\Phi_p(\nu)=\frac{\pi}{4}(s_1-s_2)+
\frac{\nu_0}{\sqrt{2}}s_2} \rightarrow \frac{\nu_0}{\sqrt{2}}s_2.
\label{2.26p}
\end{equation}
So, in the region where the LPM effect is strong the probability
Eq.(\ref{2.10p}) can written as
\begin{equation}
\frac{dW_p^c}{d\varepsilon}=\frac{\alpha (\varepsilon^2+ \varepsilon'^2)}
{\omega^2}
\left(\frac{Z^2\alpha^2 n_a}
{\pi \omega  \varepsilon \varepsilon'} L(\varrho_c)\right)^{1/2}.
\label{2.39p}
\end{equation}
This means that in this limit the spectral probability of pair
creation is proportional to the square root of the density just
as the emission probability Eq.(\ref{2.39a}).

At $\nu_0 \gg 1$ the relative contribution of the first correction
$\displaystyle{\frac{dW_p^1}{d\varepsilon}}$ is defined by
\begin{equation}
\displaystyle{r=\frac{dW_p^1}{dW_p^c}=\frac{1}{2L_c}
\left(\ln 2-C+\frac{\pi}{4} \right) \simeq \frac{0.451}{L_c}}.
\label{2.27p}
\end{equation}
In this expression the value $r$ with the accuracy up to terms
$\sim 1/L_c^2$ doesn't depend on the electron energy: $L_c \simeq
L_1+\ln(\omega/\omega_e)/2$. Hence we can find the correction to
the total probability at $\omega \gg \omega_e$. The maximal value
of the correction is attained at $\omega \sim 10\omega_e$, it is
$\sim 6\%$ for heavy elements. In this region $\nu_0 \gg 1$ or
$\omega \gg \omega_e$ the total probability of pair creation in
terms of the Bethe-Maximon total probability can be presented as
\begin{equation}
\frac{W_p}{W^{BM}_p}\simeq 2.14 \sqrt{\frac{\omega_e}{\omega}}
\left[1-0.836 \sqrt{\frac{\omega_e}{\omega}}-0.548 \frac{\omega_e}{\omega}
 + \frac{1}{4L_1}
\left(\ln \frac{\omega}{\omega_e}+0.274 \right) \right].
\label{2.33p}
\end{equation}

\subsection{Anomalous magnetic moment of the electron in a medium}

The contributions of higher orders of the perturbation theory
over the interaction with an electromagnetic field give the
electromagnetic radiative corrections to the electron mass and
lead to appearance of the anomalous magnetic moment (AMM) of the
electron \cite{S}. It is known that under influence of an external
electromagnetic field these effects in particular the AMM of
electron are changed essentially \cite{BKS1}, \cite{7}. Here we
consider how the mentioned effects modify in a medium \cite{BK1}.

The total probability of radiation $W$ is connected with imaginary
part of the radiative correction to the electron mass according to
\begin{equation}
m\Delta m=\varepsilon \Delta \varepsilon,\quad -2{\rm Im}\Delta
\varepsilon=W,\quad {\rm Im}~\Delta m = - \frac{\varepsilon}{2m}
W.
\label{2.3a}
\end{equation}
Since $T$ in Eq.(\ref{2.11}) is the analytic function of the potential
$V$ we have that
\begin{equation}
\Delta m = -\alpha m \int_{0}^{\varepsilon}
\frac{d\omega}{\varepsilon}T.
\label{2.4a}
\end{equation}
One can derive this formula considering the self-energy diagram
and the corresponding amplitude of forward scattering of electron
(see Eqs.(12.18) and (12.39) in \cite{7}).

In the presence of a homogeneous external field the Hamiltonian
${\cal H}$ (\ref{2.9}),(\ref{2.10}) acquires the linear over
coordinate term. One can find the explicit form of this term
using the Eqs.(\ref{a17}) and (\ref{a21}) of Appendix A and
carrying out the scale transformation of variables according to
Eq.(\ref{2.7}):
\begin{equation}
\Delta{\cal H}=\frac{2}{u}\mbox{\boldmath$\chi$}\mbox{\boldmath$\varrho$},
\label{2.5a}
\end{equation}
where $\mbox{\boldmath$\chi$}$ is the known parameter
characterizing quantum effects in the radiation process (see
Eq.(\ref{14})), $u=\omega/\varepsilon'$.

For polarized initial electron (see (\ref{a3a}) in Appendix A)
have for the addition to $T$ in Eq.(\ref{2.11}) depending on the
spin vector $\mbox{\boldmath$\zeta$}$:
\begin{equation}
T \rightarrow T + T_{\mbox{\boldmath$\zeta$}},\quad
T_{\mbox{\boldmath$\zeta$}}=i\frac{\omega}{\varepsilon} \left(
<0|(G^{-1}{\bf p}-{\bf p}G^{-1})|0>\times{\bf v} \right)
\mbox{\boldmath$\zeta$}.
\label{2.16a}
\end{equation}
For the radiative correction to the electron mass we have
respectively
\begin{equation}
\Delta m \rightarrow \Delta M = \Delta m + \Delta
m_{\mbox{\boldmath$\zeta$}}.
\label{2.16aa}
\end{equation}
It should be noted that expressions for $T$ (\ref{2.11}) and
(\ref{2.16a}) have universal form while the specific of the
particle motion is contained in the propagator $G^{-1}$ through
the effective potential
\begin{equation}
V_{F}(\mbox{\boldmath$\varrho$})=-iV_c(\mbox{\boldmath$\varrho$})
+\frac{2}{u}\mbox{\boldmath$\chi$}\mbox{\boldmath$\varrho$},\quad
{\cal H}_F = {\bf p}^2+V_F,\quad
 G_F= {\cal H}_F+1,
\label{2.17a}
\end{equation}
where we restrict ourselves to the main term in the decomposition
over $v$, see Eqs.(\ref{2.16}), (\ref{2.26}). This means that
result has the logarithmic accuracy over the scattering (but not
over an external field). With regard for an external field the
parameter $\varrho_c$ in Eq.(\ref{2.32a}) is defined by a set of
equations:
\begin{eqnarray}
&& \varrho_c =1\quad {\rm for}\quad 4\frac{\chi^2}{u^2}+4QL_1
\leq 1;
\nonumber \\
&& \varrho_c^4 \left[4\frac{\chi^2}{u^2}\varrho_c^2+|\nu(\varrho_c)|^2
\right]=1 \quad {\rm for} \quad 4\frac{\chi^2}{u^2}+ 4QL_1 \geq 1,
\label{2.18a}
\end{eqnarray}
where $\nu=2\sqrt{iq}$, $q=QL_c$, $L_1$, $Q$ and $L_c$ are defined
in Eqs.(\ref{2.30a}) and (\ref{2.32a}). Substituting into the
depending on the spin term in (\ref{2.16a}) we find
\begin{eqnarray}
&&\left<0|(G_F^{-1}{\bf p}-{\bf p}G_F^{-1}) |0\right>
=\frac{\mbox{\boldmath$\chi$}}{2\pi u}
\int_{0}^{\infty}\exp(-it)\frac{\varphi}{\cosh^2 \frac{\nu
t}{2}}dt,
\nonumber \\
&& \varphi \equiv \varphi(\chi, \nu, t)
=\exp\left[-\frac{4i\chi^2t}{\nu^2 u^2}\left( 1-\frac{2}{\nu
t}\tanh\frac{\nu t}{2} \right) \right].
\label{2.19a}
\end{eqnarray}
Thus, in the used approximation ($G=G_F$) we have for
$T_{\mbox{\boldmath$\zeta$}}$ Eq.(\ref{2.16a})
\begin{equation}
T_{\mbox{\boldmath$\zeta$}}=\frac{i}{2\pi
u}\frac{\omega}{\varepsilon}
\int_{0}^{\infty}\exp(-it)\frac{\varphi}{\cosh^2 \frac{\nu
t}{2}}dt (\mbox{\boldmath$\zeta$}\mbox{\boldmath$\chi$}{\bf v}),
\label{2.20a}
\end{equation}
where $(\mbox{\boldmath$\zeta$}\mbox{\boldmath$\chi$}{\bf v})=
(\mbox{\boldmath$\zeta$}\cdot(\mbox{\boldmath$\chi$}\times{\bf v}))$.
Within relativistic accuracy (up to terms of higher order over
$1/\gamma$) the combination of vectors entering in
Eq.(\ref{2.20a}) can be written as (see Eq.(12.19) in \cite{7}
and \cite{BKS3})
\begin{equation}
(\mbox{\boldmath$\zeta$}\mbox{\boldmath$\chi$}{\bf v})=
2\mu_0\frac{\mbox{\boldmath$\zeta$}{\bf H}_R}{m},
\label{3.1a}
\end{equation}
where ${\bf H}_R=\gamma ({\bf H}_{\perp}+{\bf E}\times {\bf v})$
is the magnetic field
in the electron rest frame. Here we use that
$\mbox{\boldmath$\chi$}=\chi{\bf s}$, where ${\bf s}$ is the unit
vector in the direction of acceleration (this vector is used in
\cite{7}).

In the electron rest frame one can consider the value Re~$\Delta
m_{\mbox{\boldmath$\zeta$}}$ depending on the electron spin as the
energy of interaction of the AMM of electron with the magnetic
field ${\bf H}_R$ (see Eq.(12.23) in \cite{7} and  \cite{BKS3})
\begin{equation}
{\rm Re}~\Delta m_{\mbox{\boldmath$\zeta$}}=-\mu'
\mbox{\boldmath$\zeta$}{\bf H}_R,
\label{3.2a}
\end{equation}

Taking into account Eqs.(\ref{2.4a}), (\ref{2.16a}), (\ref{2.16aa}),
(\ref{2.20a}), (\ref{3.1a}), (\ref{3.2a}) we obtain the following
general expression for the AMM of electron moving in a medium in
the presence of an external electromagnetic field \cite{BK1}:
\begin{equation}
\frac{\mu'}{\mu_0}=-\frac{\alpha}{\pi} {\rm Im}~
\int_{0}^{\infty} \frac{du}{(1+u)^3}
\int_{0}^{\infty}\exp(-it)\frac{\varphi}{\cosh^2 \frac{\nu t}{2}}dt
\label{3.3a}
\end{equation}

In the absence of scattering ($\nu \rightarrow 0$) the expression
(\ref{3.3a}) gets over into the formula for the AMM of electron in external
field (see Eq.(12.24) in \cite{7} and \cite{R}, \cite{BKS3})
\begin{equation}
\frac{\mu'}{\mu_0}=-\frac{\alpha}{\pi} {\rm Im}~
\int_{0}^{\infty} \frac{du}{(1+u)^3}
\int_{0}^{\infty}
\exp\left[-it\left(1+\frac{\chi^2 t^2}{3u^2} \right)\right]dt
\label{3.4a}
\end{equation}

In the weak external field ($\chi \ll 1, \varphi \simeq 1$) we
obtain the formula for the AMM of electron under influence of
multiple scattering
\begin{equation}
\frac{\mu'}{\mu_0}=-\frac{\alpha}{\pi} {\rm Im}~
\int_{0}^{\infty} \frac{du}{(1+u)^3}
\int_{0}^{\infty}\exp(-it)\frac{1}{\cosh^2 \frac{\nu
t}{2}}dt=\frac{\alpha}{2\pi}r,
\label{3.5a}
\end{equation}
where
\begin{eqnarray}
&& r={\rm Re}J,\quad J=2i\int_{0}^{\infty} \frac{du}{(1+u)^3}
\int_{0}^{\infty}\exp(-it)\frac{1}{\cosh^2 \frac{\nu t}{2}}dt
\nonumber \\
&& =4i\int_{0}^{\infty} \frac{du}{(1+u)^3}
\frac{1}{\nu}\left[\frac{2i}{\nu}\beta\left(\frac{i}{\nu} \right)-1 \right],
\nonumber \\
&& \beta=\frac{1}{2}\left[\psi\left(\frac{1+x}{2} \right)-
\psi\left(\frac{x}{2} \right) \right],
\label{3.6a}
\end{eqnarray}
where $\psi(x)$ is defined in Eq.(\ref{2.9b}).

The dependence of the AMM of electron on its energy $\varepsilon$ in gold
is shown in Fig. 6. It is seen that at energy $\varepsilon=500$~GeV
value of AMM is 0.85 part of standard quantity (SQ) of AMM ($r=1$) ,
at energy $\varepsilon=1$~TeV it is 0.77 part of SQ and at energy
$\varepsilon=5.5$~TeV it is 0.5 part of SQ.
Actually in all heavy elements the behavior of AMM of
electron will be quite similar, i.e. the scale of energy where the
AMM of electron deviates from SQ is of order of TeV.

In the case of weak effect of multiple scattering when
$\varepsilon \ll \varepsilon_e$, where $\varepsilon_e$ is defined
in Eq.(\ref{2.30b}) we obtain for AMM of
electron
\begin{equation}
\frac{\mu'}{\mu_0}=\frac{\alpha}{2\pi}\left(1-
\frac{\pi}{2}\frac{\varepsilon}{\varepsilon_e}\right).
\label{3.11a}
\end{equation}

At very high energy $\varepsilon \gg \varepsilon_e$
the effect of multiple scattering becomes strong. In this case the asymptotic
expression for AMM of electron is
\begin{equation}
r=\frac{2\pi}{\alpha} \frac{\mu'}{\mu_0}=\frac{\pi}{2\sqrt{2}}
\sqrt{\frac{\varepsilon_c}{\varepsilon}}\left(1-
\frac{\pi^2}{2}\frac{\varepsilon_c}{\varepsilon}\right), \quad
\varepsilon_c =\varepsilon_e\frac{L_1}{L_0},\quad
L_0=L_1+\frac{1}{2}\ln \frac{\varepsilon}{\varepsilon_e}.
\label{3.13a}
\end{equation}
Redefinition of the characteristic energy $\varepsilon_e \rightarrow
\varepsilon_c$ is connected with enlargement of the radiation cone
(in comparison with $1/\gamma$) or in another terms with increasing
of the characteristic transverse momentum transfers due to the multiple
scattering.

Let us discuss one of possibilities of experimental observation of
influence of medium on the AMM of electron. At a very high energy
where the observation becomes feasible the rotation angle
$\varphi$ of the spin vector $\mbox{\boldmath$\zeta$}$ in the
transverse to the particle velocity ${\bf v}$ magnetic field
${\bf H}$ depends only on the value of AMM and doesn't depend on
the particle energy:
\begin{equation}
\varphi =\left(\frac{m}{\varepsilon}+ \frac{\mu'}{\mu_0}
\right)\frac{e H}{m}l \simeq
r\frac{\alpha}{2\pi}\frac{H}{H_0}\frac{l}{\lambda_c}, \label{5.1a}
\end{equation}
where $l$ is the path of electron in the field, $H_0=m^2/e=4.41
\cdot 10^{13}$~Oe (see Eq.(\ref{15})). The dependence of $r$ on
energy $\varepsilon$ is found above and shown in Fig.6.

Since at radiation of hard photons in a medium the picture is
quite complicated:energy losses, cascade processes, spin flip and
depolarization, it is desirable to measure the particles which
don't radiate photons on the path $l$. The number of such
particles $N$ is determined by the total probability of radiation
in a medium found in \cite{L5}, Eqs.(3.12)-(3.14) (see
Eqs.(\ref{2.4i}) and (\ref{2.5i}) in Sec.2.2 above):
\begin{eqnarray}
&& N=N_0\exp(-\psi(\varepsilon)),\quad
\psi(\varepsilon)=W(\varepsilon)l=\frac{k(\varepsilon)\varphi_{SQ}}
{2\chi(\varepsilon_e)};\quad
k(\varepsilon)=W(\varepsilon)L_{rad}^0,
\nonumber \\
&& (L_{rad}^0)^{-1}
=\frac{\alpha}{4\pi}\frac{m^2}{\varepsilon_e},\quad
\chi(\varepsilon_e)=\frac{\varepsilon_e}{m}\frac{H}{H_0},
\label{5.2a}
\end{eqnarray}
where $\varphi_{SQ}$ is the rotation angle for standard value of
AMM in QED ($r=1$), $N_0$ is the number of initial electrons. With
energy increase the function $k(\varepsilon)$ decreases (see
Eq.(43) in \cite{L5})
\begin{equation}
k(\varepsilon \ll \varepsilon_e) \simeq \frac{4}{3}\left( \ln
\frac{\varepsilon_e}{\varepsilon}+1.96 \right),\quad
k(\varepsilon = \varepsilon_e) \simeq 3.56, \quad k(\varepsilon
\gg \varepsilon_e) \simeq
\frac{11\pi}{4\sqrt{2}}\sqrt{\frac{\varepsilon_e}{\varepsilon}}.
 \label{5.3a}
\end{equation}
The crucial part of the experiment is an accuracy of measurement
of electron polarization before target and after target. If one
supposes that spin rotation angle can be measured with accuracy
noticeably better than 1/10 then we can put that
$(1-r)\varphi_{SQ}=1/10$. In the gold we find for the energy
$\varepsilon=\varepsilon_e=2.61~$TeV and the magnetic field $H=4
\cdot 10^{5}$Oe that $1-r = 0.371$, the path of electron in the
target is $l \simeq 1~$cm and number of electron traversing the
target without energy loss is $N \simeq 3.2 \cdot 10^{-5} N_0$.
This estimates show that the measurement of the effect found in
this paper will be feasible in the not very distant future.

\subsection{Propagation of high-energy photon in a medium in presence of
an external field}

As known, the propagation of electromagnetic wave in a medium is
defined by its dielectric tensor ${\cal E}_{ik}(\omega)$. For
relatively low frequency $\omega$ (e.g. visible light) the
dielectric tensor is defined by atomic phenomena. When the
frequency of wave is much higher the atomic frequencies, the
dielectric tensor has a form
\begin{equation}
{\cal E}_{ik}(\omega)=\delta_{ik}{\cal E}(\omega),\quad {\cal
E}(\omega)= 1-\frac{\omega_0^2}{\omega^2},\quad
\omega_0^2=\frac{4\pi n_e e^2}{m},
\label{1pr}
\end{equation}
where $\omega_0$ is the plasma frequency, in any medium $\omega_0
< 100$~eV. So that for $\omega \gg \omega_0$ an influence of
atomic phenomena on propagation of electromagnetic wave in a
medium becomes small.

At very high energy the nonlinear effects of QED enter into game.
One of them is the polarization of the vacuum by a photon. In the
presence of an external electromagnetic field the polarization of
vacuum was considered first in the pioneer papers \cite{S1}. In
the strong field this effect can be essential for propagation of
high-energy photons \cite{Na}, \cite{7}.

To evaluate the polarization tensor one has to consider the
amplitude of photon scattering which included the polarization
operator \cite{BK2}. As above we use the quasiclassical operator
method \cite{7a}, \cite{7}, \cite{9}. In this method the mentioned
amplitude is described by diagram where the virtual
electron-positron pair is first created by the initial photon
with 4-momentum $k (\omega,{\bf k})$ and polarization ${\bf e}_1$
and then annihilate into final photon with 4-momentum $k$ and
polarization ${\bf e}_2$. This corresponds to use of the
non-covariant perturbation theory where at high energies ($\omega
\gg m$) the contribution of this diagram survives only. For this
energy of photon this process occurs in a rather long time (or at
a rather long distance) known as the lifetime of the virtual state
\begin{equation}
l_f=\frac{\omega}{2q_c^2}, \label{1pr.1}
\end{equation}
where $q_c \geq m$ is the characteristic transverse momentum of
the process, the system $\hbar=c=1$ is used. When the virtual
electron (or positron) is moving in a medium it scatters by atoms
and changes the velocity under influence of external
electromagnetic field. The mean square of momentum transfer to the
electron from a medium and an external field on the distance $l_f$
is
\begin{eqnarray}
&& q_f^2=q_s^2+{\bf q}_F^2, \quad q_s^2=4\pi Z^2\alpha^2n_a L
l_f,\quad L \equiv L(q_c^2)=\ln
(q_c^2 a^2), \nonumber \\
&& {\bf q}_F=e{\bf F} l_f,\quad  {\bf F}={\bf E}_{\perp}+{\bf
v}\times{\bf H} \label{1pr.2}
\end{eqnarray}
where $\alpha=e^2=1/137$, $Z$ is the charge of nucleus, $n_a$ is
the number density of atoms in the medium,  $a$ is the screening
radius of atom, ${\bf E}_{\perp}$ is the electric field strength
transverse to the velocity of particle ${\bf v} \simeq {\bf
n}={\bf k}/\omega$, ${\bf H}$ is the magnetic field strength (see
Eq.(\ref{14})).

In the case of small momentum transfer $q_f \equiv \sqrt{q_f^2}
\ll m$ the influence of a medium and an external field is weak, in
this case $q_c=m$. At high energy it is possible that $q_c \geq
m$. In this case the characteristic value of the momentum transfer
(giving the main contribution into the spectral probability) is
defined by the value of $q_f$. The self-consistency condition is
\begin{eqnarray}
&& q_c^2=q_f^2=\frac{2\pi \omega Z^2\alpha^2n_a
L(q_f)}{q_f^2}+\frac{m^6 \mbox{\boldmath$\kappa$}^2}{4q_f^4} \geq
m^2,~\mbox{\boldmath$\kappa$}=\frac{\omega}{m^3}e{\bf F}_{\gamma},
\nonumber \\
&&{\bf F}_{\gamma}={\bf E}-{\bf n}({\bf n}{\bf E})+{\bf n}\times
{\bf H}, \label{1pr.3}
\end{eqnarray}
here $\kappa=|\mbox{\boldmath$\kappa$}|$ is known parameter
characterizing the pair production process in a homogeneous
external field ${\bf F}_{\gamma}$. With $q_c$ increase the
lifetime of the virtual state Eq.(\ref{1pr.1}) decreases.

We will use the following normalization condition for the
amplitude under consideration
\begin{equation}
M=2 \omega \Delta\omega.
\label{1pr.4}
\end{equation}

The amplitude $M$ is the contraction of the tensor
$e_j^{(i)\ast}e_k^{(f)}$ (${\bf e}^{(i)}$ and ${\bf e}^{(f)}$ are
the polarization vectors of the initial and the final photons)
and the polarization tensor $M_{jk}$. We select the basic vectors
as
\begin{equation}
{\bf e}_1=\frac{{\bf F}_{\gamma}}{|{\bf F}_{\gamma}|},\quad {\bf
e}_2={\bf n}\times{\bf e}_1. \label{1pr.7}
\end{equation}
Since the tensor $M_{jk}$ is invariant under the space inversion
then in the selected basic vectors it has the diagonal form
\begin{equation}
M_{jk}=\frac{1}{2}\left[\delta_{jk}(M_{11}+M_{22})+
(e_{1j}e_{1k}-e_{2j}e_{2k})(M_{11}-M_{22}) \right] \label{1pr.8}
\end{equation}
In absence of external field it is convenient to describe the
process of photon scattering using the helicity polarization
vector ${\bf e}_{\lambda}~(\lambda = \pm 1)$ connected with
momentum transfer $\mbox{\boldmath$\Delta$}$ (see
Eq.(\ref{3.32sc}) in the next subsection). In presence of external
field and for $\mbox{\boldmath$\Delta$}=0$ we choose the
polarization vectors in the following way:
\begin{equation}
{\bf e}_{\lambda}=\frac{1}{\sqrt{2}}\left({\bf e}_1 +i\lambda{\bf
e}_2 \right),\quad ({\bf e}_{\lambda}{\bf
e}_{\lambda}^{\ast})=1,\quad ({\bf e}_{\lambda} {\bf
e}_{-\lambda}^{\ast})=0,\quad {\bf e}_{\lambda}\times {\bf
n}=i\lambda{\bf e}_{\lambda}. \label{2pr.5}
\end{equation}

In terms of helicity amplitudes $M_{++}$ and $M_{+-}$ the tensor
$M_{jk}$ and the corresponding dielectric tensor ${\cal E}_{jk}$
has a form
\begin{eqnarray}
&& M_{jk}=\delta_{jk}k_{++}^2+(e_{1j}e_{1k}-e_{2j}e_{2k})k_{+-}^2,
\nonumber \\
&& {\cal E}_{jk}=\delta_{jk}-\frac{1}{\omega^2}M_{jk},\quad
k_{++}^2 \equiv M_{++},\quad k_{+-}^2 \equiv M_{+-} \label{1pr.9}
\end{eqnarray}
The polarization tensor $k_{jk}^2$  is diagonal in the basic
vectors ${\bf e}_1$ and ${\bf e}_2$ Eq.(\ref{1pr.7}). The
corresponding mass squared are
\begin{equation}
k_1^2 \equiv k_{11}^2=k_{++}^2+k_{+-}^2, \quad k_2^2 \equiv
k_{22}^2=k_{++}^2-k_{+-}^2 \label{2pr.14}
\end{equation}

The probability of pair creation by a photon with polarization
${\bf e}$ is
\begin{equation}
W_p^F({\bf e})=-\frac{1}{\omega}{\rm Im}\left[({\bf e}{\bf
e}_1)^2k_1^2+ ({\bf e}{\bf e}_2)^2k_2^2 \right]=
-\frac{1}{\omega}{\rm Im}\left[k_{++}^2+\xi_3k_{+-}^2\right],
\label{2pr.15}
\end{equation}
where $\xi_3$ is the Stokes' parameter. For unpolarized photon
one has
\begin{equation}
W_p^F=-\frac{1}{\omega}{\rm Im}~k_{++}^2. \label{2pr.16}
\end{equation}

The method of calculation of the functions $k_{++}^2$ and
$k_{+-}^2$ which define the tensor $M_{jk}$ is similar to used in
previous subsection (see details in \cite{BK2}). The general
expressions for photon masses squared under simultaneous
influence of multiple scattering in a medium and an external
electromagnetic field has a form
\begin{eqnarray}
&& k_{++}^2=\frac{\alpha m^2}{2\pi}\int_{0}^{\omega} \frac{\omega
d\varepsilon}{\varepsilon\varepsilon'}\int_{0}^{\infty}e^{-it}
\nonumber \\
&& \times \left[s_2 \nu \varphi_p \tanh\frac{\nu
t}{2}\left(\frac{4\kappa'^2}{\nu^2}+1\right)-s_3\left(\frac{\nu}{\sinh
\nu t}\varphi_p -\frac{1}{t} \right)\right]dt.
\label{2pr.18}
\end{eqnarray}
For $k_{+-}^2$ we found respectively
\begin{equation}
k_{+-}^2= -\frac{2\alpha m^2}{\pi}\int_{0}^{\omega} \frac{\omega
d\varepsilon}{\varepsilon\varepsilon'}s_3
\kappa'^2\int_{0}^{\infty}e^{-it}\frac{\varphi_p}{\nu \sinh \nu
t}\tanh^2\frac{\nu t}{2}dt.
\label{2pr.19}
\end{equation}
Here
\begin{eqnarray}
&&
\mbox{\boldmath$\kappa'$}=\frac{\varepsilon\varepsilon'}{\omega^2}
\mbox{\boldmath$\kappa$},\quad \nu=2\sqrt{iq}, \quad q=QL_c,
\quad L_c \equiv L(\varrho_c) =\ln \frac{a_{s2}^2}{\lambda_c^2
\varrho_c^2},
\nonumber \\
&& \varphi_p \equiv \varphi(\mbox{\boldmath$\kappa'$}, \nu, t)
=\exp\left[-\frac{4i\kappa'^2t}{\nu^2}\left( 1-\frac{2}{\nu
t}\tanh\frac{\nu t}{2} \right) \right], \label{2pr.18a}
\end{eqnarray}
where $\mbox{\boldmath$\kappa$}$ is defined in Eq.(\ref{1pr.3}),
$Q$ is defined in Eq.(\ref{2.9}), the parameter $\varrho_c$ is
defined by the set of equations (compare with Eq.(\ref{2.18a})):
\begin{eqnarray}
&& \varrho_c =1\quad {\rm for}\quad
4(\mbox{\boldmath$\kappa'$}^2+QL_1) \leq 1;
\nonumber \\
&& 4\varrho_c^4 \left(\mbox{\boldmath$\kappa'$}^2\varrho_c^2+QL_c
\right)=1 \quad {\rm for} \quad 4(\mbox{\boldmath$\kappa'$}^2+
QL_1) \geq 1, \label{2pr.7}
\end{eqnarray}

In the absence of external field
($\mbox{\boldmath$\kappa'$}=0,~\varphi_p=1$) we have
\begin{eqnarray}
&& k_{++}^2=\frac{\alpha m^2}{2\pi}\int_{0}^{\omega} \frac{\omega
d\varepsilon}{\varepsilon\varepsilon'}\int_{0}^{\infty}e^{-it}
\left[s_2 \nu \tanh\frac{\nu t}{2} -\frac{s_3}{
2t}\left(\frac{\nu}{\sinh \nu t} -\frac{1}{t} \right)\right]dt
\nonumber \\
&& = -\frac{\alpha m^2}{2\pi}\int_{0}^{\omega} \frac{\omega
d\varepsilon}{\varepsilon\varepsilon'}\left[s_1\left( \ln
p-\psi\left(p+\frac{1}{2} \right) \right) +s_2 \left(\psi(p)-\ln
p +\frac{1}{2p} \right)\right],
\nonumber \\
&& k_{+-}^2=0, \label{2pr.20a}
\end{eqnarray}
where $p=i/2\nu,~ \psi(p)$ as above is the logarithmic derivative
of the gamma function. Subsisting the result obtained into formula
(\ref{2.16}) we have the probability of pair creation which
agrees with Eq.(\ref{2.10p}).

In the absence of multiple scattering $(\nu \rightarrow 0)$ we get
\begin{eqnarray}
&& k_{++}^2=\frac{\alpha m^2}{\pi}\int_{0}^{\omega} \frac{\omega
d\varepsilon}{\varepsilon\varepsilon'}\int_{0}^{\infty}e^{-it}
\left[s_2 \kappa'^2 t \varphi_F -\frac{s_3}{ 2t}\left(\varphi_F
-1 \right)\right]dt,
\nonumber \\
&& k_{+-}^2= -\frac{\alpha m^2}{2\pi}\int_{0}^{\omega}
\frac{\omega d\varepsilon}{\varepsilon\varepsilon'}s_3
\kappa'^2\int_{0}^{\infty}e^{-it}\varphi_F tdt, \label{2pr.20}
\end{eqnarray}
where
\begin{equation}
\varphi_F=\exp\left(-i\frac{\kappa'^2 t^3}{3}\right)
\label{2pr.21}
\end{equation}
For this case the expressions for $k_{1,2}^2$ (\ref{2.14}) after
substitution results of Eq.(\ref{2pr.20}) agree with masses
squared of photon calculated in an external electromagnetic field
(see \cite{7} and references therein).

In the case when the both characteristic parameters are small
($\nu_1^2 =4QL_1 \ll 1,~\kappa \ll 1$), the main terms of
decomposition of the functions $k^2$ are the sum of independent
corrections to the photon mass squared both on account of the
multiple scattering and an external field
\begin{eqnarray}
\hspace{-16mm}&& k_{++}^2=\frac{\alpha m^2}{\pi}
\Bigg[-i\frac{7}{9}\frac{\omega}{\omega_e}
\left(1-\frac{1}{21L_1} \right)+\frac{59}{225}
\left(\frac{\omega}{\omega_e}\right)^2
\nonumber \\
\hspace{-16mm}&&
-i\frac{3\sqrt{3}\pi}{16\sqrt{2}}\frac{\omega}{\omega_F}
\exp\left(-\frac{8\omega_F}{3\omega}\right)-\frac{11}{90}
\left(\frac{\omega}{\omega_F}\right)^2 \Bigg],
\nonumber \\
\hspace{-16mm}&& k_{+-}^2= \frac{\alpha m^2}{\pi} \Bigg[
i\frac{\sqrt{3}\pi}{16\sqrt{2}} \frac{\omega}{\omega_F}
\exp\left(-\frac{8\omega_F}{3\omega}\right)+\frac{1}{30}
\left(\frac{\omega}{\omega_F}\right)^2 \Bigg];
\nonumber \\
\hspace{-16mm}&&
 \omega \ll
\omega_e,\quad \omega_F=m \frac{H_0}{|{\bf F}|}, \quad \omega \ll
\omega_F, \label{2pr.22}
\end{eqnarray}
where $\omega_e$ is defined in Eq.(\ref{2.23ap}) and $H_0$ in
Eq.(\ref{15}).  The correction $\propto 1/L_1$ follows from the
first term of decomposition of the type Eq.(\ref{2.26}). Let us
remind that in gold the value $\omega_e$ is $\omega_e=10.5$~TeV,
this is the typical value for the heavy elements.

In Fig.7 the functions Re~$k_{++}^2$(curve 2) and Im~$k_{++}^2$
(curve 1) are given for the case when the influence of a medium
is taken into account only (Eq. (\ref{2pr.20a})). The both curves
are normalized to the asymptotics given by Eq. (\ref{2pr.22}) in
the limit $\omega_F \rightarrow \infty$.

It should be noted that beginning with some photon energy
$\omega=\omega_b$ the radiative correction to the value Re~${\cal
E}_{jk}$  in the absence of a field
($\mbox{\boldmath$\kappa$}=0$) becomes larger than
$\omega_0^2/\omega^2$ (see Eq.(\ref{1pr})). The estimate of
$\omega_b$ is following
\begin{equation}
\omega_b \sim \sqrt{\frac{\alpha}{\pi}}\frac{L_{rad}}{\lambda_c}
\omega_0.
\label{1pr.10}
\end{equation}
For gold one obtains $\omega_b \sim 40$~GeV.

A propagation of high-energy photons in oriented single crystals
is one of interesting applications of the result obtained in this
subsection. In this case we have both the dense matter with strong
effect of multiple scattering and high fields of crystal axes or
planes. As we saw, the LPM effect is most pronounced in the heavy
elements. The same is valid for the process under consideration.
Let the high-energy photon incident on crystal. The angle of
incidence is small and such that the distance from axis
$\mbox{\boldmath$\varrho$}$ (or the distance from plane $x$) can
be considered as a constant on the formation length of process
(the constant field approximation is applicable, see Sections
12,15 in \cite{9}).

For orientation of a crystal along an axis the ratio of density of
atoms in the vicinity of axis $n(\mbox{\boldmath$\varrho$})$ to
the mean density $n_a$ is
\begin{equation}
\xi_{ax}(\mbox{\boldmath$\varrho$})=\frac{n(\mbox{\boldmath$\varrho$})}{n_a}
=\frac{\exp(-\varrho^2/2u_1^2)}{2\pi u_1^2d n_a},
\label{3pr.1}
\end{equation}
where $u_1$ is the amplitude of thermal vibrations of atoms, $d$
is the mean distance between atoms which form the axis. This
ratio is maximal at $\mbox{\boldmath$\varrho$}=0$.  For numerical
estimates we use for definiteness the tungsten single crystal. For
the axis $<111>$ in W the ratio $\xi_{ax}(0)$=370 at the room
temperature ($T=293~K$) and $\xi_{ax}(0)$=1020 at $T=77~K$. The
effect of multiple scattering becomes strong at characteristic
photon energy $\omega_e(n_a) \simeq 11~$TeV and this value is
inversely proportional to the density. So we have that
$\omega_e(\varrho=0) \simeq 30~$GeV at $T=293~K$ and
$\omega_e(\varrho=0) \simeq 11~$GeV at $T=77~K$. It should be
noted that within logarithmic accuracy we neglect by relatively
small variation of $L_1$ due to substitution the screening radius
$a_s^2$ by the value $2u_1^2$.

It is useful to compare these estimates with known threshold
energies $\omega_t$ at which the probability of pair creation in
the field of axis is equal to the probability of the
Bethe-Heitler mechanism, see Table 12.1 in \cite{9}. For photon
energy  $\omega \geq \omega_t$ the process of pair creation in
the field of axis dominates. In W crystal for $<111>$ axis
$\omega_t=22~$GeV at T=293~K and $\omega_t=13~$GeV at T=77~K. It
is seen that for these energies the ratio $\omega/\omega_e$ which
characterize the strength the LPM effect is of the order of unity.
At $\omega \sim \omega_t$ the maximal value of the parameter
$\kappa(\mbox{\boldmath$\varrho$})$ which determines the
probability of pair creation by a photon in a field is also of
the order of unity (at $\varrho \simeq u_1$, see the mentioned
Table). So we reach the conclusion that at some energy (for axial
orientation of crystal) all the discussed effects are essential
simultaneously. The analysis in this situation will be published
elsewhere. For example, to calculate the influence of the field
of axis on the polarization tensor one have to average the
general formula (\ref{2pr.18}) over all values of
$\mbox{\boldmath$\varrho$}$ (this is integration over
$d^2\varrho$ with the weight $n_{\perp}$, where $n_{\perp}=n_a d$
is the density of axis in the perpendicular to them plane).

\subsection{An influence of multiple scattering on coherent scattering of
photon}

Investigation of the coherent photon scattering (the Delbr\"uck scattering)
has rather long history which can be found in
review (see e.g.\cite{d3}). There is a special interest to study
the process for heavy elements (the cross section $\propto Z^4$).
However, in this case the contributions of higher orders of $Z\alpha$
into the amplitude of photon scattering are very important.
This means that one needs the theory
which is exact with respect to the parameter $Z\alpha$.
The amplitudes of the coherent photon scattering valid for any $Z\alpha$
for high photon energy $\omega \gg m$ and small scattering angle
(or small momentum transfer $\Delta$) were calculated by several group of authors
(see \cite{CW},  \cite{MS1}, \cite{LM1}, \cite{LM2}, \cite{KS}).
In the last paper the process of the coherent photon scattering
was considered in frame
of the quasiclassical operator method
which appears to be very adequate for consideration of this
problem.

In the quasiclassical approximation the amplitude $M$ of the
coherent photon scattering is described by diagram where the
electron-positron pair is created by the initial photon with
4-momentum $k_1~(\omega,{\bf k}_1)$ and then annihilate into the
final photon with 4-momentum $k_2~(\omega,{\bf k}_2)$,~ so that
the photon momentum transfer is $\Delta=|{\bf k}_2-{\bf k}_1|$.
For high energy photon $\omega \gg m$ this process occurs over a
rather long distance (see Eq.(\ref{1.1})).

It is convenient to describe the process of photon scattering in
terms of helicity amplitudes. We choose the polarization vectors
with helicity $\lambda$
\begin{eqnarray}
&&{\bf e}_{\lambda}=\frac{1}{\sqrt{2}}\left({\bf e}_1
+i\lambda{\bf e}_2 \right),\quad {\bf e}_1=
\mbox{\boldmath$\nu$}=\frac{\mbox{\boldmath$\Delta$}}
{|\mbox{\boldmath$\Delta$}|},\quad {\bf e}_2={\bf n}\times
\mbox{\boldmath$\nu$},\quad {\bf n}=\frac{{\bf k}_1}{\omega},
\nonumber \\
&& \lambda=\pm 1,\quad {\bf e}_{\lambda}{\bf
e}_{\lambda}^{\ast}=1,\quad {\bf e}_{\lambda} {\bf
e}_{-\lambda}^{\ast}=0,\quad {\bf e}_{\lambda}\times {\bf
n}=i\lambda{\bf e}_{\lambda}, \label{3.32sc}
\end{eqnarray}
There are two independent helicity amplitudes:
\begin{equation}
M_{++}=M_{--},\quad M_{+-}=M_{-+},
\label{3.33sc}
\end{equation}
where the first subscript is the helicity of the initial photon
and the second is the helicity of the final photon. When the
initial photons are unpolarized the differential cross section of
scattering summed over final photons polarization contains the
combination
\begin{equation}
2[|M_{++}|^2+|M_{+-}|^2].
\label{3.34sc}
\end{equation}
The calculation of the amplitudes of photon scattering on the
separate atom appears to be most simple in the case when the
screening radius of atom $a_s$ is much smaller than the formation
length of process ($a_s \ll \lambda_c \omega/2m=l_{p,max}$
Eq.(\ref{10a})). The imaginary part of helicity amplitudes in
this case can be written in the form \cite{BK}
\begin{equation}
{\rm Im}~M_{\lambda \lambda'}=\frac{4Z^2\alpha^3 \omega}{m^2}
\int_{0}^{1}dx\int_{0}^{1}dy~ \mu_{\lambda \lambda'}f_{\lambda \lambda'},
\label{3.46sc}
\end{equation}
where
\begin{eqnarray}
&& \mu_{++}=1-2x(1-x)+4x(1-x)y(1-y),\quad \mu_{+-}=x(1-x)y^2;
\nonumber \\
&& f_{++}=\ln (ma_s)-\frac{2s^2+1}{2s\sqrt{1+s^2}} \ln
\left(s+\sqrt{1+s^2} \right) -f(Z\alpha)+\frac{41}{42},
\nonumber \\
&&f_{+-}=1-\frac{1}{s\sqrt{1+s^2}} \ln \left
(s+\sqrt{1+s^2}\right),\quad s=\frac{\Delta a_s}{2},
\label{3.47sc}
\end{eqnarray}
where the function $f(\xi)$ is defined in Eq.(\ref{2.9b}). The
important property of Eq.(\ref{3.46sc}) is that the dependence on
the screening radius $a$ originates in it from the Born
approximation. In this approximation in the case of arbitrary
screening the radius $a_s$ enters only in the combination
\begin{equation}
\frac{1}{a_s^2}+q_{\parallel}^2,\quad q_{\parallel}
=\frac{q_{m}}{x(1-x)y}, \quad q_m=\frac{m^2}{2\omega}.
\label{3.48sc}
\end{equation}
Because of this we can extend Eq.(\ref{3.47sc}) on the
case of arbitrary screening making the substitution
\begin{equation}
\frac{1}{a_s} \rightarrow \sqrt{q_{\parallel}^2+a_s^{-2}} \equiv
q_{ef},\quad s=\frac{\Delta}{2q_{ef}} \label{3.49sc}
\end{equation}

In the case $a_s \gg \omega/m^2$ (the screening radius is very
large, or in other words we consider the photon scattering in the
Coulomb field) we have to substitute in Eq.(\ref{3.47sc})
\begin{equation}
a_s \rightarrow \frac{1}{q_{\parallel}},\quad s \rightarrow
 s_c=\frac{\Delta}{2q_{\parallel}}
=\frac{\Delta \omega}{m^2}x(1-x)y.
\label{3.56sc}
\end{equation}

In the case of the complete screening ($a_s \ll \omega/m^2$) the
functions $f_{\lambda \lambda'}$ are independent of $x$ and $y$
and the corresponding integrals are
\begin{equation}
\int_{0}^{1}dx\int_{0}^{1}dy~\mu_{++}=\frac{7}{9},\quad
\int_{0}^{1}dx\int_{0}^{1}dy~\mu_{+-}=\frac{1}{18}.
\label{3.54sc}
\end{equation}
For the scattering amplitudes we have
\begin{eqnarray}
&& {\rm Im}~M_{++}=\frac{28Z^2\alpha^3\omega}{9m^2}f_{++},\quad
{\rm Im}~M_{+-}=\frac{2Z^2\alpha^3\omega}{9m^2}f_{+-},
\nonumber \\
&&{\rm Re}~M_{++}=0,\quad {\rm Re}~M_{+-}=0.
\label{3.55sc}
\end{eqnarray}
The real part of amplitudes are calculated using dispersion
relations. The photon scattering amplitude in this case for
arbitrary value of parameter $a_sm^2/2\omega$ for $\Delta=0$ was
found recently in \cite{LM2} and for arbitrary $\Delta$ in
\cite{KS}.

 When a photon is propagating in a medium it also
dissociates with probability $\propto \alpha$ into an electron-positron
pair. The virtual electron and positron interact with a medium
and can scatter on atoms. In this scattering
the electron and positron interaction with the Coulomb field
in the course of the coherent scattering of photon is involved also.
There is a direct analogue with the LPM
effect: the influence of the multiple scattering on process of the
bremsstrahlung and pair creation by a photon in a medium at high energy.
However there is the difference: in the LPM effect
the particles of electron-positron pair created by a
photon are on the mass shell while in the process
of the coherent scattering of photon this particles are off the mass shell,
but in the high energy region (this is the only region where the influence
of the multiple scattering is pronounced) the shift from the mass shell
is relatively small.
To include this scattering into consideration the amplitude
of the coherent scattering of photon
should be averaged over all possible trajectories of electron and positron.
This operation can be performed with the aid of the distribution function
averaged over the atomic positions of scatterer in the medium
in the same manner as it is done in Appendix A \cite{BK}.

The photon scattering amplitudes can be written in the form
\begin{equation}
M_{++}=M_{++}^c+M_{++}^{(1)},\quad M_{+-}=M_{+-}^c+M_{+-}^{(1)},
\label{4.16sc}
\end{equation}
where $M_{++}^c$ is the main (logarithmic) term
and $M_{++}^{(1)}$ is the first correction to the scattering amplitude.
The main terms are \cite{BK}
\begin{eqnarray}
&& M_{++}^c=\frac{\alpha m^2\omega}{2\pi n_a} \int_{}^{}
\frac{d\varepsilon}{\varepsilon \varepsilon'}
\Phi_s(\nu_s),~ M_{+-}^c=0,~
\nonumber \\
&&\displaystyle{\Phi_s(\nu_s)
=s_1\left(\ln p-\psi\left(p+\frac{1}{2}\right) \right)
+s_2\left(\psi (p) -\ln p+\frac{1}{2p}\right)},
\nonumber \\
&&\nu_s=2\sqrt{iq_s},~q_s=QL_s,~p=i/(2\nu_s),
~s_1=1,~s_2=\frac{\varepsilon^2+\varepsilon'^2}{\omega^2},
\nonumber \\
&&~Q=\frac{2\pi Z^2\alpha^2\varepsilon \varepsilon'n_a}
{m^4\omega}, L_s \equiv
L_s(\varrho_c)=\ln \frac{a_{s2}^2}{\lambda_c^2 \varrho_c^2}-
F_2\left(\frac{\beta}{2} \right),~
\nonumber \\
&&F_2(v)=\frac{2v^2+1}{v\sqrt{1+v^2}}
\ln \left(v+\sqrt{1+v^2} \right)-1,~
~\beta=a \Delta,
\label{4.17sc}
\end{eqnarray}
where $a_{s2}$ is defined in Eq.(\ref{2.16}). The structure of the
amplitude $M_{++}^c$ is quite similar to the structure of the
pair creation probability Eq.(\ref{2.10p}) under influence of the
multiple scattering (the fact that the amplitude is expressed in
terms of the probability is the common feature of the coherent
photon scattering description). This result is not occasional and
connected with origin of coherent scattering of photon as a
process which is going through dissociation of a photon into
electron-positron pair. However the expression for the basic
parameter $\nu_s$ differs from corresponding expression in
Eq.(\ref{2.10p}).

The first corrections to the amplitudes are defined by
\begin{eqnarray}
&& M_{++}^{(1)}=-\frac{\alpha m^2\omega}{4\pi n_a L_s} \int_{}^{}
\frac{d\varepsilon}{\varepsilon \varepsilon'}
F_s(\nu_s),\quad
M_{+-}^{(1)}=\frac{\alpha m^2\omega}{4\pi n_a L_s}
F_1\left(\frac{\beta}{2} \right) \int_{}^{}
\frac{d\varepsilon}{\varepsilon \varepsilon'}
F_3(\nu_s),
\nonumber \\
&& F_s(\nu_s)= \int_{0}^{\infty}\frac{dz e^{-it}}{\sinh^2z}
\left[s_1f_1(z)-2is_2f_2(z) \right],\quad
F_1(v)=1-\frac{\ln (v+\sqrt{1+v^2})}{v\sqrt{1+v^2}},
\nonumber \\
&&F_3(\nu_s)=i \nu_s s_3\int_{0}^{\infty}\frac{dz e^{-it}}{\sinh^3z} g(z)
=\frac{s_3}{2}\left[1+\frac{1}{2p}-p~\zeta(2, p) \right]
\label{4.18sc}
\end{eqnarray}
here the functions $f_1(z)$, $f_2(z)$ and $g(z)$ are defined in
Eq.(\ref{2.15p}), $\zeta(s, a)$ is the generalized Riemann zeta
function. Use of the given representations of functions
$F_3(\nu_s)$ and $G(z)$ in $f_1(z)$ simplifies the numerical
calculation. The functions $F_1(z)$ and $F_2(z)$ in
Eq.(\ref{4.17sc}) encountered in the radiation theory.

In the region of the weak effect of scattering ($|\nu_s| \ll 1,~
\varrho_c=1$) we obtain
\begin{eqnarray}
\hspace{-10mm}&& M_{++}=M_{++}^c+M_{++}^{(1)}
=i\frac{14Z^2 \alpha^3 \omega}{9m^2}
\Bigg[L_{s1}\Bigg(1+i\frac{59\omega}{175 \omega_e}\frac{L_{s1}}{L_1}
\nonumber \\
\hspace{-10mm}&&-\frac{3312}{2401}\left(\frac{\omega}{\omega_e}
\frac{L_{s1}}{L_1}\right)^2
\Bigg)-\frac{1}{21} \Bigg],
\nonumber \\
\hspace{-10mm}&& M_{+-}=M_{+-}^{(1)}=i\frac{2Z^2 \alpha^3 \omega}{9m^2}
F_1\left(\frac{\beta}{2}\right)
\left(1+i\frac{16\omega}{25 \omega_e}\frac{L_{s1}}{L_1} -
\frac{384}{245}\left(\frac{\omega}{\omega_e}\frac{L_{s1}}{L_1} \right)^2
\right),
\label{4.20sc}
\end{eqnarray}
here
\begin{equation}
 L_{s1}-\frac{1}{21}= 2\left[\ln \frac{a_{s}}{\lambda_c}
-\frac{1}{2} \left(F_2\left(\frac{\Delta a_{s}}{2}\right)+1\right)
-f(Z \alpha)+\frac{41}{42} \right],
\label{4.21sc}
\end{equation}
$\omega_e$ is defined in Eq.(\ref{2.23ap}) and $L_1$ is defined in
Eq.(\ref{2.16}). The characteristic energy $\omega_e$ encountered
in analysis of influence of the multiple scattering on the
probability of pair photoproduction \cite{L5}, in gold
$\omega_e$=10.5~TeV. The amplitudes Eq.(\ref{4.20sc}) coincide
with the formulas in absence of multiple scattering if we neglect
the terms $\propto \omega/\omega_e$ and $(\omega/\omega_e)^2$. The
terms $\propto \omega/\omega_e$ define the appearing real part of
the scattering amplitudes while the terms $\propto
(\omega/\omega_e)^2$ are the corrections to the imaginary part.

In the region where the effect of scattering is strong
($|\nu_s| \gg 1$) we obtain
\begin{eqnarray}
&& M_{++} =(i-1)\frac{3\pi Z^2 \alpha^3 \omega}{2\sqrt{2}\Delta_s^2}
\sqrt{L_{s3}}\left[
1-\frac{1}{4L_{s3}}\left(2C+\frac{1}{3}+i\frac{\pi}{2} \right) \right],
\nonumber \\
&& M_{+-} \simeq (i-1)\frac{\pi Z^2 \alpha^3 \omega}{8\sqrt{2}\Delta_s^2}
\frac{1}{\sqrt{L_{s3}}}F_1\left(\frac{\beta}{2}\right),\quad
L_{s3}=L_{s2}+\frac{1}{2} \ln L_{s2},
\nonumber \\
&& L_{s2}=L_{s1}+\frac{1}{2}\ln \frac{\omega}{\omega_e}=
2\ln(a_{s}\Delta_s)+1-F_2\left(\frac{\Delta a_{s}}{2}\right)-2 f(Z\alpha),
\nonumber \\
&& \Delta_s^4=2\pi Z^2 \alpha^2 \omega n_a,\quad
\nu_s^2=i\frac{\omega}{\omega_e}\frac{4\varepsilon \varepsilon'}{\omega^2}
\left(L_{s3}+ \frac{1}{2} \ln \frac{4\varepsilon \varepsilon'}
{\omega^2}\right).
\label{4.24sc}
\end{eqnarray}
the real and imaginary parts of the amplitudes are equal (if we
neglect the term $\propto 1/L_{s3}$ in $M_{++}$). Moreover the
amplitudes Eq.(\ref{4.24sc}) don't depend on the electron mass
$m$. In place of it we have the value $\Delta_s$.

The influence of a medium on the process of the coherent photon
scattering illustrated in Fig. 8 a and b, where Im~$M_{++}$ and
Re~$M_{++}$ as well as Im~$M_{+-}$ and Re~$M_{+-}$ are given as a
function of photon energy $\omega$ in gold. This influence is due
to the multiple scattering of electron and positron of the
virtual pair on the formation length of the process (see
Eqs.(\ref{1.1}) and (\ref{1.2}))
\begin{equation}
l_f=\frac{\omega}{2(q_s^2+m^2+\Delta^2)}.
\label{5.1sc}
\end{equation}
In the region $\Delta^2 \ll q_s^2+m^2$ this formation length
is independent of $\Delta$ and it value is coincide practically with
the formation length of pair creation by a photon $l_c$ considered in previous
subsection.
There is some difference connected with the logarithmic dependence of
$\nu_s^2$ value on $\Delta^2$:
\begin{equation}
|\nu_s^2|=\frac{q_s^2}{m^2}=\frac{4\pi Z^2 \alpha^2}{m^2}n_a l_f
\int_{q_{min}^2}^{q_{max}^2}\frac{dq^2}{q^2}
\label{5.1asc}
\end{equation}
where $q_{max}^2=m^2+q_s^2$ and $q_{min}^2=\Delta^2+a^{-2}$, $a$
is the screening radius of atom (\ref{1.2}). This defines the weak
(logarithmic) dependence of Im$M_{++}$ on $\Delta$ in the region
$\Delta < \sqrt{q_s^2+m^2}$. This can be seen in Fig. 8a where the
curves 1 and 3 represent behavior of Im~$M_{++}$ for
$\Delta$=0.4435 $m$ and $\Delta$=0.0387 $m$ respectively. For
lower value of $\Delta$ the minimal momentum transfer $q_{min}$
(\ref{1.2}) diminishes thereby the interval of contributing the
multiple scattering angles increases, so the multiple scattering
affects the photon scattering amplitude at a lower energy (and
smaller formation length). Because of this the curve 3 is shifted
to the left respect the curve 1. Note that the curve 3
($\Delta^{-1} \sim a_{s2}$) is very similar to the curve 2 in
Fig. 4  which represents the behavior of the probability of pair
photoproduction in gold vs photon energy.

The new property of influence of a medium is the appearance of
the real part of the coherent photon scattering amplitudes at
high energy $\omega$. In the region $\omega \ll \omega_e$  the
value of Re$M$ is small accordingly Eq.(\ref{4.20sc}). In the
asymptotic region $\omega \gg \omega_e$ we have -Re~$M =$ Im~$M$
according to Eq.(\ref{4.24sc}). This property is seen clearly in
Figs.8 a and b. So the value of -Re~$M$ is small at low and very
high energies of photon. At intermediate energies the value of
-Re~$M$ have the maximum at $\omega \simeq$220~TeV for
$\Delta=$0.4435 $m$ and at $\omega \simeq$80~TeV for
$\Delta=$0.0387 $m$. In Fig. 8b the same curves are shown for
amplitude $M_{+-}$. These curves are very similar to curves in
Fig. 8a. The curves in both figures are normalized to imaginary
part of the corresponding amplitude in the absence of the
multiple scattering. The ratio these imaginary parts $r$ is very
small:it is $r=0.04435$ for $\Delta$=0.4435 $m$ and $r=0.003018$
for $\Delta$=0.0387 $m$.

\subsection{The polarization of a medium and the bremsstrahlung process}

When one considers bremsstrahlung of enough soft photons $\omega
\leq \omega_0 \gamma$ (see Eq.(\ref{4})), one has to take into
account the effect of a polarization of the medium. This effect
diminishes the formation length (see Eq.(\ref{5})) as well as the
probability of radiation (see \cite{4}, the qualitative
discussion may be found in \cite{6}). For analysis we use the
general expression for the probability of radiation, see
Eq.(\ref{a1}), Appendix A. The factor in front of exponent in
this expression (see Eq.(\ref{a2})) contains two terms $A$ and
${\bf B}$, the term $A$ is not changed and the term ${\bf B}$
contains combination
\begin{equation}
{\bf v}-\frac{{\bf k}}{\omega} \simeq  \mbox{\boldmath$\vartheta$}
+{\bf n}\frac{\kappa_0^2}{2\gamma^2},\quad
\kappa_0=\frac{\omega_p}{\omega},\quad \omega_p=\omega_0 \gamma,
\label{42}
\end{equation}
and its dependence on $\kappa_0$ (term of the order $1/\gamma^2$)
may be neglected also.
So, the dependence on $\omega_0$ manifests itself in the
exponent of Eq.(\ref{a1}),
and respectively in the exponent of (\ref{a19}) only:
\begin{equation}
\displaystyle{a \rightarrow 2\frac{\omega
\varepsilon}{\varepsilon-\omega} \left(1-\frac{k}{\omega}v
-\frac{\omega_0^2}{2\varepsilon \omega}\right) \simeq a \kappa
\equiv \tilde{a}, \quad \kappa \equiv 1+\kappa_0^{'2}},\quad
\kappa_0^{'2}=\frac{\varepsilon'}{\varepsilon}\kappa_0^2.
\label{43}
\end{equation}
where the term with $\omega_0^2$ arises in the case of virtual
photon: $k^2 \equiv k_{\mu}k^{\mu}\neq 0$. Performing the
substitution $a \rightarrow \tilde{a}$ in Eq.(\ref{2.7}) we
obtain for the potential (see Eqs.(\ref{2.7}) (\ref{2.9}),
(\ref{2.9a}), (\ref{2.16}))
\begin{eqnarray}
&&\displaystyle{V(\mbox{\boldmath$\varrho$}) \rightarrow
\tilde{V}(\tilde{\mbox{\boldmath$\varrho$}}) =\tilde{Q}
\tilde{\mbox{\boldmath$\varrho$}}^2
\left(L\left(\frac{\tilde{\varrho}}{2\sqrt{\kappa}}\right)-2C\right)
=\tilde{V}_c(\tilde{\mbox{\boldmath$\varrho$}})+
\tilde{v}(\tilde{\mbox{\boldmath$\varrho$}}),\quad \tilde{\varrho}=
|\tilde{\mbox{\boldmath$\varrho$}}|=\varrho \sqrt{\kappa}},
\nonumber \\
&&\displaystyle{ \tilde{V}_c(\tilde{\mbox{\boldmath$\varrho$}})=
\tilde{q}\tilde{\mbox{\boldmath$\varrho$}}^2,\quad \tilde{q}=
\tilde{Q}\tilde{L}(\tilde{\varrho_c}),\quad
\tilde{Q}=\frac{Q}{\kappa^2},\quad
\tilde{L}(\tilde{\varrho_c})=
\ln \frac{\kappa}{\gamma^2\vartheta_2^2\tilde{\varrho_c}^2}},
\nonumber \\
&&\displaystyle{\tilde{v}(\tilde{\mbox{\boldmath$\varrho$}})=
-\frac{\tilde{q}\tilde{\mbox{\boldmath$\varrho$}}^2}{\tilde{L}}
\left(2C+\ln \frac{\tilde{\mbox{\boldmath$\varrho$}}^2}
{4\tilde{\varrho_c}^2} \right)}.
\label{43a}
\end{eqnarray}
The substitution Eq.(\ref{43}) in the expression for the
probability of radiation Eq.(\ref{2.8}) gives
\begin{equation}
\displaystyle{R_1 \rightarrow R_1,\quad R_2 \rightarrow R_2 \kappa
\equiv \tilde{R}_2}.
\label{44}
\end{equation}
The value of the parameter $\tilde{\varrho_c}$ in (\ref{43a}) is
determined by equation (compare with Eqs.(\ref{2.14}), (\ref{2.30a}) )
\begin{equation}
4\tilde{\varrho_c}^4 \tilde{Q} \tilde{L}(\tilde{\varrho_c})=1,\quad
{\rm for} \quad 4\tilde{Q}\tilde{L}(1) \geq 1.
\label{44a}
\end{equation}
In the opposite case $\tilde{\varrho_c}=1$ and
this is possible in two intervals of the photon energy $\omega$:
\begin{enumerate}
\item for $\kappa_0 \ll 1$ when the multiple scattering and effects of
the polarization of a medium are weak;
\item for $\kappa_0 \gg 1$ when effects of
the polarization of a medium become stronger then effects of the multiple
scattering ($\nu_0 < \kappa$).
\end{enumerate}
In an intermediate region we substitute $\tilde{\varrho_c}^2
\rightarrow \varrho_c^2 \kappa$ in Eq.(\ref{44a}). After it we
obtain the equation for $\varrho_c$ which coincides with
Eq.(\ref{2.14}), see also Eq.(\ref{2.30a}):
\begin{equation}
\frac{1}{\varrho_c^4}=\nu_0^2(\varrho_c),\quad \nu_0^2(\varrho_c)=
4QL(\varrho_c).
\label{44b}\end{equation}
Thus, for $\tilde{\varrho_c} < 1$ we have
\begin{equation}
\tilde{\nu_0}=\sqrt{4\tilde{Q}\tilde{L}(\tilde{\varrho_c})}=
\frac{1}{\tilde{\varrho_c}^2}=\frac{1}{\varrho_c^2 \kappa}=
\frac{\nu_0}{\kappa},\quad \tilde{L}(\tilde{\varrho_c})=L(\varrho_c),
\label{44c}\end{equation}
while for $\tilde{\nu_0} < 1$ we have
\begin{equation}
\tilde{\nu_0}=\sqrt{4\tilde{Q}\tilde{L}(1)}=\frac{2}{\kappa}\sqrt{Q}
\ln \left(\frac{a_{s2}^2 \kappa}{\lambda_c^2} \right).
\label{44d}
\end{equation}
The spectral distribution of the probability of radiation
Eq.(\ref{2.35}) with allowance for the polarization of a medium
have the form
\begin{equation}
\displaystyle{\frac{dW}{d\omega}=
\frac{\alpha}{2\pi \gamma^2} {\rm Im}~
\left[\tilde{\Phi}(\tilde{\nu})-\frac{1}{2\tilde{L}(\tilde{\varrho}_c)}
\tilde{F}(\tilde{\nu}) \right]},
\label{45}\end{equation}
where
\[
\displaystyle{\tilde{\Phi}(R_1,R_2)=\Phi(R_1, \tilde{R}_2),\quad
\tilde{F}(R_1,R_2)=F(R_1, \tilde{R}_2)},
\]

In the case $\nu_0 \ll \kappa_0^2$, the characteristic momentum
transfer in the used units ($\zeta_c$) are defined by value
$\kappa_0^2 (\tilde{\varrho}_c^2=1)$, one can use asymptotic
expansion Eq.(\ref{2.33})  and we have for the spectral
distribution of the probability of radiation
\begin{equation}
\displaystyle{\frac{dW}{d\omega}=\frac{16}{3}\frac{Z^2\alpha^3n_a}
{m^2\omega \kappa_0^2} \left(L_{p}+\frac{1}{12}-f(Z\alpha)
\right)= \frac{4}{3\pi}\frac{Z\alpha^2\omega}{m\gamma^2}
\left(L_{p}+\frac{1}{12}-f(Z\alpha) \right)},
\label{47}\end{equation} where $f(Z\alpha)$ is defined in
(\ref{2.9b}), $\displaystyle{L_{p}=\ln \left(183Z^{-1/3}\kappa_0
\right)}$. The results obtained agree with given in \cite{4}
where calculations are fulfilled within a logarithmic accuracy and
without Coulomb corrections. It is seen that the dependence of
spectral distribution on photon energy ($\omega d\omega$) differs
essentially from the Bethe-Maximon one ($d\omega/\omega$), and the
probability is independent on the density $n$.

\section{Impact of a medium for a target of finite thickness}

\setcounter{equation}{0}

\subsection{General consideration}

In the case under consideration probability of radiation is
defined not only by the relative time $\tau=t_2-t_1$ as in Sec.
2.1. The used radiation theory is formulated in terms of two
times (see Eqs.(\ref{a1}) - (\ref{a3}) of Appendix A). Proceeding
from this formulation we can obtain more general expression for
the radiation probability. With allowance for the polarization of
a medium we have for the spectral distribution of the probability
of radiation (compare with Eq.(\ref{2.8}))
\begin{eqnarray}
&&\displaystyle{\frac{dw}{d\omega}=\frac{4\alpha}{\omega}
{\rm Re}\int_{-\infty}^{\infty}dt_2
\int_{-\infty}^{t_2}dt_1 \exp \left(-i\int_{t_1}^{t_2}\mu(t)dt \right)}
\nonumber \\
&&\displaystyle{\times \left[r_1 \varphi_0 (0, t_2, t_1)-ir_2
\mbox{\boldmath$\nabla$} \mbox{\boldmath$\varphi$}(0, t_2, t_1) \right]},
\label{50}
\end{eqnarray}
where
\begin{equation}
\mu(t)=1+\kappa_0^{'2} g(t),\quad \kappa_0^{'2}
=\frac{\varepsilon'}{\varepsilon}
\kappa_0^2,\quad
r_1 = \frac{\omega^2}{\varepsilon^2},\quad
r_2=1+\frac{\varepsilon'^2}{\varepsilon^2},
\label{51}
\end{equation}
here the function $g(t)$ describes the density of atoms on the
projectile trajectory, $\kappa_0$ is defined in Eq.(\ref{42}). The
functions $\varphi_{\mu}(\mbox{\boldmath$\varrho$}, t_2,
t_1),\quad (\varphi_{\mu}=(\varphi_0,
\mbox{\boldmath$\varphi$}))$ satisfy the equation (\ref{2.9}),
but now the potential $V$ depends on time
\begin{eqnarray}
&& \displaystyle{i\frac{\partial \varphi_{\mu}}{\partial t} =
H(t) \varphi_{\mu},\quad H (t)={\bf p}^2-
iV(\mbox{\boldmath$\varrho$})g(t)};
\nonumber \\
&& \displaystyle{\varphi_0(\mbox{\boldmath$\varrho$},t_1,t_1)=
\delta(\mbox{\boldmath$\varrho$}),\quad
\mbox{\boldmath$\varphi$}(\mbox{\boldmath$\varrho$},t_1,t_1)={\bf p}
\delta(\mbox{\boldmath$\varrho$})}.
\label{52}
\end{eqnarray}
So, the functions $\kappa_0^{'2}$ and $V(\mbox{\boldmath$\varrho$})$
in Eqs.(\ref{51}) and (\ref{52}) depend now on mean density of atoms
on the projectile trajectory while the function $g(t)$
describes the modulation of the density as a function of time
\[
g(t)=\frac{n_a(t)}{\overline{n}}
\]
Using an operator form of a solution of Eq.(\ref{52}) (compare
with Eq.(\ref{2.10})) we can present the probability Eq.(\ref{50})
in the form
\begin{eqnarray}
&&\hspace{-10mm}\frac{dw}{d\omega}= \frac{4\alpha}{\omega}{\rm Re}
\int_{-\infty}^{\infty}dt_2
\int_{-\infty}^{t_2}dt_1 \exp \left(-i\int_{t_1}^{t_2}\mu(t)dt \right)
\nonumber \\
&&\hspace{-10mm}\times \left<0|r_1 S(t_2,t_1)+r_2 {\bf p} S(t_2,
t_1) {\bf p}|0 \right>,\quad S(t_2, t_1)={\rm T} \exp
\left[-i\int_{t_1}^{t_2}  H(t) dt \right], \label{53}
\end{eqnarray}
where the symbol ${\rm T}$ means the chronological product. Note,
that in Eqs.(\ref{50}) and (\ref{53}) it is implied that
subtraction is made at $V=0$,~$\mu(t)=1$.

\subsection{Boundary effects for a thick target}

For the homogeneous target of finite thickness $l$ the radiation
process in a medium depends on interrelation between $l$ and
formation length $l_{f}$ Eq.(\ref{5}). In the case when $l \gg
l_{f}$ we have the thick target where radiation on the boundary
should be incorporated. In the case when $l \ll l_{f}$ we have
the thin target where the mechanism  of radiation is changed
essentially and in the case when $l \sim l_{f}$ we have
intermediate thickness. Integrals over time in Eq.(\ref{53}) we
present as integrals over four domains:
\begin{enumerate}
\item $t_1 \leq 0,~0 \leq t_2 \leq T$;
\item $ 0 \leq t_1 \leq T,~0 \leq t_2 \leq T$;
\item $0 \leq t_1 \leq T,~ t_2 \geq T$;
\item $t_1 \leq 0,  t_2 \geq T$;
\end{enumerate}
where
\begin{equation}
\mu(t)=\vartheta(-t)+\vartheta(t-T)+\kappa\vartheta(t)\vartheta(T-t),\quad
T=\frac{l}{l_{f0}}=\frac{l\omega m^2}{2\varepsilon \varepsilon'},\quad
\kappa=1+\kappa_0^{'2}.
\label{51a}
\end{equation}
In two more domains $t_{1,2} \leq 0$ and $t_{1,2} \geq T$ an
electron is moving entirely free and there is no radiation. We
consider in this subsection the case, when the thickness of a
target $l$ is much larger than formation length $l_f$ Eq.(\ref{5})
or $(\nu_0+\kappa)T \gg 1$. In this case domain 4. doesn't
contribute. The contributions of other domains are
\begin{eqnarray}
&&\hspace{-14mm}I_1 \simeq \int_{-\infty}^{0}dt_1 \int_{0}^{\infty}dt_2
\exp \left(i(t_1-\kappa t_2) \right) \exp \left(-iHt_2 \right)
\exp \left(i H_0 t_1 \right)=-\frac{1}{H+\kappa}
\frac{1}{H_0+1},
\nonumber \\
&&\hspace{-14mm}I_2=\int_{0}^{T} dt_2 \int_{0}^{t_2} dt_1 \exp\left(-i
(H+\kappa)(t_2-t_1) \right) \simeq T \int_{0}^{\infty} d\tau
\exp \left(-i(H+\kappa)\tau \right)
\nonumber \\
&&\hspace{-14mm}- \int_{0}^{\infty} \tau d\tau
\exp \left(-i(H+\kappa)\tau \right)
=-i\frac{T}{H+\kappa}+\frac{1}{(H+\kappa)^2},\quad
I_3 \simeq - \frac{1}{ H_0+1}\frac{1}{H+\kappa},
\label{54}
\end{eqnarray}
where $ H_0= {\bf p}^2$. The term in $I_2$:~$-iT/(H+\kappa)$
describes the probability of radiation considered in previous
subsections. All other terms define the probability of radiation
of boundary photons \footnote[1] {Radiation of boundary photons
in an inhomogeneous electromagnetic field was considered in
\cite{18a}.}. So, making mentioned subtraction we have for the
spectral distribution of the probability of radiation of boundary
photons
\begin{eqnarray}
&&\displaystyle{\frac{dw_b}{d\omega}=\frac{4\alpha}{\omega}
{\rm Re}\left<0|r_1 M+ r_2 {\bf p} M {\bf p}|0 \right>},
\nonumber \\
&& M=\left(\frac{1}{H+\kappa}-\frac{1}{{\bf p}^2+1} \right)^2
\label{55}
\end{eqnarray}

In the case when both the LPM effect and effect of polarization
of a medium are weak one can decompose combination in $M$ in
Eq.(\ref{55})
\begin{equation}
\frac{1}{H+\kappa}-\frac{1}{{\bf p}^2+1} \simeq \frac{1}{{\bf
p}^2+1}(iV-\kappa_0^{'2})\frac{1}{{\bf p}^2+1}. \label{51b}
\end{equation}
Using this decomposition one can find an estimate for probability
of radiation
\begin{equation}
\frac{dw_b}{d\omega} \simeq \frac{\alpha}{\pi}\left(-c_1
\nu_1^4+c_2\kappa_0^{'4}\right), \label{51c}
\end{equation}
where $c_1$ and $c_2$ are some positive coefficients. So, in the
case under consideration the probability of boundary photon
radiation is negligible small.

In the case when $\nu_1 \ll \kappa_0^{'2} \geq 1$ one can omit the
potential $V$ in Eq.(\ref{51b}), so that the effect of
polarization of a medium is essential, then
\begin{eqnarray}
&&\frac{dw_b}{d\omega} =
\frac{4\alpha}{\omega(2\pi)^2}\int_{}^{}\left<0|r_1 M+ r_2 {\bf
p} M {\bf p}|0 \right>d^2p \nonumber \\
&& =\frac{\alpha}{\pi
\omega}\left\{r_1\left(1+\frac{1}{\kappa}-\frac{2}{\kappa-1}\ln
\kappa \right)+r_2\left[\left(1+\frac{2}{\kappa-1}\right)\ln
\kappa -2 \right]\right\} \label{51d}
\end{eqnarray}
This result is the quantum generalization of the transition
radiation probability \cite{z10}.

Let us estimate the probability of radiation Eq.(\ref{55}) in the
case when both the LPM effect and effect of polarization of a
medium are strong $(\nu_0+\kappa \gg 1)$. In this case we can
neglect the term $1/(H+\kappa)$ in the term $\propto r_1$ in $M$
Eq.(\ref{55}). In the term $\propto r_2$ one can put
$V(\mbox{\boldmath$\varrho$}) \simeq \nu_0^2
\mbox{\boldmath$\varrho$}^2 \sim \nu_0^2/{\bf p}^2$ so that
\begin{equation}
\int_{}^{} {\bf p}^2\left(\frac{1}{H+\kappa}- \frac{1}{{\bf
p}^2+1} \right)^2 d^2p \simeq \pi \ln (\nu_0+\kappa). \label{51e}
\end{equation}
Substituting these results into Eq.(\ref{55}) we have up to
logarithmic accuracy
\begin{equation}
\frac{dw_b}{d\omega} =
\frac{\alpha}{\omega \pi}\left[r_1+r_2 \ln (\nu_0+\kappa) \right].
\label{51g}
\end{equation}
In the case $\nu_0 \ll \kappa$ this result agrees with Eq.(\ref{51d}).
The case $\nu_0 \geq \kappa$ will be considered below (see Eqs.(\ref{z24})
and (\ref{z25})).

\subsection{A thin target}

We consider now a situation when the formation length of
radiation is much larger than the thickness $l$ of a target
\cite{13}
\begin{equation}
l \ll l_f =\frac{l_{f0}}{\zeta},\quad \zeta =
1+\gamma^2\vartheta^2,\quad T =\frac{l}{l_{f0}} \ll
\frac{1}{\zeta},
\label{69}
\end{equation}
where $l_f,~l_{f0}$ are defined in Eqs.(\ref{5}), (\ref{6}). In
the case $\kappa T \ll 1$ the radiated photon is propagating in
the vacuum and one can neglect the polarization of a medium. The
spectral distribution of the probability of radiation from a thin
target is
\begin{equation}
\frac{dw_{th}}{d\omega}= \displaystyle{\frac{\alpha}{\pi^2
\omega} \int_{}^{} d^2\varrho \left[r_1 K_0^2(\varrho)+r_2
K_1^2(\varrho) \right] \left(1-\exp (-V(\varrho)T) \right)},
\label{72}
\end{equation}
where $V(\varrho)$ is defined in Eqs.(\ref{2.9}), (\ref{2.9a}),
$K_n$ is the modified Bessel function. If $V(\varrho=1)T \ll 1$
one can expand the exponent (the contribution of the region
$\varrho \gg 1$ is exponentially damped because in this region
$K_{0,1}(\varrho) \propto \exp (-\varrho)$). In the first order
over $VT$ using the explicit expression for the potential
Eq.(\ref{2.9a}) one obtains in this case the Bethe-Maximon formula
with the Coulomb corrections Eq.(\ref{2.35}).

In the opposite case when the multiple scattering of a particle
traversing a target is strong ($V(\varrho=1)T \gg 1$, the mean
square of multiple scattering angle $\vartheta_s^2 \gg
1/\gamma^2$). We present the function $V(\varrho)T$ (see
Eqs.(\ref{2.9}), (\ref{2.9a}) and (\ref{2.18}) as
\begin{eqnarray}
&&\displaystyle{V(\varrho)T= \frac{\pi Z^2 \alpha^2 n l}{m^2}\varrho^2
\left(\ln \frac{4a_{s2}^2}{\lambda_c^2 \varrho^2}-2C \right)=
A\varrho^2 \ln \frac{\chi_t}{\varrho^2}=
A\varrho^2 \left( \ln \frac{\chi_t}{\varrho_t^2} -
\ln \frac{\varrho^2}{\varrho_t^2} \right)}
\nonumber \\
&&\displaystyle{=k\varrho^2\left(1-\frac{1}{L_t}\ln \frac{\varrho^2}
{\varrho_t^2}\right);\quad A\varrho_t^2 \ln \frac{\chi_t}{\varrho_t^2}=1,\quad
L_t=\ln \frac{\chi_t}{\varrho_t^2} =
\ln \frac{4a_{s2}^2}{\lambda_c^2 \varrho_t^2}-2C},
\label{73}
\end{eqnarray}
where $\varrho_t$ is the lower boundary of values contributing into the
integral over $\varrho$.
Substituting this expression into (\ref{72}) we have the integral
\begin{equation}
\displaystyle{2\pi\int_{0}^{\infty}\varrho d\varrho K_1^2(\varrho)
\left\{1-\exp \left[-k\varrho^2 \left(1-\frac{1}{L_t}
\ln \frac{\varrho^2}{\varrho_t^2} \right) \right] \right\} \equiv \pi J}.
\label{74}
\end{equation}
In this integral we expand the exponent in the integrand over $1/L_t$ keeping
the first term of the expansion.
It is convenient to substitute $z=k\varrho^2$.
We find
\begin{eqnarray}
&&\displaystyle{J_1=\frac{1}{k}\int_{0}^{\infty}
\left[K_0\left(\sqrt{\frac{z}{k}}\right)K_2\left(\sqrt{\frac{z}{k}}\right)
-K_1^2\left(\sqrt{\frac{z}{k}}\right) \right]\exp(-z) zdz},
\nonumber \\
&&\displaystyle{J_2=-\frac{1}{kL_t}\int_{0}^{\infty}
K_1^2\left(\sqrt{\frac{z}{k}}\right) \exp(-z) \ln z zdz},\quad
J=J_1+J_2.
\label{76}
\end{eqnarray}
Expanding the modified Bessel functions $K_n(x)$ at $x \ll 1$ and taking the
integrals in the last expression
we have
\begin{eqnarray}
&&\displaystyle{J=J_1+J_2=\left(1+\frac{1}{2k}\right)\left(\ln 4k -C\right)+
\frac{1}{2k}-1 +\frac{C}{L_t}},
\nonumber \\
&&\displaystyle{k=\frac{\pi Z^2\alpha^2}{m^2}
n_a l L_t}.
\label{77}
\end{eqnarray}
In the term with $K_0^2$ in Eq.(\ref{72}) the region $\varrho \sim
1$ contributes. So we have
\begin{equation}
\displaystyle{J_3=2 \int_{0}^{\infty} K_0^2(\varrho)
\left(1-\exp (-VT) \right)\varrho d\varrho \simeq
2 \int_{0}^{\infty} K_0^2(\varrho)
\varrho d\varrho=1}.
\label{78}
\end{equation}
Substituting found $J$ and $J_3$ into Eq.(\ref{72}) we obtain for
the spectral distribution of the probability of radiation in a
thin target at conditions of the strong multiple scattering
\begin{equation}
\frac{dw_{th}}{d\omega}=\frac{\alpha}{\pi \omega}
\left(r_1+r_2 J\right).
\label{79}\end{equation}
The logarithmic term in this formula is well known in theory of the
collinear photons radiation at scattering of a radiating particle on angle
much larger than characteristic angles of radiation $\sim 1/\gamma$.
It is described with logarithmic accuracy in a quasi-real electron
approximation (see \cite{20a}, Appendix B2).

The formula (\ref{72}) presents the probability of radiation in
the case when the formation length $l_f \gg l$. It is known,see
e.g. \cite{7}, that in this case a process of scattering of a
particle is independent of a radiation process and a differential
probability of radiation at scattering with the momentum transfer
${\bf q}$ can be presented in the form
\begin{equation}
dW_{\gamma}=dw_s({\bf q})dw_r({\bf q}, {\bf k}),
\label{80}\end{equation} where $dw_s({\bf q})$ is the
differential probability of scattering with the momentum transfer
${\bf q}$ which depends on properties of a target. The function
$dw_r({\bf q}, {\bf k})$ is the probability of radiation of a
photon with a momentum ${\bf k}$ when an emitting electron
acquires the momentum transfer ${\bf q}$. This probability has a
universal form which is independent of properties of a target.
For an electron traversing an amorphous medium this fact is
reflected in  Eq.(\ref{72}). Indeed, passing on to a momentum
space we have
\begin{eqnarray}
&&\hspace{-16mm}\displaystyle{dw_r({\bf q}, {\bf k})=
\frac{\alpha d\omega}{\pi^2 \omega} \int_{}^{} d^2\varrho
\left[r_1 K_0^2(\varrho)+r_2 K_1^2(\varrho) \right]
\left(1-\exp (-i{\bf q}\mbox{\boldmath$\varrho$}) \right)}
\nonumber \\
&&\hspace{-16mm}\displaystyle{
=\frac{\alpha d\omega}{\pi \omega}\left[r_1 F_1\left
(\frac{q}{2}\right)+r_2 F_2\left(\frac{q}{2}\right) \right]},
\label{81}
\end{eqnarray}
where the functions $F_1(z)$ and $F_2(z)$ are defined in
Eqs.(\ref{4.18sc}) and (\ref{4.17sc}). Remind that $q$ is
measured in electron mass. The probability of radiation in this
form was found in \cite{18b}. For a differential probability of
scattering (here we consider the multiple scattering) there is a
known formula (cp~Eqs.(\ref{a12}), (\ref{a17}), (\ref{2.9}) and
(\ref{2.9a}))
\begin{eqnarray}
&&dw_s({\bf q})=F_s({\bf q})d^2q,\quad F_s({\bf q})=
\frac{1}{(2\pi)^2}\int_{}^{}d^2\varrho
\exp \left(-i{\bf q} \mbox{\boldmath$\varrho$}\right)
\exp \left(-V_s(\varrho) l\right),
\nonumber \\
&&V_s(\varrho)=
n\int_{}^{}d^2q\left(1-\exp(-i{\bf q}\mbox{\boldmath$\varrho$}) \right)
\sigma({\bf q}),
\label{82}
\end{eqnarray}
where $\sigma({\bf q})$ is the cross section of single
scattering.

The formula (\ref{82})  is the exact solution of the kinetic
equation for the pure scattering problem (without radiation). Let
us consider scattering of relativistic electrons in the case when
scattering angles $\vartheta_s \ll 1$ and the condition
$\vartheta_s \gg \vartheta_1=\displaystyle{\frac{1}{a_s
\varepsilon}}$ is fulfilled (the angle $\vartheta_1$ is defined
in Appendix A, Eq.(\ref{a12})). For combination $V_s(\varrho)l$
in Eq.(\ref{82}) one can use Eq.(\ref{73}). Let us remind that the
impact parameters $\varrho$ measured in the Compton wavelengths
$\lambda_c$, which is conjugate to space of the transverse
momentum transfers $q$ measured in the electron mass $m$.
\begin{eqnarray}
&& V_s(\varrho)l \simeq V(\varrho)T =
\frac{\varrho^2}{\varrho_t^2}-\frac{1}{L_t}\frac{\varrho^2}{\varrho_t^2}\ln
\frac{\varrho^2} {\varrho_t^2},
\nonumber \\
&& L_t=l_t+L_t,\quad \frac{1}{\varrho_t^2}=AL_t
\nonumber \\
&& l_t=\ln (A\chi_t)=\ln (4\pi Z^2 \alpha^2 n_al
a_s^2)+1-2C-2f(Z\alpha).
\label{82a}
\end{eqnarray}
Substituting Eq.(\ref{82a}) into Eq.(\ref{82}) and passing to the
variable $\nu =q\varrho_t/2$ we obtain
\begin{eqnarray}
&&dw_s(\nu)=f(\nu)\nu d\nu,
\nonumber \\
&& f(\nu)=\int_{0}^{\infty}u J_0(\nu u)\exp
\left(-\frac{u^2}{4}+\frac{u^2}{4L_t}\ln \frac{u^2}{4}\right)du.
\label{82b}
\end{eqnarray}
Solution of scattering problem in angular variables is given in
the end of Appendix A (see Eq.(\ref{a23})).

On adopted assumptions ($\vartheta_s^2/\vartheta_1^2=A\chi_t \gg
1,~l_t \gg 1,~L_t \gg1 $) one can expands  in Eq.(\ref{82b}) over
powers of $1/L_t$:
\begin{eqnarray}
&& f(\nu)=\sum_{n=0}^{\infty}\frac{1}{n!}\int_{0}^{\infty} u
J_0(\nu u)\exp
\left(-\frac{u^2}{4}\right)\left(\frac{u^2}{4L_t}\ln
\frac{u^2}{4}\right)^n du
\nonumber \\
&& =f^{(0)}(\nu) + \frac{1}{L_t}f^{(1)}(\nu)+
\frac{1}{L_t^2}f^{(2)}(\nu) + \ldots
\label{82c}
\end{eqnarray}

It is instructive to compare these results with the classical
paper Bethe on multiple scattering \cite{B1}. The obtained
formula Eq.(\ref{82c}) is consistent with Eq.(25) of \cite{B1},
However there is difference in evaluation of the Coulomb
corrections entering into the expression for the function $l_t$
Eq.(\ref{82a}) (see $e^b$ in Eq.(22) in \cite{B1}). In the last
formula in \cite{B1} the Coulomb correction is contained in the
form $-\ln(1+3.34 Z^2\alpha^2)$ (it should be noted that a) Bethe
was referring to the numerical coefficient at $(Z\alpha)^2$
calculated by Moli\`ere \cite{Mo1} as "only approximate"; b) the
Coulomb corrections to the cross sections of bremsstrahlung and
pair creation by photon which contain the famous function
$f(Z\alpha)$ defined in Eq.(\ref{2.9b}) were found after
publication of \cite{B1}) while in Eq.(\ref{82a}) enters different
correction: $-2f(Z\alpha)$. To compare these results numerically
let us consider case $Z\alpha \ll 1$. In this case the correction
in \cite{B1} is -$3.34 Z^2\alpha^2$ while for Eq.(\ref{82a}) we
have the correction $-2\zeta(3)Z^2\alpha^2=-2.404 Z^2\alpha^2$.
Besides, for accounting the inelastic scattering on electrons the
use of Eq.(\ref{40a}) for the potential $V_s(\varrho)$ gives more
accurate results than the simple substitution $Z^2 \rightarrow
Z(Z+1)$ done in \cite{B1}.

The main term of the expansion in Eq.(\ref{82c}) has the Gaussian
form
\begin{equation}
\displaystyle{f^{(0)}(\nu)=2 e^{-\nu^2}}
\label{82d}
\end{equation}
The next terms contains the corrections to the Gaussian form.
They were analyzed in detail in \cite{B1}.

Using the formula (\ref{81}) one can easily obtain to within
logarithmic accuracy expressions (\ref{79}),(\ref{51g}). Both a
radiation of boundary photons and a radiation in a thin target
may be considered as a radiation of collinear photons (see e.g.
\cite{20a}) in the case when an emitting particle deviates at
large angle ($\vartheta_s \gg 1/\gamma, q \gg 1$). Using
Eq.(\ref{81}) at $x \gg 1$ we find
\begin{eqnarray}
&& dw_r(q) \simeq \frac{\alpha d\omega}{\pi \omega}\left[
r_1+r_2\left(\ln q^2 -1 \right) \right];
\nonumber \\
&&\int_{}^{}d^2q dw_r(q) F_s({\bf q})
\simeq \frac{\alpha d\omega}{\pi \omega}
\left[r_1+r_2\left(\ln \overline{q^2} -1 \right) \right].
\label{83}
\end{eqnarray}
For a thin target value of $\overline{q^2}$ is defined by mean
square of multiple scattering angle on a thickness of a target
$l$, and for boundary photons is the same but on the formation
length $l_f$. However, if we one intends to perform computation
beyond a logarithmic accuracy, the method given in this subsection
has advantage since there is no necessity to calculate $F_s({\bf
q})$. Radiation from a thin target using the path integral method
was considered in \cite{SF} and \cite{SF1}.

\subsection{A target of intermediate thickness}

In the case when a target has intermediate thickness ($l \sim l_f$)
the separation of contributions on photon emission inside target and
boundary photon emission becomes senseless. We consider
this case \cite{14} neglecting by
the correction term $v(\mbox{\boldmath$\varrho$})$.
We present the spectral probability of radiation as
\begin{eqnarray}
&&\hspace{-6mm} \frac{dw}{d\omega}=\frac{\alpha}{\pi \omega}
\sum_{k=1}^{5}\left[-r_1 {\rm Im}~J_k^{(1)} +
r_2 {\rm Re}~J_k^{(2)}\right]; \nonumber \\
&&\hspace{-6mm} J_1^{(m)} = J_3^{(m)} =\int_{0}^{\infty} dt_1 \int_{0}^{T} dt_2
{\rm e}^{-i\left(t_1+\kappa t_2 \right)}\left[\left(t_1+t_2\right)^{-m}
-N_1^m \right], \nonumber \\
&&\hspace{-6mm} J_2^{(m)} = \int_{0}^{T} dt (T-t){\rm e}^{-i \kappa t}
\left[t^{-m}-N_2^m \right], \nonumber \\
&& J_4^{(m)} =\int_{0}^{\infty} dt_1 \int_{0}^{\infty} dt_2
{\rm e}^{-i\left(t_1+t_2+\kappa T \right)}\left[\left(t_1+t_2+T\right)^{-m}
-N_4^m \right],
\nonumber \\
&&\hspace{-6mm} J_5^{(m)}(T)=2 \int_{0}^{\infty}dt_1\int_{0}^{T}dt_2
\frac{{\rm e}^{-it_1}}
{\left(t_1+t_2 \right)^m}\left({\rm e}^{-it_2}-{\rm e}^{-i\kappa t_2} \right)
+\int_{0}^{T}dt \frac{(T-t)}{(t-i0)^{m}}
\left({\rm e}^{-it}-{\rm e}^{-i\kappa t} \right) \nonumber \\
&&\hspace{-6mm} + \int_{0}^{\infty}dt_1\int_{0}^{\infty}dt_2
\frac{{\rm e}^{-i\left(t_1+t_2 \right)}}
{\left(t_1+t_2 +T \right)^m}\left({\rm e}^{-iT}-{\rm e}^{-i\kappa T} \right).
\label{z12}
\end{eqnarray}
where $T$ is defined in (\ref{51}), the functions $N_k$ are
\begin{eqnarray}
&&\hspace{-8mm} N_1=\frac{\nu}{\sinh \nu t_2 + \nu t_1 \cosh \nu t_2},\quad
N_2 = \frac{\nu}{\sinh \nu (t-i0) }, \nonumber \\
&&\hspace{-8mm} N_4=\frac{\nu}{\left(1+\nu^2t_1t_2 \right)\sinh
\nu T + \nu \left(t_1+t_2 \right) \cosh \nu T},\quad \nu =
\frac{1+i}{\sqrt{2}}\nu_0,
\label{z11a}
\end{eqnarray}
here $\kappa$ and $\nu_0$ are defined in Eq.(\ref{43}) and in
Eq.(\ref{2.30b}).

We confine ourselves to the case $\varepsilon \ll \varepsilon_e$
when the LPM effect manifests itself for soft emitted photons
($\omega \ll \varepsilon$). So we can neglect the terms with $r_1$
in (\ref{z12}).

Let us consider first the case when LPM effect is weak ($\nu_1 \ll
1$). We assume here that condition $\nu_1(\omega_p) \geq 1$
(definition of $\omega_p$ see in Eq.(\ref{5})) is fulfilled, that
is in the region where $\nu_1 \ll 1$ one has $\omega \gg \omega_p$
and effects of the polarization of a medium are negligible. This
is true for high energies ($\varepsilon \geq 10~GeV$). Then for
thickness $T \ll 1/\nu_1$ the transverse shift of the projectile
due to the multiple scattering in a target as a whole have no
influence on coherent effects defined by the phase $\phi=\omega
l(1-{\bf nv})$ in the factor $\exp(-i\phi)$. Indeed, for the
projectile traversing a target in the case $\nu_1T \ll 1$ an
increment of the phase $\phi$ is small
\begin{equation}
\Delta \phi \sim \omega l \vartheta_s^2 \sim
\omega l \frac{\nu_1^2T}{\gamma^2} \sim \nu_1^2 T^2 \ll 1
\label{z22a}
\end{equation}
The angle of multiple scattering $\vartheta_s$ is small also
comparing with an characteristic angle of radiation
$1/\gamma~(\gamma^2\vartheta_s^2 =\nu_1^2T \ll 1)$. So, in the
case $\nu_1 \ll 1,~ \nu_1T \ll 1$ the radiation originates on
separate atoms of a target and an interference on target
boundaries is defined by the value $\omega l (1-v)=T$. At $T \ll
1$ this interference is weak, while at $T \gg 1$ there is a
damping of the interference terms due to integration over photon
emission angles. Expanding over $\nu_1$ in Eq.(\ref{z12}) we
obtain ($\kappa=1$):
\begin{eqnarray}
&& J^{(2)}(T) \equiv {\rm Re}~\sum_{k=1}^{4}J_k^{(2)}(T) \simeq
\frac{\nu_1^2T}{3}\left[1-3T\int_{1}^{\infty}\frac{(x-1)^2}{x^3}
\sin (xT) dx \right]  \nonumber \\
&& =\frac{\nu_1^2T}{3} \left[1+3T\left(\left(1-\frac{T^2}{2} \right)
{\rm si}(T)-2T {\rm ci}(T) +\frac{3}{2} \sin T -
\frac{T}{2} \cos T \right) \right].
\label{z22b}
\end{eqnarray}
For case $T \ll 1$
\begin{equation}
J^{(2)}(T) \simeq \frac{\nu_1^2T}{3} \left[1-\frac{3\pi}{2}T
+6T^2\left(\ln \frac{1}{T}+1-C \right) \right],
\label{z22c}
\end{equation}
and for case $T \gg 1$
\begin{equation}
J^{(2)}(T) \simeq \frac{\nu_1^2T}{3} \left(1+6
\frac{\cos T}{T^2} \right).
\label{z22d}
\end{equation}
Thus, in the case $\nu_1 \ll 1,~ \nu_1T \ll 1$ the probability of
radiation is defined by Bethe-Maximon formula both for $T \ll 1$ and
for $T \gg 1$. However, for $T \sim 1$ the interference on the target
boundaries is essential. If we present as above $J^{(2)}$ as product
of $\nu_1^2 T/3$ (Bethe-Maximon formula) and some interference
factor, then this factor attains $0.53$ at $T=0.32$ (minimum of the
interference factor) and $1.33$ at $T=1.84$ (maximum of the
interference factor).

When the parameter $\nu_1T$ is large ($\nu_1 \ll 1,~ \nu_1T \gg
1$) the radiation is formed inside a target and the interference
terms are damped exponentially. In this case formulae derived for
thick target in Sec.2.1 are applicable. Taking into account the
contribution of boundary photons (see Sec.3.2) we have
\begin{equation}
J^{(2)}(\infty) = \frac{\nu_1^2T}{3} \left(1+\frac{1}{6L_1}
-\frac{16 \nu_1^4}{21}\right) - \frac{2\nu_1^4}{21},
\label{z22f}
\end{equation}
where $L_1$ and $\nu_1$ are defined in Eqs.(\ref{2.16}) and
(\ref{2.30a}).

We consider now the case when the LPM effect is strong ($\nu_0 \gg 1$)
and the parameter $T \ll 1$ while the value which characterize the
thickness of a target $\nu_0 T \sim 1$.
We find for $\kappa T \ll 1$ ($r_2 \simeq 2$)
\begin{eqnarray}
&& \frac{dw}{d\omega}=\frac{2\alpha}{\pi \omega}
\left(J^{(2)}+J_5^{(2)}\right),~J_5^{(2)}
\simeq \frac{(\kappa-1)^2}{2}\left(\ln \frac{1}{T}+
\frac{1}{2}-C \right), \nonumber \\
&&J^{(2)} \simeq
{\rm Re}~\Bigg\{ \ln (\nu \sinh \nu T)-1-C-
\kappa \frac{\pi T}{4} + \nonumber \\
&& \frac{2i}{\nu \tanh \nu T}\left[\ln (\nu \tanh \nu T)+1-C-
\frac{i\pi}{2} \right]+i\kappa T\left(\ln \frac{\cosh \nu T}{\tanh \nu T}-
\frac{\nu T}{\tanh \nu T} \right) \nonumber \\
&& + \frac{i\kappa}{\nu}\int_{0}^{\nu T} dt \left(\frac{4t}{\sinh 2t}-
\frac{t^2}{\sinh^2 t} \right)\Bigg\}, \quad
\nu = \exp\left(i\frac{\pi}{4}\right)\nu_0.
\label{z24}
\end{eqnarray}

For a relatively thick target ($\nu_0T \gg 1$) we have from
Eq.(\ref{z24})
\begin{eqnarray}
&& J^{(2)} \simeq \ln \nu_0 -1 -C -
\ln 2+\frac{\sqrt{2}}{\nu_0}\left(\kappa\frac{\pi^2}{24}+
\ln \nu_0 +1-C+\frac{\pi}{4} \right) \nonumber \\
&& + \frac{\nu_0T}{\sqrt{2}}\left(1-\frac{\pi \kappa}{2\sqrt{2}\nu_0} \right)
\label{z25}
\end{eqnarray}
Here the terms without $T$ are the contribution of boundary
photons (see Eq.(4.14) in \cite{13}) while the term $\propto T$
gives the probability of radiation inside target (with correction
$\sim \kappa/\nu_0$ but without corrections $\sim 1/L$). The
relative value of the last corrections at $\nu_0 \gg 1$ is given
by Eq.(\ref{2.40}).

In the limiting case when a target is very thin and $\nu_0 T \ll
1$ but when $\nu_0^2 T \gg 1$ we have from Eq.(\ref{z24})
\begin{eqnarray}
&& J^{(2)} \simeq \left(1+\frac{2}{\nu_0^2T} \right)
\left[\ln (\nu_0^2T) +1 -C \right] - 2 +\delta, \nonumber \\
&& \delta= \frac{(\nu_0 T)^4}{180}+\frac{2(\nu_0 T)^2}{45}T\left(
\ln \nu_0^2 T - C\right)-\frac{\kappa T}{6}(\nu_0 T)^2.
\label{z26}
\end{eqnarray}
The terms without $\delta$ in this expression coincide
with Eq.(\ref{79}) (up to terms $\propto~C/L_t$).

In the photon energy region where $\nu_0T \ll 1$ the contribution
of the terms $J_k^{(m)}$~($k$=1,2,3) is very small ($\sim
\delta$) and decreases with photon energy reduction ($\propto
\omega$), so that in the spectral distribution of radiation only
the terms $J_4^{(m)}, J_5^{(m)}$ contribute. We obtain for the
function $J_4^{(2)}$ in the case when $(1+\nu_0)T \ll 1$ and when
the parameter $\nu_0^2T$, which characterizes the mean square
angle of the multiple scattering in a target as a whole, has an
arbitrary value
\begin{eqnarray}
&& \hspace{-4mm} J_4^{(2)}{\rm e}^{i\kappa T} = \int_{0}^{\infty}dx x
\int_{0}^{\infty}\frac{dt_1}{t_1^2}
\int_{0}^{\infty}\frac{dt_2}{t_2^2} \exp \left(-\left(1+
\frac{x}{t_1t_2}\right)(t_1+t_2) \right)
\left[1-\exp \left(-x\nu_0^2 T \right) \right] \nonumber \\
&& \hspace{-4mm} = 4 \int_{0}^{\infty}dx K_1^2\left(2\sqrt{x} \right)
\left[1-\exp \left(-x\nu_0^2 T \right) \right] =2\int_{0}^{\infty}d\varrho
\varrho K_1^2(\varrho)\left[1-\exp \left(-k\varrho^2 \right)
\right], \nonumber \\
&& \hspace{-4mm} 4k=\nu_0^2T,
\label{z29}
\end{eqnarray}
where $K_1(\varrho)$ is the modified Bessel function. Formula
(\ref{z29}) corresponds at $\kappa=1$ to result for a thin target
obtained above (see Eq.(\ref{72})) without terms $\propto 1/L$.
Since the dependence on the parameter $\kappa$ is contained in
Eq.(\ref{z29}) as a common phase multiplier $\exp (-i\kappa T)$,
one can write more accurate expression for $J_4^{(2)}$ (with
terms $\propto 1/L$) using the found results (see Eq.(\ref{73})):
\begin{equation}
J_4^{(2)}=2{\rm e}^{-i\kappa T} \int_{0}^{\infty}d\varrho \varrho
K_1^2(\varrho)\left[1-\exp \left(-V(\varrho)T \right) \right].
\label{z30}
\end{equation}
For the case $\nu_0^2T \gg 1$ it has the form
\begin{eqnarray}
&& {\rm e}^{i\kappa T}~J_4^{(2)} = \left(1+\frac{1}{2k} \right)
\left[\ln 4k +1 -C \right] - 2 +\frac{C}{L_t}, \nonumber \\
&& k= \frac{\pi Z^2 \alpha^2 n l}{m^2}L_t, \quad L_t= \ln
\frac{4a_{s2}^2}{\lambda_c^2 \varrho_t^2}-2C,\quad
k(\varrho_t)\varrho_t^2=1
\label{ad31}
\end{eqnarray}
In the case when parameter $k$ is not very high one has to use an
exact expression Eq.(\ref{72}). For $k \ll 1$ one can expand the
exponent in the integrand of Eq.(\ref{z30}). Then we find
\begin{equation}
{\rm e}^{i\kappa T}J_4^{(2)} = \frac{\nu_1^2T}{3}
\left(1+\frac{1}{6L_1} \right),\quad \nu_1^2T= \frac{4\pi Z^2
\alpha^2nl}{m^2}L_1.
\label{ad32}
\end{equation}
At $\kappa T \ll 1$ the spectral distribution of probability is
\begin{equation}
\frac{dw}{d\omega}=\frac{2\alpha}{3\pi \omega}\nu_1^2 T\left(1+
\frac{1}{6L_1} \right)\left(1-\frac{\omega}{\varepsilon}\right).
\label{ad33}
\end{equation}
This is the Bethe-Maximon formula for not very hard photons (terms
$\displaystyle{\propto
\left(\frac{\omega}{\varepsilon}\right)^2}$ are omitted).

When a photon energy decreases, the parameter $\kappa$ increases
as well as the combination $\kappa T \propto 1/\omega$, while the
value $(\nu_0 T)^2$ decreases $\propto \omega$. Just this value
defines an accuracy of Eq.(\ref{z30}). Using Eq.(\ref{z12}) at $T
\ll 1, \kappa T \geq 1$ we find for the probability of transition
radiation ($J_5^{(2)}$) the following expression
\begin{eqnarray}
&& \frac{dw_{tr}}{d\omega} \simeq \frac{2\alpha}{\pi}
\Bigg\{\left(1+\frac{2}{\kappa-1} \right)
\Bigg[\ln \kappa -{\rm ci}(\kappa T) + \cos (\kappa T)(\ln T+C) \nonumber \\
&& + \frac{\pi}{2} \sin (\kappa T) \Bigg] + \kappa T {\rm si}(\kappa T)
-4 \sin^2 \frac{\kappa T}{2} \Bigg\}.
\label{z34}
\end{eqnarray}
In the limiting case $\kappa T \gg 1$ the probability
Eq.(\ref{z34}) turns into standard probability of the transition
radiation (\ref{51d}) with oscillating additions
\begin{eqnarray}
&&\frac{dw_{tr}}{d\omega}=\frac{2\alpha}{\pi}\left[J_{tr}+\cos
(\kappa T) \left(\ln T +C+1 \right)+\frac{\pi}{2}\sin (\kappa T)
\right],
\nonumber \\
&& J_{tr}=\left(1+\frac{2}{\kappa-1}\right)\ln
\kappa-2.
\label{z34a}
\end{eqnarray}
Note, that there is a qualitative difference in a behaviors of
interference terms in Eqs.(\ref{z22d}) and (\ref{z34a}). In the
former an amplitude of oscillation with $\omega$ increase
decreases as $1/\omega^2$ while in the latter the corresponding
amplitude weakly (logarithmically) increases with $\omega$
decrease.

From the above analysis follows that in the case when
$\nu_0 T \ll 1$ ($\nu_0 \gg 1$)
the spectral distribution of probability of
radiation with the polarization of a medium
taken into account has the form
\begin{equation}
\frac{dw}{d\omega} = \frac{dw_{tr}}{d\omega} +\cos (\kappa T)
\frac{dw_{th}}{d\omega},
\label{z35}
\end{equation}
where $dw_{th}/d\omega$ is the spectral distribution of
probability of radiation in a thin target without regard for the
polarization of a medium. In the case $4k = \nu_0^2 T \gg 1$ the
probability $dw_{th}/d\omega$ is defined by Eqs.(\ref{77}),
(\ref{79}) and for the case $k \ll 1$ it is defined by
Eq.(\ref{z22f}). More accurate representation of the probability
of radiation $dw_{th}/d\omega$ may be obtained using
Eq.(\ref{z30}). It follows from Eqs.(\ref{z34a}) and (\ref{z35})
that if we make allowance for multiple scattering at $\kappa T
\gg 1$ this results in decreasing of oscillations of the
transition radiation probability by magnitude of the
bremsstrahlung probability in a thin target.

 The radiation of the boundary photons with regard for the
multiple scattering was considered in \cite{17b} (for $\omega \ll
\varepsilon$), the effect of polarization of a medium  was
included into consideration of the problem in \cite{18b} and
\cite{18c}. In these papers the probability of radiation of
boundary photons (under condition of applicability of
Eqs.(\ref{z25}) and (\ref{z26})) was analyzed also to within the
logarithmic accuracy (see Eq.(20) in \cite{18b} and Eq.(15) in
\cite{18c}). This accuracy is insufficient for parameters
connected with experiment \cite{12}-\cite{14}. For example, for
$\varepsilon=25$~GeV and in heavy elements the value $\nu_0$
equates $\kappa$ for $\nu_0 \sim 20$. One can see from
Eqs.(\ref{z25}) and (\ref{z26})) that in this case $\ln \nu_0$ is
nearly completely compensated by constant terms. Our results,
which are consistent with obtained \cite{18b}, are more accurate
and presented in more convenient for application form and the
Coulomb corrections are included.

\subsection{Multiphoton effects in energy loss spectra}

It should be noted that in the experiments \cite{10}-\cite{12}
the summary energy of all photons radiated by a single electron
is measured. This means that besides mentioned above effects
there is an additional "calorimetric" effect due to the multiple
photon radiation. This effect is especially important in
relatively thick used targets. Since the energy loss spectrum of
an electron is actually measured, which is not coincide in this
case with the spectrum of photons radiated in a single
interaction, one have to consider the distribution function of
electrons over energy after passage of a target \cite{15}. As it
is known, this distribution function is the solution of the
corresponding kinetic equation. However, the problem can be
simplified essentially if one is interested in the soft part of
the energy loss spectrum. Just this situation was in the
experiments \cite{10}-\cite{12}, since the measurements were
performed in the region of the photon energies of one to five
orders of magnitude lower than the electron energy.

We consider first the spectral distribution of the energy loss.
The probability of the successive radiation of $n$ soft photons
with energies \newline $\omega_1, \omega_2, \ldots \omega_n$ by a
particle with energy $\varepsilon~(\omega_k \ll \varepsilon, k=1,2
\dots n)$ on the length $l$ in the energy intervals $d\omega_1
d\omega_2 \ldots d\omega_n$ is given by expression
\begin{eqnarray}
&& dw(\omega_1, \omega_2, \ldots \omega_n)=
A\int_{0}^{l}dW(\omega_1)dl_1 \int_{0}^{l_1}dW(\omega_2)dl_2 \ldots
\int_{0}^{l_{n-1}}dW(\omega_n)dl_n \nonumber \\
&&=\frac{A}{n!} dw(\omega_1)dw(\omega_2) \ldots dw(\omega_n),
\label{y1}
\end{eqnarray}
where $A$ is the normalization constant, $dW(\omega)/d\omega$ is
the differential probability of the photon radiation per unit
length, $dw(\omega)/d\omega=ldW(\omega)/d\omega$ is the
differential probability of the photon radiation per length $l$.
If the probability $dw/d\omega$ doesn't depend on the particle
energy $\varepsilon$ then integrating Eq.(\ref{y1}) over all
photon energies we obtain
\begin{equation}
w_n=\frac{A}{n!}w^n,\quad w=\int_{}^{}\frac{dw}{d\omega}d\omega.
\label{y2}
\end{equation}
The value $A$ is defined by the condition that probability of all
the possible events with radiation of any number of photons or without
photon radiation is equal to unit.
\begin{equation}
\sum_{n=0}^{\infty}w_n=A\sum_{n=0}^{\infty}\frac{1}{n!}w^n=A\exp w=1,\quad
A=\exp(-w).
\label{y3}
\end{equation}
Using Eqs.(\ref{y1}),(\ref{y3}) we can write the expression for
the differential distribution of the energy loss in the form
\begin{equation}
\frac{1}{\omega}\frac{d\varepsilon}{d\omega}=
\sum_{n=1}^{\infty}\frac{1}{n!}\exp (-w)\int_{}^{}\frac{dw}{d\omega_1}
\frac{dw}{d\omega_2} \cdots \frac{dw}{d\omega_n}
\delta\left(\sum_{k=1}^{n}\omega_k-\omega\right)d\omega_1 \ldots d\omega_n.
\label{y4}
\end{equation}
Here on the right-hand side we have the sum of the probabilities
of radiation of $n$ photons with summary energy $\omega$.

Using standard parametrization of $\delta$-function
\[
\delta\left(\sum_{k=1}^{n}\omega_k-\omega\right)
=\frac{1}{2\pi}\int_{-\infty}^{\infty}
\exp \left(is\left(\omega-\sum_{k=1}^{n}\omega_k\right)\right)ds
\]
we obtain
\begin{eqnarray}
&& \frac{d\varepsilon}{d\omega}=\frac{\omega}{2\pi}
\int_{-\infty}^{\infty}\exp\left(is\omega-w \right)
\sum_{1}^{\infty}\frac{1}{n!}\left(\int_{}^{}\frac{dw}{d\omega_1}
\exp (-is\omega_1)d\omega_1 \right)^n ds \nonumber \\
&& =\frac{\omega}{2\pi}
\int_{-\infty}^{\infty}\exp\left(is\omega \right)
\exp\left\{-\int_{0}^{\infty}\frac{dw}{d\omega_1}
\left[1-\exp(-is\omega_1) \right] d\omega_1\right\}ds \nonumber \\
&& =\frac{1}{\pi}~{\rm Re} \int_{0}^{\infty}\exp\left(ix \right)
\exp\left\{-\int_{0}^{\infty}\frac{dw}{d\omega_1}
\left[1-\exp\left(-ix\frac{\omega_1}{\omega}\right)\right]
d\omega_1\right\}dx
\label{y5}
\end{eqnarray}
The formula (\ref{y5}) was derived by Landau \cite{y9a} (see also
\cite{1}) as solution of the kinetic equation under assumption
that energy loss are much smaller than particle's energy (the
paper \cite{y9a} was devoted to the ionization energy loss). Let
us notice the following. The energy loss are defined by the hard
part of the radiation spectrum. In the soft part of the energy
loss spectrum Eq.(\ref{y5}) the probability of radiation of one
hard photon only is taken into account accurately. To calculate
the probability of the emission of two and more hard photons one
has to take into account step by step the recoil in previous acts
of the photon emission. The probability of radiation of two and
more hard photons is of the order $(l/L_{rad})^2$ and so on.
Thus, the formula (\ref{y5}) is applicable for the thin targets
and has accuracy $l/L_{rad}$. If we want to improve accuracy of
Eq.(\ref{y5}) and for the case of thick targets $l \geq L_{rad}$
one has to consider radiation of an arbitrary number of hard
photons. This problem is solved in Appendix of \cite{15} for the
case when hard part of the radiation spectrum is described by the
Bethe-Maximon formula. In this case the Eq.(\ref{y5}) acquires
the additional factor. As a result we extend this formula on the
case thick targets.

We will analyze first the interval of photon energies where the
Bethe-Maximon formula is valid. We write it in the form (within
the logarithmic accuracy)
\begin{equation}
\omega \frac{dw}{d\omega}=\frac{l}{L_{rad}}\left[\frac{4}{3}
\left(1-\frac{\omega}{\varepsilon} \right)+\frac{\omega^2}{\varepsilon^2}
\right] \vartheta(\varepsilon-\omega)=
\beta\left(1-x_1+ \frac{3}{4}x_1^2 \right)\vartheta(1-x_1),
\label{y7}
\end{equation}
where $l$ is the thickness of the target, $L_{rad}$ is the radiation
length,
\[
\beta=\frac{4l}{3L_{rad}},
\quad x_1=\frac{\omega}{\varepsilon}.
\]
The integral in the curly brackets in Eq.(\ref{y5}) is
\begin{equation}
J \simeq \beta \left(\ln x+\mu+i\frac{\pi}{2} \right),\quad
\mu=\ln \frac{\varepsilon}{\omega}-\frac{5}{8}+C
\label{y10}
\end{equation}
Substituting this result into Eq.(\ref{y5}) we have
\begin{equation}
\frac{d\varepsilon}{d\omega}=
\beta \frac{\exp (-\beta \mu)}{\Gamma(1+\beta)}
\label{y12}
\end{equation}
where $\Gamma(z)$ is the Euler gamma function. If we consider
radiation of the one soft photon, we have from Eq.(\ref{y5})
$d\varepsilon/d\omega=\beta$. Thus, the formula (\ref{y12}) gives
additional "reduction factor" $f_{BH}$ which characterizes the
distortion of the soft Bethe-Maximon spectrum due to multiple
photons radiation. The derivation of Eq.(\ref{y12}) is based on
expression (\ref{y5}) in which the statistical independence of
photon radiation acts is assumed (the Poisson distribution). This
statistical independence is broken when two or more hard photons
(energies of which are of the order of the energy of the initial
electron) are emitted. Since the probability of hard photon
radiation is of the order of $\beta$, Eq.(\ref{y12}) is
applicable, strictly speaking, at $\beta \ll 1$ and it has the
accuracy up to factor ($1+O(\beta^2)$). However, the contribution
of arbitrary number of hard photons radiated in the Beth-Maximon
region is exactly calculated. This contribution is contained in
the factor $g(\beta)$ (see Appendix of \cite{15}, Eqs.(A.15),
(A.16)). So we obtain the reduction factor valid for $\beta \geq
1$
\begin{equation}
f_{BH}=g(\beta)\frac{\exp (-\beta \mu)}{\Gamma(1+\beta)}
=\left(\frac{\omega}{\varepsilon}\right)^{\beta}(1+\beta)^{1/4}
\left(1+\frac{\beta}{2}\right)^{3/4}.
\label{y13}
\end{equation}
In Fig. 9 the function $f_{BH}(\omega)$ is given for electron with
the energy $\varepsilon$ = 25~GeV. One can see that for $\omega$
= 100~MeV the difference between the value $f_{BH}(\omega)$ and
unit is 5-6 times greater than value $\beta$. Let us discuss this
circumstance. The emission of accompanying photons with energy
much less or of the order of $\omega$ (we consider in this figure
the situation connected with SLAC experiment \cite{10}-\cite{12}
where $\omega \ll \varepsilon,~\beta \ll 1 $) changes the
spectral distribution on quantity order of $\beta$. However, if
one photon with energy $\omega_r > \omega$ is emitted, at least,
then photon with energy $\omega$ is not registered at all in the
corresponding channel of the calorimeter. Since mean number of
photons with energy larger than $\omega$ is determined by the
expression (see Eqs.(\ref{y1})-(\ref{y5}))
\begin{equation}
\sum_{n=0}^{\infty} n w_n =w_{\omega}=\int_{\omega}^{\varepsilon}
\frac{dw}{d\omega}d\omega,
\label{y13a}
\end{equation}
the probability of the event when no photon with energy $\omega_r
> \omega$ is radiated is defined by $\exp \left(-w_{\omega}
\right)$. This is just the main factor in the expression for the
reduction factor $f_{BH}$ (\ref{y13}). In the case, when
radiation is described by the Bethe-Maximon formula the value
$w_{\omega}$ increases as a logarithm with $\omega$ decrease
($w_{\omega} \simeq \beta \ln \varepsilon/\omega$) and for large
ratio $\varepsilon/\omega$ the value $w_{\omega}$ is much larger
than $\beta$. Thus, amplification of the effect is connected with
a large interval of the integration ($\omega - \varepsilon$) at
evaluation of the radiation probability.

The Bethe-Maximon formula becomes inapplicable for the photon
energies $\omega \leq \omega_c, (\omega_c$ is defined in Eq.
(\ref{2.31b})), where LPM effect starts to manifest itself (see
Eqs.(\ref{2.25}), (\ref{2.41})). Calculating the integral in
Eq.(\ref{y5}) we find for the distribution of the spectral energy
loss
\begin{equation}
\frac{d\varepsilon}{d\omega}=3\beta \sqrt{\frac{\omega}{2\omega_c}}f_{LPM},
\quad f_{LPM} = g(\beta)\left(1+\frac{3\pi}{2\sqrt{2}}\beta
\sqrt{\frac{\omega}{\omega_c}} \right) \exp (-w_c),
\label{y26}
\end{equation}
where $f_{LPM}$ is the reduction factor in the photon energy range
where the LPM effect is essential,
\begin{equation}
w_c=\beta\left(\ln \frac{\varepsilon}{\omega_c}+C_2 \right),\quad
C_2 \simeq 1.959.
\label{y21}
\end{equation}
In this expression the terms $\sim 1/L$ (see Sec.2.1) are not
taken into account. In Fig. 10 the function $f_{LPM}(\omega)$ is
shown for $\omega <$100~MeV ($\varepsilon$ = 25~GeV, $\omega_c$ =
228~MeV). It is seen in Fig. 10 that the reduction factor
$f_{LPM}$ changes appreciably in the region of high photon
energies solely. This is due to the fact that here the total
probability of photon radiation is finite (in contract to the
Bethe-Maximon formula) and the integral which defines this
probability converges at $\omega \rightarrow 0$.

In the above analysis we neglected an influence of the
polarization of a medium on the bremsstrahlung. This is correct if
$\kappa \ll \nu_0$ (see Eqs.(\ref{42}), (\ref{43}),(\ref{2.30b}),
Sec. 2.6) where
\begin{equation}
\kappa=1+\frac{\omega_p^2}{\omega^2} \equiv 1+ \kappa_0^2,\quad
\omega_p=\gamma \omega_0,\quad
\nu_0^2=\frac{\omega_c}{\omega}.
\label{y27}
\end{equation}
We assume, for definiteness, that $\omega_p \ll \omega_c$. This
is true in any case for a dense matter if electron energy
$\varepsilon \geq 10~GeV$. In the opposite case when $\kappa \gg
\nu_0 \gg 1$ there is an additional suppression of the
bremsstrahlung (see Sec. 2.6):
\begin{equation}
\omega \frac{dw}{d\omega} \simeq \frac{3l}{4\kappa L_{rad}}
= \frac{\beta}{\kappa}
\label{y28}
\end{equation}
and this contribution into reduction factor (\ref{y26}) can be neglected.

The main contribution into effect considered for the photon
energies such that $\omega \ll \omega_p$ gives the transition
radiation Eqs.(\ref{51d}), (\ref{z34a}). Let us note, that for
the transition radiation the given above derivation of the
formula (\ref{y5}) is directly inapplicable. However, for
application of formula (\ref{y5}) it is enough that acts of
radiation of soft photons are statistically independent. Taking
integrals in Eq.(\ref{y5}) and taking into account that $\eta \ll
1$ (but it may be that $\eta \sigma \sim 1$) we obtain
\begin{eqnarray}
&& \frac{d\varepsilon}{d\omega}=\frac{g(\beta)}{\pi} \exp(-w_c)
\exp \left[-\eta\left(\sigma^2
-\frac{3\pi^2}{4}+\psi^2(1-2\eta \sigma)+\psi'(1-2\eta \sigma)
\right) \right] \nonumber \\
&& \times \sin \left[2\pi \eta \left(\sigma+\psi(1-2\eta \sigma)
\right)\right] \Gamma(1-2\eta \sigma),
\label{y38}
\end{eqnarray}
where $\psi(z)=d \ln \Gamma(z)/dz$, $g(\beta)$ is defined in
Eq.(\ref{y13}),
\begin{equation}
\eta=\frac{2\alpha}{\pi},\quad \sigma=\ln \frac{\omega_p}{\omega}+C-1.
\label{y34}
\end{equation}
In the case $\eta \sigma \ll 1$ we have
\begin{eqnarray}
&& \frac{d\varepsilon}{d\omega}=
2\eta \left(\ln \frac{\omega_p}{\omega}-1\right)g(\beta) \exp(-w_c) f_{tr},
\nonumber \\
&& f_{tr}=\exp \left[-\eta\left(\ln \frac{\omega_p}{\omega}-1\right)^2
\left(1+\frac{\eta \pi^2}{3} \right)+\frac{\eta \pi^2}{4} \right].
\label{y39}
\end{eqnarray}

The case of a thin target, where the photon formation length
\begin{equation}
l_f = \frac{2\gamma^2}{\omega}\left(\kappa + \nu_0 \right)
\geq l,
\label{y40}
\end{equation}
$l$ is the target thickness, should be analyzed separately. We
consider situation when effects of the polarization of a medium
are weak ($\kappa_0^2 < \nu_0$) In this case the scattering takes
place on a target as a whole during the radiation process and the
spectral probability $dw/d\omega$ in Eqs.(\ref{y2})-(\ref{y5})
depends on the momentum transfer $q=\gamma \vartheta$ (see
Eq.(\ref{81}))
\begin{equation}
\frac{dw_r}{d\omega}=\frac{2\alpha}{\pi} \frac{d\omega}{\omega}
F_2\left(\frac{q}{2}\right) \equiv \beta_1(q)\frac{d\omega}{\omega}.
\label{y41}
\end{equation}
On the final step one have to average $d\varepsilon_r/d\omega$ over
${\bf q}$ with the distribution function $F_s({\bf q})$ Eq.(\ref{82}).

Since the expression for the spectral distribution (\ref{y41}) has
the same form as in the Bethe-Maximon case, one can use Eq.
(\ref{y12}) for calculation of the reduction factor. As a result we
have for the contribution of the thin target region
\begin{equation}
f_{th}=\frac{1}{\left<\beta_1(q)\right>} \left<\beta_1(q)
\exp \left[-\beta_1(q) \ln \frac{\omega_{th}}{\omega}\right]\right>
\simeq 1-\frac{2\alpha}{\pi}
\displaystyle{\frac{\left<F_2^2\left(\frac{q}{2}\right) \right>}
{\left<F_2\left(\frac{q}{2}\right) \right>}\ln \frac{\omega_{th}}{\omega}},
\label{y42}
\end{equation}
where $\omega_{th}$ is the boundary energy starting from which the target
becomes thin. The lower limit of the applicability of Eq.(\ref{y42}) is
determined from the condition $l_f=l$. The expression (\ref{y42})
depends essentially on the mean square of the momentum transfer $q_s^2$
associated with the target thickness. In the case $q_s^2 \ll 1$ when
an influence of the multiple scattering is weak
($\vartheta_s^2 \ll 1/\gamma^2$)
\begin{equation}
F_s(q)=\frac{4Z^2\alpha^2}{\left(q^2+\alpha_{s}^2 \right)^2}\frac{nl}{m^2},
\quad \alpha_s=\frac{\lambda_c}{a_{s2}},
\label{y43}
\end{equation}
where $a_{s2}$ is the effective screening radius  Eq.(\ref{2.18}).
When the multiple scattering is strong ($q_s^2 \gg 1,~
\vartheta_s^2 \gg 1/\gamma^2$), the reduction factor can be calculated
using the Gaussian distribution for $F_s(q)$
\begin{equation}
F_s(q)=\frac{1}{\pi q_s^2} \exp \left(-\frac{q^2}{q_s^2} \right),~
q_s^2=\frac{4\pi Z^2\alpha^2}{m^2} nl \int_{0}^{q_s^2+1}
\frac{q^2dq^2}{\left(q^2+\alpha_{s}^2 \right)^2} \simeq
\frac{4\pi Z^2\alpha^2}{m^2} nl \ln \frac{q_s^2+1}{\alpha_s^2}.
\label{y44}
\end{equation}
In the first case ($q_s^2 \ll 1$) one has $\left<F_2^2 \right> \ll
\left<F_2 \right>$ and one can neglect the correction to unit in
Eq.(\ref{y42}). In the opposite case ($q_s^2 \gg 1$) the main
contribution into entering mean values of $F_2$ gives values $q
\gg 1$ where
\begin{equation}
F_2(q) \simeq \ln q^2 - 1 + \frac{2}{q^2}.
\label{y45}
\end{equation}
In this case the effect under consideration may be noticeable.

From the above analysis one can find an approximate expression
for the reduction factor in the general form. One can present the
integral over $\omega_1$ in Eq.(\ref{y5}) as
\begin{equation}
\int_{0}^{\infty}\frac{dw}{d\omega_1}
\left[\exp\left(-ix\frac{\omega_1}{\omega}\right)-1 \right] d\omega_1
\simeq -\int_{\omega}^{\infty}\frac{dw}{d\omega_1}d\omega_1
+\int_{0}^{\omega}\frac{dw}{d\omega_1}
\left[\exp\left(-ix\frac{\omega_1}{\omega}\right)-1 \right] d\omega_1.
\label{y3.1}
\end{equation}
The second integral on the right-hand side of Eq.(\ref{y3.1}) (as
well as terms omitted in the first integral) has an order $\omega
dw/d\omega$. For the case $\beta \ll 1$ this value is small
\begin{eqnarray}
&& \omega \frac{dw}{d\omega} \leq
\frac{4l}{3L_{rad}} \equiv \beta \quad (\omega \geq \omega_p);
\nonumber \\
&& \omega \frac{dw}{d\omega} \simeq \frac{4\alpha}{\pi}
\left(\ln \frac{\omega_p}{\omega}-1 \right)=2\eta(\sigma-C),\quad
\eta=\frac{2\alpha}{\pi} \quad (\omega \ll \omega_p).
\label{y3.2}
\end{eqnarray}
Let us note that a contribution of the transition radiation can be
enlarged $n$ times if one makes a target as a collection of $n$ plates
conserving the total thickness $l$ provided that definite conditions are
fulfilled for the plate thicknesses and gaps between plates. In that
case $\eta \rightarrow n \eta$ and above formulae are valid if
$n \eta \leq 1$.

Expanding the exponential in Eq.(\ref{y5}) with the second
integral of Eq.(\ref{y3.1}) and integrating over $x$ we have
\begin{eqnarray}
&& \frac{1}{2\pi} \int_{-\infty}^{\infty}\exp (ix) dx
\int_{0}^{\omega}\frac{dw}{d\omega_1}
\left[\exp\left(-ix\frac{\omega_1}{\omega}\right)-1 \right] d\omega_1
\nonumber \\
&&=\int_{0}^{\omega}\frac{dw}{d\omega_1}\delta\left(1-\frac{\omega_1}
{\omega} \right)d\omega_1 = \omega \frac{dw}{d\omega}
\label{y3.4}
\end{eqnarray}
Thus, the spectral distribution of the energy loss and reduction
factor $f$ have the following general form
\begin{equation}
\frac{d\varepsilon}{d\omega}=\omega \frac{dw}{d\omega} f,\quad
f= \exp \left[-\int_{\omega}^{\infty}\frac{dw}{d\omega_1}d\omega_1 \right]
\left(1+O\left(\omega \frac{dw}{d\omega} \right) \right)
\label{y3.5}
\end{equation}

\subsection{Radiation from structured target}

The radiation from several plates for the relatively hard part of
the spectrum in which the bremsstrahlung in the condition of the
strong LPM effect dominates was investigated recently in
\cite{x16b}. A rather curious interference pattern in the
spectrum of the radiation was found which depends on a number
(and a thickness) of plates and the distance between plates (the
polarization of a medium was neglected). In the present subsection
the probability of radiation in a radiator consisting of $N$
plates is considered \cite{x15}. The transition radiation
dominates in the soft part of the considered spectrum, while the
bremsstrahlung under influence of the strong LPM effect dominates
in the hard part. The intermediate region of the photon energies
where contributions of the both mentioned mechanisms are of the
same order is of evident interest. We analyze this region in
detail. In this region effects of the polarization of the medium
are essential.

Here we consider the case when the target consists of N identical
plates of thickness $l_1$ with the equal gaps $l_2$ between them.
The case $l_1 \ll l_f$ will be analyzed where $l_f$ is the
characteristic formation length of radiation in absence of a
matter ($\omega_0 = 0$), see Eqs.(\ref{4})-(\ref{6})
\begin{equation}
l_f=\frac{l_{f0}}{1+\gamma^2\vartheta_c^2}=\frac{l_0}{1+p_c^2},\quad
T_{1,2}=\frac{l_{1,2}}{l_{f0}},\quad \frac{l_1}{l_f}=(1+p_c^2)T_1 \ll 1,
\quad T_1+T_2=T,
\label{x5}
\end{equation}
where $\vartheta_c$ is the characteristic angle of radiation.

One can split the spectral distribution of the probability of
radiation into two parts
\begin{equation}
 \frac{dw}{d\omega} = \frac{dw_{br}}{d\omega} + \frac{dw_{tr}}{d\omega},
\label{x23}
\end{equation}
where $dw_{br}/d\omega$ is the spectral distribution of the
probability of bremsstrahlung with allowance for the multiple
scattering and polarization of a medium and $dw_{tr}/d\omega$ is
the probability of the transition radiation
for the $N$-plate target.
For $dw_{br}^{(N)}/d\omega$ one has:
\begin{eqnarray}
&& \frac{dw_{br}^{(N)}}{d\omega}= \frac{\alpha}{\pi \omega}{\rm Re}
\sum_{n_1,n_2=0}^{N-1}\int_{}^{}d\tau_2
\int_{}^{}d\tau_1 \exp \left[-i\left(\tau_1+\tau_2+\kappa T_1+
(n-1)\overline{\kappa}T \right) \right]
\nonumber \\
&& \times \left[r_1 \left(G_n-G_n(0) \right)+
r_2 \left(G_n^2-G_n^2(0)  \right)  \right],
\label{x27}
\end{eqnarray}
where
\begin{eqnarray}
&& G_n^{-1}= \beta_n\left[\left(d_n-d_{n-1} \right)^2-1 \right]+
i\left(d_n-d_{n-1} \right)\left(\tau_1+\tau_2\right)-
\beta_n^{-1}\tau_1\tau_2;
\nonumber \\
&& N \geq n=n_2-n_1+1 \geq 1;\quad n_1=0,~0 \leq \tau_1 < \infty; \quad
n_2=N-1,~0 \leq \tau_2 < \infty; \nonumber \\
&&  n_1 \geq 1,~ ~0 \leq \tau_1 \leq T;\quad
n_2 \leq N-2,~0 \leq \tau_2 \leq T,
\nonumber \\
&&\overline{\kappa}=1+\overline{\kappa_0^2},~
\overline{\kappa_0^2}=\kappa_0^2 \frac{l_1}{l_1+l_2}
\label{x28}
\end{eqnarray}
Note for the subtraction procedure one has that when
$q \rightarrow 0$, the function $b,~\beta_n \rightarrow \infty$.
The function $d_n$ can be presented in the form
\begin{equation}
d_n=\frac{\sinh n\eta}{\sinh \eta},\quad
\sinh \eta = 2\sqrt{iqT_1T\left(1+iqT_1T \right)}.
\label{x18}
\end{equation}
For the case $\eta \ll 1$ one has
\begin{equation}
\eta \simeq 2\sqrt{iqT_1T} \equiv 2T\sqrt{i\overline{q}},\quad
\overline{q}=q\frac{T_1}{T}=q\frac{l_1}{l_1+l_2}.
\label{x19}
\end{equation}
In above formulae
\begin{equation}
T=T_1+T_2,\quad b=\frac{1}{4qT_1},\quad \beta_n=\frac{b}{d_n},\quad
G_n(0)=G_n(q=0)
\label{xd1}
\end{equation}

For one plate ($N=1$) we have $n_1=n_2=0,~n=1$ and
\begin{equation}
G_1^{-1}=i(\tau_1+\tau_2)-\frac{1}{b}\tau_1\tau_2,\quad
G_1^{-1}(0)=i(\tau_1+\tau_2).
\label{x29}
\end{equation}
In the integral (\ref{x27}) we rotate the integration contours over
$\tau_1,~\tau_2$ on the angle $-\pi/2$ and substitute variables
$\tau_{1,2} \rightarrow -ix_{1,2}$. Then we carry out change of
variables $x=x_1+x_2,~x_2=zx$. We have
\begin{eqnarray}
\hspace{-8mm}&& \frac{dw_{br}^{(1)}}{d\omega}=
\frac{\alpha}{\pi \omega} \cos \kappa T_1
\int_{0}^{\infty} \exp (-x)dx\int_{0}^{1}\left[r_1\left(1-
\frac{1}{g(x,z)} \right)+r_2\left(1-\frac{1}{g^2(x,z)} \right) \right]dz
\nonumber \\
\hspace{-8mm}&& = \frac{\alpha b}{\pi \omega}
\cos (\kappa T_1) \int_{0}^{\infty}
\exp (-bx) \left[r_1F_1\left(\sqrt{\frac{x}{4}} \right) +
r_2F_2\left(\sqrt{\frac{x}{4}} \right)\right],\nonumber \\
&& g(x,z)=1+\frac{x}{b}z(1-z),
\label{x30}
\end{eqnarray}
where the functions $F_1(x)$ and $F_2(x)$ are defined in (\ref{4.18sc})
and (\ref{4.17sc}).

The special case is when a target consists of two plates ($N=2$).
Since the formation length is enough long for the soft photons
($\omega \ll \varepsilon$) only, we consider the term with $r_2$
in Eq.(\ref{x27}) as far as $r_1=\omega^2/\varepsilon^2 \ll 1$.
For the case ($N=2$) the sum in Eq.(\ref{x27}) consists of three
terms: $1) n_1=n_2=0;~2) n_1=n_2=1;~3) n_1=0, n_2=1$. For the two
first $n=1$ and we have from Eq.(\ref{x29})
\begin{eqnarray}
&& \frac{dw_{br1}^{(2)}}{d\omega}= \frac{dw_{br2}^{(2)}}{d\omega}=
\frac{\alpha r_2}{\pi \omega}~ {\rm Re}~\Bigg[ \exp (-i\kappa T_1)
\int_{0}^{\infty} d\tau_2\int_{0}^{T} \exp (-i(\tau_1+\tau_2))
\nonumber \\
&& \times \left[\frac{1}{(\tau_1+\tau_2)^2}-
\frac{1}{(\tau_1+\tau_2+i\tau_1\tau_2/b)^2} \right]d\tau_1 \Bigg].
\label{x31}
\end{eqnarray}
For the third term $n=2,~d_2=2\cosh \eta=2(1+2iqT_1T)$ (see
Eq.(\ref{x18})) and we have
\begin{eqnarray}
&& G_2^{-1}=iT+i\left(1+\frac{iT}{b} \right)(\tau_1+\tau_2)
-\frac{2}{b}\left(1+\frac{iT}{2b} \right)\tau_1\tau_2
\nonumber \\
&& = i(T+\tau_1+\tau_2)-\frac{2}{b}\tau_1\tau_2 +\frac{iT}{b}
\left[i(\tau_1+\tau_2)-\frac{\tau_1\tau_2}{b} \right],
\label{x32}
\end{eqnarray}
so that
\begin{eqnarray}
\hspace{-14mm}&& \frac{dw_{br3}^{(2)}}{d\omega}=
\frac{\alpha r_2}{\pi \omega}~ {\rm Re}
~\Bigg[ \exp (-i(\overline{\kappa} T+\kappa T_1))
\int_{0}^{\infty} dx_1\int_{0}^{\infty} \exp (-(x_1+x_2))
\nonumber \\
\hspace{-14mm}&& \times \left[\frac{1}{(iT+x_1+x_2)^2}-
\frac{1}{\left[iT+x_1+x_2+\frac{2}{b}x_1x_2 +\frac{iT}{b}
\left(x_1+x_2+\frac{x_1x_2}{b}  \right)\right]^2} \right]dx_2 \Bigg].
\label{x33}
\end{eqnarray}
Here we rotate the integration contours over
$\tau_1,~\tau_2$ on the angle $-\pi/2$ and substitute variables
$\tau_{1,2} \rightarrow -ix_{1,2}$.

In the case of the weak multiple scattering ($b \gg 1$)
neglecting the effect of the polarization of a medium ($\kappa =
1$) and expanding the integrand in Eqs.(\ref{x31}) and (\ref{x33})
over $1/b$ we have for the probability of radiation
\begin{eqnarray}
&& \frac{dw_{br}^{(2)}}{d\omega}=2 \frac{dw_{br1}^{(2)}}{d\omega}+
\frac{dw_{br3}^{(2)}}{d\omega} \simeq \frac{2\alpha r_2}{3\pi \omega b}~
\left[1-\frac{3}{10b}+\frac{1}{b}G(T) \right],
\nonumber \\
&& G(T)=T^2 \int_{0}^{\infty} \frac{x^3 \exp (-x)}{(x^2+T^2)^2}
\Bigg[\left(1-\frac{3x^2}{10T^2} \right)
\cos\left(T-4\arctan\frac{x}{T} \right)
\nonumber \\
&& +\frac{2x}{T}
\sin\left(T-4\arctan\frac{x}{T} \right) \Bigg].
\label{x34}
\end{eqnarray}
Note that the main term of the decomposition in Eq.(\ref{x34}) is
the Bethe-Maximon probability of radiation from two plates which is
independent of the distance between plates. This means that in the case
considered  we have independent radiation from each plate
without interference in the main order over $1/b$. The interference effects
appear only in the next orders over $1/b$.

In the case of the strong multiple scattering we have $b \ll 1,~p_c^2 \gg 1$.
When the formation length is much longer than the target thickness
as a whole ($T \ll b$) we find
for the probability of radiation
\begin{equation}
\frac{dw_{br}^{(2)}}{d\omega} \simeq \frac{\alpha r_2}{\pi \omega}
\left[(1+b)\left(\ln \frac{2}{b}+1- C\right)-2 \right] \cos (2\kappa T_1)
\label{x40}
\end{equation}
In the opposite case $b \ll T \ll 1$
neglecting the effect of the polarization of a medium
($\kappa T_1 \ll 1$) we have
\begin{equation}
\frac{dw_{br}^{(2)}}{d\omega} \simeq \frac{\alpha r_2}{\pi \omega}
\left[(1+2b)\left(\ln \frac{T}{b^2}+1- C\right)-2 \right]
\label{x41}
\end{equation}

The case of the strong multiple scattering ($b \ll 1$) for the
photon energies where the value $T \geq 1$ is of the special interest.
In this case we can neglect the polarization of a medium
$\kappa=\overline{\kappa}=1$, and disregard the terms $\propto \kappa T_1$
in the exponent of the expressions (\ref{x31}) and (\ref{x33}) since
$T_1 \ll 1$. We obtain as a result
\begin{eqnarray}
&& \frac{dw_{br}^{(2)}}{d\omega}=2\frac{dw_{br1}^{(2)}}{d\omega}+
\frac{dw_{br3}^{(2)}}{d\omega} \simeq 2\frac{dw_{br}^{(2)}}{d\omega}-
\frac{dw_{br3}^{(2)}}{d\omega},\quad \frac{dw_{br3}^{(2)}}{d\omega}=
\frac{\alpha r_2}{\pi \omega}~F(T),
\nonumber \\
&&{\rm Re}~F(T)=\int_{0}^{\infty}d\tau_2 \int_{T}^{\infty} d\tau_1
\frac{1}{(\tau_1+\tau_2)^2} \exp(-i(\tau_1+\tau_2))=
\int_{T}^{\infty} \frac{d\tau}{\tau^2}\exp(-i\tau) (\tau-T)
\nonumber \\
&&=-({\rm ci}(T) + T{\rm si}(T) + \cos T),
\label{x42a}
\end{eqnarray}
where ${\rm si}(z)$ is the integral sine and ${\rm ci}(z)$
is the integral cosine.
At $T \gg 1$ we have
\begin{equation}
F(T) = -\frac{\cos T}{T^2}.
\label{x42b}
\end{equation}

Let us turn to the case of large $N$ ($N \geq 3$). If the
formation length of the bremsstrahlung is shorter than the
distance between plates ($T > 1$) the interference of the
radiation from neighboring plates takes place. Using the
probability of radiation from two plates (\ref{x34}) we obtain in
the case of weak multiple scattering ($b \gg 1$)
\begin{equation}
\frac{dw_{br}^{(N)}}{d\omega}
\simeq \frac{N\alpha r_2}{3\pi \omega b}~
\left[1-\frac{3}{10b}+2\frac{N-1}{Nb}G(T) \right],
\label{x43}
\end{equation}
where for $T \gg 1$ the function $G(T)$ is defined in
Eq.(\ref{x34}). In the case of strong multiple scattering ($b \ll
1$) and large $T$ we have (compare with Eqs.(\ref{x42a}) and
(\ref{x42b}))
\begin{equation}
\frac{dw_{br}^{(N)}}{d\omega}
\simeq N\left(\frac{dw_{br}^{(1)}}{d\omega} +
\frac{\alpha r_2}{\pi \omega b}~
\frac{N-1}{Nb}\frac{\cos T}{T^2} \right).
\label{x44}
\end{equation}

The behavior of the spectral distribution $\displaystyle{\omega
\frac{dw}{d\omega}}$ can be discussed using as an example the case
of two plates with the thickness $l_1$ and the distance between
plates $l_2 \geq l_1$ which was analyzed in detail above. For
plates with the thickness $l_1 \geq 0.2\% L_{rad}$ and in the
energy interval $\omega > \omega_p$, in which the effects of the
polarization of a medium can be discarded, the condition
(\ref{x5}) is fulfilled only for enough high energy $\varepsilon$
so that $\omega_p \ll \omega_c$ (see Eqs.(\ref{5}),
(\ref{2.31b})). We study the situation when the LPM suppression
of the intensity of radiation takes place for relatively soft
energies of photons: $\omega \leq \omega_c \ll \varepsilon$.

For the hard photons $\omega_c \ll \omega < \varepsilon$ the
formation length $l_{f0}$ Eq.(\ref{6}) is much shorter than the
plate thickness $l_1~(T_1 \gg 1)$, the radiation intensity is the
incoherent sum of radiation from two plates and it is independent
of the distance between plates. In this interval the
Bethe-Maximon formula is valid.

For $\omega \leq \omega_c$ the LPM effect turns on, but when
$\omega = \omega_c$ the thickness of plate is still larger
than the formation length $l_0$
\begin{equation}
\frac{l_1}{l_0(\omega_c)} = T_1(\omega_c) \equiv T_c
\simeq \frac{2\pi}{\alpha}\frac{l_1}{L_{rad}} > 1,
\label{x47}
\end{equation}
so that the formation of radiation takes place mainly inside each of plates.
With $\omega$ decreasing we get over to the region where
the formation length $l_c > l_1$, but effects of the polarization
of a medium are still weak ($\omega > \omega_p$).
Within this interval (for $\omega < \omega_{th}$) the main condition
(\ref{x5}) is fulfilled. To estimate the value $\omega_{th}$ we have to take
into account the characteristic radiation angles ($p_c^2$ in Eq.(\ref{x5})),
connected with mean square angle of the multiple scattering. Using
the definition of the parameter $b=1/(4qT_1)$ in
Eq.(\ref{xd1})) we have
\begin{eqnarray}
&& p_c^2 \leq \frac{1}{b}=\frac{2\pi}{\alpha}\frac{l_1}{L_{rad}}
\left(1-\frac{\ln \varrho_b}{\ln (a_{s2}/\lambda_c)} \right)
\simeq T_c \left(1+\frac{\ln T_c}{2\ln (a_{s2}/\lambda_c)} \right);
\nonumber \\
&& \omega_{th}=\frac{\omega_c}{T_c\left(1+p_c^2 \right)}
\geq \omega_b=\frac{\omega_c}{T_c(1+T_c)},
\label{x48}
\end{eqnarray}
where the parameter $\varrho_b$ is defined by a set of equations (we
rearranged terms in
(\ref{73})):
\begin{eqnarray}
&& \varrho_b =1,\quad L_b=L_1\quad {\rm for}\quad 4QL_1T_1=\frac{2\pi}{\alpha}
\frac{l_1}{L_{rad}} \leq 1,\quad q=QL_1;
\nonumber \\
&& 4QL(\varrho_b)T_1\varrho_b^2=\frac{2\pi}{\alpha}
\frac{l_1}{L_{rad}} \varrho_b^2\left(1-\frac{\ln \varrho_b^2}{L_1} \right)
=1,\quad {\rm for}\quad 4QL_1T_1 > 1,
\label{x13}
\end{eqnarray}
here $L_1=\ln (a_{s2}/\lambda_c)^2$. One can show (see discussion
after Eq.(3.6) in \cite{14}) that $\omega_{th} \simeq 4\omega_b$.
Naturally, $\omega_{th} < \omega_c/T_c$, and when $\omega
=\omega_c/T_c$ one has $T_1=1,~ l_1=l_0$. It is seen from
Eq.(\ref{x48}) that when the value $l_1$ decreases, the region of
applicability of results of this subsection grows.

So, when $\omega < \omega_{th}$ the formation length is longer
than the thickness of the plate $l_1$ and the coherent effects
depending on the distance between plates $l_2$ turn on.
For the description of these effects for $T=\left(1+l_2/l_1 \right) T_1
\geq 1$ one can use Eq.(\ref{x42a}). For $T \geq \pi \gg 1$
one can use the asymptotic expansion (\ref{x42b}) and it is seen
that at $T=\pi$ the spectral curve has minimum. Let us note an accuracy
of formulas is better when $\omega$ decreases, and the description is
more accurate for $T \gg T_1~(l_2 \gg l_1)$.

With further decreasing of the photon energy $\omega$ the value $T$
diminishes and the spectral curve grows until $T \sim 1$. When $T < 1$
the spectral curve decreases $\propto \ln T$
according with Eq.(\ref{x41}). So, the
spectral curve has maximum for $T \sim 1$.
The mentioned decreasing continues until the photon energy $\omega$
for which $(1+2/b)T \sim 1$. For smaller $\omega$ the thickness of the target
is shorter than the formation length.

The results of numerical calculations are given in Fig. 11. The
formulas (\ref{x31}), (\ref{x33}) were used. The spectral curves
of energy loss were obtained for the case of two gold plates with
the thickness $l_1=11.5~\mu m$ with different gaps $l_2$ between
plates. The initial energy of electrons is $25$~GeV. The
characteristic parameters for this case are:
\begin{eqnarray}
&& \omega_c \simeq 240~{\rm MeV},\quad T_c \simeq 2.9,\quad b^{-1} \simeq 3.3,
\quad \omega_{th} \simeq 80~{\rm MeV},
\nonumber \\
&& \omega_{p} \simeq 3.9~{\rm MeV},\quad \omega_{1} \simeq 30~{\rm keV},
\quad \frac{T}{T_1} \equiv k =\frac{l_1+l_2}{l_1} = 3,5,7,9,11.
\label{x58}
\end{eqnarray}
At $\omega > 80~$MeV the radiation process occurs independently
from each plate according with theory of the LPM effect given
above. The interference pattern appears at $\omega < 80~$MeV
where the formation length is longer than the thickness of one
plate and the radiation process depends on the distance between
plates $T$. According to Eqs.(\ref{x42b}), (\ref{x43}) the curves
1-5  have minimums at $\omega \simeq \pi \omega_{th}/k~(T=\pi)$
which are outside of Fig. 11 and will be discussed below. In
accord with the above analysis the spectral curves in Fig. 11 have
the maximums at photon energies $\omega \simeq
\omega_{th}/k~(T=1)$. These values (in MeV) are $\omega \simeq
27, 16, 11, 9, 7$ for curves $1, 2, 3, 4, 5$ respectively. At
further decrease of $\omega(T)$ the spectral curves diminish
according to Eq.((\ref{x41})) and attain the minimum at
$\omega_{min} = \omega_{th}/(k(1+2/b))~(T(1+2/b)=1)$. The
corresponding values (in MeV) are $\omega \simeq 3.5, 2, 1.4, 1.2$
for curves $1, 2, 3, 4$. The least value of $\omega_{min} \simeq
1~$MeV has the curve $5$. However, one has to take into account
that at $\omega \leq 1.5~$MeV ($\kappa_0^2=p_c^2 \simeq 2/b$) the
contribution of the transition radiation becomes significant.
Starting from $\omega \leq 0.6~$MeV the contribution of the
transition radiation dominates.

The approach presented in this subsection is applicable in the
interval of photon energies where effects of the polarization of
a medium are essential. It includes also the soft part of the LPM
effect. For the case given in Fig. 11 our results are given up to
$\omega_{max} \sim 20~$MeV. On the other hand, in \cite{x16b} the
hard part of the LPM effect spectrum was analyzed where one can
neglect the effects of the polarization of a medium ($\omega >
5~$MeV for the mentioned case). Although here (as in \cite{x15})
in and in \cite{x16b} the different methods are used, the results
obtained in overlapping regions are in a quite reasonable
agreement among themselves.

It is interesting to discuss behavior of the spectral curves
obtained in \cite{x16b} from the point of view of our results. We
consider the low value $l_1~(T_c$ is not very large) and a
situation when corrections to the value $b$ in Eq.(\ref{x48})
which neglected in \cite{x16b} are less than $20\%$. This leads
to the difference in results less than $10\%$. We concentrate on
the case of gold target with the total thickness
$Nl_1=0.7\%~L_{rad}$. The case $N=1$ where
$T_c=5.8,~b^{-1}=7.3,~\omega_c \simeq 240~$Mev, $\omega_{th}=
4\omega_c/(T_c(1+T_c)) \simeq 24~$MeV was considered in detail in
\cite{x16b} (see Sec. 2). The curves in figures in \cite{x16b}
are normalized on the Bethe-Heitler probability of radiation (no
the Coulomb corrections), i.e. they measured in units $\alpha r_2
T_c/(3\pi)$. In the region where our results are applicable
$\omega < \omega_{th} \simeq 24~$MeV in Fig. 2 of \cite{x16b}
$~(G=0)$ one can see plateau the ordinate of which is $10\%$ less
than calculated according with Eq.(\ref{x30}). The case of two
plates ($T_c \simeq b^{-1} \simeq 3,~\omega_c \simeq
240~$MeV,~$\omega_{th} \simeq \omega_c/T_c = 80~$MeV) is given in
Fig. 3 of \cite{x16b}. The lengths of the gaps are the same as in
our Fig. 11, except $k=9$. The positions and ordinates of the
minimums and the maximums in the characteristic points ($\omega
\simeq \pi \omega_{th}/k$ for minimums and $\omega \simeq
\omega_{th}/k~(T=1)$ for maximums, see above) as well as behavior
of the spectral curves is described quite satisfactory by our
formulas (see e.g. asymptotic Eqs.(\ref{x40})-(\ref{x42b})).

\subsection{Comparison of the results obtained in different papers}

As it was said above, Migdal have calculated the probability of
the bremsstrahlung and pair creation in the logarithmic approximation
\cite{2}.This result, now considered as classic one,
was confirmed in the most of subsequent calculation.

We developed the method of calculation \cite{13, 14, 15, x15}
which goes beyond the logarithmic approximation. The formula
(\ref{2.25}) gives the spectral distribution of the probability
of radiation which coincides formally with the probability
calculated by Migdal (Eq.(49) in \cite{2}). However, here Coulomb
corrections are included into parameter $\nu$ in contrast to
\cite{2}, \cite{3}. As it was noted, the parameter $\varrho_c$
entering into the parameter $\nu$ (see Eqs.(\ref{2.24}),
(\ref{2.16}) and (\ref{2.14})) is defined up to the factor $\sim
1$, what is inherent in the logarithmic approximation. However,
we calculated also the next term of the decomposition over
$v(\mbox{\boldmath$\varrho$})$ (an accuracy up to the "next to
leading logarithm") and this permits to obtain the result which
is independent of the parameter $\varrho_c$. It was shown that
the definition of the parameter $\varrho_c$ minimizes corrections
to Eqs.(\ref{2.25}) and  (\ref{2.10p}) practically for all values
of the parameter $\varrho_c$. It should be emphasized also that
here the Coulomb corrections are included into the parameter
$\nu$ in contrast to \cite{3}. Moreover, the procedure of fine
tuning presented above (see \cite{13}) permits one (after adding
the term $dW_1/d\omega$ (\ref{2.30})) to obtain more accurate
expression for the probability of radiation, which has power
accuracy in the limit of weak multiple scattering, see
(\ref{2.35}), while the probability calculated by Migdal \cite{2}
has logarithmic accuracy only in the mentioned limit. The
interrelation between the main term $dW_c/d\omega$ and the first
correction $dW_1/d\omega$ depends essentially on fine tuning
procedure which was chosen, as it was just said, in a such way
that minimize the value of the first correction. For example, one
can see that the maximal value of $dW_1/d\omega$ is of the order
7~\% of the main term for all four spectral curves shown in the
Fig. 3. This means that in our case there are more rigid
requirements on fine tuning procedure than used by Migdal.

In the approach used by Blankenbecler and Drell (BD) \cite{16}
the scattering process was considered in the second approximation
of the eikonal formalism and an analysis of scattering in medium
was based on the model of a random medium. The radiation was
calculated for all possible paths and averaged. One advantage of
this approach is that it treats naturally finite thickness
targets. There are a few short-comings in the BD approach. The
Gaussian distribution of scattered electrons was obtained by
Migdal \cite{2} as a solution of the Fokker-Planck equation,
while in BD approach this is an input. Actually there are
deviations from the Gaussian distribution, see end of Sec.3.3. The
calculation of the probability $dW/d\omega$ is quite unwieldy in
BD approach and general formula for $dW/d\omega$ (Eq.(5.32) in
\cite{16})is written down in an implicit form only, and the
functions $\eta(z_2,z_1,l)$ and $\lambda(z_2,z_1,l)$ in the
mentioned formula have to be calculated in the each case
separately. This means that comparison of BD probability with
that calculated by Migdal can be done in limiting cases only. In
the case of weak multiple scattering the probability in \cite{16}
coincides with the Bethe-Heitler probability taken with
logarithmic accuracy (just as in Migdal paper \cite{2}). However,
in the limit of the strong LPM effect, where the probability of
radiation is given by Eqs.(\ref{2.25}) and (\ref{2.39}) above and
which coincides with obtained by Migdal Eq.(52) in \cite{2} (if
one omits the Coulomb corrections included into the parameter
$\nu_0$), the probability calculated by BD (the term with $R_2$)
differs from (\ref{2.39}) by factor $\sqrt{3\pi/8}$=1.0854. The
origin of this discrepancy was not analyzed in \cite{16}.

In the paper of R.Baier, Dokshitzer, Mueller, Peigne, and Schiff
(BDMPS) \cite{19} the multiple scattering of high-energy
electrons off a large number of scatterers (presented as a
screened Coulomb potential) is analyzed within eikonal formalism.
Radiation of soft photons only is included using the classical
current approach, although this approach gives no significant
simplification for the problem under consideration. All the
results are obtained within logarithmic accuracy. No general
formula of the type of Eq.(\ref{2.41}) is written down. In the
"Bethe-Heitler limit" the result of \cite{19} does not completely
match the standard Bethe-Heitler formula. Since the method used
consists in direct summation over scatterers it naturally
included the consideration of a target of finite thickness. For
thin target ($T \leq l_{f0}$) the result of \cite{19} is similar
to logarithmic term in Eq.(\ref{77}) where $r_2=2$ for soft
photons. The explicit formula for radiation spectrum is given in
\cite{19} in the limit of strong LPM effect where it is similar to
Eq.(\ref{2.39a}) if $\omega \ll \varepsilon$ and the Coulomb
corrections are omitted.

Note that in Migdal papers \cite{2}, \cite{3} and in our papers
the kinetic equation is the basic element of the used methods
while in the both papers \cite{19} and \cite{16} the
consideration does not included the kinetic equation. Due to this
feature the calculations in the both papers \cite{19} and
\cite{16} are more cumbersome and no general formulas for
radiation spectra are obtained.

A path integral method is applied  successfully for analysis of
multiple scattering (see e.g. \cite{LMNS}). Zakharov \cite{17},
\cite{18} is used this method to study of the LPM effect. Using a
transverse Green function based on a path integral he succeeded in
reduction of the problem to the two dimensional Schr\"{o}dinger
equation, i.e. he re-derived equation a la Eq.(\ref{2.9}), which
was obtained earlier in \cite{8}. The probability of radiation
was calculated (see Eq.(19) in \cite{17}) within logarithmic
accuracy (as in Migdal papers \cite{2} and \cite{3}) although it
is written down in a more cumbersome form. The procedure of fine
tuning so important for Migdal calculation \cite{2} is
oversimplified in \cite{17}. For earlier attempt to do study of
the LPM effect in a such way see \cite{LMNS}. The formalism used
in \cite{17} has the same region of validity as in Migdal's
papers \cite{2}, \cite{3} and in our papers \cite{8}, \cite{13}
and \cite{14} in spite of claims about "rigorous treatment of the
LPM effect". The developed formalism allows naturally to consider
the LPM effect in finite thickness targets \cite{18}. As an
example the calculation of spectral intensity of radiation is
given for electron energy $\varepsilon=25$~GeV and gold target
with thickness $l=0.7\% L_{rad}$. This calculation was carried out
numerically with the complete potential (see Eq.(\ref{2.9}))
outside of scope of logarithmic approximation. The Coulomb
corrections were discarded as well as a multiple photon
radiation. After arbitrary diminishing of calculated value by
7$\%$ it was found very good agreement with a portion of data (in
the region of the LPM drop up to beginning of plateau). Actually
this diminishing should be attributed to Coulomb corrections
found in \cite{13} (for details see \cite{14} and below in Sec.
3.10). More late results of Zakharov see in \cite{Za3}.

\subsection{Qualitative behavior of the spectral intensity of radiation}

We consider the spectral intensity of radiation for the energy of
the initial electrons when the LPM suppression of the intensity of radiation
takes place for relatively soft energies of photons:
$\omega \leq \omega_c \ll \varepsilon$:
\begin{equation}
\displaystyle{\nu_0(\omega_c)=1,\quad \omega_c=\frac{16\pi Z^2 \alpha^2}{m^2}
\gamma^2 n_a \ln \frac{a_{s2}}{\lambda_c}},
\label{100}\end{equation}
see Eqs.(\ref{2.9}), (\ref{2.16}), (\ref{2.32a})-(\ref{2.31b}).
This situation corresponds to
the \newline experimental conditions \cite{10, 11, 12, E2}.

The ratio of the thickness of the target $l$ and the formation length of
radiation $l_{f}(\omega)$ (\ref{5})
is an important characteristics of the process.
In Eq.(\ref{5}) both  the multiple scattering and the polarization
of a medium are taken into account. This ratio may be written as
\begin{eqnarray}
&&\displaystyle{\beta(\omega) \equiv \frac{l}{l_{f}(\omega)}=
T\left(\nu_0+\kappa \right)\simeq T_c\left[
\frac{\omega}{\omega_c}+\sqrt{\frac{\omega}{\omega_c}}+
\frac{\omega_p^2}{\omega \omega_c} \right]},
\nonumber \\
&&\displaystyle{T=\frac{l \omega}{2\gamma^2},\quad
\omega_p=\omega_0 \gamma,\quad
T_c \equiv T(\omega_c)
\simeq \frac{2\pi}{\alpha}\frac{l}{L_{rad}}},
\label{101}
\end{eqnarray}
where we put that $\displaystyle{\nu_0
\simeq \sqrt{\frac{\omega_c}{\omega}}}$.
Below we assume that $\omega_c \gg \omega_p$ which is true under
the experimental conditions.

If $\beta(\omega_c) \gg 1~(T_c \gg 1)$ then at $\omega \geq \omega_c$
a target is thick and one has the LPM suppression for $\omega \leq \omega_c$.
There are two opportunities depending on the minimal value of the parameter
$\beta$:
\begin{equation}
\displaystyle{\beta_m \simeq \frac{3}{2}T_c
\sqrt{\frac{\omega_1}{\omega_c}},\quad
\omega_1=\omega_p\left(\frac{4\omega_p}{\omega_c}\right)^{1/3},\quad
\beta_m \simeq 2T_c\left(\frac{\omega_p}{\omega_c} \right)^{2/3}}.
\label{102}
\end{equation}
If $\beta_m \ll 1$ then for photon energies $\omega > \omega_1$ it will
be $\omega_b$ such that
\begin{equation}
\beta(\omega_b)=1,\quad \omega_b \simeq \frac{\omega_c}{T_c^2}
\label{103}\end{equation} and for $\omega < \omega_b$ the
thickness of a target becomes smaller than the formation length
of radiation so that for $\omega \ll \omega_b$ the spectral
distribution of the radiation intensity is described by Eqs.
(\ref{z24}), (\ref{z30}). Under these conditions for
$4k=\nu_0^2T=T_c \gg 1$ the spectral curve has a plateau
\begin{equation}
\displaystyle{\frac{dI}{d\omega}=\frac{2\alpha J}{\pi}={\rm const}}
\label{103a}
\end{equation}
in accordance with Eq.(\ref{79}). Under conditions $\kappa T \ll
1, \omega < \omega_b$ the spectral intensity of radiation is
independent of photon energy $\omega$. It should be noted that
due to smallness of the coefficients in expression for $\delta$
Eq.(\ref{z26}), such behaviors of the spectral curve begins at
$\omega < \omega_{th} =4\omega_b \simeq 4\omega_c/T_c^2$. It will
continue until photon energies where one has to take into account
the polarization of a medium and connected with it a contribution
of the transition radiation.

At $\beta_m \gg 1$ a target remains thick for all photon energies
and radiation is described by formulae of Sections 2.2 and 2.6.
In this case at $\omega \ll \omega_c~(\nu_0 \gg 1)$ and
$\displaystyle{\omega \gg \left(\omega_p/\omega_c
\right)^{1/3}\omega_p}$ ~$(\nu_0 \gg \kappa)$ the spectral
intensity of radiation formed inside a target is given by
Eqs.(\ref{2.25}) and (\ref{2.30}) and the contribution of the
boundary photons is given by Eq.(\ref{51g}).

It is important to include also the contribution of boundary
photons. Since the contribution into the spectral intensity of
radiation from a passage of the electron inside the target
($\propto T$) is diminishing and a contribution of the boundary
photons is increasing with $\omega$ decrease, the spectral curve
has a minimum at $\omega=\omega_m$. The value of $\omega_m$ may
be estimated from equations (see (\ref{2.39}),
(\ref{z24})-(\ref{z25})))
\begin{eqnarray}
&&\displaystyle{\frac{d}{d\omega}\left(\frac{\nu_0 T}{\sqrt{2}}+\ln \nu_0
+ \frac{\pi^2 \sqrt{2}}{24}\frac{\kappa}{\nu_0} \right)=0,\quad
\frac{\nu_0T}{\sqrt{2}} \simeq 1+\frac{\pi^2 (\kappa-1)}{4\sqrt{2}\nu_0}},
\nonumber \\
&&\displaystyle{T_c \simeq \left(\frac{2 \omega_c}{\omega_p} \right)^{1/2}
\sqrt{x}+\frac{\pi^2}{4}x^2,\quad x=\frac{\omega_p}{\omega}}.
\label{104}
\end{eqnarray}
When the value of $T_c$ is high enough, the solution of
Eq.(\ref{104}) does not satisfy the condition $\nu_0 \gg \kappa$
and in this case the equation (\ref{104}) ceases to be valid. For
determination of $\omega_m$ in this case we use the behavior of
the spectral intensity of radiation at $\kappa \gg \nu_0$. In
this case the contribution into radiation from inside passage of
the target is described by Eq.(\ref{47}) while the radiation of
the boundary photons reduces to the transition radiation and its
contribution is given by Eq.(\ref{51d}). Leaving the dominant
terms ($\nu_0^2T$ is $\omega$ independent) we have
\begin{equation}
\displaystyle{\frac{d}{d\omega}\left(\frac{\nu_0^2T}{3\kappa}
+\ln \kappa \right)=0,\quad \frac{\nu_0^2T}{3\kappa} = 1,\quad
\kappa_m=\frac{T_c}{3},\quad \omega_m \simeq
\sqrt{\frac{3}{T_c}}\omega_p}.
\label{105}
\end{equation}
Since the value $\pi^2/12 \simeq 0.8$ is of the order of unity,
the solution of (\ref{104}) at $\kappa_m \gg \nu_0$ differs only
slightly from $\omega_m$. Because of this, if the condition
$\displaystyle{2T_c(\omega_p/\omega_c)^{2/3} \gg 1}$ is
fulfilled, the position of the minimum is defined by
Eq.(\ref{104}). The formulas (\ref{104}) and (\ref{105}) show the
position of a minimum of the spectral intensity curve which is
the sum of contributions of radiation inside a target and boundary
radiation. Due to the LPM effect the intensity of radiation
inside a target diminishes with $\omega$ decreasing. However at
low $\omega$ the contribution of the boundary radiation is
sharply increasing and this results in appearance of a minimum of
spectral curve. Eq.(\ref{104}) gives its position at small $T_c$
while Eq.(\ref{105}) gives minimum position at large $T_c$.

If $\beta(\omega_c)=2T_c \ll 1$ then at $\omega=\omega_c$ a
target is thin and the Bethe-Maximon spectrum of radiation which
is valid at $\displaystyle{\omega \gg \omega_c
\left(\frac{dI(\omega)}{d\omega}={\rm const}\right)}$ will be
also valid  at $\omega \leq \omega_c$ in accordance with
Eq.(\ref{z30}) since $4k=\nu_0^2T=T_c \ll 1$. This behavior of the
spectral curve will continue with $\omega$ decrease until photon
energies where a contribution of the transition radiation become
essential. In this case the spectral distribution of radiation
has the form (\ref{z35}) for all $\omega$
\begin{equation}
\frac{dw}{d\omega} = \frac{dw_{tr}}{d\omega} +\cos (\kappa T)
\frac{dw_{BM}}{d\omega},
\label{z38}
\end{equation}
Since for soft photons ($\omega \ll \varepsilon$)
\begin{equation}
\frac{dI}{d\omega} = \frac{2\alpha}{\pi}\left[J_5^{(2)} +
\frac{T_c}{3}\left(1+\frac{1}{6L_1} \right)\cos (\kappa T) \right]
\label{z39}
\end{equation}
and $T_c/3 \ll 1$ a contribution of the transition radiation
become visible already at $\kappa T \ll 1$. For $\omega >
\omega_c~(T_c \ll 1)$ the probability of radiation is defined by
Eqs.(\ref{z22b})-(\ref{z22d}). In this case a considerable
distinction from Bethe-Maximon formula will be in the region
$\omega \sim \omega_c/T_c$.

As well known, for soft photons ($\omega \ll \varepsilon$) the
Bethe-Maximon formula for the spectral intensity of radiation
doesn't depend on a photon energy as well. So, the ratio of these
spectral intensities (see Eqs. (\ref{z30}) and (\ref{ad33})) is an
important characteristics of the phenomenon under consideration:
\begin{equation}
R=\frac{dI_{th}}{d\omega}/\frac{dI_{BM}}{d\omega}=
\frac{6}{T_c}\left(1+\frac{1}{6L_1} \right)^{-1}
\int_{0}^{\infty}d\varrho
\varrho K_1^2(\varrho)\left[1-\exp \left(-V(\varrho)T \right)
\right]
\label{z41c}
\end{equation}
For $T_c \ll 1$ one has $R=1$ and for $T_c \gg 1$ one has using
expression (\ref{77})
\begin{equation}
R \simeq \frac{3}{T_c} \left(1+\frac{1}{2k} \right)
\left[\ln 4k +1 -C \right] - 2 +\frac{C}{L_t}
\label{z41d}
\end{equation}
For estimates one can put with a good accuracy $4k \simeq T_c$
and  $L_t \simeq L_1$.

There is, in principle, an opportunity to measure the electron
energy (in region of high energies) using the LPM effect. For this one can
measure the spectral curve on a target with thickness a few percent of
$L_{rad}$ and compare the result with the theory prediction.

Existence of the plateau of the spectral curve in a region of
photon energies where a target is thin was first found in
\cite{18b} within Migdal approach (quantum theory). Recently this
item was discussed in \cite{z11} (in classical theory), \cite{16}
and \cite{18}.

\subsection{Experimental investigation of the LPM effect}

Bremsstrahlung or pair creation suppression can be studied with high
energy electron or photon beams. Because pair creation suppression
requires photons with $\omega>\omega_e$ (see Eq.(\ref{2.23ap})),
beyond the reach of current
accelerators, pair creation has been studied only with cosmic rays,
with consequently very limited statistics.  The best suppression
studies have used electron beams at accelerators.  Besides the LPM
effect, these beams have been used to study effect of 
polarization of a medium and boundary photon emission.

The first tests of LPM suppression came shortly after Migdal's appeared.
These were cosmic rays experiments with high energy photons
($\omega > 1$~TeV) and studied the depth of pair conversion in a dense target
\cite{CRE1}-\cite{CRE4}, see also more recent \cite{CRE5}.
All of the air shower experiments suffered from the poor statistics. 
Uncertainties in the photon spectrum complicated
the analysis. So, these experiments are at best qualitative verification
of the LPM effect.

The first accelerator based study of LPM effect was done in Serpukhov
\cite{E1} using 40~GeV electrons. Photons
with energy 20~MeV $< \omega <$ 70~MeV emitted from carbon,
aluminum, lead and tungsten targets were detected in an sodium 
iodide calorimeter. It were several limitations due to substantial 
background. The results agreed with Migdal's predictions.

A precision measurement of LPM suppression, and study of effect of 
polarization of a medium was performed by the E-146 collaboration 
at SLAC using 25~GeV and 8~GeV electrons. The experimental setup 
was conceptually similar to the Serpukhov experiment, but heavily 
optimized to minimize background. The target used are listed in Table 1.
Photons with energy 200~keV $< \omega <$ 500~MeV  were detected in a
BGO calorimeter.The logarithmic binning of photon spectrum was used
with 25 bins per decade. A very large data set was collected \cite{10} - \cite{E2}.
A detailed, high statistics Monte Carlo simulation was used by E-146.
Its main purpose  was to understand multi-photon emission (see Sec.3.5).

Recently new study of LPM effect at higher energies of electrons
($\varepsilon=$149, 207 and 287~GeV), where the effect has influence upon much 
wider part of spectrum comparing with $\varepsilon=$25~GeV, 
was performed in the H2 beam line of the CERN SPS \cite{HU}.
The experimental setup was also conceptually similar to the Serpukhov experiment.
Photons with energy 2~GeV $< \omega < \varepsilon $   were detected in a
lead glass calorimeter. The logarithmic binning of photon spectrum was used
with 25 bins per decade as at SLAC.

\subsection{Discussion of theory and experiment}

\subsubsection{SLAC E-146}

Here we consider the experimental data 
\cite{10}-\cite{12},\cite{E2} from the
point of view of the above analysis. It was shown that the
mechanism of radiation depends strongly on the thickness of the
target. First, we estimate the thickness of used target in terms
of the formation length. From Eq.(\ref{101}) we have that
\begin{equation}
T_c=\frac{2\pi l}{\alpha L_{rad}} \geq 20 \quad {\rm at}
\quad \frac{l}{L_{rad}} \geq 2~\%.
\label{D1}
\end{equation}
The minimal value of the ratio of the thickness of a target to the formation
length is given by Eq.(\ref{102})($\beta_m \simeq 2T_c
(\omega_p/\omega_c)^{2/3}$). For defined value of $T_c$ this ratio is least
of all for the heavy elements. Indeed, the value of $\omega_p=
\omega_0 \gamma$ depends weakly on nucleus charge
$Z~(\omega_0=30 \div 80)$~eV,
while $\displaystyle{\omega_c\frac{4\pi \gamma^2}{\alpha L_{rad}} \propto
Z^2}$. Furthermore, the ratio $\omega_p/\omega_c$ decreases with
the energy increase. Thus, among all the targets with thickness
$l \geq 2~\% L_{rad}$ the minimal value of $\beta_m$ is attained for the heavy
elements (Au, W, U) at the initial energy $\varepsilon=25$~GeV. In this case
one has $\omega_c \simeq 250$~MeV, $\omega_p \simeq 4$~MeV,
$\beta_m \geq 2.5$. Since the parameter $T_c$ is energy independent
and the ratio $\omega_p/\omega_c \propto 1/\varepsilon$, the minimal value
$\beta_m \geq 5$ is attained  at the initial energy $\varepsilon=8$~GeV
for all the targets with thickness
$l \geq 2~\% L_{rad}$. So, all such targets can be considered
as thick targets at both energies.

As an example we calculated the spectrum of the energy losses in
the tungsten target with the thickness $l=0.088$~mm (=$2.7~\%
L_{rad}$) for both initial energies shown in Fig. 12 (a) and (b).
The characteristic parameters for this case are given in Table 2.
We calculated the main (Migdal type) term Eq.(\ref{2.25}), the
first correction term Eq.(\ref{2.30}) taking into account an
influence of the polarization of a medium according to
Eq.(\ref{45}), as well as the Coulomb corrections entering into
parameter $\nu_0$ Eq.(\ref{2.32a}) and value $L(\varrho_c)$
Eq.(\ref{2.16}). The contribution of an inelastic scattering of a
projectile on atomic electrons was not included into the
numerical calculation (it is $\sim 1\%$), although it can be done
using Eq.(\ref{40a}). We calculated also the contribution of
boundary photons (see Eq.(4.12) in \cite{13}). Here in the soft
part of the spectrum $\omega < \omega_d~(\omega_d \simeq 2$ MeV
for $\varepsilon=25$~GeV) the transition radiation term
Eq.(\ref{51d}) dominates, while in the harder part of the
boundary photon spectrum $\omega > \omega_d$ the terms depending
on both the multiple scattering and the polarization of a medium
give the contribution; for $\varepsilon=8$~GeV one has $\omega_d
\simeq 700$ keV. All the mentioned contribution presented
separately in Fig. 12. Under conditions of the experiment the
multiphoton reduction of the spectral curve is very essential.
The curve "T" in the Fig. 12 involves the reduction factor which
was constructed as the interpolation of Eqs.(\ref{y13}),
(\ref{y26}) and (\ref{y39}) with the accuracy up to $1\% $. The
curves "T" in Fig. 12 (a) and (b) are the final theory prediction
in units $2 \alpha /\pi$.

Experimental data are taken from \cite{12} and recalculated
according with given in mentioned paper procedure:
\begin{equation}
\left(\frac{d\varepsilon}{d\omega}\right)_{exp}=
\frac{1}{L_{rad}}\frac{N_{exp}}{k},\quad  N_{exp}\equiv \left(
\frac{dN}{d\ln\omega}\right)_{exp}. 
\label{D2}
\end{equation}
Since photon energies were histogrammed logarithmically,
using 25 bins per decade of energy, one has for the coefficient $k$  
\begin{eqnarray}
&&k_{h}=\frac{\omega_{max}-\omega_{min}}{\omega_{min}}
=\displaystyle{e^s-1}=0.096,
\nonumber \\
&&k_{m}=2\frac{\omega_{max}-\omega_{min}}{\omega_{max}+\omega_{min}}
=2\frac{e^s-1}{e^s+1}=0.092,
\nonumber \\
&&k_{l}=\frac{\omega_{max}-\omega_{min}}{\omega_{max}}=1-e^{-s}=0.88,
\nonumber \\
&&s=\ln\frac{\omega_{max}}{\omega_{min}}=\frac{\ln 10}{25}=0.092,
\label{D3}
\end{eqnarray} 
depending on the normalization point within bin.
In the review \cite{E2} 
it was indicated that $k=0.096=k_{h}$. In more earlier paper
\cite{12} the value $k=0.09$ was given and we used it in our paper
\cite{13} and \cite{14}. The date recalculated with $k=0.096$ are
also given in Fig. 12. It is seen that there is a perfect
agreement of the curves T with data for both energies.

The gold targets with thickness $l=0.7\% L_{rad}$ and $l=0.1\%
L_{rad}$ are an exception and can't be considered as thick
targets. We calculated energy losses spectra in these targets for
both the initial electron energies $\varepsilon=25~GeV$ and
$\varepsilon=8~GeV$. The characteristic parameters of radiation
for these cases are given in Table 3.

In Fig. 13(a) results of calculations are given for target with a
thickness
\newline $l=0.7\%L_{rad}$ at $\varepsilon=25~GeV$. The curves 1,2,3,4
present correspondingly the functions $J_1^{(2)}, J_2^{(2)},
J_3^{(2)}, J_4^{(2)}$ (\ref{z12}). At $\omega=500~MeV$ the value
$\displaystyle{\nu_1 T = \sqrt{\frac{\omega}{\omega_c}}T_c = 8.4
\gg 1}$, the interference terms are exponentially small and one
can use formulae for a thick target. In this case the parameter
$\nu_1=0.69$ and contribution of boundary photons
($J_b=J_1^{(2)}+J_3^{(2)}+J_4^{(2)}$) is small
($\displaystyle{J_b \simeq - \frac{2\nu_1^4}{21}}$, see \cite{8},
Eq.(4.16)) and distinction Bethe-Maximon formula
($J_{BM}=T_c/3=1.94$) from ${\rm Re}~J^{(2)}={\rm
Re}~(J_1^{(2)}+J_2^{(2)}+ J_3^{(2)}+J_4^{(2)})$ is of the order
10\% according with asymptotic damping factor
$\displaystyle{\left(1-\frac{16\nu_1^4}{21}\right)}$.

At $\omega < \omega_{th} \simeq 30~MeV$ for the case $T_c \gg 1$
and $\beta_m < 1$ the spectral curve turns into plateau according
with discussion in previous subsection. In this photon energy
region the parameter $\nu_0 > 3$ and Eq.(\ref{z24}) for a target
with intermediate thickness describes the spectral probability of
radiation with a good accuracy. With the further photon energy
decrease one can use limiting formula (\ref{z26}) where $\nu_0^2
T =4k \simeq 7.4$. For this case the ratio of ordinate of the
plateau to ordinate of the Bethe-Maximon intensity (see
Eq.(\ref{z41d})) is $R \simeq 0.57$. Note, that in formulae for
$J_1^{(2)} \div J_4^{(2)}$ the potential
$V_c(\mbox{\boldmath$\varrho$})$ Eq.(\ref{2.16}) is used which
doesn't include corrections $\sim
1/L~(v(\mbox{\boldmath$\varrho$}))$. These corrections were given
above for both thin and thick targets (see Eq.(\ref{2.40})). In
our case ($\nu_0 \gg 1,~ \nu_0 T \gg 1$) the expressions with
corrections $\sim 1/L$ are given in Eqs.(\ref{2.40}) and
(\ref{ad31}) for a thick target and a thin target respectively.
Taking into account behaviors of correction in the region $\nu_0
\leq 1$ (see curve 2 in Fig. 12(a)) we construct an interpolation
factor (taking into account the term $\sim 1/L$) with accuracy of
order 1\%. The summary curve ($T$) in Fig. 13(a) contains this
factor.

The transition radiation contributes in the region $\omega \leq
\omega_p$ (function ${\rm Re}~J_5^{(2)}$, curve 5 in Fig. 13(a)).
When $\kappa T \ll 1$ this curve is described by asymptotic of
${\rm Re}~J_5^{(2)}$ Eq.(\ref{z24}). The contribution of the
transition radiation increases with $\omega$ decrease and for
$\kappa T \sim 1$ it describes by Eq.(\ref{z34}). The contribution
of the multiple scattering diminishes due to interference factor
$\cos (\kappa T)$ in Eq.(\ref{z35}) (at $\omega=0.2~MeV,~\kappa
T=1.9$). The curve $T$ in Fig. 13(a) gives the summary
contribution of the multiple scattering (the curve $S$, where
factor $(1-\omega/\varepsilon)$ is included) and the transition
radiation (the curve 5).

In Fig. 13(b) results of calculations are given for target with a
thickness
\newline $l=0.7\%L_{rad}$ at $\varepsilon=8~GeV$.
The notations are the same as in Fig. 13(a). In this case the
characteristic photon energy $\omega_c$ is one order of magnitude
lower than for $\varepsilon=25~GeV$, so that at $\omega=500~MeV$
the parameter $\nu_1$ is small ($\nu_1^2 \simeq 1/20$). Because
of this the right part of the curve $S$ coincides with a good
accuracy with the Bethe-Maximon formula (the Coulomb corrections
are included). Note, that for this electron energy the effect of
recoil (factor $(1-\omega/\varepsilon)$) is more essential.
Strictly speaking, a target with a thickness $0.7\%L_{rad}$ at
$\varepsilon=8~GeV$ is not thin target for any photon energy
($\beta_m=1.6$). However, for bremsstrahlung this target can be
considered as a thin one for $\omega < \omega_{th}=3~MeV$. Since
the polarization of a medium becomes essential in the same region
($\omega_p=1.25~MeV$), the interference factor $\cos (\kappa T)$
in Eq.(\ref{z35}) causes an inflection of the spectral curve $S$
at $\omega \sim 1~MeV$. The transition radiation grows from the
same photon energy $\omega$ and because of this the total spectral
curve $T$ has a minimum at $\omega \simeq 1~MeV$. As far as there
is some interval of energies between $\omega_p$ and $\omega_{th}$
($\omega_{th}-\omega_p \sim 3~MeV$), this minimum is enough wide.
Moreover, the value of its ordinate coincide with a good accuracy
with ordinate of the plateau of the spectral curve $S$ in Fig.
18(a) because bremsstrahlung on a thin target is independent of
electron energy (\ref{ad31}).It should be mentioned also that for
this thickness the reduction factor due to multiple photon
emission (see Eq.(\ref{y26})) $f_{LPM} \simeq 0.94$ (so this is
the effect of the same scale) but this reduction is compensated
nearly exactly because of more precise definition of data (change
of a bin width $\Delta k/k = 0.09$ in \cite{10} (used in
\cite{18}) to a bin width $\Delta k/k = 0.096$ in \cite{12},
\cite{E2}, see discussion in Sec.3 of \cite{15}).

In Fig.14(a) results of calculations are given for target with a
thickness
\newline $0.1\%L_{rad}$ at $\varepsilon=25~GeV$.
The notations are the same as in Fig. 13(a). For this thickness
$T_c=0.96$ and one has a thin target starting from $\omega \leq
\omega_c$. So, we have here very wide plateau. The left edge of
the plateau is defined by the contribution of transition
radiation ($\omega \sim \omega_p$). Since in this case
$4k=\nu_1^2 T=T_c \simeq 1$ (see Eqs.(\ref{z30})-(\ref{ad33})),
one has to calculate the ordinate of the plateau using the exact
formula for a thin target (\ref{z30}). For this case the ratio of
ordinate of the plateau to ordinate of the Bethe-Maximon
intensity (see Eq.(\ref{z41d})) is $R \simeq 0.85$. The same
ordinate has the plateau for electron energy $\varepsilon=8~GeV$
(Fig. 13(b)). However, a width of the plateau for this electron
energy is more narrow ($1~MeV \div 20~MeV$) due to diminishing of
the interval between $\omega_p$ and $\omega_c$. For $\omega >
\omega_c$ the formation length of radiation becomes shorter than
target thickness ($T=T_c\omega/\omega_c > 1$) and the parameter
$\nu_1$ decreases. The value $\nu_1 T=T_c\sqrt{\omega/\omega_c}$
increases with $\omega$ growth. A target becomes thick and the
spectral curves is described by the Bethe-Maximon formula in
Fig. 14(b) starting from photon energy $\omega \sim 100~MeV$. In
Fig. 14(b) the contributions of separate terms into ${\rm
Re}~J^{(2)}$ are shown as well. Their behavior at $\omega >
\omega_c$ is described quite satisfactory by Eqs.
(\ref{z22b})-(\ref{z22f}) (see also discussion at their
derivation).

We compared our calculations with experimental data \cite{12}.
The curves $T$ in Fig. 13, 14 give theory prediction (no fitting
parameters !) in units $2\alpha/\pi$. It should be mentioned that
in our papers \cite{13} and \cite{14} we put $k$=0.09 according
with the instruction given in paper \cite{12}. But the
multiphoton effects were not included. However, as was stated
above one has to use $k$=0.096. So, from the one side for the
gold target with the thickness $l=0.7~\%~L_{rad}$ and energy
$\varepsilon$=25~GeV the reduction factor $f \simeq 0.94$ for
photons with energy $\omega < 10~$ MeV (plateau region). From the
other side, use of the coefficient $k$ = 0.096 in (\ref{D2})
instead of $k$ = 0.09 lowers data upon $\sim 6\%$ and this is
imitate inclusion of the reduction factor. As a result, the
excellent agreement of the theory and data noted in \cite{14} is
not broken. It is seen that in Fig. 13(a) there is a perfect
agreement of the theory and data. In Fig. 13(b) there is an overall
difference: data is order of $10\%$  higher than theory curve.
For photon energy $\omega=500~MeV$ the theory coincides with
Bethe-Maximon formula (with the Coulomb corrections) applicable
for this energy. Note that just for this case it was similar
problem with normalization of data matching with the Migdal Monte
Carlo simulation ($+12.2\%$, see Table II in \cite{12}).

For thickness $l=0.1\%~L_{rad}$ there is a qualitative difference between
our theory prediction and Monte Carlo simulation in \cite{12}. There was a
number of experimental uncertainties associated with this target.
Nevertheless, we show data for $\varepsilon=25~GeV$ which are
lying higher than theory curve.

\subsubsection{CERN SPS experiment}

We consider now the data \cite{HU} from the
point of view of the above analysis \cite{BK3}. 
Energy loss spectra were measured
in iridium target of 0.128 mm thickness (4.36 \% $L_{rad}$). Basing on 
Eq.(\ref{D1}) and following analysis we see that this 
is a thick target. Since the photons with the energy 
$\omega \geq 2~$GeV were measured, no boundary effects were observed. 
For Ir the characteristic electron energy Eq.(\ref{2.30b}) 
$\varepsilon_e$=2.27~TeV and the characteristic photon energy
Eq.(\ref{2.31b})  $\omega_c(\varepsilon)$ for which the LPM effect well 
manifests itself is $\omega_c$(287~GeV)=32~GeV. 
The result of calculation for the initial electron 
energy $\varepsilon$=287~GeV
is shown in Fig. 15. The curves 1, 2, 3 are
calculated using Eqs.(\ref{2.35}), (\ref{2.30}), (\ref{2.25})
correspondingly. It should be mentioned
that the prediction of our theory (curve 4) in the hard end of
spectrum coincides with the Bethe-Maximon curve within the accuracy 
better than $10^{-3}$. For used thickness 
of target the multi-photon effects are very essential. In this 
case it is convenient to calculate the reduction factor $f$ using 
Eq.(\ref{y3.5}), where the mentioned spectrum Eq.(\ref{2.25}) 
is used. The result obtained is in a good agreement
with Eqs.(\ref{y12}), (\ref{y26}). The reduction factor for this
energy is presented by curve 3 in Fig.18.
The final prediction with the reduction factor taken 
into account for the used target thickness is
presented by the curve $T$. The data are recalculated according with
Eqs.(\ref{D2}), (\ref{D3}) using the coefficient $k_m$.
The result of calculation for the initial electron 
energy $\varepsilon$=207~GeV is shown in Fig. 16 and 
the result of calculation for the initial electron 
energy $\varepsilon$=149~GeV is shown in Fig. 17.
The reduction factors for these energies are presented by 
the curves 2 and 1 in Fig.18 correspondingly.
From Fig.18 it follows that if the electron energy decreases 
the reduction factor $f$ diminishes in agreement 
with Eqs.(\ref{y12}), (\ref{y26}).
It is seen that for energy 287 GeV there is the quite satisfactory 
agreement of theory with data, for energy 207 GeV the 
agreement is somewhat less satisfactory, and
for energy 149 GeV data are 10-15\% below the theory. The difference
is the same in hard part of spectrum ($\omega >$ 40 GeV) where 
the Bethe-Maximon intensity spectrum is valid.

The decrease of relative energy losses of electron 
$\displaystyle{\frac{I}{\varepsilon}}$
in consequence of the LPM effect can be important in 
electromagnetic calorimeters operating in detectors on colliders 
in TeV range \cite{L5}.
This item was discussed in Sec.2.2 using the main (Migdal) term,
$\displaystyle{\frac{I}{\varepsilon}L_{rad}^0}$ in gold is given
in Fig. 4 (curve 1).  An analysis with better accuracy is of
evident interest. The relative energy loss of electron
per unit time including the contribution of the 
correction term Eq.(\ref{2.25}) in terms
of the Bethe-Maximon radiation length $L_{rad}$ Eq.(\ref{2.2i})
for iridium ($\varepsilon_e$=2.27~TeV)
curve 1 and for lead ($\varepsilon_e$=4.385~TeV) curve 2
is shown in Fig.19. The increase of effective radiation length
is 13\% in Ir and 7.5\% in Pb for the electron energy 1 TeV,
is 38\% in Ir and 24.5\% in Pb for the electron energy 5 TeV,
and is 58\% in Ir and 38.5\% in Pb for the electron energy 10 TeV.
For both media the corrections attain the maximal value of 
order 5\% at energy $\varepsilon \sim \varepsilon_e$.

\section{Effects in colliding electron-positron beams}

\setcounter{equation}{0}

\subsection{Mechanisms of radiation}

In this section we consider the radiation at head-on collision of
high energy electron and positron beams. The properties of photon
emission process from a particle are immediately connected with
details of its motion. It is convenient to consider the motion
and radiation from particles of one beam in the rest frame of
other beam (the target beam). In this case the target beam is an
ensemble of the Coulomb centers. The radiation takes place at
scattering of a particle from these centers. If the target
consists of neutral particles forming an amorphous medium, a
velocity of particle changes (in a random way) only at small
impact distances because of screening. In the radiation theory
just the random collisions are the mechanism which leads to the
incoherent radiation. For colliding beams significant
contributions into radiation give the large impact parameters
(very small momentum transfers) due to the long-range character
of the Coulomb forces. As a result, in the interaction volume,
which is determined also by the formation length $l_f$
Eqs.(\ref{5}), (\ref{6}) (in the longitudinal direction), it may
be large number of target particles. Let us note that in the case
when the contribution into the radiation is given by impact
parameters comparable with the transverse size of target, the
number of particles in the interaction volume is determined by
the ratio of the radiation formation length to the mean
longitudinal distance between particles.

However, not all cases of momentum transfer should be interpreted
as a result of random collisions. One have to exclude the
collisions, which are macroscopic certain events. For elaboration
of such exclusion we present the exact microscopic momentum
transfer to the target particle in the form: ${\bf q}=<{\bf
q}>+{\bf q}_s$. Here $<{\bf q}>$ is the mean value of momentum
transfer calculated according to standard macroscopic
electrodynamics rules with averaging over domains containing many
particles. The longitudinal size of these domains should be large
with respect to longitudinal distances between target particles
and simultaneously small with respect to the radiation formation
length. The motion of particle in the averaged potential of
target beam, which corresponds to the momentum transfer $<{\bf
q}>$, determines the coherent radiation ({\it the beamstrahlung}).
While the term ${\bf q}_s$ describes the random collisions which
define the process of incoherent radiation ({\it the
bremsstrahlung}). Such random collisions we will call
``scattering'' since $<{\bf q}_s>=0$.

One of principal characteristics of particle motion defining the
properties of coherent radiation is the ratio of variation of its
transverse momentum to the mass during the whole time of passage
across the opposite beam $T$
\begin{equation}
\frac{\Delta {\bf p}_{\perp}}{m} \sim \frac{eE_{\perp}\sigma_z}{m}
\sim \frac{2\alpha N_c \lambda_c}{\sigma_x+\sigma_y} \equiv
\delta, \label{c2.1}
\end{equation}
where $N_c$ is the number of particles in the opposite beam,
$\sigma_x$ and $\sigma_y$ is its transverse dimensions ($\sigma_y
\leq \sigma_x$), $\sigma_z$ is the longitudinal size of opposite
beam. The dispersion of particle momentum during time $T$ is
small comparing with $m$. It attains the maximum for the coaxial
beams:
\begin{equation}
\frac{\left<\textbf{q}_s^2\right>}{m^2} = \gamma^2
\left<\vartheta_s^2\right> \sim
\frac{8\alpha^2N_c\lambda_c^2}{\sigma_x\sigma_y}L \ll 1,
\label{c2.2}
\end{equation}
here $\vartheta_s^2$ is the square of mean angle of multiple
scattering, $L$ is the characteristic logarithm of scattering
problem ($L \sim 10$). This inequality permits one to use the
perturbation theory for consideration of bremsstrahlung, and to
analyze the beamstrahlung independently from the
bremsstrahlung\footnote{Actually more soft condition should be
fulfilled:
\[
\left<\textbf{q}_s^2(l_f)\right>/m^2=\left<\textbf{q}_s^2\right>/m^2~l_f/\sigma_z
\ll1
\]}.

Another important characteristics of motion is the relative
variation of particle impact parameter during time $T$
\begin{equation}
\frac{\Delta\varrho_i}{\varrho_i} \sim
\frac{eE_{\perp}\sigma_z^2}{\varepsilon \sigma_i} \sim
\frac{2\alpha N_c \lambda_c \sigma_z}{\gamma(\sigma_x+\sigma_y)
\sigma_i} \equiv D_i, \label{c2.3}
\end{equation}
here  $i$ is $x$ or $y$. When the disruption parameter $D_i \ll
1$, the collision doesn't change the beam configuration and the
particle crosses the opposite beam on the fixed impact parameter.
If in addition the parameter $\delta \ll 1$ (this situation is
realized in colliders with relatively low energies) then the
beamstrahlung process can be calculated using the dipole
approximation. The main contribution into the beamstrahlung give
soft photons with an energy
\begin{equation}
\frac{\omega}{\varepsilon} \leq \frac{\gamma \lambda_c}{\sigma_z}
\ll 1. \label{c2.4}
\end{equation}

In the opposite case $\delta \gg 1$ the main part of
beamstrahlung is formed when the angle of deflection of particle
velocity is of the order of characteristic radiation angle
$1/\gamma$ and the radiation formation length $l_m$ is defined by
\begin{equation}
\frac{eE_{\perp}l_m}{m}\sim \frac{2\alpha N_c \lambda_c
l_m}{(\sigma_x+\sigma_y)\sigma_z}=1,\quad l_m =
\frac{\sigma_z}{\delta}; \label{c2.5}
\end{equation}
and the characteristic photon energy is
\begin{equation}
\omega \sim \omega_m=\frac{\gamma^2}{l_m} =\varepsilon
\chi_m,\quad \chi_m \equiv 2\alpha N_c
\gamma\frac{\lambda_c^2}{(\sigma_x+\sigma_y)\sigma_z} (\chi_m \ll
1). \label{c2.6}
\end{equation}
Here $\chi$ is the invariant parameter Eq.(\ref{14}) which defines
properties of magnetic bremsstrahlung in the constant field
approximation (CFA). For applicability of CFA it is necessary
that relative variation of~ ${\bf E}_{\perp}$  was small on the
radiation formation length $l_m$. As far $l_m$ is shorter than
$\sigma_z$ in $\delta \gg 1$ times  the characteristic parameter
becomes
\begin{equation}
D_{mi}=D_i\frac{l_m}{\sigma_z}=\frac{D_i}{\delta}=
\frac{\sigma_z}{\gamma \sigma_i},~ (i=x,y) \label{c2.8}
\end{equation}
to that extent. The condition $D_{mi} \ll 1$ is fulfilled in all
known cases. The mean number of photons emitted by a particle
during the whole time of passage across the opposite beam $T$ is
$N_{\gamma} \sim \alpha\delta$, it include the electromagnetic
interaction constant. Using the estimate (\ref{c2.6}) we get an
estimate of relative energy loss
\begin{equation}
\frac{\Delta\varepsilon}{\varepsilon} \sim \alpha\delta
\chi_m~(\chi_m \ll 1) \label{c2.9}
\end{equation}
In the case $\chi_m \ll 1$ (this condition is satisfied in all
existing facilities and proposed collider projects) the soft
photons with energy $\omega \sim \omega_m =\varepsilon\chi_m \ll
\varepsilon$ are mainly emitted. For $\omega \gg \omega_m$ the
emission probability is exponentially suppressed. So, such
photons are emitted in the bremsstrahlung  process only. The
boundary photon energy $\omega_b$, starting from which the
bremsstrahlung  process dominates, depends on particular
parameters of facility. If $\chi_m \sim 1/10$ the energy is
$\omega_b \sim \varepsilon$. The formation length for $\omega \gg
\omega_m$ is much shorter than $l_m$. On this length the particle
deflection angle is small comparing with $1/\gamma$ and one can
neglect the variation of transverse beam dimensions (see
Eq.(\ref{c2.8})). This means that all calculations of
bremsstrahlung characteristics can be carried out in adiabatic
approximation using local beam characteristics $\sigma_{x,y}(t),
{\bf v}(t)$ etc, with subsequent averaging of radiation
characteristics over time. Note that actually we performed a
covariant analysis and the characteristic parameters are defined
in a laboratory frame.

As an example we consider the situation when the configuration of
beams doesn't change during the beam collision (the disruption
parameter $D \ll 1$), and the total particle deflection angle
during intersection of whole beam is small comparing with the
characteristic radiation angle $1/\gamma$ (the dipole case). The
target beam in its rest frame is the ensemble of classical
potentials centers with coordinates ${\bf r}_a~({\bf x}_a,~z_a)$
and the transverse coordinate of emitting particle is ${\bf
r}_{\perp}$. In the perturbation theory the total matrix element
of the radiation process can be written as
\begin{equation}
\textbf{M}(\textbf{r}_{\perp})=\sum_{a=1}^{N_c} \textbf{m}
(\textbf{r}_{\perp}-\textbf{x}_a )\exp(iq_{\parallel} z_a)
\label{c1.1}
\end{equation}
We represent the combination $M_iM_j^{\ast}$ in the form
\begin{eqnarray}
\hspace{-7mm}&& M_iM_j^{\ast}=\sum_{a=b}
m_i(\textbf{r}_{\perp}-\textbf{x}_a)
m_j(\textbf{r}_{\perp}-\textbf{x}_b)
\nonumber \\
\hspace{-7mm}&& +\sum_{a \neq b}
m_i(\textbf{r}_{\perp}-\textbf{x}_a)
m_j(\textbf{r}_{\perp}-\textbf{x}_b)
\exp(-iq_{\parallel}\left(z_a-z_b)\right). \label{c1.2}
\end{eqnarray}
In the expression Eq.(\ref{c1.2}) we have to carry out averaging
over position of scattering centers. We will proceed under
assumption that there are many scattering centers within the
radiation formation length $l_f=1/q_{\parallel}$
\begin{equation}
N_f=n_zl_f \gg 1, \label{c1.3}
\end{equation}
where for the Gaussian distribution
\begin{equation}
n_z =\frac{N_c}{\sqrt{2\pi}\sigma_z}
\exp\left(-\frac{z^2}{2\sigma_z^2} \right), \label{c1.4}
\end{equation}
here $N_c$ and $\sigma_z$ are introduced in Eq.(\ref{c2.1}). Note
that in the situation under consideration $\varrho_{max}
=|\textbf{r}_{\perp}-\textbf{x}_a|_{max} \geq \sigma_t$, where $
\sigma_t$ is the characteristic transverse size of target beam.
Let us select terms with approximately fixed phase
$q_{\parallel}(z_a-z_b)=\phi_{ab}$ in the sum with $a \neq b$ in
Eq.(\ref{c1.2}). If the condition (\ref{c1.3}) is fulfilled,
there are many terms for which the phase variation is small
($\Delta\phi_{ab} \ll 1$). For this reason one can average over
the transverse coordinates (${\bf x}_a, {\bf x}_b$) of target
particles in Eq.(\ref{c1.2}) without touching upon the
longitudinal coordinates ($z_a, z_b$)
\begin{eqnarray}
\hspace{-7mm}&& M_iM_j^{\ast}=N_c \left\langle
m_im_j\right\rangle_{\perp} +\left\langle
m_i\right\rangle_{\perp} \left\langle m_j\right\rangle_{\perp}
\sum_{a \neq b} \exp(-iq_{\parallel}\left(z_a-z_b)\right)
\nonumber \\
\hspace{-7mm}&&=N_c\Big( \left\langle m_im_j\right\rangle_{\perp}
-\left\langle m_i\right\rangle_{\perp} \left\langle
m_j\right\rangle_{\perp} \Big) +\left\langle
m_i\right\rangle_{\perp} \left\langle m_j\right\rangle_{\perp}
\left|\sum_{a}\exp(iq_{\parallel}z_a)\right|^2, \label{c1.5}
\end{eqnarray}
where
\begin{eqnarray}
\hspace{-7mm}&& \left\langle m_i\right\rangle_{\perp} =\int
m_i(\textbf{r}_{\perp}-\textbf{x} ) w_c(\textbf{x})d^2x,
\nonumber \\
\hspace{-7mm}&& \left\langle m_im_j\right\rangle_{\perp} =\int
m_i(\textbf{r}_{\perp}-\textbf{x} )
m_j(\textbf{r}_{\perp}-\textbf{x} )w_c(\textbf{x})d^2x,
\label{c1.6}
\end{eqnarray}
here $w_c(\textbf{x})$ is the probability density of target
particle distribution over the transverse coordinates normalized
to unity. In Eq.(\ref{c1.5}) in the sum with $a\neq b$ we add and
subtract the terms with $a=b$. The first term (proportional to
$N_c$) on the right-hand side of Eq.(\ref{c1.5}) is the
incoherent contribution to  radiation ({\it the bremsstrahlung}).
The second term gives the coherent part of radiation. For Gaussian
distribution  Eq.(\ref{c1.4}) performing averaging over the
longitudinal coordinate $z_a$ one has
\begin{equation}
\left|\sum_{a}\exp(iq_{\parallel}z_a)\right|^2 \rightarrow
\left|\int_{-\infty}^{\infty} n_z\exp(iq_{\parallel}z)dz\right|^2=
N_c^2\exp(-q_{\parallel}^2\sigma_z^2). \label{c1.7}
\end{equation}

\subsection{The beam-size effect in the bremsstrahlung}

It was pointed in Sec.1.3 that the external factors act
differently on the radiating particle and on the recoil particle.
This leads to a specific effect in the bremsstrahlung process  at
electron-electron(positron) collision. The point is that in the
rest frame of recoil particle the formation length is longer by
$\varepsilon/m$ times than in the laboratory frame. Just this
length determines the maximal impact parameters at scattering
(e.g. the screening radius). The are a few factors which could
act on the recoil electron. One of them is the presence of an
external magnetic field in the region of particle collision
\cite{20} (see Sec.1.3). Another effect can appear due to
smallness of the linear interval $l$ where the collision occurs
in comparison with $l_v$ Eq.(\ref{6}). This was pointed in
\cite{22}.

The  special experimental study of  bremsstrahlung was performed
at the electron-positron colliding beam facility VEPP-4 of
Institute of Nuclear Physics, Novosibirsk \cite{exp1}. The
deviation of the bremsstrahlung spectrum from the standard QED
spectrum was observed at the electron energy
$\varepsilon=1.84~$GeV. The effect was attributed to the smallness
of the transverse size of the colliding beams. In theory the
problem was investigated in \cite{cBKS1}, where the bremsstrahlung
spectrum at the collision of electron-electron(positron) beams
with the small transverse size was calculated to within the power
accuracy (the neglected terms are of the order
$1/\gamma=m/\varepsilon$). After the problem was analyzed in
\cite{BD}, and later on in \cite{KPS} where some of results for
the bremsstrahlung found in \cite{cBKS1} were reproduced. It
should be noted that in \cite{cBKS1} (as well as in all other
papers mentioned above) an incomplete expression for the
bremsstrahlung spectrum was calculated. One has to perform the
subtraction associated with the extraction of pure fluctuation
process as it was shown above. An analogous problem encountered in
the analysis of incoherent processes in the oriented crystals
\cite{cBKS6},\cite{BG2} where the necessity of the subtraction
procedure was indicated. Without the subtraction the results for
the incoherent processes in oriented crystals would be
qualitatively erroneous.

The correction to photon emission probability due to the small
transverse dimensions of colliding beam for unpolarized electrons
and photon was calculated in \cite{cBK2} basing on subtraction
procedure as in Eq.(\ref{c1.5}). It is obtained after integration
over the azimuth angle of the emitted photon
\begin{equation}
dw_1=\frac{\alpha^3}{\pi
m^2}\frac{\varepsilon'}{\varepsilon}\frac{d
\omega}{\omega}U(\zeta) F(\omega, \zeta) d\zeta,\quad \zeta =
1+\gamma^2 \vartheta^2, \label{c3.11}
\end{equation}
where $\vartheta$ is the photon emission angle,
$\varepsilon'=\varepsilon-\omega$,
\begin{eqnarray}
\hspace{-7mm}&& U(\zeta)=v-\frac{4(\zeta-1)}{\zeta^2},~
v=\frac{\varepsilon}{\varepsilon'}
 +\frac{\varepsilon'}{\varepsilon},\quad
F(\omega, \zeta)=F^{(1)}(\omega, \zeta)-F^{(2)}(\omega, \zeta),
\nonumber \\
\hspace{-7mm}&& F^{(1)}(\omega, \zeta) =\frac{\eta^2}{\zeta^2}\int
\left[K_0(\eta \varrho)K_2(\eta \varrho)-K_1^2(\eta
\varrho)\right]\varrho
\frac{d\Phi(\mbox{\boldmath$\varrho$})}{d\varrho}d^2\varrho
\nonumber \\
\hspace{-7mm}&& F^{(2)}(\omega, \zeta)=\frac{2\eta^2}{\zeta^2}\int
\left(\int K_1(\eta
\varrho)\frac{\mbox{\boldmath$\varrho$}}{\varrho}w_c({\bf
x}-\mbox{\boldmath$\varrho$})d^2\varrho \right)^2w_r({\bf x})d^2x,
\label{c3.12}
\end{eqnarray}
here
\begin{equation}
\eta=q_{min}\zeta,~ q_{min}=m^3\omega/4\varepsilon^2
\varepsilon',~ \Phi(\mbox{\boldmath$\varrho$})=\int w_r({\bf
x}+\mbox{\boldmath$\varrho$}) w_c({\bf x})d^2x, \label{c3.15}
\end{equation}
where $w_c({\bf x})$ is defined in (\ref{c1.6}), $w_r({\bf x})$
is the same but for the radiating beam,  value $q_{min}$ is
defined in c.m.frame of colliding particles. The term
$F^{(2)}(\omega, \zeta)$ is the subtraction term. The total
probability is $dw_{\gamma}=dw_0+dw_1$, where $dw_0$ is standard
QED probability. The analysis in \cite{cBK2} was based on
Eqs.(\ref{c3.11})-(\ref{c3.12}).

We considered in \cite{cBK2} the actual case of the Gaussian
beams. The Fourier transform was used
\begin{eqnarray}
\hspace{-9mm}&&w({\bf x})=\frac{1}{(2\pi)^2}\int d^2q \exp(-i{\bf
q x})w({\bf q});
\nonumber \\
\hspace{-9mm}&& w_r({\bf
q})=\exp\left[-\frac{1}{2}(q_x^2\Delta_x^2+q_y^2\Delta_y^2)\right],~
w_c({\bf
q})=\exp\left[-\frac{1}{2}(q_x^2\sigma_x^2+q_y^2\sigma_y^2)\right],
\label{c4.1}
\end{eqnarray}
where as above the index $r$ relates to the radiating beam and the
index $c$ relates to the target beam, $\Delta_y$ and $\Delta_x$
($\sigma_y$ and $\sigma_x$) are the vertical and horizontal
transverse dimensions of radiating (target) beam. Substituting
Eq.(\ref{c4.1}) into Eq.(\ref{c3.15}) we find
\begin{equation}
\Phi(\mbox{\boldmath$\varrho$})=
\frac{\Sigma_x\Sigma_y}{\pi}\exp[-\varrho_x^2\Sigma_x^2-\varrho_y^2\Sigma_y^2];~
\Sigma_x^2=\frac{1}{2(\sigma_x^2+\Delta_x^2)},~
\Sigma_y^2=\frac{1}{2(\sigma_y^2+\Delta_y^2)}. \label{c4.2}
\end{equation}
Using the relation $d\sigma_1=\Phi^{-1}(0) dw_1$ the following
expression for the correction to spectrum was found in
\cite{cBK2} starting from Eq.(\ref{c3.11})
\begin{eqnarray}
\hspace{-2mm}&& d\sigma_1^{(1)}=\frac{2\alpha^3}{m^2}
\frac{\varepsilon'}{\varepsilon}
\frac{d\omega}{\omega}f^{(1)}(\omega),~f(s) =
\frac{\sqrt{\pi}}{2s}(v-8s^2){\rm
erfc}(s)+4\displaystyle{e^{-s^2}}+2{\rm Ei}(-s^2),
\nonumber \\
\hspace{-2mm}&& f^{(1)}(\omega)= -\frac{1}{\pi\Sigma_x
\Sigma_y}\int_{0}^{2\pi}\frac{d\varphi}{\Sigma_x^{-2}\cos^2\varphi+
\Sigma_y^{-2}\sin^2\varphi} \int_{0}^{\infty}F_2(z)f(s)s ds,
\label{c4.9} \\
\hspace{-2mm}&& z^2=\frac{s^2
q_{min}^{-2}}{\Sigma_x^{-2}\cos^2\varphi+
\Sigma_y^{-2}\sin^2\varphi},~ F_2(x)=\frac{2x^2+1}{x\sqrt{1+x^2}}
\ln(x+\sqrt{1+x^2})-1, \nonumber
\end{eqnarray}
where Ei($x$) is the exponential integral function and erfc($x$)
is the error function. This formula is quite convenient for the
numerical calculations.

The subtraction term ($F^{(2)}(\omega, \zeta)$ in
Eq.(\ref{c3.12})) gives for coaxial beams
\begin{equation}
d\sigma_1^{(2)}=-\frac{2\alpha^3}{m^2}
\frac{\varepsilon'}{\varepsilon}
\frac{d\omega}{\omega}J^{(2)}(\omega), \label{c4.19}
\end{equation}
where
\begin{eqnarray}
&& J^{(2)}(\omega)= \frac{\sqrt{a b}}{\Sigma_x \Sigma_y}
\int_{0}^{\infty} ds_1 \int_{0}^{\infty} ds_2  g\left(
\frac{q_{min}\sqrt{s}}{2} \right)G(s_1, s_2),
\nonumber \\
&& G(s_1, s_2)=\frac{1}{2} \left( \frac{a_1a_2 b_1 b_2}{A
B}\right)^{1/2}\left[\frac{a_1a_2}{A}+\frac{b_1b_2}{B}\right]
\label{c4.20}
\end{eqnarray}
Here the function $g(q)$ is:
\begin{equation}
g(q)= \left( v
-\frac{2}{3}\right)\displaystyle{e^{-q^2}}-2q^2\left[
\frac{\sqrt{\pi}}{2q}\left(v-\frac{8}{3}q^2\right){\rm
erfc}(q)+\frac{4}{3}\displaystyle{e^{-q^2}}+{\rm Ei}(-q^2)\right].
\label{c4.21}
\end{equation}
In Eq.(\ref{c4.20}) we introduced the following notations
\begin{eqnarray}
&& a=\frac{1}{2\Delta_x^2}, \quad b=\frac{1}{2\Delta_y^2},\quad
 a_{1,2}=\frac{1}{s_{1,2}+2\sigma_x^2},\quad
b_{1,2}=\frac{1}{s_{1,2}+2\sigma_y^2},
\nonumber \\
&& A=a_1+a_2+a, \quad B=b_1+b_2+b,\quad s=s_1+s_2. \label{c4.22}
\end{eqnarray}

In the case of narrow beams one has  $q_{min}/(\Sigma_x+\Sigma_y)
\ll 1$. In this case of coaxial beams $d\sigma_{\gamma}=
d\sigma_0 + d\sigma_1$ is
\begin{equation}
d\sigma_{\gamma}= \frac{2\alpha^3}{m^2}
\frac{\varepsilon'}{\varepsilon}
\frac{d\omega}{\omega}\Bigg\{\left(v-\frac{2}{3}\right)\Bigg[2\ln
\frac{m}{\Sigma_x+\Sigma_y} +C +2
  -J_{-}\Bigg]+\frac{2}{9}\Bigg\},
 \label{c5.7}
\end{equation}
where
\begin{equation}
 J_{-}=\frac{\sqrt{a b}}{\Sigma_x
\Sigma_y} \int_{0}^{\infty} ds_1 \int_{0}^{\infty} ds_2 G(s_1,
s_2) \label{c5.8}
\end{equation}

The dimensions of beams in the experiment \cite{exp1} were
$\sigma_y=\Delta_y=24~\mu m, \sigma_x=\Delta_x=450~\mu m$, so
this is the case of flat beams. The estimate for this case gives
$J_{-} \simeq (4/3\sqrt{3})\pi \sigma_y/\sigma_x \ll 1$. This
term is much smaller than other terms in Eq.(\ref{c5.7}). This
means that for this case the correction to the spectrum
calculated in \cite{cBKS1} is very small.

The result of calculation and VEPP-4 (INP, Novosibirsk) data are
presented in Fig.20 where the bremsstrahlung intensity spectrum
$\omega d\sigma/d\omega$ is given in units $2\alpha r_0^2$ versus
the photon energy in units of initial electron energy
($x=\omega/\varepsilon$)\cite{cBK2}. The upper curve is the
standard QED spectrum, the three close curves below are
calculated using Eqs.(\ref{c4.9}) and (\ref{c4.19}) for the
different vertical dimensions of colliding beams (equal for both
colliding beams $\sigma=\sigma_z=\Delta_z$):$\sigma=20~\mu m$
(bottom), $\sigma=24~\mu m$ (middle), $\sigma=27~\mu m$ (top)
(this is just the 1-sigma dispersion for the beams used in the
experiment). All the theoretical curves are calculated to within
the relativistic accuracy (the discarded terms are of the order
$m/\varepsilon$). It is seen that the effect of the small
transverse dimensions is very essential in soft part of spectrum
(at $\omega/\varepsilon = 10^{-4}$ the spectral curve diminishes
in 25 \%), while for $\omega/\varepsilon > 10^{-1}$ the effect
becomes negligible. The data measured in \cite{exp1} are
presented as circles (experiment in 1980) and as triangles
(experiment in 1981) with 6 \% systematic error as obtained in
\cite{exp1} (while the statistical errors are negligible). This
presentation is somewhat different from \cite{exp1}. It is seen
that the data points are situated systematically below the theory
curves but the difference is not exceed the 2-sigma level
\cite{exp1}. It should be noted that this is true also in the
hard part of spectrum where the beam-size effect is very small.

One more measurement of beam-size effect was performed at the HERA
electron-proton collider (DESY, Germany) \cite{P}. The electron
beam energy was $\varepsilon$=27.5~GeV, the proton beam energy was
$\varepsilon_p$=820~GeV. The parameters of beam in this
experiment were (in our notation): $\sigma_y=\Delta_y=(50 \div
58) \mu m$, $\sigma_x=\Delta_x=(250 \div 290) \mu m$. Since the
ratio of the vertical and the horizontal dimensions is not very
small, the contribution of subtraction term (Eq.(\ref{c4.19})) is
essential (more than 10\%). The bremsstrahlung intensity spectrum
$\omega d\sigma/d\omega$ in units $2\alpha r_0^2$ versus the
photon energy in the units of initial electron energy
($x=\omega/\varepsilon$) for the HERA experiment is given in
Fig.21. For details of  comparison of experimental data
\cite{exp1}, \cite{P} with theory see \cite{cBK2}, where we
discussed also possible use of beam-size effect for linear
collider tuning. It should be noted that for linear collider the
condition of strong beam-size effect $\sigma_y q_{min} \ll
1~(\sigma_y \ll \sigma_x)$ is fulfilled for the whole spectrum.
This can be seen in Fig.22, where the lower curve is calculated
using Eq.(\ref{c4.9}) and the subtraction term is very small
since $\sigma_y/\sigma_x < 0.01$. As far as the narrow beams are
considered in Fig.22, the lower curve is consistent also with
Eq.(\ref{c5.7}). This curve depends on the energy and the
transverse sizes of beams. It will be instructive to remind that
the analysis in \cite{cBK2} (see Eq.(2.8)) and here is valid if
$\chi_m/u \ll 1$ (see Eq.(\ref{c2.6}), $u=\omega/\varepsilon'$).
The parameter $\chi_m$ depends also on number of particles $N_c$
and the longitudinal beam size. So, for low $N_c$ Fig.22 is valid
for any $x$, but for TESLA project ($\chi_m$=0.13) it holds in
hard part of spectrum only. In fact, the probability of
incoherent radiation becomes larger than the probability of
coherent radiation only at $x > 0.7$ where the lower curve in
Fig.22 is certainly applicable.

\subsection{Coherent radiation}

The particle interaction at beam-beam collision in linear
colliders occurs in an electromagnetic field provided by the
beams. As a result, 1)the phenomena induced by this field turns
out to be very essential, 2)the cross section of the main QED
processes are modified comparing to the case of free particles.
These items were considered by V.M.Strahkhovenko and authors
\cite{coBKS1}.

The magnetic bremsstrahlung mechanism dominates and its
characteristics are determined by the value of the quantum
parameter $\chi(t)$ Eq.(\ref{14}) dependent on the strength of the
incoming beam field at the moment $t$ (the constant field
limit(CFA)).

In the CFA the spectral probability of radiation from an electron
per unit time has the form \cite{7}, Eq.(10.25)  (see also
Eq.(4.24) in \cite{9} or Eq.(90.23) in \cite{16a} )
\begin{eqnarray}
\hspace{-7mm}&& \frac{dw_{\gamma}}{dt} \equiv dW_{\gamma}(t)=
\frac{\alpha}{2\sqrt{3}\pi \gamma^2}
\int_{-\infty}^{\infty}\Phi_{\gamma}(t)dt d\omega,
\nonumber \\
\hspace{-7mm}&&\Phi_{\gamma}(t)=\left(\frac{\varepsilon}{\varepsilon'}+
\frac{\varepsilon'}{\varepsilon}
\right)K_{2/3}(z)-\int_{z}^{\infty}K_{1/3}(y)dy, \label{co1}
\end{eqnarray}
where $K_{\nu}(z)$ is the Macdonald functions, $z=2u/3\chi(t)$,
$u=\omega/\varepsilon'$. The corresponding probability for
polarized electrons and photons is given in \cite{coBK3}.

For the Gaussian beams
\begin{equation}
\chi(t)=\chi_0(x,y)\exp(-2t^2/\sigma_z^2), \label{co2.3}
\end{equation}
here the function $\chi_0(x,y)$ depends on transverse coordinates.

It turns out that for the Gaussian beams the integration of the
spectral probability over time can be carried out in a general
form:
\begin{eqnarray}
&& \frac{dw_{\gamma}}{du}=\frac{\alpha m \sigma_z}{\pi\gamma
\sqrt{6}} \frac{1}{(1+u)^2}\Bigg[\left(1+u+\frac{1}{1+u} \right)
\nonumber \\
&&\times \int_{1}^{\infty}K_{2/3}\left(ay \right)
\frac{dy}{y\sqrt{\ln y}}-2a\int_{1}^{\infty}K_{1/3}\left(ay
\right) \sqrt{\ln y}~dy\Bigg], \label{co2.4}
\end{eqnarray}
where $a=2u/3\chi_0$. In the case when $a \gg 1$ the main
contribution into the integral (\ref{co2.4}) gives the region
$y=1+\xi, \xi \ll 1$. Taking the integrals over $\xi$ we obtain
\begin{equation}
\frac{dw_{\gamma}^{CF}}{du} \simeq \frac{\sqrt{3}\alpha m
\sigma_z}{4\gamma} \frac{1+u+u^2}{u(1+u)^3} \chi_0 \exp\left(
-\frac{2u}{3\chi_0}\right) \label{co2.5}
\end{equation}

For round beams the integration over transverse coordinates is
performed with the density
\begin{equation}
n_{\perp}(\mbox{\boldmath$\varrho$})=\frac{1}{2\pi
\sigma_{\perp}^2} \exp\left(-\frac{\varrho^2}{2\sigma_{\perp}^2}
\right) \label{co2.6}
\end{equation}
The parameter $\chi_0(\varrho)$ we present in the form
\begin{eqnarray}
&&\chi_0(\varrho)=\chi_{rd}\frac{f(x)}{f_0},\quad
x=\frac{\varrho}{\sigma_{\perp}}, \quad
f(x)=\frac{1}{x}\left(1-\exp(-x^2/2) \right),
\nonumber \\
&&\chi_{rd}=0.720 \alpha N_c \gamma
\frac{\lambda_c^2}{\sigma_z\sigma_{\perp}},~
 f'(x_0)=0,~f_0=f(x_0)=0.451256,
\label{co2.7}
\end{eqnarray}
where $N_c$ is the number of electron in the bunch.

The Laplace integration of Eq.(\ref{co2.5}) gives for radiation
intensity $dI/du=\varepsilon u/(1+u)dW/du$
\begin{equation}
\frac{dI_{as}}{du} \simeq \alpha m^2 \sigma_z
\frac{3}{4}\sqrt{\frac{\pi}{|f''_0|}}
\frac{1+u+u^2}{\sqrt{u}(1+u)^4}
 f_0^{3/2}\chi_{rd}^{3/2}\exp\left(-\frac{2u}{3\chi_{rd}}\right),
\label{co2.8}
\end{equation}
where $f''_0=f''(x_0)=-0.271678$.

Integration of Eq.(\ref{co2.4}) over transverse coordinates gives
the final result for the radiation intensity. For the round beams
it is shown in Fig.23 for $\chi_{rd}=0.13$, the curve attains the
maximum at $u \simeq 0.02$. The right slope of the curve agrees
with the asymptotic intensity (\ref{co2.8}) and the left slope of
the curve agrees with the standard classical intensity.
\begin{equation}
\frac{dI_{cl}}{du}=\frac{e^2
m^2}{\pi}3^{1/6}\Gamma(2/3)\chi_{rd}^{2/3}u^{1/3} \label{co2.8a}
\end{equation}

It will be instructive to compare the spectrum in Fig.23, found
by means of integration over the transverse coordinates with
intensity spectrum which follows from Eq.(\ref{co2.4})
(multiplied by $\omega$) with averaged over the density
Eq.(\ref{co2.6}) value $\chi_0$ Eq.(\ref{co2.7}):
$\overline{\chi_0}=\chi_{rd} \cdot 0.8135 $. The last spectrum
reproduces the spectrum given in Fig.23, in the interval $10^{-3}
\leq u/\overline{\chi_0} < 1$ with an accuracy better than 2\%
(near maximum better than 1\%) while for $ u/\overline{\chi_0}
\leq 10^{-3}$ one can use the classic intensity (\ref{co2.8a})
and for $ u/\overline{\chi_0} \gg 1$ the asymptotics
(\ref{co2.8}) is applicable.

For the flat beams ($\sigma_x \gg \sigma_y$) the parameter
$\mbox{\boldmath$\chi$}_0(\mbox{\boldmath$\varrho$})$ takes the
form
\begin{equation}
\mbox{\boldmath$\chi$}_0=\frac{2\gamma {\bf E}_{\perp}}{H_0}=
\chi_m e^{-v^2} \left[\textbf{e}_y{\rm erf}\left(w \right)
 -i \textbf{e}_x{\rm erf}\left(iv \right) \right],\quad
 \chi_m= \frac{2N_c\alpha\gamma \lambda_c^2}{\sigma_z\sigma_x},
\label{co2.9}
\end{equation}
here $v=x/\sqrt{2}\sigma_x,~w=y/\sqrt{2}\sigma_y$, ${\rm
erf}(z)=2/\sqrt{\pi}\int_{0}^{z}\exp(-t^2)dt$, $\textbf{e}_x$ and
$\textbf{e}_y$ are the unit vectors along the corresponding axes.
The formula (\ref{co2.9}) is consistent with given in \cite{N}.
In \cite{coBKS1} the term with $\textbf{e}_x$ was missed. Because
of this the numerical coefficients in results for the flat beams
are erroneous.

To calculate the asymptotics of radiation intensity for the case
$u \gg \chi_m$ one has to substitute
\begin{equation}
\chi_0=|\mbox{\boldmath$\chi$}_0|=\chi_m e^{-v^2} \left[ {\rm
erf}^2\left(w \right) -{\rm erf}^2\left(iv \right) \right]^{1/2}
\label{co2.10}
\end{equation}
into Eq.(\ref{co2.5}) and take integrals over transverse
coordinates $x,y$ with the weight
\begin{equation}
n_{\perp}(x,y)=\frac{1}{2\pi \sigma_x
\sigma_y}\exp\left(-\frac{x^2}{2\sigma_x^2}
-\frac{y^2}{2\sigma_y^2} \right). \label{co2.11}
\end{equation}
Integral over $x$ can be taken using the Laplace method, while
for integration over $y$ it is convenient to introduce the
variable
\begin{equation}
\eta=\frac{2}{\sqrt{\pi}}\int_{w}^{\infty}\exp(-t^2)dt,\quad
w=\frac{y}{\sqrt{2}\sigma_y}. \label{co2.12}
\end{equation}
As a result we obtain for the radiation intensity in the case of
flat beams
\begin{equation}
\frac{dI_{fl}}{du}=\frac{9}{8\sqrt{2\left(1-2/\pi \right)}}
\alpha m^2 \sigma_z \chi_m^{5/2}
 \frac{1+u+u^2}{u^{3/2}(1+u)^4}\exp\left( -\frac{2u}{3\chi_m}\right).
\label{co2.13}
\end{equation}
It is interesting to compare the high-energy end of intensity
spectrum at collision of flat beams Eq.(\ref{co2.13}) with
intensity spectrum of incoherent radiation with regard for
smallness of the transverse beam sizes considered above. For
calculation we use the project TESLA parameters \cite{T}:
~$\varepsilon=250$~GeV,~$\sigma_x=553$~nm,
$\sigma_y=5$~nm,~$\sigma_z=0.3$~mm, $N=2\cdot 10^{10}$,
$\chi_m$=0.13. The result is shown in Fig.24, where the curves
calculated according to Eq.(\ref{co2.13}), and Eq.(\ref{c5.7}).
It is seen that for $x=\omega/\varepsilon > 0.7$ the incoherent
radiation dominates. For $x > 0.7$ the incoherent radiation may
be used for a tuning of beams \cite{cBK2}.

Along with the spectral characteristics of radiation the total
number of photons emitted by an electron is of evident interest as
well as the relative energy losses. We discuss an actual case of
flat beams and the parameter $\chi_m \ll 1$. In this case one can
use the classic expression for intensity (bearing in mind that
starting from $\chi_m \sim 1/10$ the quantum effects become
substantial). In classical limit the relative energy losses are
\begin{equation}
\left(\frac{\Delta
\varepsilon}{\varepsilon}\right)_{cl}=\frac{2\alpha
m^2}{3\varepsilon} \int \chi^2(t, x, y) n_{\perp}(x,y)dt dx dy.
\label{co2.14}
\end{equation}
Using Eqs.(\ref{co2.3}), (\ref{co2.10}), and (\ref{co2.11}) we get
\begin{equation}
\left(\frac{\Delta
\varepsilon}{\varepsilon}\right)_{cl}=\frac{2}{9}\sqrt{\frac{\pi}{3}}r
\chi_m^2, \quad r=\frac{\alpha \sigma_z}{\gamma \lambda_c}.
\label{co2.15}
\end{equation}
For mean number of emitted by an electron photons we find
\begin{equation}
n_{\gamma}^{cl}=\frac{5\alpha m^2}{2\sqrt{3}\varepsilon} \int
\chi(t, x, y) n_{\perp}(x,y)dt dx dy =0.59275
\frac{5}{2}\sqrt{\frac{\pi}{6}}r \chi_m =1.072 r \chi_m.
\label{co2.16}
\end{equation}
If $\chi_m > 1/10$ one have to use the quantum formulas. For the
energy losses one can use the  approximate expression (the
accuracy is better than 2\% for any $\chi$) \cite{coBKS5}
\begin{equation}
d\varepsilon/dt=2/3 \alpha m^2 \chi^2
\left[1+4.8(1+\chi)\ln(1+1.7\chi)+2.44\chi^2 \right]^{-2/3}.
\label{co2.17}
\end{equation}
Here $\chi$ is the local value, so this expression for
$d\varepsilon/dt$ should be integrated over time and averaged
over the transverse coordinates. For mean number of photons
emitted by an electron there is the approximate expression (the
accuracy is better than 1\% for any $\chi_0$)
\begin{equation}
n_{\gamma}(\mbox{\boldmath$\varrho$}) =\frac{1.81\chi_0
r}{[1+1.5(1+\chi_0)\ln(1+3\chi_0)+0.3\chi_0^2]^{1/6}}
\label{co2.18},
\end{equation}
where the expression should be averaged over the transverse
coordinates. For the project TESLA one gets for $(\Delta
\varepsilon/\varepsilon)_{cl}=4.3\%$ according to
Eq.(\ref{co2.15}), while the correct result from Eq.(\ref{co2.17})
is $\Delta \varepsilon/\varepsilon=3.2\%$. For mean number of
photons emitted by an electron we have correspondingly
$n_{\gamma}^{cl} \simeq 1.6$ while the correct result is
$n_{\gamma} \simeq 1.5$.

\subsection{Mechanisms of $e^+e^-$- pair creation}

The probability of pair creation by a photon in the external
field can be find from formulas (\ref{co1}) using the
substitution rule (see e.g.\cite{16a}, \cite{9}): $\varepsilon
\rightarrow -\varepsilon, \quad \omega \rightarrow -\omega, \quad
\omega^2d\omega \rightarrow - \varepsilon^2d\varepsilon$.
Performing these substitutions we obtain (see also Eq.(3.50) in
\cite{9})
\begin{eqnarray}
\hspace{-7mm}&& \frac{dw_{e}}{dt} \equiv dW_{e}(t)=\frac{\alpha
m^2}{2\sqrt{3}\pi \omega^2} \Phi_{e},
\nonumber \\
\hspace{-7mm}&&\Phi_{e}(t)=\left(\frac{\varepsilon}{\varepsilon'}+
\frac{\varepsilon'}{\varepsilon}
\right)K_{2/3}(y)+\int_{y}^{\infty}K_{1/3}(x)dx,
~y=\frac{2\omega^2}{3\varepsilon\varepsilon'\kappa},~
\kappa=\frac{\omega F}{m H_0} \label{co4}
\end{eqnarray}
here $\varepsilon$ is the energy of the created electron,
$\varepsilon'=\omega-\varepsilon$ is the energy of created
positron.

There are different mechanisms of electron-positron pair creation
\begin{enumerate}
\item Real photon radiation in the field and pair creation
by this photon in the same field of the opposite beam. This
mechanism dominates in the case $\chi \geq 1$.
\item Direct electroproduction of electron-positron pair in the
field through virtual photon. This mechanism is also essential in
the case $\chi \geq 1$.
\item Mixed mechanism(1):photon is radiated in the bremsstrahlung
process, i.e. incoherently, and the pair is produced by photon in
an external field.
\item Mixed mechanism(2): photon is radiated via the beamstrahlung
mechanism, and pair is produced in the interaction
of this photon with individual particles of oncoming beam, i.e. in
interaction with potential fluctuations.
\item Incoherent electroproduction of pair.
\end{enumerate}

In the actual case $\chi \ll 1$ mixed and incoherent mechanisms
mostly contribute. We start with the mixed mechanism (2). In the
case $\chi_m \ll 1$ the parameter $\kappa$ Eq.(\ref{co4})
containing the energy of emitted photon is also small and the
incoherent cross section of pair creation by a photon is weakly
dependent on the photon energy. To logarithmic accuracy
\begin{equation}
\sigma_p=28/9\alpha^3 \lambda_c^2 \ln(\sigma_y/\lambda_c)
\left(1+ 396\kappa^2/1225 \right). \label{co3.1}
\end{equation}
If the term $\propto\kappa^2 $ being neglected, the pair creation
probability is factorized (we discuss coaxial beams of identical
configuration). The number of pairs created by this mechanism
(per one initial electron) is
\begin{eqnarray}
\hspace{-7mm}&&n_p^{(2)}=\int_{-\infty}^{\infty}
W_{\gamma}(\mbox{\boldmath$\varrho$}, t)dt
\int_{t}^{\infty}2\sigma_p n_{\perp}(\mbox{\boldmath$\varrho$},
t')dt'd^2\varrho
\nonumber \\
\hspace{-7mm}&&=\frac{35}{9}\sqrt{\frac{\pi}{6}}N r \alpha^3
\lambda_c^2 \ln \frac{\sigma_y}{\lambda_c} \int
\chi_0(\mbox{\boldmath$\varrho$}) n_{\perp}^2
(\mbox{\boldmath$\varrho$})d^2\varrho
=\frac{35}{9}\sqrt{\frac{\pi}{6}}N r \chi_m \alpha^3
\frac{\lambda_c^2}{2\pi\sigma_x\sigma_y }\frac{4}{\pi}
\nonumber \\
\hspace{-7mm}&& \times \int_{0}^{\infty}\exp(-3v^2-2w^2) \big[
{\rm erf}^2\left(w \right)-{\rm erf}^2\left(iv \right)
\big]^{1/2} dw dv
\nonumber \\
\hspace{-7mm}&&=0.1167\alpha^3 N r \chi_m
\frac{\lambda_c^2}{\sigma_x\sigma_y } \ln
\frac{\sigma_y}{\lambda_c}. \label{co3.2}
\end{eqnarray}
So for the project  TESLA parameters the total number of produced
pairs by this mechanism is $Nn_p^{(2)} \sim 1.4 \cdot 10^4$ per
bunch in one collision.

Now we turn over to discussion of incoherent electroproduction of
pairs when both intermediate photons are virtual. To within the
logarithmic accuracy for any $\chi$ and $\kappa$ one can use the
method of equivalent photons
\begin{equation}
\frac{d\sigma_v}{d\omega}=n(\omega)\sigma_p(\omega), \label{co3.7}
\end{equation}
where
\begin{eqnarray}
\hspace{-7mm}&& n(\omega)=\frac{2\alpha}{\pi \omega
}\ln\frac{\Delta}{q_m}, ~\Delta=m(1+\kappa)^{1/3},
\nonumber \\
\hspace{-7mm}&& q_m=m\frac{\omega}{\varepsilon}\left(1+
\frac{\varepsilon \chi}{\omega}\right)^{1/3} +
\frac{1}{\sigma_y}\equiv q_F +q_\sigma. \label{co3.8}
\end{eqnarray}
Taking into account that the cross section of pair
photoproduction is
\begin{equation}
\sigma_p(\omega)=\frac{28
\alpha^3}{9m^2}\ln\frac{m}{q_m(\omega')},~
\omega'=\frac{m^2}{\omega}~(\kappa \ll 1);~\sigma_p \propto
\kappa^{-2/3}~ (\kappa \gg 1), \label{co3.9}
\end{equation}
we obtain in the main logarithmic approximation for the cross
section of the pair electroproduction
\begin{equation}
\sigma(2e \rightarrow 4e)= \frac{56 \alpha^4}{9m^2}
\int_{\omega_{min}}^{\omega_{max}}
\ln\frac{m}{q_m(\omega)}\ln\frac{m}{q_m(\omega')}\frac{d\omega}{\omega},
\label{co3.10}
\end{equation}
where
$\omega_{max}=\varepsilon/(1+\chi),~\omega_{min}=m^2/\omega_{max}$.
If we put $\chi=0,~\sigma_y=\infty$ we obtain the standard
Landau-Lifshitz cross section $\sigma_{LL}$ (see e.g. \cite{16a},
\cite{7})
\begin{equation}
\sigma_{LL}= \frac{28 \alpha^4}{27m^2}\ln^3\gamma^2.
\label{co3.11}
\end{equation}
With regard for the bounded transverse dimensions of beam and
influence of an external field the equivalent photon spectrum
changes substantially. For $\chi \sim 1$ we have
$q(\omega)=m(\omega/\varepsilon)^{2/3}+1/\sigma_y$ and under
condition $\gamma^{2/3}\lambda_c/\sigma_y \geq 1$ we find
\begin{equation}
\sigma_{v}= \frac{28 \alpha^4}{3m^2}\ln\left(\gamma^{4/3}
\frac{\lambda_c}{\sigma_y}\right)\ln^2\frac{\sigma_y}{\lambda_c}.
\label{co3.12}
\end{equation}
For TESLA parameters $\gamma^{2/3}\lambda_c/\sigma_y \simeq 1$
then
\begin{equation}
\sigma_{v}= \frac{28 \alpha^4}{81m^2}\ln^3\gamma^2. \label{co3.13}
\end{equation}
This cross section is three times smaller than standard
$\sigma_{LL}$. For the project TESLA parameters the number of
pairs produced by this mechanism is $n_v= L\sigma_{v}/2 \simeq
1.5 \cdot 10^4$ in one direction per bunch in one collision,
$L=1/(4\pi \sigma_x \sigma_y )$ is the geometrical luminosity per
bunch. So the both discussed mechanisms give nearly the same
contribution for this project. It should be noted that the above
analysis was performed under assumption that the configuration of
beams doesn't changed during collision, although in the TESLA
project the disruption parameter $D_y > 1$.

So we see that the beam-size effect  and an external field
effects are very essential for pair creation mechanism also.

\vspace{0.7cm}

{\bf Acknowledgments}

\vspace{0.5cm}

The authors are indebted to the Russian Foundation for Basic
Research supported in part this research by Grants 00-02-18007,
03-02-16154.

We would like to thank S. Klein for discussions and U.Uggerhoj
for information about experiment and data. 

\newpage

\setcounter{equation}{0}
\Alph{equation}

\appendix

\section{Appendix}

{\large {\bf Basic equations}} \vskip5mm The quasiclassical
operator method developed by authors is adequate for
consideration of the electromagnetic processes at high energy.
The probability of the radiation process has a form (see
\cite{9}, p.63, Eq.(2.27)); the method is given also in
\cite{16a}, for classical limit see \cite{5a}
\begin{equation}
dw=\frac{e^2}{(2\pi)^2} \frac{d^3k}{\omega}
\int_{}^{}dt_2\int_{}^{}dt_1 R^{\ast}(t_2) R(t_1)
\exp \left[-\frac{i\varepsilon}{\varepsilon'}
\left(kx(t_2)-kx(t_1)\right) \right],
\label{a1}
\end{equation}
where $k=(\omega, {\bf k})$ is the 4-momentum of the radiated
photon, $k^2=0$, $x(t)=(t, {\bf r}(t))$, $t$ is the time,
and ${\bf r}(t)$ is the particle location on a classical trajectory,
$kx(t)=\omega t- {\bf kr}(t)$,
$\varepsilon'=\varepsilon-\omega$, let us remind that
we employ units such that $\hbar=c=1$.
The matrix element $R(t)$ is defined by the structure of a current.
For an electron (spin 1/2 particle) one has
\begin{eqnarray}
&& R(t)=\frac{m}{\sqrt{\varepsilon \varepsilon'}}
\overline{u}_{s_{f}}({\bf p'})\hat{e}^{\ast}u_{s_{i}}({\bf p})
=\varphi_{s_{f}}^{+}\left(A(t)+i\mbox{\boldmath$\sigma$}{\bf B}(t) \right)
\varphi_{s_{i}}, \nonumber \\
&& A(t)=\frac{{\bf e}^{\ast}{\bf p}(t)}{2\sqrt{\varepsilon \varepsilon'}}
\left[\sqrt{\frac{\varepsilon'+m}{\varepsilon+m}}+
\sqrt{\frac{\varepsilon+m}{\varepsilon'+m}} \right]
\simeq \frac{1}{2}\left(1+\frac{\varepsilon}{\varepsilon'} \right)
{\bf e}^{\ast}\mbox{\boldmath$\vartheta$}, \nonumber \\
&& {\bf B(t)}=\frac{1}{2\sqrt{\varepsilon \varepsilon'}}
\left[\sqrt{\frac{\varepsilon'+m}{\varepsilon+m}} \left({\bf e}^{\ast}
\times {\bf p}(t) \right)
- \sqrt{\frac{\varepsilon+m}{\varepsilon'+m}} \left({\bf e}^{\ast}
\times ({\bf p}(t)-\hbar {\bf k}) \right)\right] \nonumber \\
&&\simeq \frac{\hbar \omega}{2\varepsilon'}\left({\bf e}^{\ast} \times
\left(\frac{{\bf n}}{\gamma}- \mbox{\boldmath$\vartheta$}\right) \right),
\label{a2}
\end{eqnarray}
here ${\bf p,~p'=p-k}$ is the initial and final momentum of an
electron, ${\bf e}$ is the vector of the polarization of a photon
(the Coulomb gauge is used),the four-component spinors $u_{s_f},
u_{s_i}$ and the two-component spinors $\varphi_{s_f},
\varphi_{s_i}$ describe the initial ($s_i$) and final ($s_f$)
polarization of the electron, {\bf v}={\bf v}(t) is the electron
velocity, $\mbox{\boldmath$\vartheta$} = v^{-1}\left({\bf v}-{\bf
n}({\bf n}{\bf v}) \right) \simeq {\bf v}_{\perp}$, ${\bf
v}_{\perp}$ is the component of particle velocity perpendicular
to the vector ${\bf n}={\bf k}/|{\bf k}|$, $\gamma=\varepsilon/m$
is the Lorentz factor. The final expressions in (\ref{a2}) are
given for radiation of ultrarelativistic electrons, they are
written down with relativistic accuracy (terms $\sim 1/\gamma$
are neglected) and in the small angle approximation. If we are
not interested in the initial and final particles polarization,
then
\begin{equation}
R^{\ast}(t_2) R(t_1) \rightarrow \frac{1}{2\varepsilon'^2}
\left[\frac{\omega^2}{\gamma^2}+\left(\varepsilon^2+\varepsilon'^2 \right)
\mbox{\boldmath$\vartheta$}\mbox{\boldmath$\vartheta$}' \right]
=\frac{1}{2}
{\cal L}(\mbox{\boldmath$\vartheta$}',\mbox{\boldmath$\vartheta$}),
\label{a3}
\end{equation}
where we have used the notation $\mbox{\boldmath$\vartheta$}'=
\mbox{\boldmath$\vartheta$}(t_2),~\mbox{\boldmath$\vartheta$}=
\mbox{\boldmath$\vartheta$}(t_1)$. For polarized initial electron
we have
\begin{equation}
R^{\ast}(t_2)R(t_1) = \frac{\varepsilon}{2\varepsilon'\gamma^2}
\left\{\frac{\omega^2}{\varepsilon\varepsilon'}+
\left(\frac{\varepsilon}{\varepsilon'}+\frac{\varepsilon'}{\varepsilon}
\right){\bf p}{\bf p}'+i\frac{\omega}{\varepsilon} (({\bf
p}'-{\bf p})\times{\bf v}) \mbox{\boldmath$\zeta$} \right\},
\label{a3a}
\end{equation}
where $\mbox{\boldmath$\zeta$}$ is the vector describing the
initial polarization of the electron (in its rest frame), ${\bf
p}= \gamma \mbox{\boldmath$\vartheta$}(t_1)$ and ${\bf p}'=\gamma
\mbox{\boldmath$\vartheta$}(t_2)$.

If the particle moves along a definite trajectory, then
substituting the classical values ${\bf r}(t)$ and ${\bf v}(t)$ in
(\ref{a1})-(\ref{a3}), we obtain the desired probability for this
process. When there is scattering, Eq.(\ref{a1}) must be averaged
over all possible particle trajectories. This operation is
performed with the aid of the distribution function, averaged
over atomic positions in the scattering medium and satisfying the
kinetic equation. We consider here general case with the external
field (particle acceleration) taken into account. The emission
probability per unit time is then (in this Appendix we follow the
paper \cite{8})
\begin{eqnarray}
&& dW= \left<\frac{dw}{dt} \right> = \frac{e^2}{(2\pi)^2}\frac{d^3k}{\omega}
{\rm Re}\int_{0}^{\infty} d\tau \exp \left(-i\frac{\varepsilon}{\varepsilon'}
\omega \tau \right)\int_{}^{}d^3 v d^3 v'd^3 r d^3 r'
{\cal L}(\mbox{\boldmath$\vartheta$}',\mbox{\boldmath$\vartheta$})
\nonumber \\
&& \times F_i({\bf r}, {\bf v}, t)
F_f({\bf r}', {\bf v}', \tau; {\bf r}, {\bf v})
\exp \left(-i\frac{\varepsilon}{\varepsilon'}
{\bf k}({\bf r}'-{\bf r}) \right)
\label{a4}
\end{eqnarray}
The distribution function $F$ in (\ref{a4}) satisfies the kinetic
equation
\begin{eqnarray}
&& \frac{\partial F({\bf r}, {\bf v}, t)}{\partial t}+
{\bf v}\frac{\partial F({\bf r}, {\bf v}, t)}{\partial {\bf r}}+
{\bf w}\frac{\partial F({\bf r}, {\bf v}, t)}{\partial {\bf v}} \nonumber \\
&& =n\int_{}^{}\sigma({\bf v},{\bf v}')\left[F({\bf r}, {\bf v}', t)-
F({\bf r}, {\bf v}, t) \right]d^3v',
\label{a5}
\end{eqnarray}
where $n$ is the number density of atoms in the medium,
and $\sigma({\bf v},{\bf v}')$ is the scattering cross section.
The normalization condition
\begin{equation}
\int_{}^{}d^3r\int_{}^{}d^3v F({\bf r}, {\bf v}, t)=1
\label{a6}
\end{equation}
should be also satisfied, as well as the initial condition for $F_f$:
\[
F_f({\bf r}', {\bf v}', 0; {\bf r}, {\bf v})=\delta({\bf r}-{\bf r}')
\delta({\bf v}-{\bf v}').
\]
In Eq.(\ref{a4}), we integrate over $d^3r d^3(r-r')$,
taking advantage of the fact that
$F_f({\bf r}', {\bf v}', \tau; {\bf r}, {\bf v})$ can only depend on the
coordinate difference ${\bf r}'-{\bf r}$:
\begin{equation}
dW= \frac{e^2}{(2\pi)^2}\frac{d^3k}{\omega}
{\rm Re}\int_{0}^{\infty} d\tau
\int_{}^{}d^3 v d^3 v'
{\cal L}(\mbox{\boldmath$\vartheta$}',\mbox{\boldmath$\vartheta$})
F_i({\bf v}, t) F_k({\bf v}', \tau; {\bf v}),
\label{a7}
\end{equation}
where
\begin{eqnarray}
\hspace*{-15mm}&& F_i({\bf v}, t)=\int_{}^{}d^3r F_i({\bf r}, {\bf v}, t)
\nonumber \\
\hspace*{-15mm}&& F_k({\bf v}', \tau; {\bf v})=
\exp \left(-i\frac{\varepsilon}{\varepsilon'}
\omega \tau \right)\int_{}^{}d^3 r'
\exp \left(-i\frac{\varepsilon}{\varepsilon'}
{\bf k}({\bf r}'-{\bf r}) \right)
F_f({\bf r}', {\bf v}', \tau; {\bf r}, {\bf v})
\label{a8}
\end{eqnarray}
We can put
\begin{eqnarray}
&& F_k({\bf v}', \tau; {\bf v})=
U(\mbox{\boldmath$\vartheta$}',\mbox{\boldmath$\vartheta$}, \tau)
\delta(|{\bf v}'|-|{\bf v}|)
\nonumber \\
&& U(\mbox{\boldmath$\vartheta$}',\mbox{\boldmath$\vartheta$}, 0)=
\delta(\mbox{\boldmath$\vartheta$}'-\mbox{\boldmath$\vartheta$}),
\label{a9}
\end{eqnarray}
if we take into account that in ultrarelativistic limit
${\bf w}{\bf v} \sim O(1/\gamma^3)$, ${\bf w} \simeq {\bf w}_{\perp}$,
and the scattering cross section is
\begin{equation}
\sigma({\bf v},{\bf v}')=\delta(|{\bf v}'|-|{\bf v}|)
\sigma(\mbox{\boldmath$\vartheta$}',\mbox{\boldmath$\vartheta$}).
\label{a10}
\end{equation}
Making use of these results and Eq.(\ref{a5}) for $F_f$,
we obtain the following equation for
$U(\mbox{\boldmath$\vartheta$}',\mbox{\boldmath$\vartheta$}, \tau)$
\begin{eqnarray}
&& \frac{\partial U}{\partial \tau}+
i\frac{\omega \varepsilon}{2\varepsilon'}\left(\frac{1}{\gamma^2}+
\mbox{\boldmath$\vartheta$}'^2 \right)U +
{\bf w}\frac{\partial U}{\partial \mbox{\boldmath$\vartheta$}'}
\nonumber \\
&& =n\int_{}^{} d^2\vartheta''
\sigma(\mbox{\boldmath$\vartheta$}',\mbox{\boldmath$\vartheta$}'')
\left[U(\mbox{\boldmath$\vartheta$}'',\mbox{\boldmath$\vartheta$}, \tau)
-U(\mbox{\boldmath$\vartheta$}',\mbox{\boldmath$\vartheta$}, \tau)\right],
\label{a11}
\end{eqnarray}
where for scattering in the screened Coulomb potential the cross
section in the Born approximation is
\begin{equation}
\sigma(\mbox{\boldmath$\vartheta$}',\mbox{\boldmath$\vartheta$})
=\frac{4Z^2\alpha^2}{\varepsilon^2
((\mbox{\boldmath$\vartheta$}'-\mbox{\boldmath$\vartheta$})^2+
\vartheta_1^2)^2},
\label{a12}
\end{equation}
here $\displaystyle{\vartheta_1=\frac{1}{\varepsilon a_s}}$, $a_s$ is the
screening radius ($a_s=0.81a_BZ^{-1/3}$, $a_B$ is the Bohr radius).
A similar equation for the $\mbox{\boldmath$\vartheta$}$-dependence of $U$
can be obtained by letting $\mbox{\boldmath$\vartheta$}' \rightarrow
\mbox{\boldmath$\vartheta$}$ and ${\bf w} \rightarrow -{\bf w}$.

If the final state of the charged particle is of no interest, the
probability (\ref{a7}) must be integrated over $d^3v'$. The resulting
emission probability per unit time, normalized to a single particle
moving at a speed ${\bf v}$ is then
\begin{equation}
dW= \frac{e^2}{(2\pi)^2}\frac{d^3k}{\omega}
{\rm Re}\int_{0}^{\infty} d\tau \exp \left(-i\frac{a\tau}{2}
\right) \left[\frac{\omega^2}{\varepsilon'^2 \gamma^2}
V_0(\mbox{\boldmath$\vartheta$}, \tau)+
\left(1+\frac{\varepsilon^2}{\varepsilon'^2} \right)
\mbox{\boldmath$\vartheta$}
{\bf V}(\mbox{\boldmath$\vartheta$}, \tau) \right],
\label{a13}
\end{equation}
where $\displaystyle{a=\frac{\omega m^2}{\varepsilon \varepsilon'}}$,
and $V_{\mu}=(V_0, {\bf V})$ satisfies the equation
\begin{eqnarray}
&& \frac{\partial V_{\mu}}{\partial \tau}+ i\frac{b}{2}
\mbox{\boldmath$\vartheta$}^2 V_{\mu} - {\bf w}\frac{\partial
V_{\mu} }{\partial \mbox{\boldmath$\vartheta$}}
\nonumber \\
&& =n\int_{}^{} d^2\vartheta'
\sigma(\mbox{\boldmath$\vartheta$},\mbox{\boldmath$\vartheta$}')
\left[V_{\mu}(\mbox{\boldmath$\vartheta$}', \tau)
-V_{\mu}(\mbox{\boldmath$\vartheta$}, \tau)\right],
\label{a14}
\end{eqnarray}
here $\displaystyle{b=\frac{\omega \varepsilon}{\varepsilon'}}$.
The initial conditions for $V_{\mu}$ are
\begin{equation}
V_0(\mbox{\boldmath$\vartheta$}, 0)=1,\quad
{\bf V}(\mbox{\boldmath$\vartheta$}, 0)=\mbox{\boldmath$\vartheta$}.
\label{a15}
\end{equation}

If the scattering cross section
$\sigma(\mbox{\boldmath$\vartheta$},\mbox{\boldmath$\vartheta$}')$
depends solely on the angle difference
$\mbox{\boldmath$\vartheta$}-\mbox{\boldmath$\vartheta$}'$,
as Eq.(\ref{a12}),
then Eq.(\ref{a14}) is most conveniently solved by Fourier transforming
with respect to the variable $\mbox{\boldmath$\vartheta$}$:
\begin{eqnarray}
&& \varphi_{\mu}({\bf x}, \tau)=\frac{1}{(2\pi)^2}\int_{}^{}d^2\vartheta
\exp (i\mbox{\boldmath$\vartheta$} {\bf x})
V_{\mu}(\mbox{\boldmath$\vartheta$}, \tau),
\nonumber \\
&& V_{\mu}(\mbox{\boldmath$\vartheta$}, \tau)
 =\int_{}^{} d^2x \exp (-i\mbox{\boldmath$\vartheta$} {\bf x})
\varphi_{\mu}({\bf x}, \tau).
\label{a16}
\end{eqnarray}
From Eq.(\ref{a14}), with the initial conditions (\ref{a15}), we have
\begin{eqnarray}
&& \frac{\partial \varphi_{\mu}({\bf x}, \tau)}{\partial \tau}-
i\frac{b}{2} \Delta_x \varphi_{\mu}({\bf x}, \tau) +
i{\bf wx}\varphi_{\mu}({\bf x}, \tau)
\nonumber \\
&& = (2\pi)^2 n \left[\Sigma({\bf x})-\Sigma(0) \right]
\varphi_{\mu}({\bf x}, \tau),\quad \Sigma({\bf x})=
\int_{}^{} d^2\vartheta
\exp (i\mbox{\boldmath$\vartheta$} {\bf x})
\sigma(\mbox{\boldmath$\vartheta$});
\nonumber \\
&& \varphi_{0}({\bf x}, 0)=
\delta({\bf x}),\quad \mbox{\boldmath$\varphi$}({\bf x}, 0)
=-i\mbox{\boldmath$\nabla$}\delta({\bf x})
\label{a17}
\end{eqnarray}
If the angular distribution of the radiation is of no interest,
Eq.(\ref{a13}) must be integrated over the photon emission angles
$\mbox{\boldmath$\vartheta$}$. Bearing in mind that $d^3 k=\omega^2 d\omega
d^2 \vartheta$, and that
\begin{eqnarray}
&& \frac{1}{(2\pi)^2} \int_{}^{} V_{0}(\mbox{\boldmath$\vartheta$}, \tau)
d^2 \vartheta = \varphi_{0}(0, \tau),
\nonumber \\
&& \frac{1}{(2\pi)^2} \int_{}^{} \mbox{\boldmath$\vartheta$}
{\bf V}(\mbox{\boldmath$\vartheta$}, \tau)
d^2 \vartheta = -i\mbox{\boldmath$\nabla$}
\mbox{\boldmath$\varphi$}({\bf x}, \tau) |_{{\bf x}=0},
\label{a18}
\end{eqnarray}
we obtain the following expression for the spectral distribution
of emission probability per unit time
\begin{equation}
\frac{dW}{d\omega}= \alpha \omega
{\rm Re}\int_{0}^{\infty} d\tau \exp \left(-i\frac{a\tau}{2}
\right) \left[\frac{\omega^2}{\varepsilon'^2 \gamma^2}
\varphi_{0}(0, \tau) -
i\left(1+\frac{\varepsilon^2}{\varepsilon'^2} \right)
\mbox{\boldmath$\nabla$}
\mbox{\boldmath$\varphi$}(0, \tau)\right].
\label{a19}
\end{equation}
The equation like (\ref{a17}) in the classical limit and with ${\bf w}=0$
was discussed in \cite{4}, Appendix V.

For the cross section (\ref{a12}) we have for $\Sigma({\bf x})$ (\ref{a17})
\begin{equation}
\Sigma(x) = \frac{Z^2 \alpha^2}{\pi \varepsilon^2} \frac{x}{\vartheta_1}
K_1(x\vartheta_1),
\label{a20}
\end{equation}
where $K_1(x\vartheta_1)$ is the modified Bessel function.
Bearing in mind that contributions to the cross section come from
$x \sim \vartheta_{ef}^{-1} \ll \vartheta_{1}^{-1}$, and expanding
$K_1(x\vartheta_1)$ as a power series in $x\vartheta_1$ we obtain
the following equation for $\varphi_{\mu}({\bf x}, \tau)$ from
Eq.(\ref{a17}), to power-series accuracy:
\begin{eqnarray}
&& \frac{\partial \varphi_{\mu}({\bf x}, \tau)}{\partial \tau}-
i\frac{b}{2} \Delta_x \varphi_{\mu}({\bf x}, \tau) + i{\bf
wx}\varphi_{\mu}({\bf x}, \tau)= u(x)
\nonumber \\
&& u(x)= \frac{2\pi n Z^2 \alpha^2}{\varepsilon^2}x^2\left( \ln
\frac{x\vartheta_1}{2}+C-\frac{1}{2} \right), \label{a21}
\end{eqnarray}
where $C=0.577216...$ is the Euler constant.

The "potential" $u(x)$ is calculated in the Born approximation
since Eq.(\ref{a12}) was used. The Coulomb correction to the
potential $u(x)$ were calculated in Appendix A of \cite{13}. The
eikonal approximation was use which is valid for arbitrary value
$Z\alpha$. Substituting this correction we have
\begin{equation}
u(x)= \frac{2\pi n Z^2 \alpha^2}{\varepsilon^2}x^2\left( \ln
\frac{x\vartheta_1}{2}+C-\frac{1}{2}+f(Z\alpha) \right),
\label{a22}
\end{equation}
where the function $f(Z\alpha)$ is defined in Eq.(\ref{2.9b}). In
absence of radiation and an external field (b={\bf w}=0)
Eqs.(\ref{a21}) and (\ref{a16}) at the initial condition
$\varphi({\bf x}, 0)= (2\pi)^{-2}$ give the solution of pure
scattering problem (the initial condition for the distribution
function $F(\mbox{\boldmath$\vartheta$}, t)$ is
$F(\mbox{\boldmath$\vartheta$}, 0)= \delta
(\mbox{\boldmath$\vartheta$})$):
\begin{eqnarray}
&& \varphi(x, t)=\frac{1}{(2\pi)^2}e^{u(x)t},
\nonumber \\
&& F(\mbox{\boldmath$\vartheta$}, t)=\frac{1}{(2\pi)^2}\int
\exp(-i\mbox{\boldmath$\vartheta$}{\bf x}+u(x)t)d^2x
\nonumber \\
&& dw(\vartheta, t)=\left[\int_{0}^{\infty}J_0(\vartheta
x)e^{u(x) t}xdx\right]\vartheta d\vartheta
 \label{a23}
\end{eqnarray}
For another form of solution see Eq.(\ref{82b}).

\newpage

{\bf Figure captions}

\vspace{15mm}
\begin{itemize}

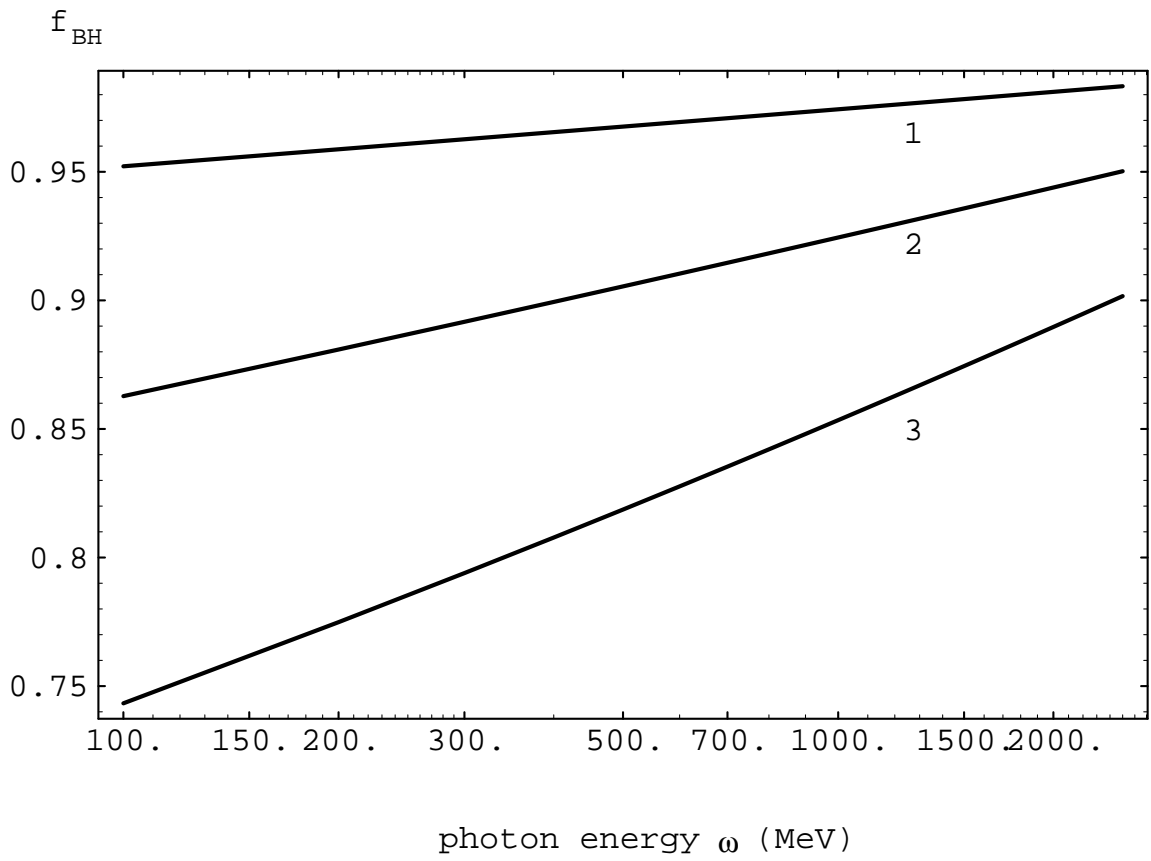
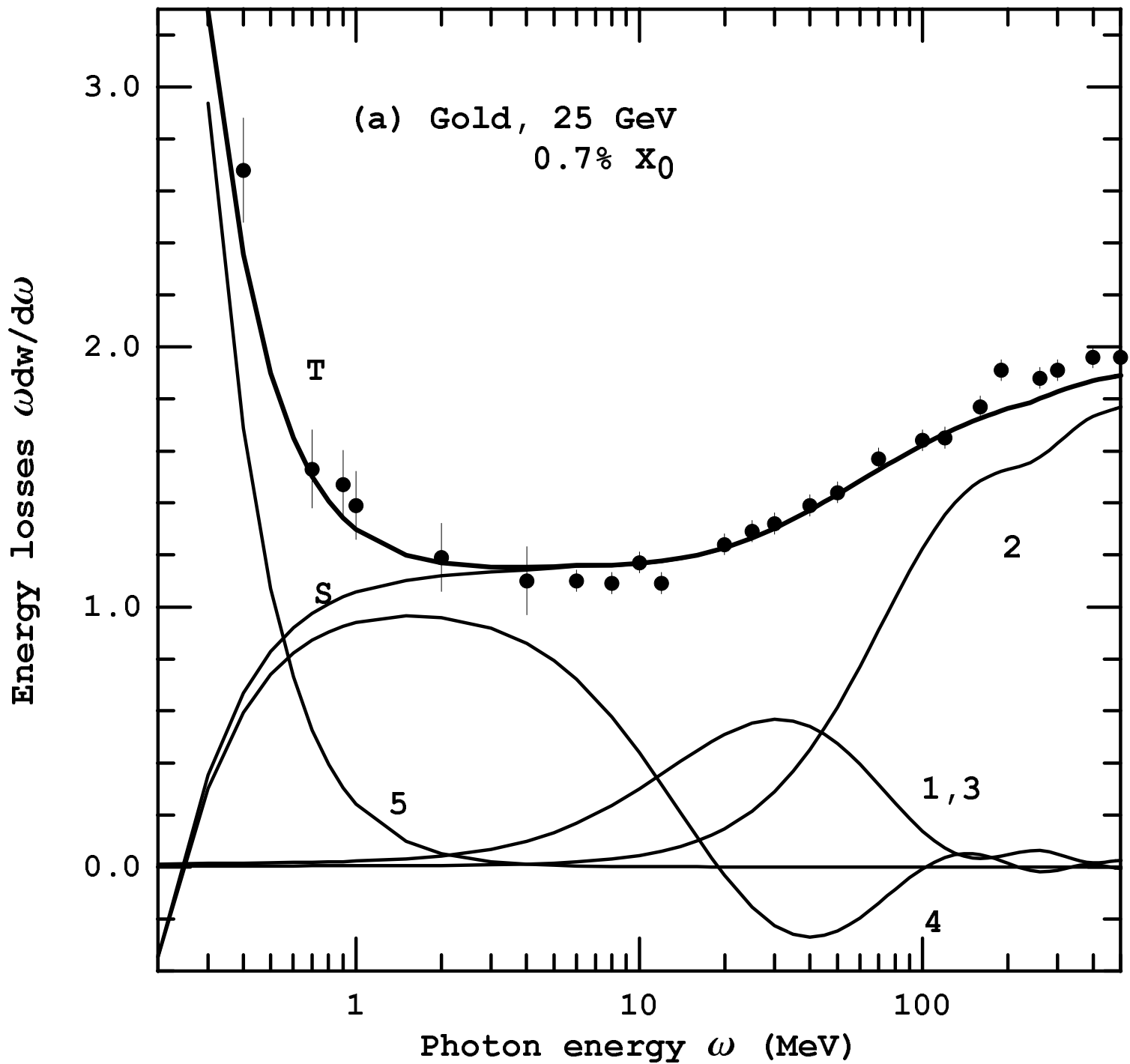
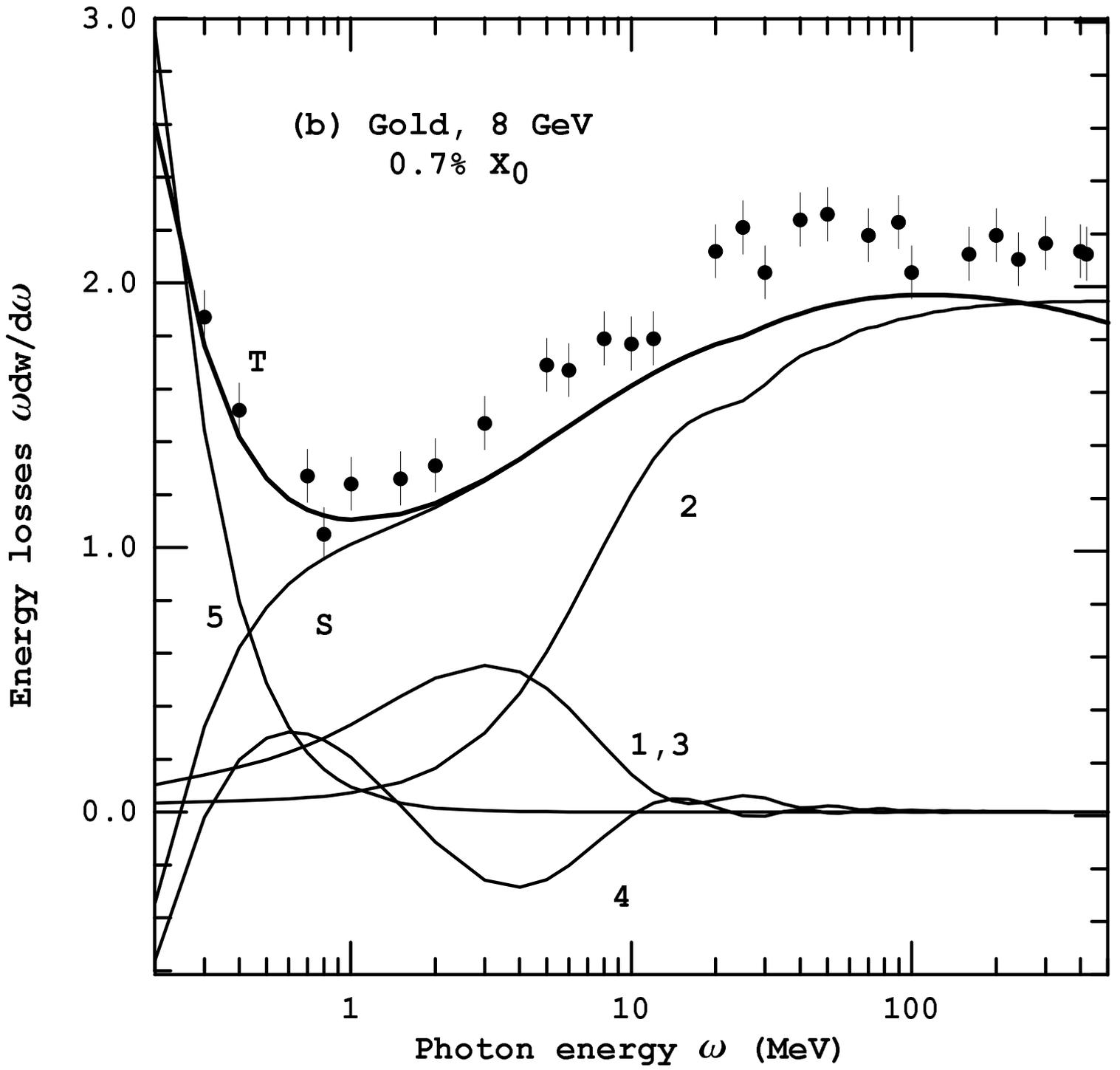
\item {\bf Fig.1} The functions $G(s)$ (curve 1) and $\phi(s)$
(curve 2) in Eq.(\ref{2.42}).

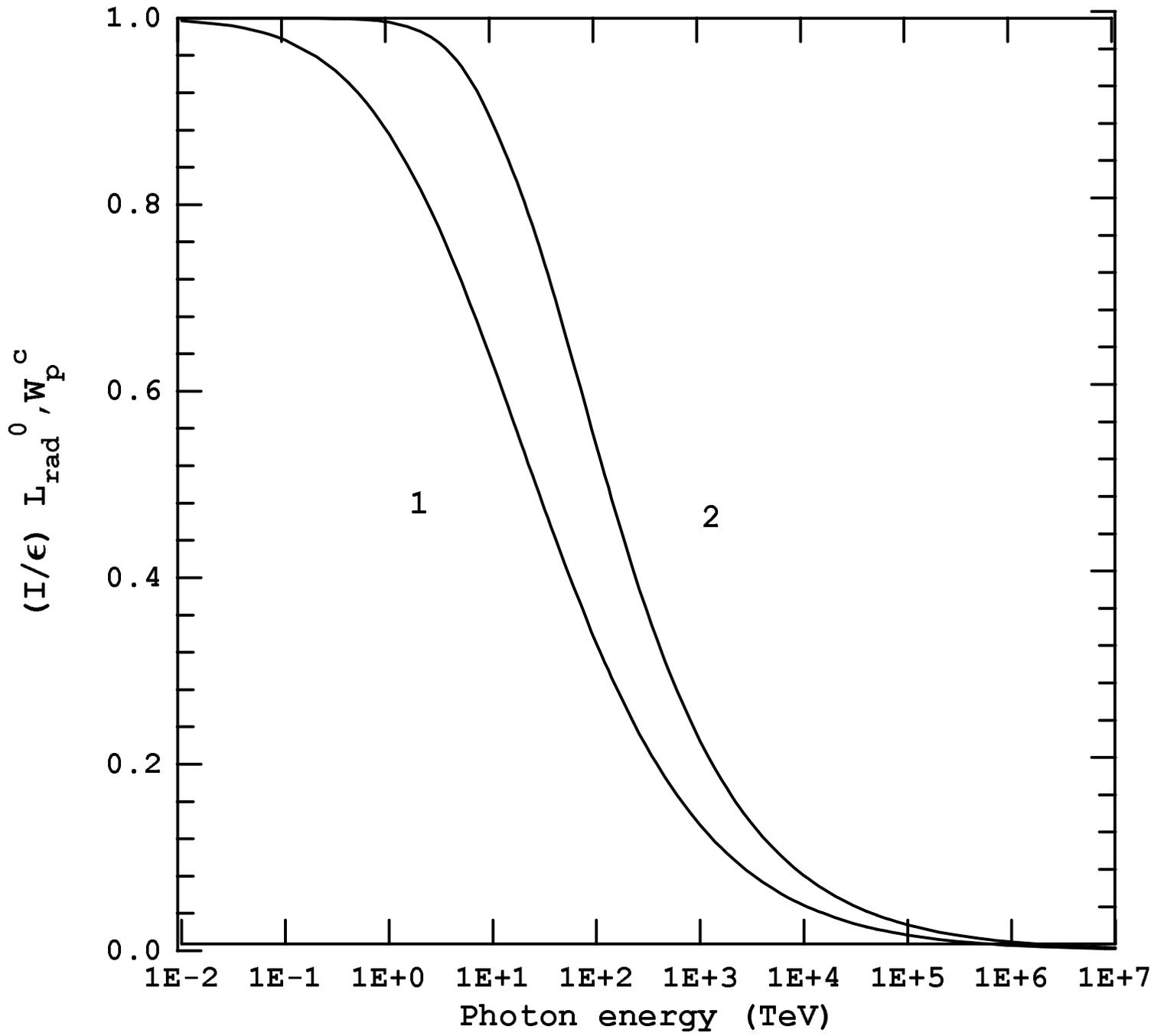
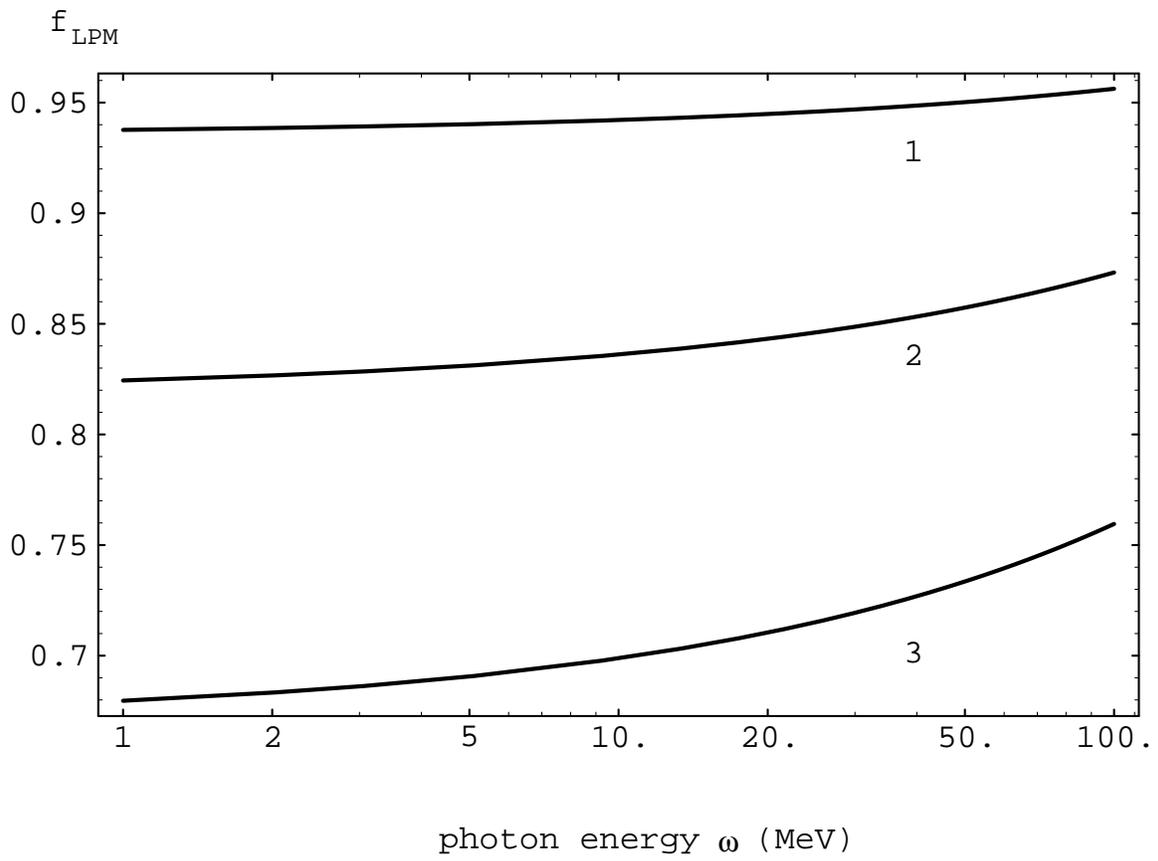
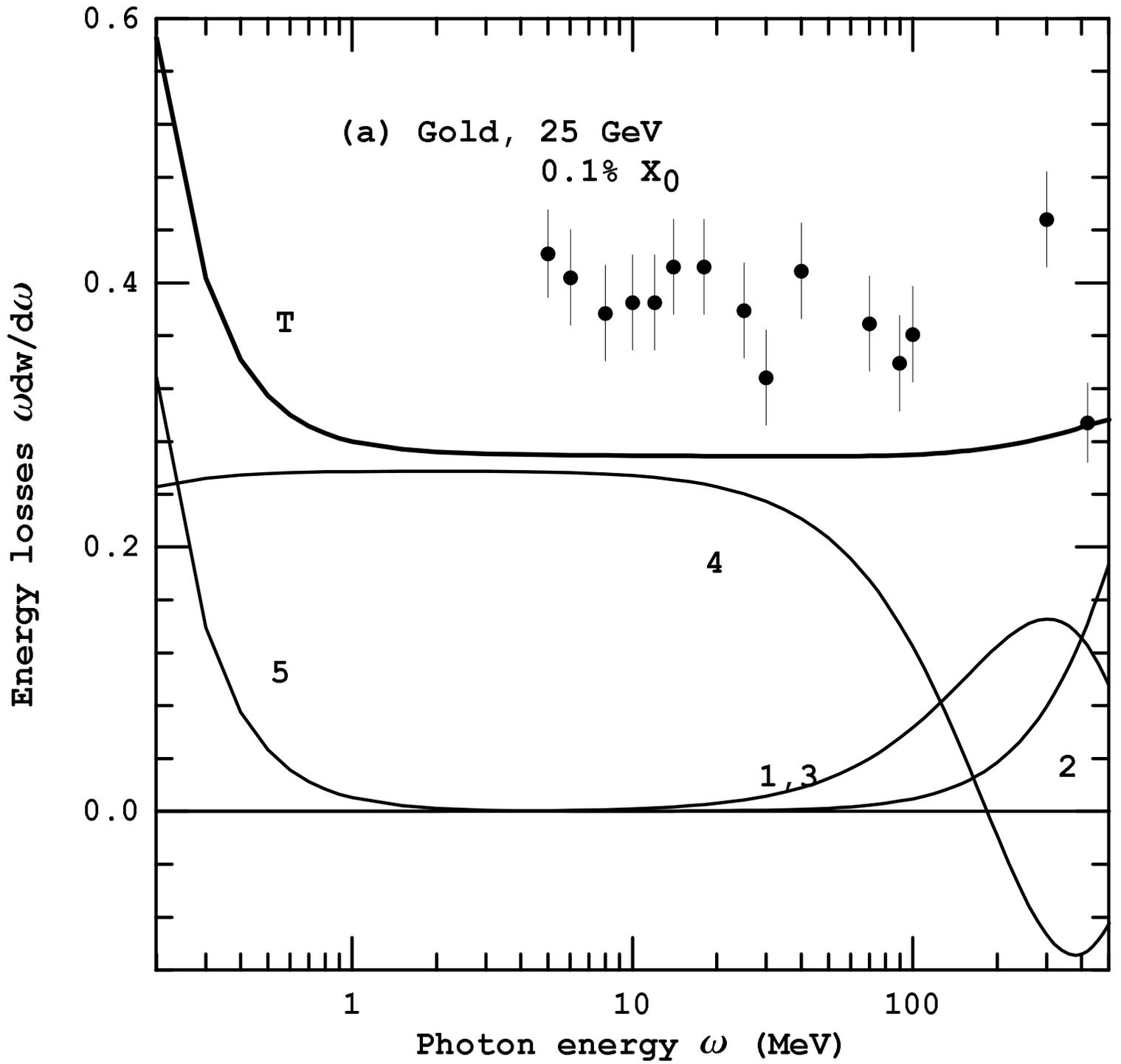
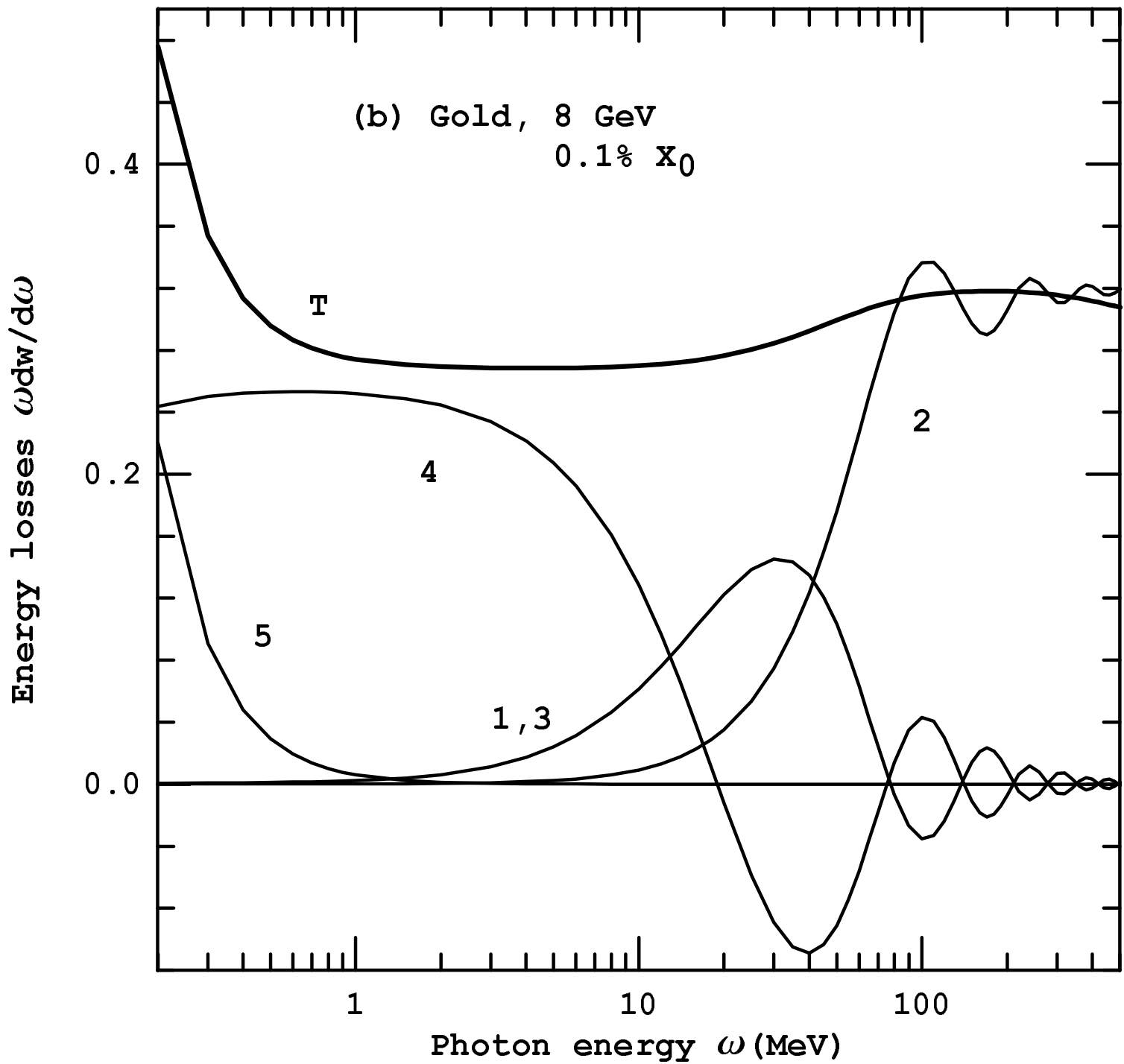
\item {\bf Fig.2} The functions $D_1(\nu_0)$ (curve 1) and $D_2(\nu_0)$
(curve 2) in Eq.(\ref{2.30}).

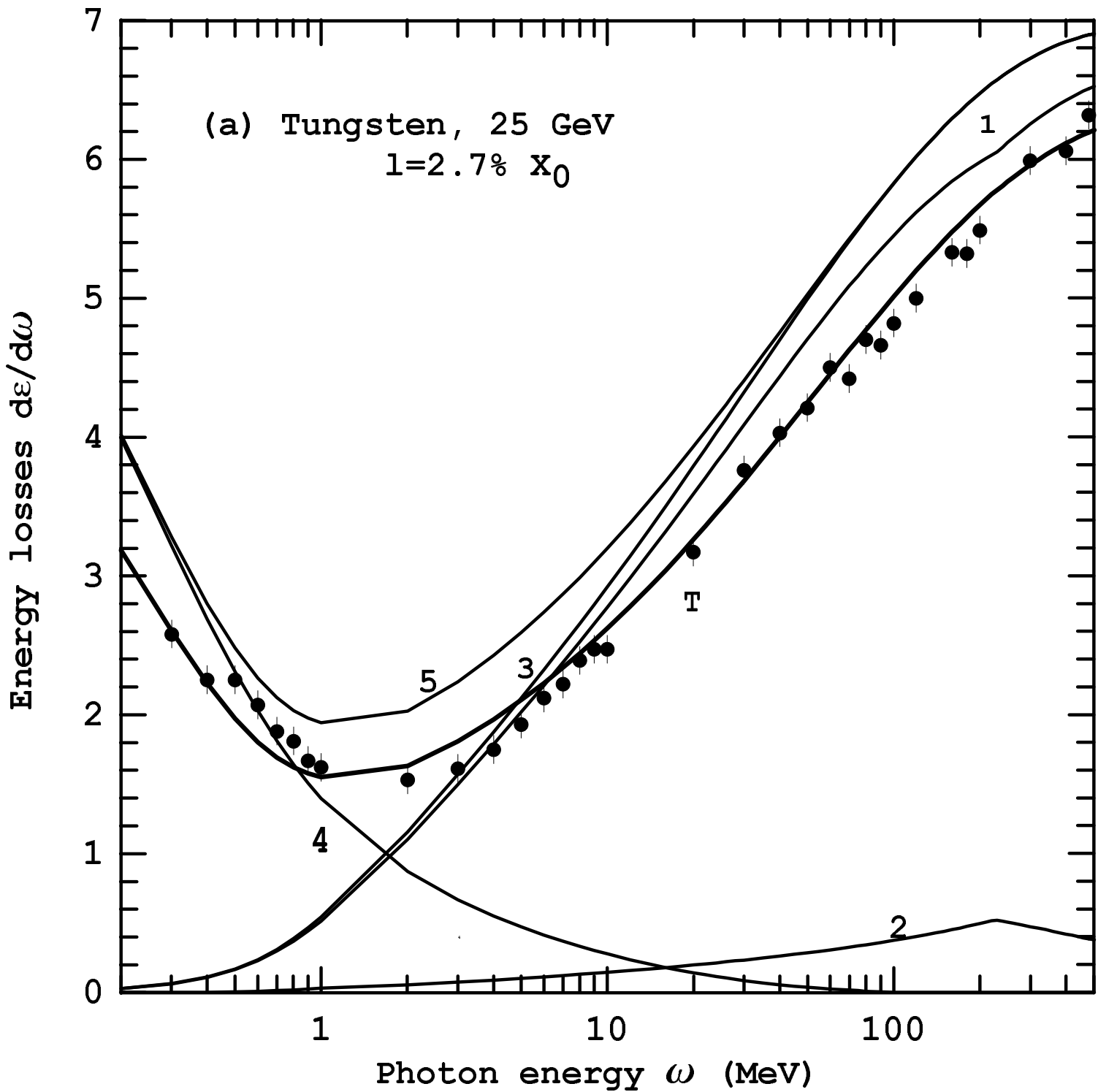
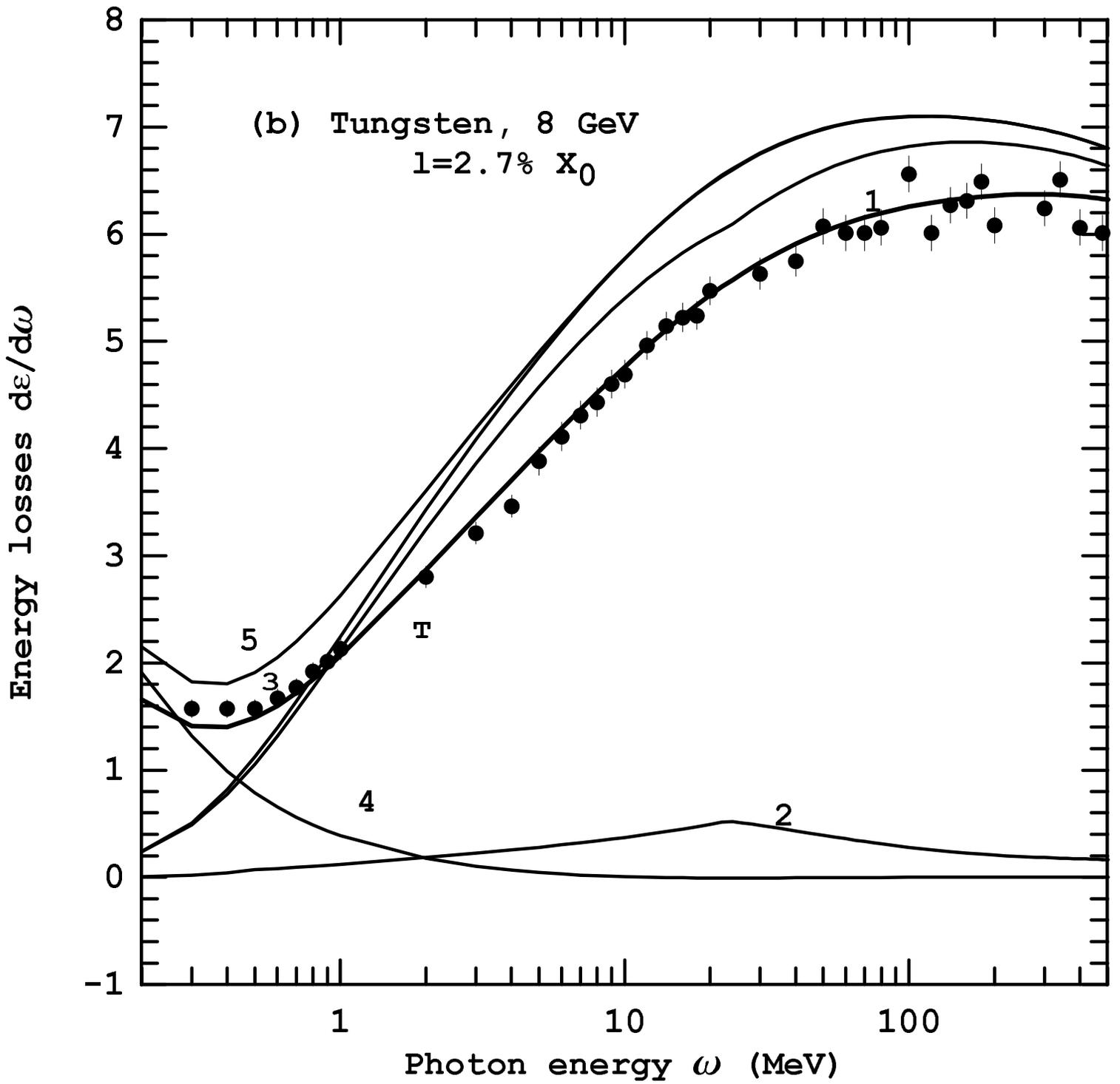
\item {\bf Fig.3} The spectral intensity of radiation
$\displaystyle{\omega \frac{dW}{d\omega}= x\frac{dW}{dx},~x=
\frac{\omega}{\varepsilon}}$
in gold in terms of $3 L_{rad}$ taken
with the Coulomb corrections (see Eq.(\ref{2.2i})).
\begin{itemize}
\item Curve BH is the Bethe-Maximon spectral intensity
(see Eq.(\ref{2.35}));
\item curve 1 is the logarithmic approximation
$\omega dW_c/d\omega$ Eq.(\ref{2.25}), curve c1 is the first
correction to the spectral intensity $\omega dW_1/d\omega$
Eq.(\ref{2.30}) and curve T1 is the sum of the previous
contributions for the electron energy $\varepsilon=25$~GeV;
\item curve 2 is the logarithmic approximation
$\omega dW_c/d\omega$ Eq.(\ref{2.25}), curve c2 is the first
correction to the spectral intensity $\omega dW_1/d\omega$
Eq.(\ref{2.30}) and curve T2 is the sum of the previous
contributions for the electron energy $\varepsilon=250$~GeV;
\item curves 3, c3, T3 are the same for the electron energy
$\varepsilon=2.5$~TeV;
\item curves 4, c4, T4 are the same for the electron energy
$\varepsilon=25$~TeV.
\end{itemize}

\item {\bf Fig.4} The relative energy losses of electron
per unit time in terms
of the Bethe-Maximon radiation length $L_{rad}^0$:
$\displaystyle{\frac{I}{\varepsilon}L_{rad}^0}$ in gold vs the initial energy
of electron
(curve 1) and the total pair creation probability
per unit time $W_p^c$ (see Eq.(\ref{2.23bp}))in terms
of the Bethe-Heitler total probability of pair creation $W_{p0}^{BH}$
(see Eq.(\ref{2.23p}))
in gold vs the initial energy of photon (curve 2).

\item {\bf Fig.5} The pair creation spectral probability
$\displaystyle{\frac{dW_p}{dx}, x=\frac{\varepsilon}{\omega}}$ in
gold in terms of the exact total Bethe-Maximon probability taken
with the Coulomb corrections (see Eq.(\ref{2.23p})).
\begin{itemize}
\item Curve BH is the Bethe-Maximon spectral probability
(see Eq.(\ref{2.22p}));
\item curve T1 is the total contribution (the sum of the
logarithmic approximation $dW_p^c/d\varepsilon$ (\ref{2.10p}) and
the first correction to the spectral probability $dW_p^1/d\varepsilon$
(\ref{2.15p})) for the photon energy $\omega=2.5$~TeV;
\item curve 2 is the logarithmic approximation
$dW_p^c/d\varepsilon$ (\ref{2.10p}), curve c2 is the first correction
to the spectral probability $dW_p^1/d\varepsilon$
(\ref{2.15p})) and curve T2 is the
sum of the previous contributions for the photon energy $\omega=25$~TeV;
\item curves 3, c3, T3 are the same for the photon energy $\omega=250$~TeV;
\item curves 4, c4, T4 are the same for the photon energy $\omega=2500$~TeV;
\end{itemize}

\item {\bf Fig.6} Anomalous magnetic moment(AMM) of electron in units
$\alpha/2\pi$ ($r$ in Eq.(\ref{3.6a}))~in gold vs electron energy
in TeV.

\item {\bf Fig.7} The functions Re~$k_{++}^2$(curve 2) and Im~$k_{++}^2$ (curve 1)
versus the photon energy taken in units $\omega_e$ (because of
this the curves are universal) for the case when the influence of
a medium is taken into account only (Eq.(\ref{2pr.20a})). The both
curves are normalized to the asymptotics given by
Eq.(\ref{2pr.22}) in the limit $\omega_F \rightarrow \infty$.

\item {\bf Fig.8}(a) The amplitude $M_{++}$ of the coherent photon scattering
in gold under influence of the multiple scattering at the different
momentum transfer to the photon $\Delta$ in terms of the amplitude
Im$M_{++}$ (\ref{3.55sc}) calculated for the screened Coulomb
potential.
\begin{itemize}
\item Curve 1 is Im$M_{++}$ for $\Delta=$0.4435 $m$.
\item Curve 2 is Re$M_{++}$ for $\Delta=$0.4435 $m$.
\item Curve 3 is Im$M_{++}$ for $\Delta=$0.0387 $m$.
\item Curve 4 is Re$M_{++}$ for $\Delta=$0.0387 $m$.
\end{itemize}

\item {\bf Fig.8}(b) The same as in Fig.1 but for
the amplitude $M_{+-}$ in terms of the amplitude
Im$M_{+-}$ (\ref{3.55sc}) calculated for the screened Coulomb
potential.

\item {\bf Fig.9} The reduction factor $f_{BH}$ for the energy of
the initial electron $\varepsilon=25$~GeV. The curves 1,2 and 3 are
respectively for $\beta$ =0.01, 0.03 and 0.06.

\item{\bf Fig.10} The reduction factor $f_{LPM}$ for energy of the initial
electron $\varepsilon$=25~GeV in tungsten ($\omega_c$=228~MeV).
The curves 1, 2 and 3 are respectively for
$\beta$=0.01, 0.03 and 0.06.

\item {\bf Fig.11} The energy losses spectrum
$\displaystyle{\frac{d\varepsilon}{d\omega}}$
in units $\displaystyle{\frac{2\alpha}{\pi}}$,
in the target consisting
of two gold plates with thickness $l_1=11.5~\mu m$
for the initial electrons energy $\varepsilon$=25~GeV .
\begin{itemize}
\item Curve 1 is for distance between plates $l_2=2l_1$;
\item Curve 2 is for distance between plates $l_2=4l_1$;
\item Curve 3 is for distance between plates $l_2=6l_1$;
\item Curve 4 is for distance between plates $l_2=8l_1$;
\item Curve 5 is for distance between plates $l_2=10l_1$.
\end{itemize}

\item{\bf Fig.12} The energy losses
$\displaystyle{\frac{d\varepsilon}{d\omega}}$
in tungsten with thickness $l=0.088~mm$ in units
$\displaystyle{\frac{2\alpha}{\pi}}$,
((a) is for the initial electrons energy $\varepsilon$=25~GeV and (b)
is for $\varepsilon$=8~GeV).
The Coulomb corrections and the polarization of a medium are included.
\begin{itemize}
\item Curve 1 is the contribution of the main term describing LPM effect;
\item curve 2 is the correction term;
\item curve 3 is the sum of the previous contributions;
\item curve 4 is the contribution of the boundary photons;
\item curve 5 is the sum of the previous contributions;
\item curve T is the final theory prediction with regard for
the reduction factor (the multiphoton effects).
\end{itemize}
Experimental data from Fig.9  of \cite{12}.

\item {\bf Fig.13} The energy losses
$\displaystyle{\omega \frac{dW}{d\omega}}$
in gold with thickness $l=0.023~mm$ in units
$\displaystyle{\frac{2\alpha}{\pi}}$,
((a) is for the initial electrons energy $\varepsilon$=25~GeV and (b)
is for $\varepsilon$=8~GeV).
The Coulomb corrections and the polarization of a medium are included.
\begin{itemize}
\item Curve 1 is the contribution of the term ${\rm Re}~J_1^{(2)}=
{\rm Re}~J_3^{(2)}$;
\item curve 2 is the contribution of the term ${\rm Re}~J_2^{(2)}$;
\item curve 4 is the contribution of the term ${\rm Re}~J_4^{(2)}$, all
(\ref{z12});
\item curve S is the sum of the previous contributions ${\rm Re}~J^{(2)}$;
\item curve 5 is the contribution of the boundary photons, see Eq.(2.20), \cite{14};
\item curve T is the total prediction for the radiation energy losses .
\end{itemize}
Experimental data from Fig.12 of \cite{12}.

\item {\bf Fig.14} The energy losses
$\displaystyle{\omega \frac{dW}{d\omega}}$
in gold with thickness $l=0.0038~mm$ in units
$\displaystyle{\frac{2\alpha}{\pi}}$,
((a) is for the initial electrons energy $\varepsilon$=25~GeV and (b)
is for $\varepsilon$=8~GeV).
The Coulomb corrections and the polarization of a medium are included.
\begin{itemize}
\item Curve 1 is the contribution of the term ${\rm Re}~J_1^{(2)}=
{\rm Re}~J_3^{(2)}$;
\item curve 2 is the contribution of the term ${\rm Re}~J_2^{(2)}$;
\item curve 4 is the contribution of the term ${\rm Re}~J_4^{(2)}$, all
(\ref{z12});
\item curve S is the sum of the previous contributions ${\rm Re}~J^{(2)}$;
\item curve 5 is the contribution of the boundary photons see Eq.(2.20), \cite{14};
\item curve T is the total prediction for the radiation energy losses .
\end{itemize}
Experimental data from Fig.13 of \cite{12}.

\item{\bf Fig.15} The energy loss
$\displaystyle{\frac{d\varepsilon}{d\omega}}$
in iridium target with thickness $l=0.128~mm$ in units
of inverse radiation length $L_{rad}^0$  (Eq.(\ref{2.2i}))
for the initial electrons energy $\varepsilon$=287~GeV. 
The Coulomb corrections are included.
\begin{itemize}
\item curve 1 is the the Bethe-Maximon intensity spectrum;
\item curve 2 is the contribution of the main (Migdal) term describing LPM effect;
\item curve 3 is the correction term;
\item curve 4 is the sum of the previous contributions;
\item curve T is the final theory prediction with regard for
the reduction factor (the multiphoton effects).
\end{itemize}
Experimental data from Fig.2  of \cite{HU}.

\item{\bf Fig.16} The same as in Fig.15 but 
for the initial electrons energy $\varepsilon$=207~GeV.  

\item{\bf Fig.17} The same as in Fig.15 but 
for the initial electrons energy $\varepsilon$=149~GeV.  

\item{\bf Fig.18} The reduction factor for iridium target 
with thickness $l=0.128~mm$ (4.36 \% $L_{rad}$) versus 
$x=\omega/\varepsilon$. The curves 1, 2, 3 are for energies
149~GeV, 207~GeV and 287~GeV correspondingly.

\item{\bf Fig.19} The relative energy losses of electron
per unit time including the contribution of the 
correction term Eq.(\ref{2.25}) in terms
of the Bethe-Maximon radiation length $L_{rad}$ Eq.(\ref{2.2i}):
$\displaystyle{\frac{I}{\varepsilon}L_{rad}}$ in iridium (curve 1)
and in lead (curve 2) vs the initial energy
of electron.

\item {\bf Fig.20} The bremsstrahlung intensity spectrum $\omega
d\sigma/d\omega$ in units $2\alpha r_0^2$ versus the photon
energy in units of initial electron energy
($x=\omega/\varepsilon$) for VEPP-4 experiment. The upper curve is
the standard QED spectrum, the three close curves below are
calculated  for the different vertical dimensions of colliding
beams (equal for two colliding beams
$\sigma=\sigma_z=\Delta_z$):$\sigma=20~\mu m$ (bottom),
$\sigma=24~\mu m$ (middle), $\sigma=27~\mu m$ (top). The data
measured in \cite{exp1} are presented as circles (the experiment
in 1980) and as triangles (the experiment in 1981) with 6 \%
systematic error as obtained in \cite{exp1}.

\item {\bf Fig.21} The bremsstrahlung intensity spectrum $\omega
d\sigma/d\omega$ in units $2\alpha r_0^2$ versus the photon
energy in units of initial electron energy
($x=\omega/\varepsilon$) for the HERA experiment. The upper curve
is the standard QED spectrum, the two close curves below are
calculated  with the beam-size effect taken into account: the
bottom curve is for $\sigma_z=\Delta_z=50~\mu m,
\sigma_y=\Delta_y=250~\mu m$; while the top curve is for
$\sigma_z=\Delta_z=54~\mu m, \sigma_y=\Delta_y=250~\mu m$. The
data taken from Fig.5c in \cite{P}

\item {\bf Fig.22} The bremsstrahlung intensity spectrum $\omega
d\sigma/d\omega$ in units $2\alpha r_0^2$ versus the photon
energy in units of initial electron energy
($x=\omega/\varepsilon$) for linear collider with beam energy
$\varepsilon=250~$GeV. The upper curve is the standard QED
spectrum. The curve below is calculated  with the beam-size
effect taken into account for $\sigma_x=553nm, \sigma_y=5nm$.

\item {\bf Fig.23} Spectral intensity of radiation of round beams in units
 $\alpha m^2 \sigma_z$ for
 $\chi_{rd}$=0.13 calculated according to
 Eqs.(\ref{co2.4}),(\ref{co2.7}).

\item {\bf Fig.24} The spectral radiation intensity $dI/d\omega$
of coherent radiation (fast falling with $x=\omega/\varepsilon$
increase curve) and of incoherent radiation (the curve which is
almost constant) in units $N\alpha^2\lambda_c/\sigma_x$ for beams
with dimensions $\sigma_x=553$~nm, $\sigma_y=5$~nm, for
$\chi_m$=0.13.

\end{itemize}

\newpage

\begin{table}
\begin{center}
{\sc TABLE 1}~
{The characteristics of the LPM effect}\\
{Listed are the charge of nucleus $Z$, the density $\varrho$, the
radiation length $ L_{rad}$},{\newline $\varepsilon_e $ (see
Eq.(\ref{2.30b})), $\omega_e $(see Eq.(\ref{2.23ap})), $\omega_c$
Eq.(\ref{2.31b})), $\omega_p$ Eq.(\ref{42}))}, {\newline $l$ is
the target thickness in percent of the radiation length $
L_{rad}$} {\newline (all photon energies $\omega_c$ and $\omega_p$
are for the electron energy $\varepsilon = 25~$GeV)}
\end{center}
\begin{center}
\begin{tabular}{*{9}{|c}|}
\hline Material& Z& $\varrho (g/cm^3)$& $L_{rad} (cm)$
&$\varepsilon_e$(TeV) & $\omega_e$ (TeV)& $\omega_c$(MeV) & $\omega_p$(MeV) & $l$ \\
\hline C   & 6 & 2.2  & 18.8& 144 &  580 & 4.3 & 1.5 & 2.1, 6 \\
\hline Al   & 13& 2.7  & 8.9 & 68  &  270 & 9.2 & 1.6 & 3.5, 6 \\
\hline Fe     & 26& 7.87 & 1.76& 14  &  56  & 46  & 2.7 & 2.8, 6.1\\
\hline Pb     & 82& 11.35& 0.56& 4.3 &  17.2& 150 & 3.0 & 2.7     \\
\hline W      & 74& 19.3 & 0.35& 2.7 &  10.8& 230 & 3.9 & 2.7, 6.4\\
\hline U      & 92& 18.95& 0.32& 2.5 &  10  & 250 & 3.8 & 2.2, 4.2 \\
\hline Au     & 79& 19.32& 0.33& 2.5 &  10  & 240 & 3.9 & 0.1, 0.7 \\
\hline Water    & - &  1   & 36.1& 277 &  1100& 2.6 & -  & -  \\
\hline Std.rock & 11 & 2.65& 10.0& 77  &  300 & 8.1 & -  & -  \\
\hline
\end{tabular}
\end{center}
\end{table}

\begin{table}
\begin{center}
{\sc TABLE 2}~
{Characteristic parameters of the radiation process in}\\
{tungsten with the thickness $l = 2.7\% L_{rad}$}
\end{center}
\begin{center}
\begin{tabular}{*{7}{|c}|}
\hline $\varepsilon~(GeV)$ & $\omega_c~(MeV)$ & $\omega_p~(MeV)$&
$T_c$& $\omega_1~(MeV)$& $\beta_m$&$\omega_m~(MeV)$ \\ \hline 25
& 230  &  3.9& 21.3&  1.6 &  2.7&  2 \\ \hline
 8 & 23&  1.3& 21.3&  0.76&  5.7& 0.5\\ \hline
\end{tabular}
\end{center}
\end{table}

\begin{table}
\begin{center}
{\sc TABLE 3}~
{Characteristic parameters of the radiation process}\\
{in gold with the thickness $l = 0.7\% L_{rad}$ and $l = 0.1\% L_{rad}$}
{\newline (all photon energies $\omega$ are in MeV)}
\end{center}
\begin{center}
\begin{tabular}{*{9}{|c}|}
\hline $\varepsilon~(GeV)$ & $\omega_c~$ & $\omega_p~$&
$T_c(0.7)$& $T_c(0.1)$&$\omega_1(0.7)~$&
$\beta_m(0.7)$&$\beta_m(0.1)$& $\omega_{th}~$ \\ \hline 25 & 240 &
3.9& 5.82& 0.96&  1.6 & 0.75 & 0.12& 28 \\ \hline
 8 & 25&  1.3 & 5.82& 0.96& 0.76 & 1.6  & 0.25& 3.0 \\ \hline
\end{tabular}
\end{center}
\end{table}


\newpage

\begin{thebibliography}{99}

\bibitem{1} L. D. Landau and I. Ya. Pomeranchuk, Dokl.Akad.Nauk SSSR
{\bf 92} (1953) 535, 735. See in English in {\em The
Collected Papers of L. D. Landau}, Pergamon Press, 1965.
\bibitem{2} A. B. Migdal, Phys. Rev. {\bf 103} (1956) 1811.
\bibitem{3} A. B. Migdal, Sov. Phys. JETP {\bf 5} (1957) 527.
\bibitem{4} M. L. Ter-Mikaelian, {\em High Energy Electromagnetic
Processes in Condensed Media}, John
Wiley \& Sons, 1972.
\bibitem{5} E. L. Feinberg and I. Ya. Pomeranchuk, Nuovo Cimento,
Supplement to Vol. {\bf 3} (1956) 652.
\bibitem{6} V. M. Galitsky and I. I. Gurevich, Nuovo Cimento
{\bf 32} (1964) 396.
\bibitem{7a} V. N. Baier and V. M. Katkov,
Sov. Phys. JETP, {\bf 26} (1968) 854; {\bf 28} (1969) 807.
\bibitem{7} V. N. Baier, V. M. Katkov and V. S. Fadin,
{\em Radiation from Relativistic Electrons} (in Russian) Atomizdat,
Moscow, 1973.
\bibitem{8} V. N. Baier, V. M. Katkov and V. M. Strakhovenko,
Sov. Phys. JETP {\bf 67} (1988) 70.
\bibitem{9} V. N. Baier, V. M. Katkov and V. M. Strakhovenko,
{\em Electromagnetic Processes at High Energies in Oriented
Single Crystals}, World Scientific Publishing Co, Singapore, 1998.
\bibitem{16a} V. B. Berestetskii, E. M. Lifshitz and L. P. Pitaevskii,
{\em Quantum Electrodynamics} Pergamon Press, Oxford, 1982.
\bibitem{10} P. L. Anthony, R. Becker-Szendy, P. E. Bosted {\em et al},
Phys. Rev. Lett. {\bf 75} (1995) 1949.
\bibitem{11} P. L. Anthony, R. Becker-Szendy, P. E. Bosted {\em et al},
Phys. Rev. Lett. {\bf 76} (1996) 3550.
\bibitem{12} P. L. Anthony, R. Becker-Szendy, P. E. Bosted {\em et al},
Phys.Rev.{\bf d56} (1997) 1373.
\bibitem{E2} S. Klein,
Rev. Mod. Phys. {\bf 71} (1999) 1501.
\bibitem{HU} H. D. Hansen, U. I. Uggerhoj, C.C.Biino {\em et al},
Phys.Rev.Lett. {\bf 91} (2003) 014801.
\bibitem{13} V. N. Baier and V. M. Katkov,
Phys.Rev. {\bf D57} (1998) 3146.
\bibitem{14} V. N. Baier and V. M. Katkov,
{\it Quantum Aspects of Beam Physics}, ed.P. Chen,
World Scientific PC, Singapore, 1998, p.525.
\bibitem{15} V. N. Baier and V. M. Katkov,
Phys.Rev. {\bf D59} (1999) 056003.
\bibitem{x15} V. N. Baier and V. M. Katkov,
Phys.Rev. {\bf D60} (1999) 076001.
\bibitem{L5} V. N. Baier and V. M. Katkov,
Phys.Rev. {\bf D62} (2000) 036008.
\bibitem{16} R. Blancenbeckler and S. D. Drell, Phys.Rev.
{\bf D53} (1996) 6265.
\bibitem{17} B. G. Zakharov,
Pis'ma v ZhETF {\bf 63 } (1996) 906.
\bibitem{18} B. G. Zakharov,
Pis'ma v ZhETF {\bf 64 } (1996) 781.
\bibitem{19} R. Baier, Yu. L. Dokshitzer, A. H. Mueller, S. Peigne,
and D. Schiff, Nucl. Phys. {\bf B478} (1996) 577.
\bibitem{19b} V. N. Baier and V. M. Katkov,
Phys.Rev. {\bf D63} (2001) 116008.
\bibitem{20} V. N. Baier, and V. M. Katkov,
Sov. Phys. Dokl. {\bf 17} (1973) 1068.
\bibitem{21}  V. M. Katkov, and V. M. Strakhovenko,
Sov. J. Nucl.Phys. {\bf 25} (1977) 660.
\bibitem{S} J. Schwinger, Phys.Rev. {\bf 73} (1948) 416; {\bf 76}
(1949) 790.
\bibitem{BKS1} V. N. Baier, V. M. Katkov and V. M. Strakhovenko,
Lett. Nuovo Cimento {\bf 15} (1976) 149; Sov. J. Nuc. Phys. {\bf
24} (1976) 197.
\bibitem{BK1} V. N. Baier, and V. M. Katkov,
Phys.Lett. A 280 (2001) 275.
\bibitem{BK2} V. N. Baier, and V. M. Katkov,
Phys.Lett. A 286 (2001) 299.
\bibitem{BKS3} V. N. Baier, V. M. Katkov and V. M. Strakhovenko,
Sov. Phys. Doklady {\bf 16} (1971) 230.
\bibitem{R} V. I. Ritus,
Sov. Phys. JETP {\bf 30} (1970) 1181.
\bibitem{S1} J. Schwinger, Phys.Rev. {\bf 75} (1949) 1912; {\bf 82}
(1951) 664.
\bibitem{Na} N. B. Narozhny, Zh. Eksp.Teor.Fiz., {\bf
55} (1968) 714.
\bibitem{d3} A. I. Milstein and M. Schumacher,
Phys. Rep.C243 (1994) 183.
\bibitem{CW} M. Cheng and T. T. Wu,
Phys.Rev. {\bf 182} (1969) 1873; D {\bf 2} (1970) 2444;
\newline D {\bf 5} (1972) 3077.
\bibitem{MS1} A. I. Milstein and V. M. Strakhovenko,
Sov. Phys. JETP {\bf 58} (1983) 8.
\bibitem{LM1} R. N. Lee and A. I. Milstein,
Phys.Lett. A 198 (1995) 217.
\bibitem{LM2} R. N. Lee, A. I. Milstein, and V. M. Strakhovenko,
Jour. Exper. Theor. Phys. {\bf 90} (2000) 66.
\bibitem{KS} V. M. Katkov and V. M. Strakhovenko,
Jour. Exper. Theor. Phys. {\bf 92} (2001) 561.
\bibitem{BK} V. N. Baier, V. M. Katkov,
Phys.Rev. {\bf D63} (2001) 116008, hep-ph/0005215.
\bibitem{B1} H. A. Bethe, Phys.Rev. {\bf 89} (1953) 1256.
\bibitem{Mo1} G. Moli\`ere, Z.Naturforsch.{\bf 2a} (1947) 133.
\bibitem{22}  V. M. Katkov, and V. M. Strakhovenko,
Sov. Phys. Dokl. {\bf 21} (1976) 658.
\bibitem{22c} Y.-S. Tsai, Rev. Mod. Phys. {\bf 46} (1974), 815.
\bibitem{18a} V. N. Baier, V. M. Katkov and V. M. Strakhovenko,
Nucl. Phys. {\bf B328} (1989) 387.
\bibitem{16c} R. Blankenbecler,
Phys.Rev {\bf D55} (1997) 2441.
\bibitem{20a} V. N. Baier, V. S. Fadin, V. A. Khoze and E.A.
Kuraev, Phys. Rep. {\bf 78} (1981) 293.
\bibitem{17b} I. I. Gol'dman,
Sov. Phys. JETP {\bf 11} (1960) 1341.
\bibitem{18b} F. F. Ternovskii,
Sov. Phys. JETP {\bf 12} (1960) 123.
\bibitem{18c}V. E. Pafomov,
Sov. Phys. JETP {\bf 20} (1965) 253.
\bibitem{SF} N. F. Shul'ga and S. P. Fomin,
Sov. Phys. JETP {\bf 86} (1998) 32.
\bibitem{SF1} N. F. Shul'ga and S. P. Fomin,
Nucl.Instr.and Meth. B {\bf 145} (1998) 73.
\bibitem{z8a} I. S. Gradshtein and I. M. Ryzhik,
{\em Table of Integrals, Series, and Products},
Academic Press, New York, 1965.
\bibitem{z9} X. Artru, G. B. Yodth and G. Mennesier,
Phys.Rev. {\bf D12} (1975) 1289.
\bibitem{z10} V. N. Baier and V. M. Katkov,
Phys.Lett. A 252 (1999) 263.
\bibitem{z11} N. F. Shul'ga and S. P. Fomin,
JETP Lett. {\bf 63} (1996) 873.
\bibitem{y9a} L. D. Landau, J.Phys. USSR {\bf 8} (1944) 201.
\bibitem{x16b} R. Blankenbecler,
Phys.Rev {\bf D55} (1997) 190.
\bibitem{x14} V. N. Baier and V. M. Katkov,
 Nucl. Instr. Meth. A 439 (2000) 189.
\bibitem{LMNS} N. V. Laskin, A. S. Mazmanishvili, N. N. Nasonov,
N. F. Shul'ga, Jour. Exper. Theor. Phys. {\bf 89} (1985) 763.
\bibitem{Za3} V. G. Zakharov, Phys. Atom. Nucl.{\bf 62} (1999) 1075.
\bibitem{BK3} V. N. Baier, and V. M. Katkov,
``Variation of radiation length due to LPM effect'' (to be published)
\bibitem{CRE1} M. Miezowitz, O. Stanisz and W. Wolther,
Nuovo Cimento {\bf 5} (1957) 513.
\bibitem{CRE2} P. H. Fowler, D. H. Perkins and K. Pinkau,
Phil.Mag. {\bf 4} (1959) 1030.
\bibitem{CRE3} A. Varfolomeev, I. Gerasimova, I. I. Gurevitch et al.,
Sov.Phys. JETP {\bf 11} (1960) 23.
\bibitem{CRE4} E. Lohrmann,
Phys.Rev {\bf 122} (1961) 1908.
\bibitem{CRE5} S. C. Strautsz et al.,
in {\em Proc.of the 22th Intl.Cosmic Ray Conf.},Dublin, Ireland, V.4, p.233.
\bibitem{E1} A. Varfolomeev, et al.,
Sov.Phys. JETP {\bf 42} (1976) 218.
\bibitem{5a} L. D. Landau and E. M. Lifshitz, {\em Classical Theory of Fields}
4th English Ed., Pergamon Elmsford, New York, 1975.
\bibitem{exp1} A. E. Blinov, A. E. Bondar, Yu. I. Eidelman et al.,
{\it Preprint INP 82-15, Novosibirsk 1982}; Phys.Lett. {\bf B113}
(1982) 423.
\bibitem{cBKS1} V. N. Baier, V. M. Katkov, V. M. Strakhovenko,
{\it Preprint INP 81-59, Novosibirsk 1981}; Sov.J.Nucl.Phys, {\bf
36} (1982) 95.
\bibitem{BD} A. I. Burov and Ya. S. Derbenev,
{\it Preprint INP 82-07, Novosibirsk 1982}.
\bibitem{KPS} G. L. Kotkin, S. I. Polityko, and V. G. Serbo,
 Yad.Fiz., {\bf 42} (1985) 692, 925.\\
G. L. Kotkin et al.,  Z.Phys., {\bf C39} (1988) 61.
\bibitem{cBK2} V. N. Baier and V. M. Katkov,
Phys. Rev. {\bf D66} (2002) 053009.
\bibitem{P} K. Piotrzkowski,
{\it Z.Phys.} {\bf C67} (1995) 577.
\bibitem{coBKS1} V. N. Baier, V. M. Katkov,and  V. M. Strakhovenko,
Particle Accelerators, {\bf 30} (1990) 43.
\bibitem{N} R. J. Noble,  Nucl.Instr.Meth.,
{\bf A256} (1987) 427.
\bibitem{T} O. Napoly,
{\it Proceedings of the 2001 Particle Accelerator Conference},
Chicago, (2001) p.402.
\bibitem{coBKS5} V. N. Baier, V. M. Katkov, and  V. M. Strakhovenko,
Phys. Lett. {\bf  A229} (1992) 429.
\bibitem{BG2} V. A. Bazylev, V. V. Golovoznin, and A. V. Demura,
 Zhurn.Teor.i Eksp. Fiziki, {\bf 92} (1987) 1921.
\bibitem{cBKS6} V. N. Baier, V. M. Katkov, V. M. Strakhovenko,
 phys.stat.sol.(b) {\bf 149} (1988) 403.
\bibitem{coBK3} V. N. Baier and V. M. Katkov,
{\it On coherent radiation in electron-positron colliders},
Proceedings of {\it Quantum Aspects of Beam Physics}, ed.P. Chen,
World Scientific PC, Singapore, 2003 (in print).



\end{thebibliography}
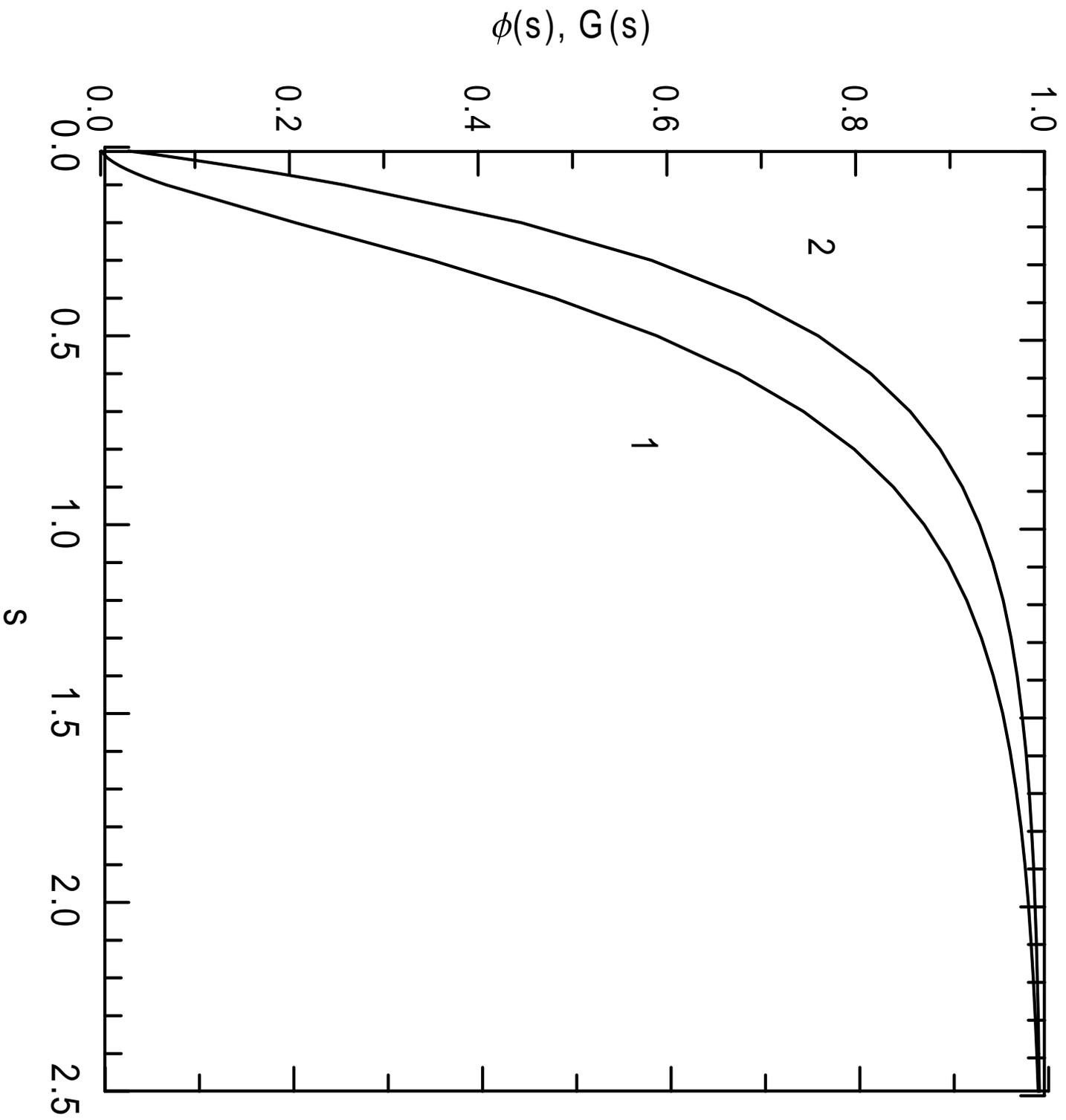
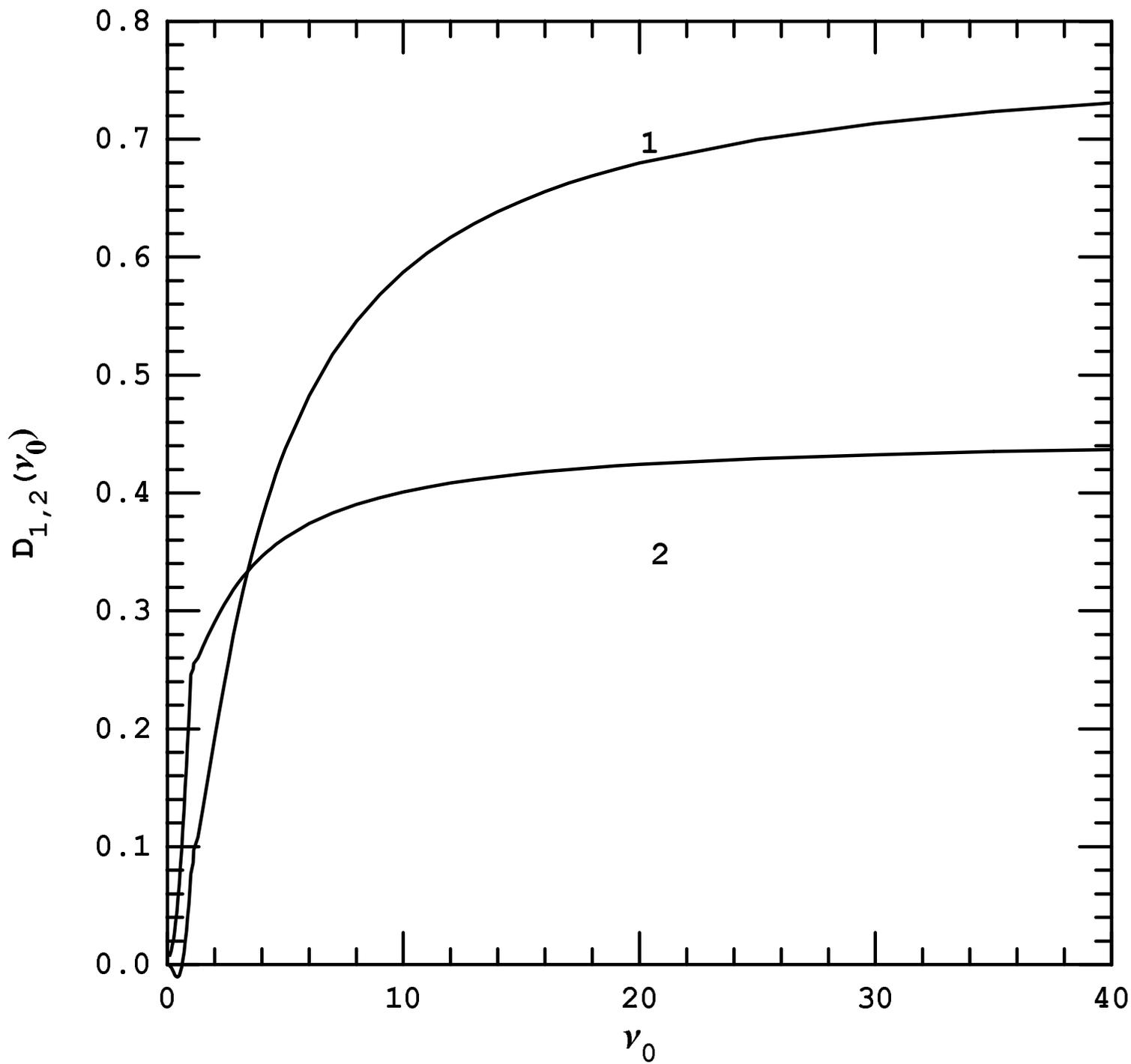
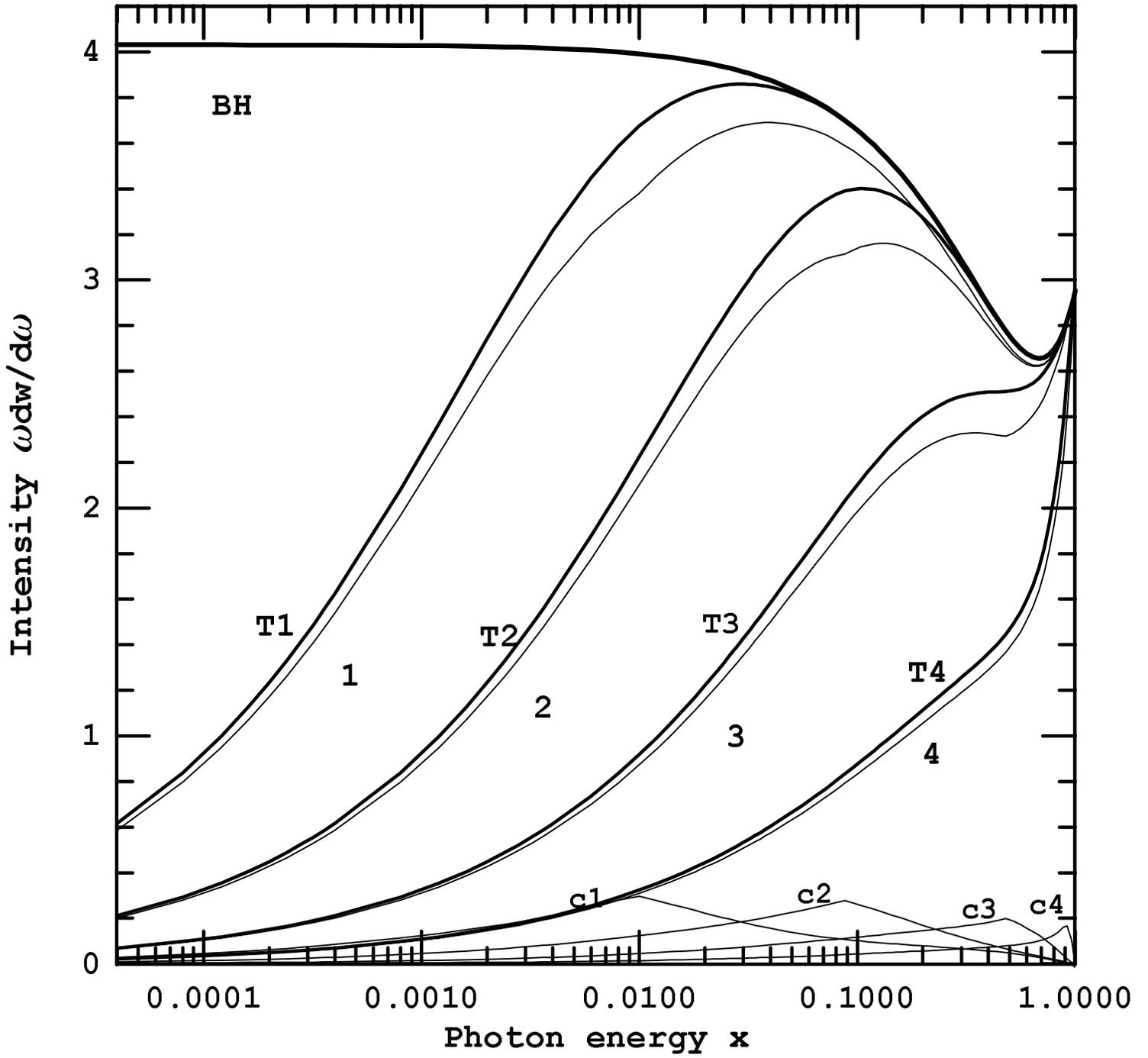
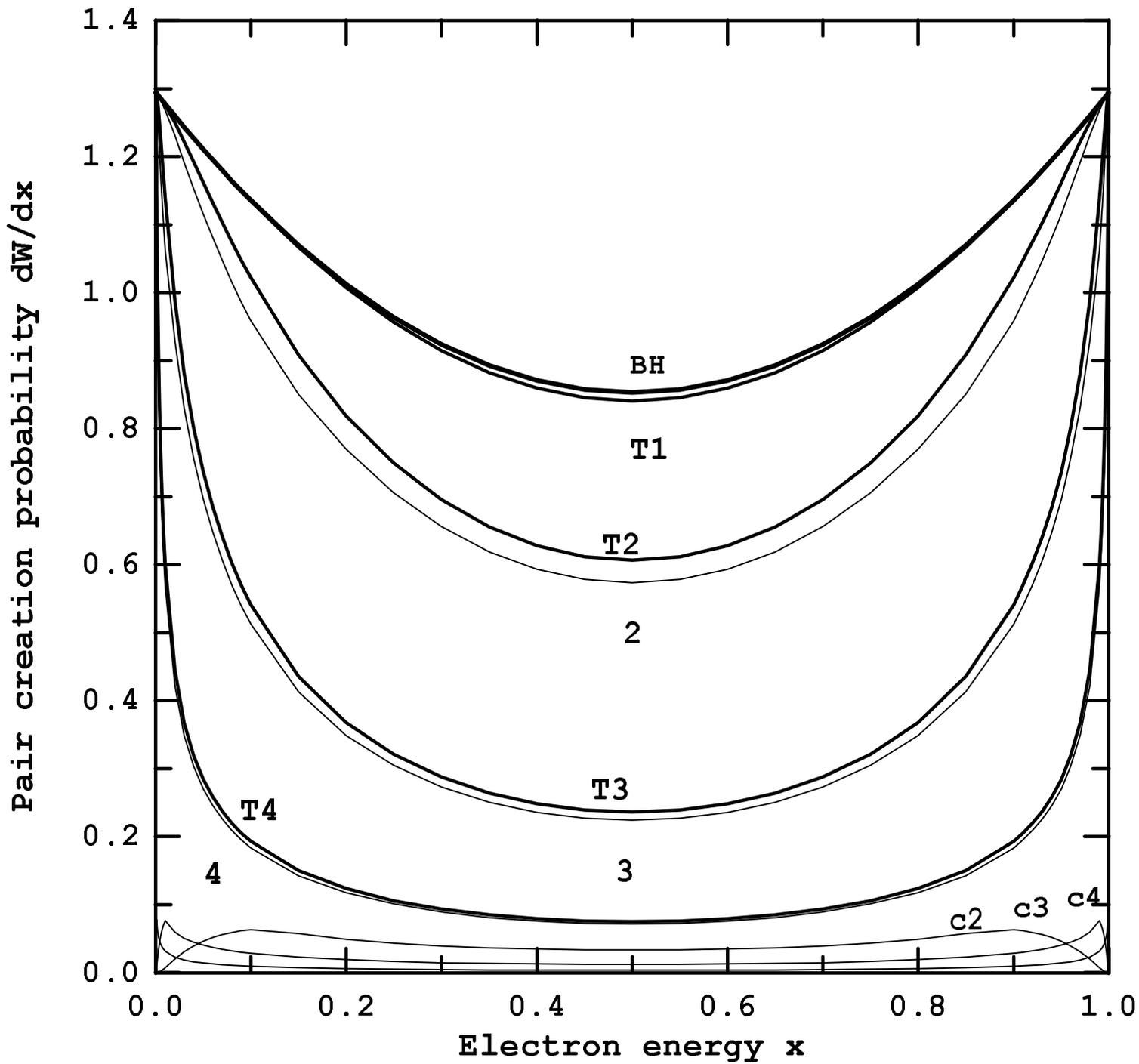
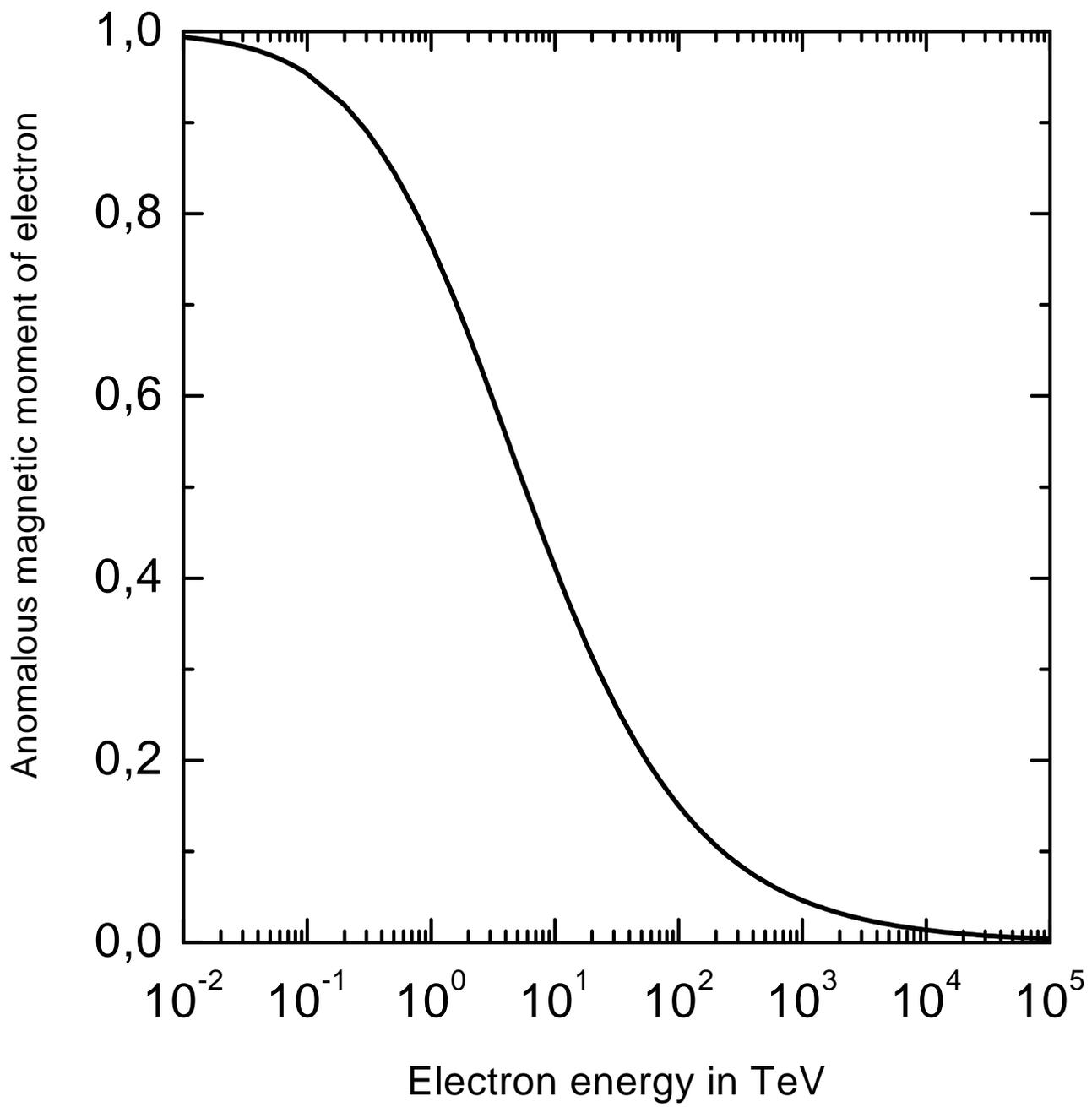
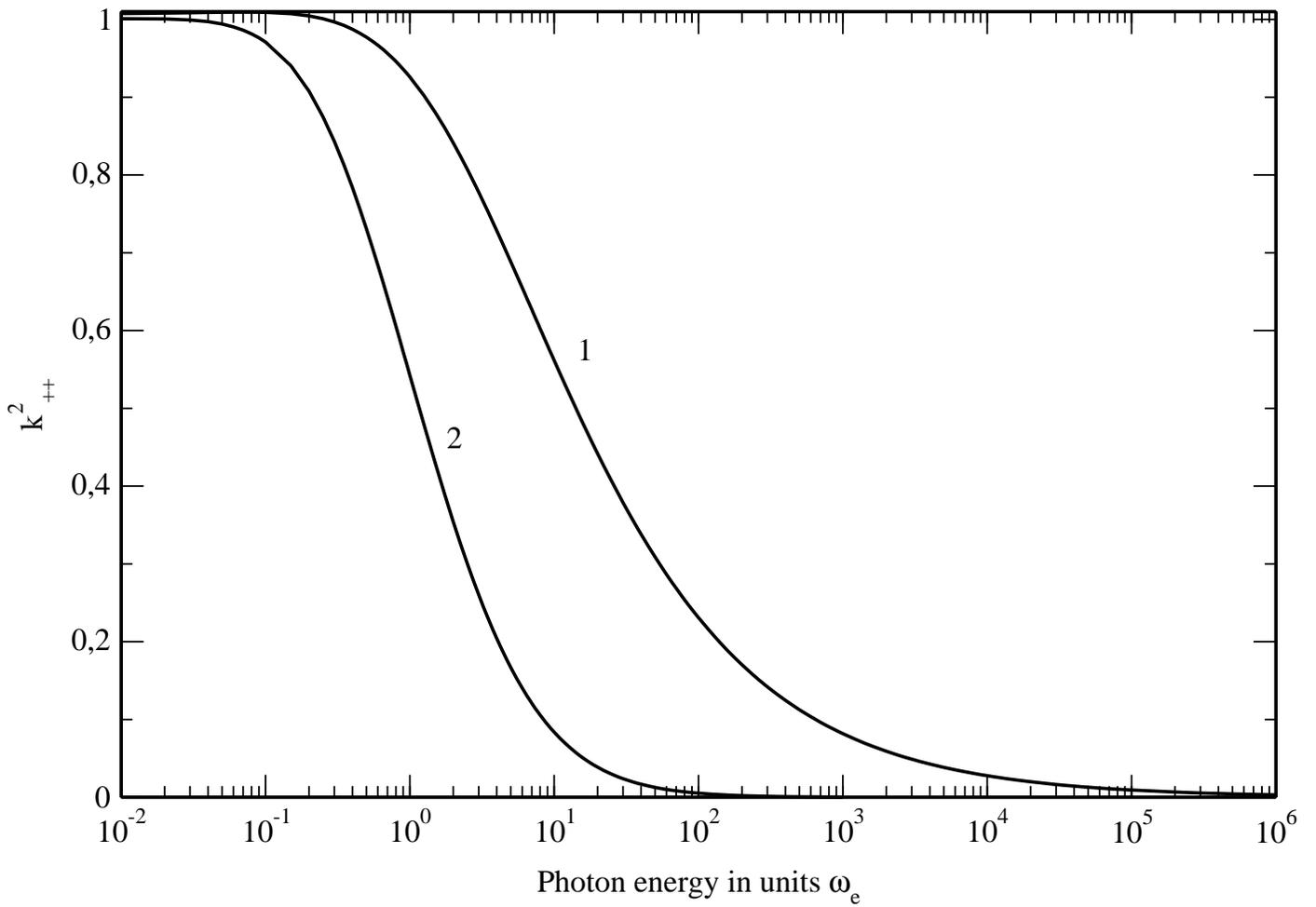
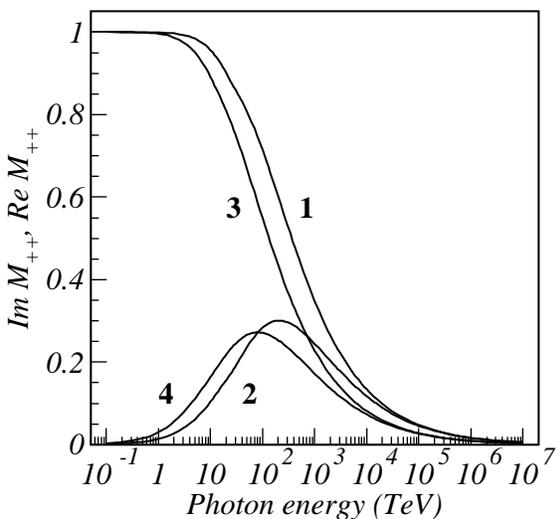
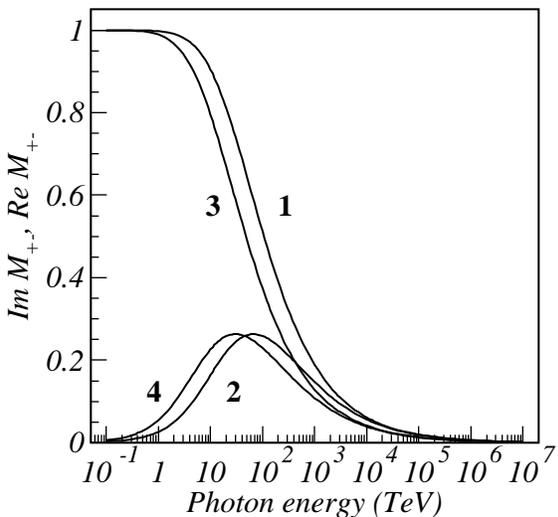
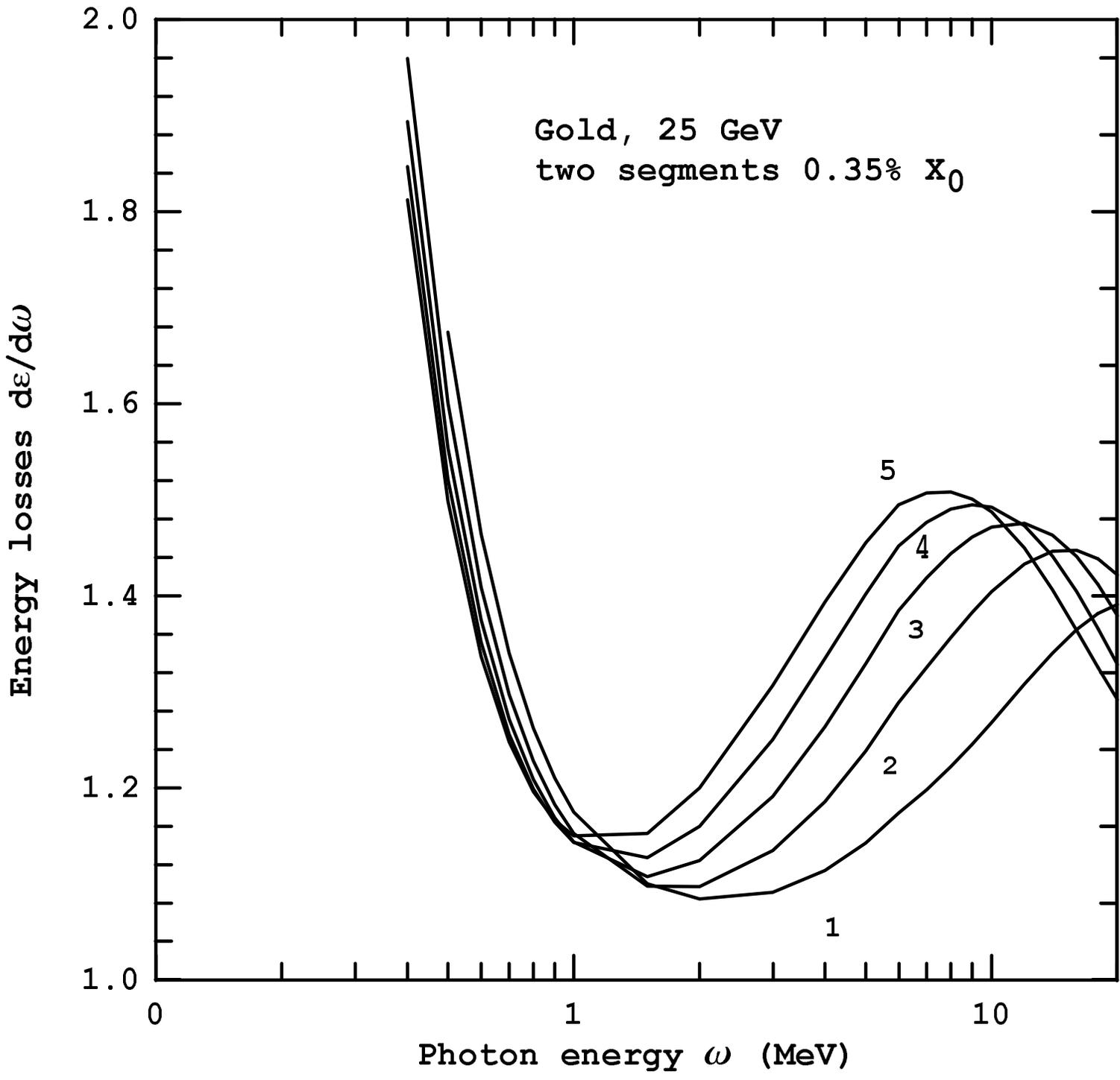
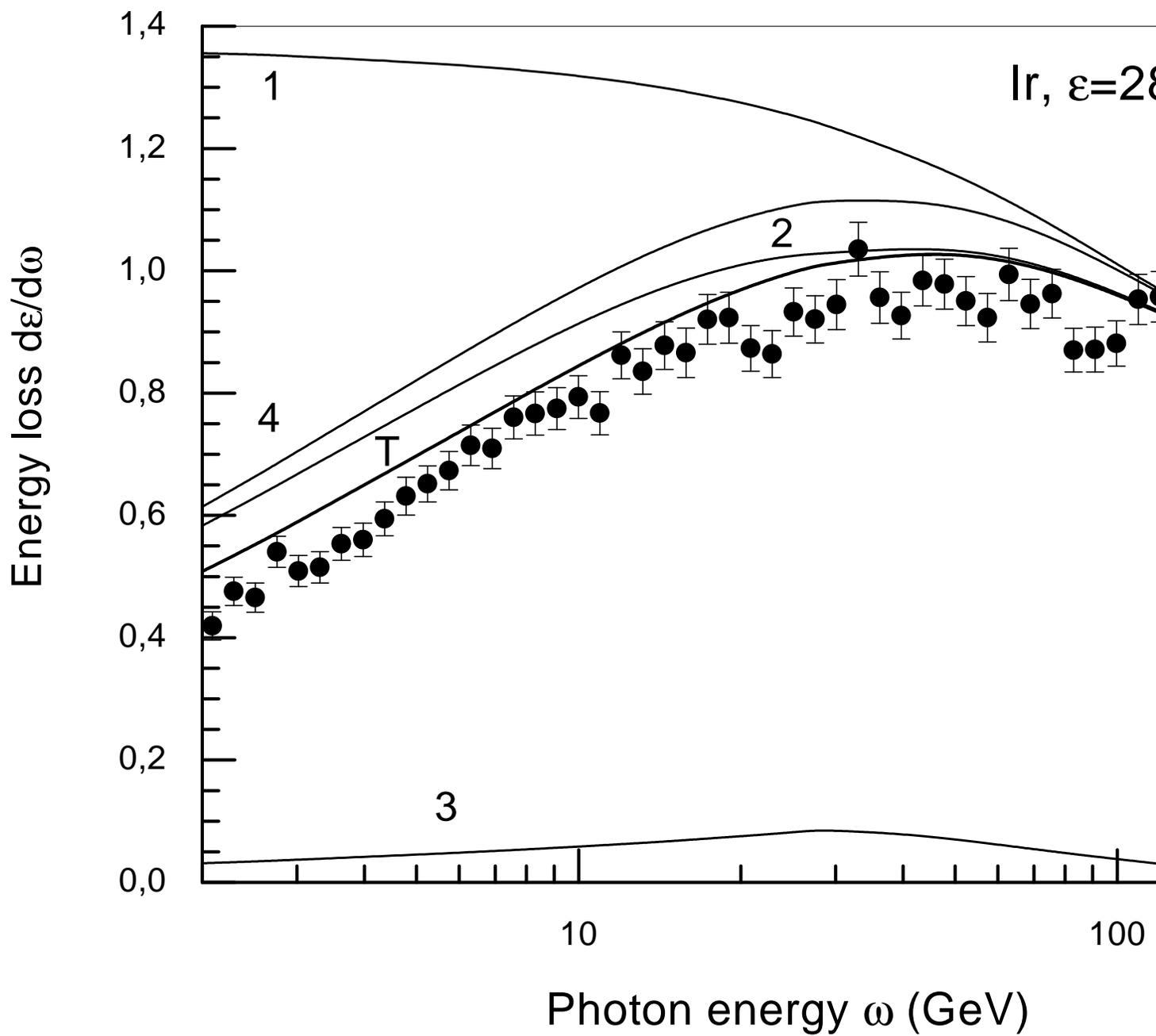
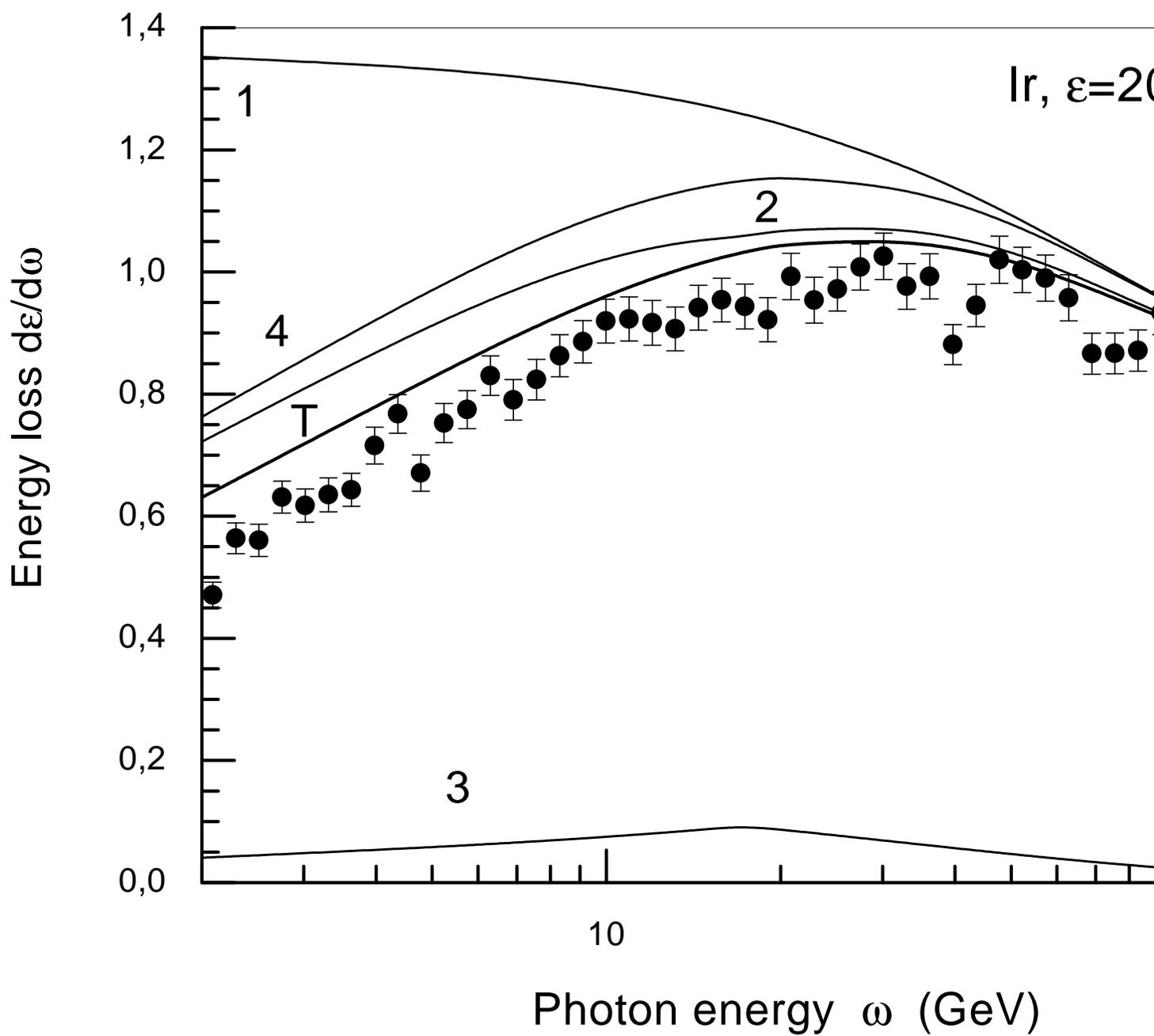
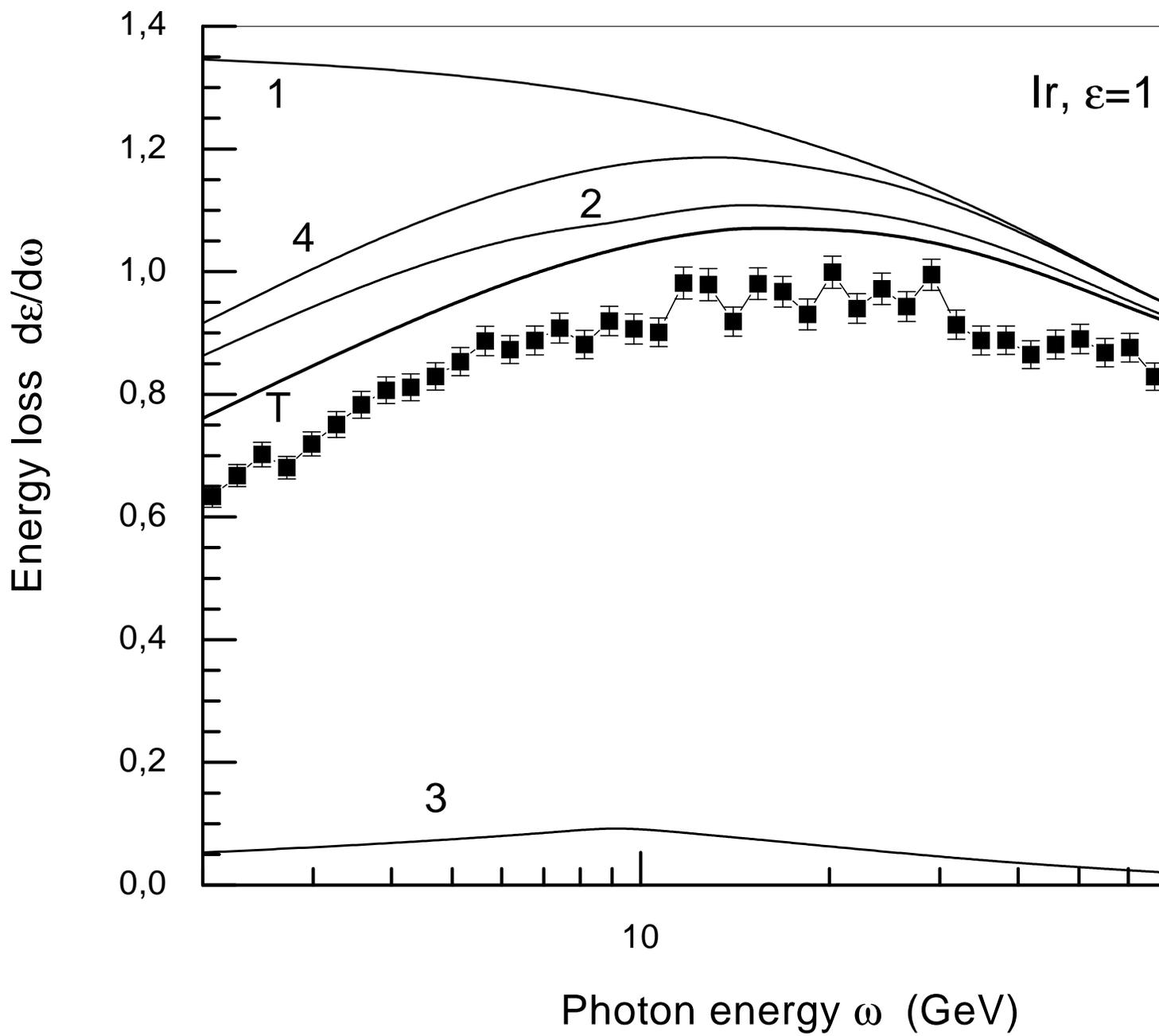
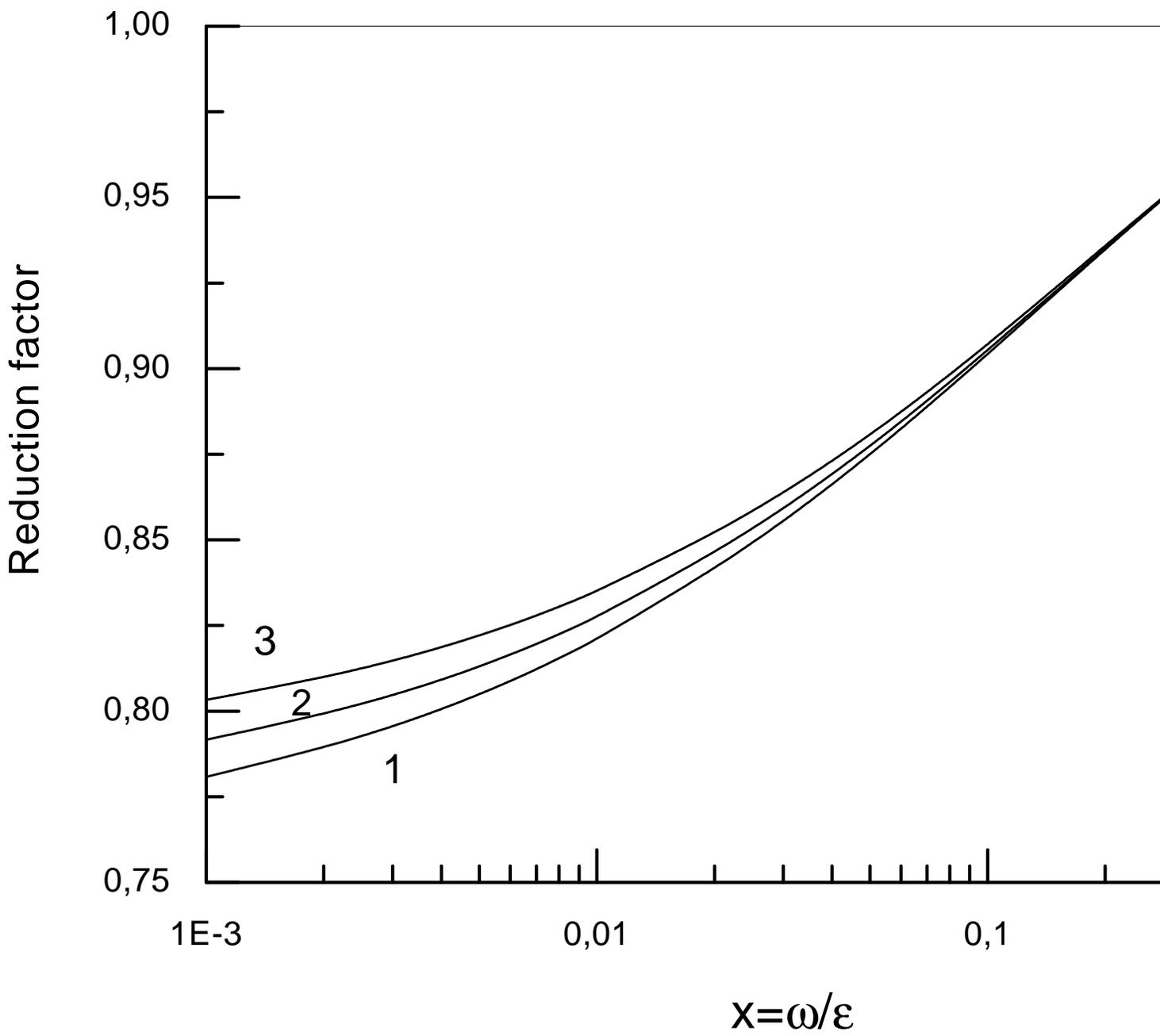
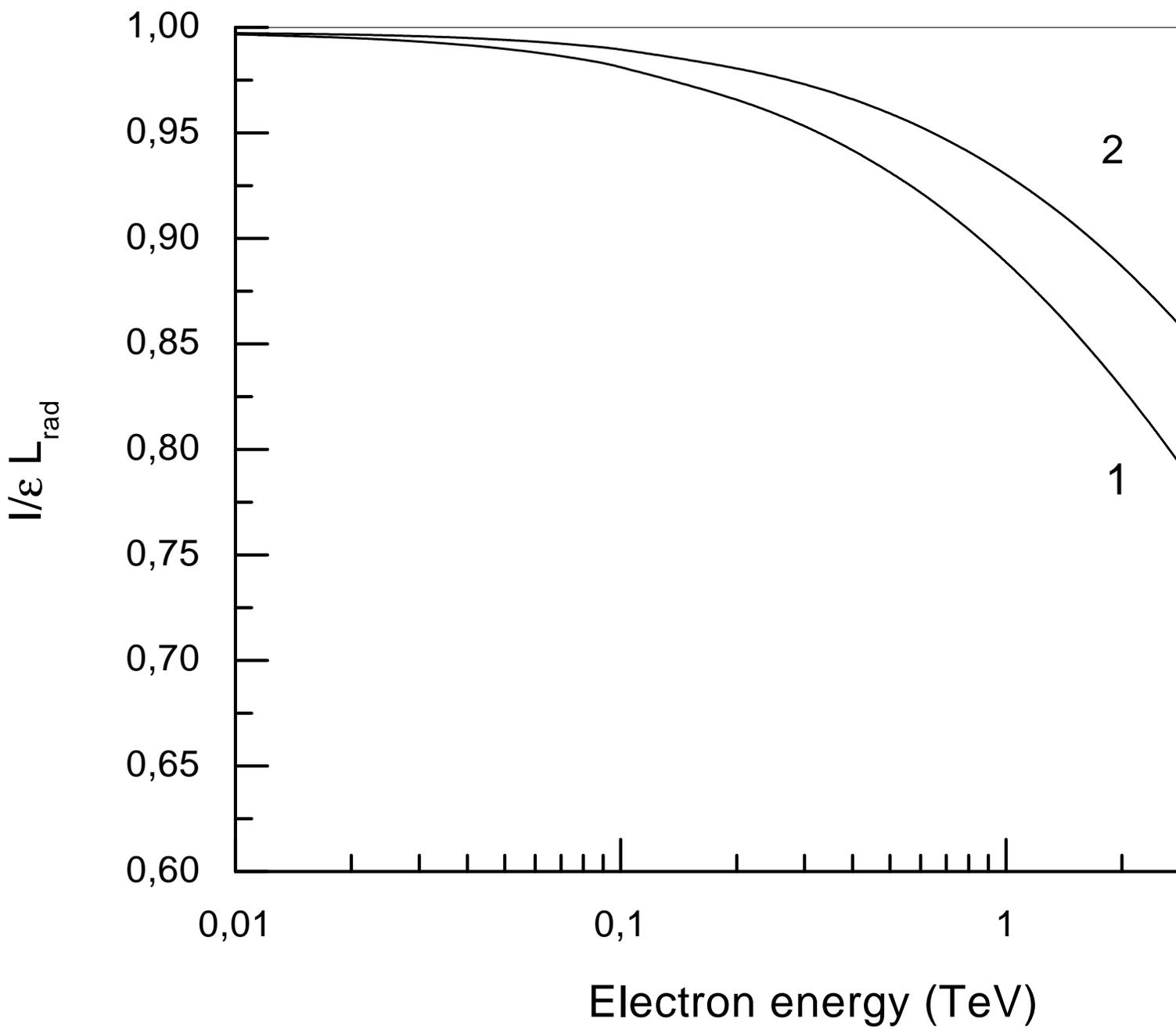
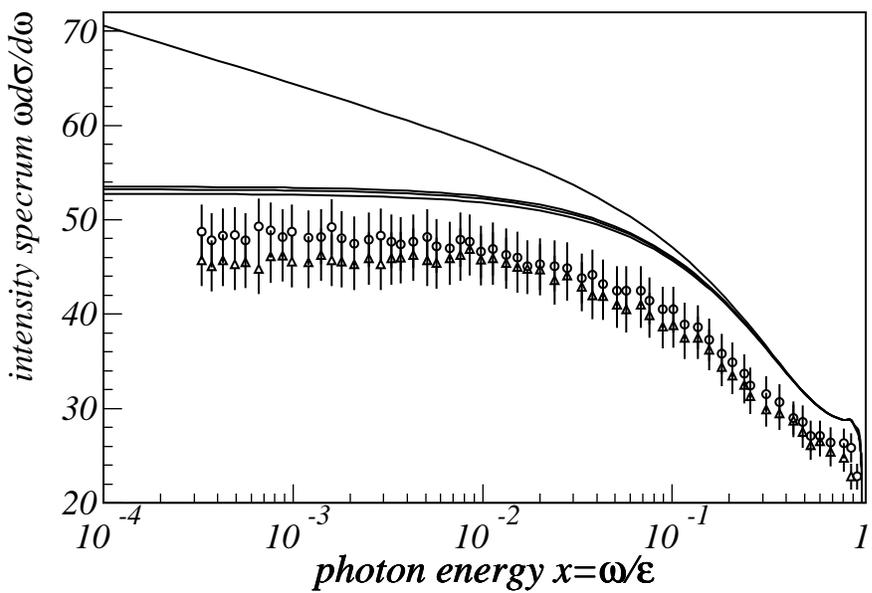
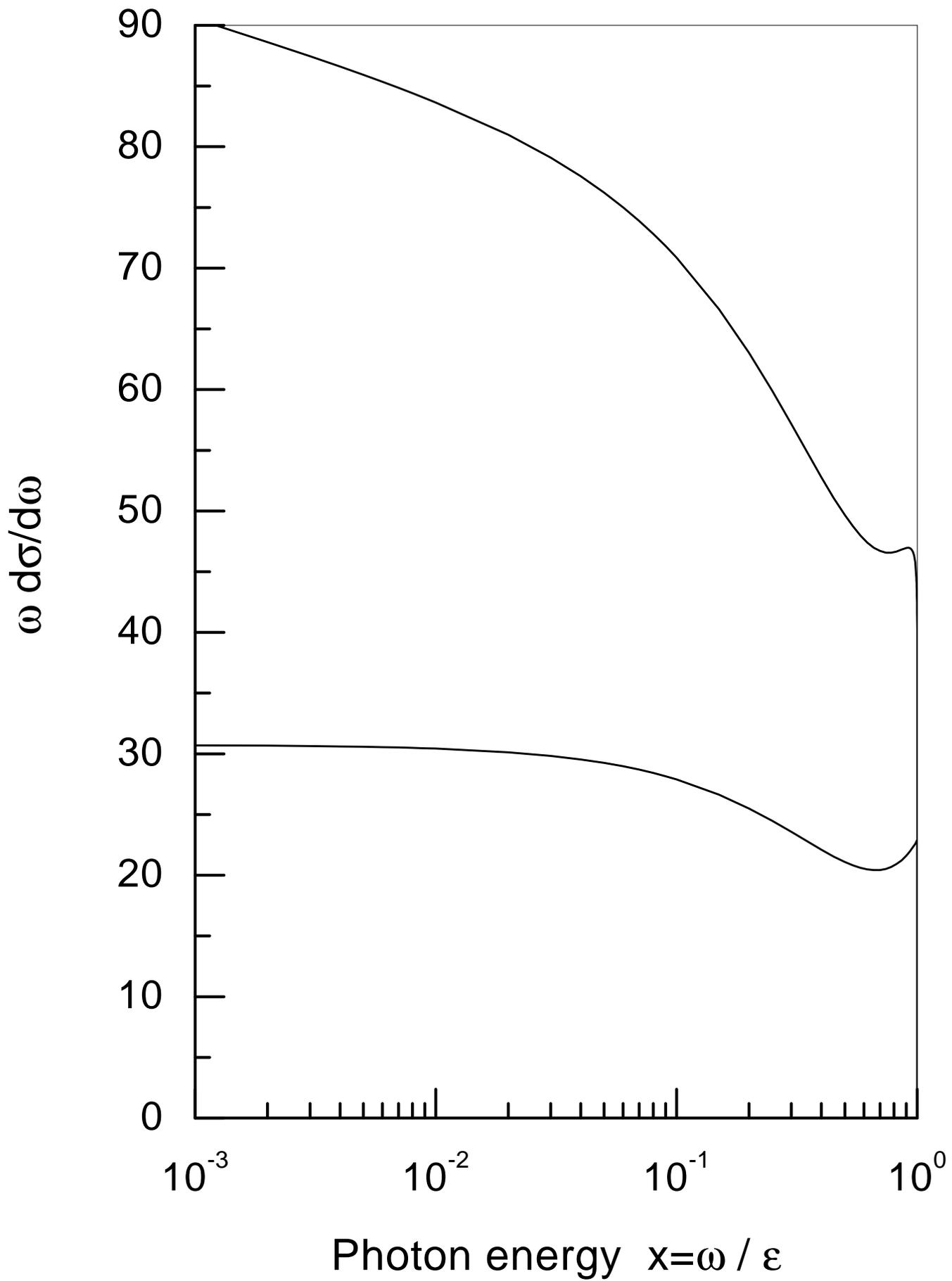
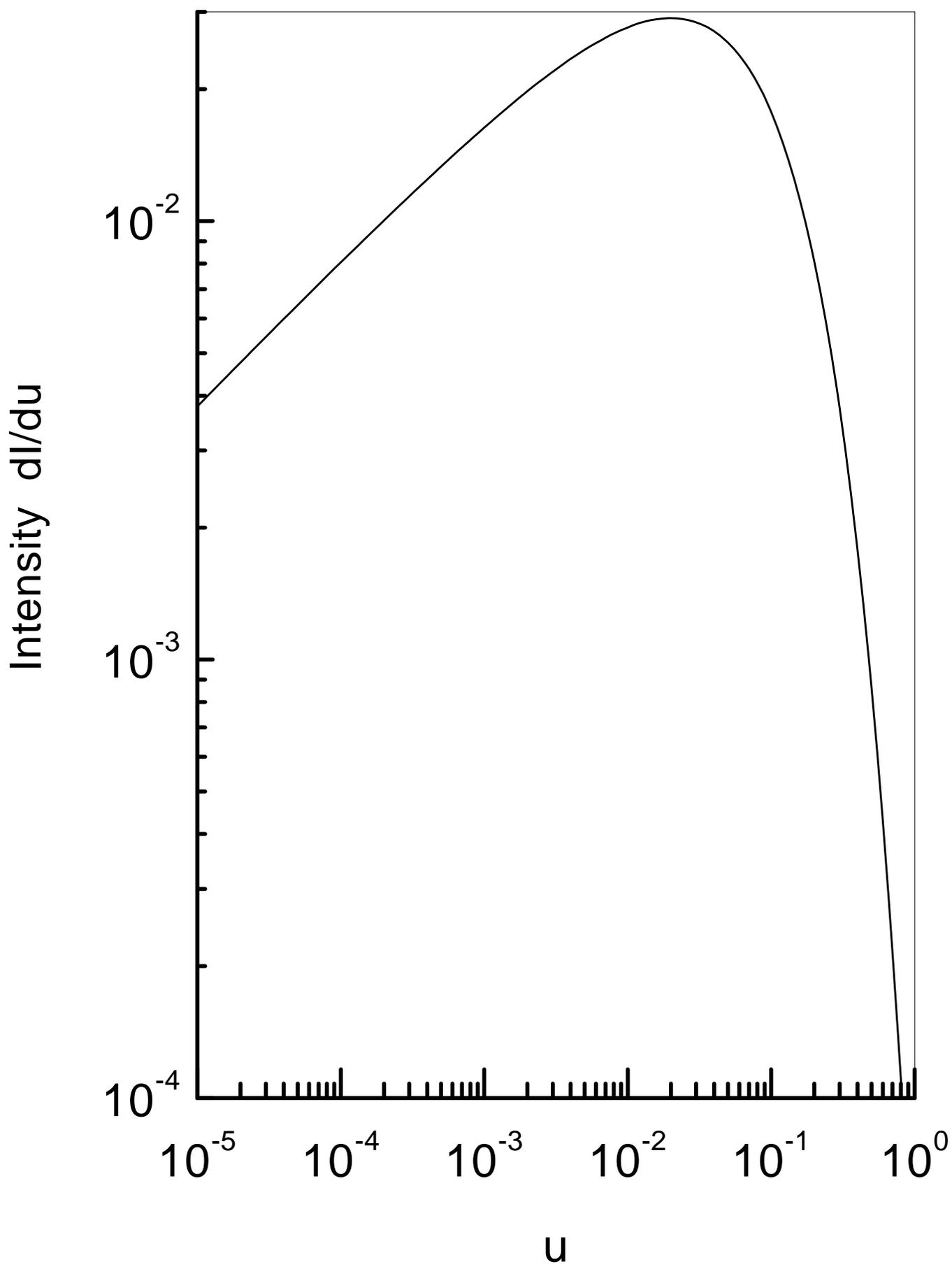
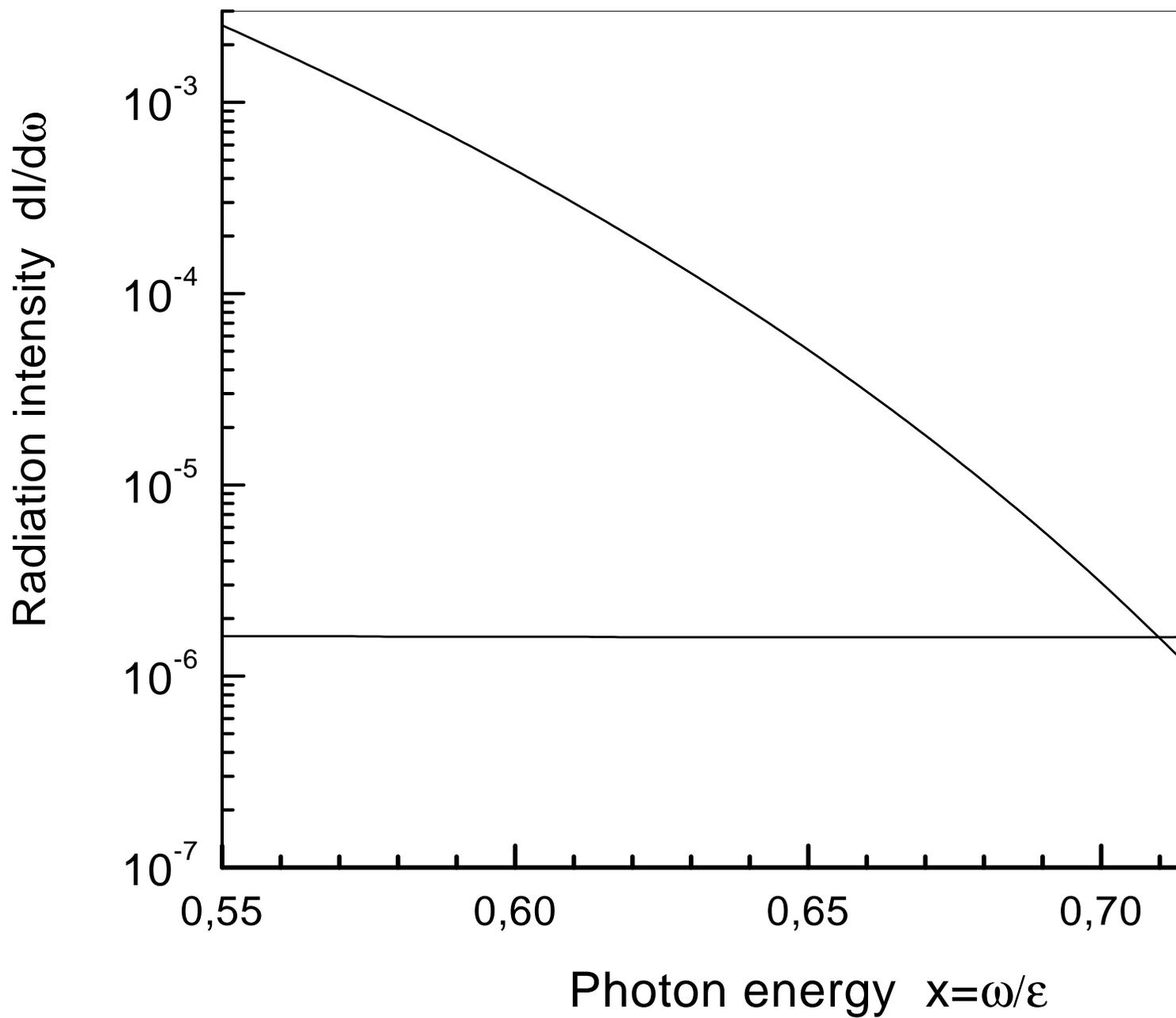
\end{document}